\documentclass[12pt,a4paper,british,openany]{book}
\usepackage[top=2.5cm, bottom=2cm, left=2cm, right=2.2cm]{geometry}
\usepackage{titlesec}
\titleformat{\chapter}{\LARGE\bfseries}{\thechapter}{1em}{}
\titleformat{\section}{\Large\bfseries}{\thesection}{0.6em}{}
\titleformat{\subsection}{\large\bfseries}{\thesubsection}{0em}{}
\titleformat{\subsubsection}{\bfseries}{\thesubsubsection}{0em}{}
\usepackage[font=small,labelfont=bf]{caption}
\usepackage[toc,page]{appendix}
\usepackage{amsmath,amssymb,amsfonts,braket,mathtools}
\usepackage{mathrsfs,bbm}
\usepackage{cite}

\usepackage[colorlinks=true,linkcolor=black, citecolor=black,
urlcolor=black]{hyperref}
\usepackage{fancyhdr}

\usepackage{graphicx}
\usepackage{booktabs}
\usepackage{subfig}
\usepackage{tikz}
\usetikzlibrary{plotmarks,calc,decorations, decorations.pathmorphing,trees,
decorations.markings}

\tikzstyle dynkin node=[very thick,shape=circle,draw,inner sep=0pt,minimum size=5mm]
\tikzstyle directed=[postaction={decorate,decoration={markings,
    mark=at position .65 with {\arrow[draw=cyan]{>}}}}]
\tikzstyle dynkin line=[very thick]
\tikzstyle inverse line=[gray,line width=1.46pt,line cap=round, dash pattern=on 0pt off 2\pgflinewidth]
\tikzstyle red phase=[red,decoration={snake,amplitude=0.15mm,segment length=1.6mm},decorate]
\tikzstyle blue phase=[blue,decoration={snake,amplitude=0.1mm,segment length=0.9mm},decorate]
\tikzstyle S-mat=[circle,thick,draw=black,fill=yellow!20!white,inner sep=1pt]
\tikzstyle spin=[circle,thick,draw=black,fill=yellow!15!white,,inner sep=1pt]
\tikzstyle spinex=[circle,thick,draw=black,fill=yellow!50!white,inner sep=1pt]
\tikzstyle spin1=[circle,thick,draw=black,fill=cyan!20!white,inner sep=1pt]
\tikzstyle spin2=[circle,thick,draw=black,fill=red!20!white,inner sep=1pt]

\pgfdeclarelayer{background}
\pgfdeclarelayer{foreground}
\pgfsetlayers{background,main,foreground}

\newcommand\KinNC{\mathtt{K}}
\newcommand\KinCC{\colorbox{yellow}{$\mathtt{K}$}}
\newcommand\DynNC{\mathtt{D}}
\newcommand\DynCC{\colorbox{yellow}{$\mathtt{D}$}}

\newlength\vertdist
\newlength\hordist
\newcommand{\fixedspaceL}[2]{\mathrlap{#2}\phantom{#1}}
\newcommand{\fixedspaceR}[2]{\phantom{#1}\mathllap{#2}}

\DeclareMathAlphabet{\mathsfit}{\encodingdefault}{\sfdefault}{m}{sl}
\def\eps{\varepsilon}
\def\Xint#1{\mathchoice
  {\XXint\displaystyle\textstyle{#1}}%
  {\XXint\textstyle\scriptstyle{#1}}%
  {\XXint\scriptstyle\scriptscriptstyle{#1}}%
  {\XXint\scriptscriptstyle\scriptscriptstyle{#1}}%
  \!\int}
\def\XXint#1#2#3{{\setbox0=\hbox{$#1{#2#3}{\int}$}
    \vcenter{\hbox{$#2#3$}}\kern-.5\wd0}}
\def\pint{\;\Xint-}

\def\a{\alpha}
\def\b{\beta}

\def\d{\delta}

\def\N{\mathcal{N}}
\def\AdS{\text{AdS}}
\def\CFT{\text{CFT}}

\def\S{\text{S}}
\def\T{\text{T}}
\def\Smat{\mathbf{S}}
\def\Id{\mathbf{I}}
\def\de{\text{d}}

\def\so{\mathfrak{so}}
\def\vac{0}
\def\su{\mathfrak{su}}
\def\psu{\mathfrak{psu}}
\def\u{\mathfrak{u}}
\def\o{\mathfrak{o}}
\def\gl{\mathfrak{gl}}
\def\sl{\mathfrak{sl}}
\def\bigL{\text{L}}
\def\bigR{\text{R}}
\def\L{\smallL}
\def\R{\smallR}
\def\d21a{\mathfrak{d}(2,1;\a)}

\def\tr{\text{tr}}
\def\str{\text{str}}

\def\Ham{\mathbf{H}}
\def\IN{\text{in}}
\def\OUT{\text{out}}

\def\am{\text{am}}
\def\dn{\text{dn}}
\def\cn{\text{cn}}
\def\sn{\text{sn}}
\def\K{\text{K}}

\newcommand{\gen}[1]{\mathfrak{#1}}

\newcommand{\I}{\text{I}}
\newcommand{\II}{\text{II}}

\newcommand{\order}{\mathcal{O}}

\newcommand{\CDD}{\scriptscriptstyle\text{CDD}}

\renewcommand{\Re}{\mathop{\operatorname{Re}}}
\renewcommand{\Im}{\mathop{\operatorname{Im}}}

\DeclareMathOperator{\sign}{sign}

\newcommand{\prodsigma}{\sigma^+}
\newcommand{\ratiosigma}{\sigma^-}
\newcommand{\sumtheta}{\theta^+}
\newcommand{\difftheta}{\theta^-}
\newcommand{\BES}{\text{BES}}
\newcommand{\BLMMT}{{\scriptscriptstyle\text{BLMMT}}}
\newcommand{\AFS}{\text{AFS}}
\newcommand{\HL}{\text{HL}}
\newcommand{\newq}{\mathcal{Q}}
\newcommand{\curvearrowurl}{\curvearrowleft}
\newcommand{\curvearrowdlr}{\rotatebox[origin=c]{180}{$\curvearrowleft$}}
\newcommand{\circlearrowrot}{\rotatebox[origin=c]{-30}{$\circlearrowleft$}}
\newcommand{\inturl}{\;{\int \negthickspace \negthickspace \negthickspace\negthickspace \negthickspace \curvearrowurl}\mbox{ }\,} 
\newcommand{\intdlr}{\;{\int \negthickspace \negthickspace \negthickspace \negthickspace \curvearrowdlr}\mbox{ }\,} 
\newcommand{\ointc}{\;{\int \negthickspace \negthickspace \negthickspace \negthinspace \negthinspace \circlearrowrot}\mbox{ }\,}

\newcommand{\Ael}{\mathsf{A}}
\newcommand{\Bel}{\mathsf{B}}
\newcommand{\Cel}{\mathsf{C}}
\newcommand{\Del}{\mathsf{D}}
\newcommand{\Eel}{\mathsf{E}}
\newcommand{\Fel}{\mathsf{F}}

\newcommand{\smallL}{\mbox{\tiny L}}
\newcommand{\smallR}{\mbox{\tiny R}}

\newcommand{\smallLL}{\mbox{\tiny LL}}
\newcommand{\smallLR}{\mbox{\tiny LR}}
\newcommand{\smallRL}{\mbox{\tiny RL}}
\newcommand{\smallRR}{\mbox{\tiny RR}}

\newcommand{\comm}[2]{\big[#1,#2\big]}
\newcommand{\acomm}[2]{\big\{#1,#2\big\}}

\def\etc{\textit{etc.}}
\def\ie{\textit{i.e.}}
\def\eg{\textit{e.g.}}

\begin{document}
\frontmatter
\tikzset{
time/.style={draw=black,thick, postaction={decorate},
    decoration={markings,mark=at position 1 with {\arrow[draw=black]{>}}}},
byarrow/.style={draw=black,thick, postaction={decorate},
    decoration={markings,mark=at position 1 with {\arrow[draw=black]{>}}}},
partic1/.style={draw=blue,thick, postaction={decorate},
    decoration={markings,mark=at position .5 with {\arrow[draw=blue]{>}}}},
partic2/.style={draw=red,thick, postaction={decorate},
    decoration={markings,mark=at position .5 with {\arrow[draw=red]{>}}}},
particle/.style={draw=blue,line width=1, postaction={decorate},
    decoration={markings,mark=at position .5 with {\arrow[draw=blue]{>}}}},
antiparticle/.style={draw=blue,line width=1, postaction={decorate},
    decoration={markings,mark=at position .5 with {\arrow[draw=blue]{<}}}},
particlecross/.style={draw=red,line width=1,dashed, postaction={decorate},
    decoration={markings,mark=at position .5 with {\arrow[draw=red]{>}}}},
antiparticlecross/.style={draw=red,line width=1,dashed, postaction={decorate},
    decoration={markings,mark=at position .5 with {\arrow[draw=red]{<}}}},
nopart/.style={draw=none}
 }
\pagestyle{empty}
\begin{titlepage}
\begin{flushright}
\texttt{HU-Mathematik-2014-14}\\
\texttt{HU-EP-14/24}
\end{flushright}
\vspace*{1cm}
\begin{center}
{\LARGE\bf Towards integrability for $\AdS_{\mathbf{3}}/\CFT_{\mathbf{2}} $}\\[2cm]
{\Large Alessandro Sfondrini}\\[1cm]
{\textit{Institut f\"ur Mathematik und Institut f\"ur Physik, Humboldt-Universit\"at zu Berlin\\
IRIS Geb\"aude, Zum Grossen Windkanal 6, 12489 Berlin, Germany}}\\[1cm]
{\texttt{Alessandro.Sfondrini@physik.hu-berlin.de}\\[2cm]}
{\bf Abstract}\\[0.4cm]
\parbox{0.95\textwidth}{
We review the recent progress towards applying worldsheet integrability techniques to the $\AdS_3/\CFT_2 $ correspondence to find its all-loop S~matrix and Bethe-Yang equations.
We study in full detail the massive sector of $\AdS_3\times\S^3\times\T^4 $ superstrings supported by pure Ramond-Ramond (RR) fluxes. The extension of this machinery to accommodate massless modes, to the $\AdS_3\times\S^3\times\S^3\times\S^1 $ pure-RR background and to backgrounds supported by mixed background fluxes is also reviewed.
While the results discussed here were found elsewhere, our presentation sometimes deviates from the one found in the original literature in an effort to be pedagogical and self-contained. 
  }
\end{center}

\end{titlepage}
\cleardoublepage
\pagestyle{plain}

\tableofcontents

\mainmatter
\pagestyle{fancy}
\fancyhf{}
\fancyhead[LE,RO]{\thepage}
\fancyhead[LO]{\nouppercase{\rightmark}}
\fancyhead[RE]{\nouppercase{\leftmark}}
\renewcommand{\headrulewidth}{0.5pt}

\chapter{Introduction}
The holographic conjecture~\cite{'tHooft:1993gx} is a major advance in theoretical Physics. Its study, which is mainly performed in the framework of string theory~\cite{Susskind:1994vu, Maldacena:1997re,Witten:1998qj,Gubser:1998bc}, has generated an incredible number of results. Our focus here is on a particular instance of holography, that is the duality between gravity (superstring) theories on backgrounds involving three-dimensional anti-de Sitter space ($\AdS_3 $) and supersymmetric two-dimensional conformal field theories ($\CFT_2 $).
The interest of this case is evident, due to the special properties of $\AdS_3 $ gravity and $\CFT $s in two dimensions.

This particular duality has been considered since the early days of holography and of AdS/CFT, and has been investigated by several techniques in the past fifteen years. It is however a recent realisation that $\AdS_3/\CFT_2 $ may be amenable to the integrability approach that proved very successful especially in the case of  $\AdS_5/\CFT_4 $. More specifically, in ref.~\cite{Babichenko:2009dk} Babichenko, Stefa{\'n}ski and Zarembo have shown that the non-linear $\sigma$ model (NLSM) action that describes the dynamics of free strings on maximally supersymmetric $\AdS_3 $ backgrounds supported by Ramond-Ramond (RR) fluxes yields an integrable classical field theory. This prompted a rapid progress in adapting the S-matrix integrability techniques that worked so well for $\AdS_5/\CFT_4 $ to this lower-dimensional and less supersymmetric case. Our main aim here is to review this progress in a self-contained and accessible way. 

Before starting with our review, we will first briefly overview some well-established facts about the $\AdS_3/\CFT_2 $ duality. We will then sketch the general aspects of the spectral problem in AdS/CFT, as well as briefly present the historical development of the S-matrix integrability approach in this context. The reader who is familiar with these topics may want to skip this introductory discussion, and jump to the end of this chapter where we present the plan of the review.

\section{\texorpdfstring{$\AdS_3 $}{AdS3} gravity and holography}
Gravity on $\AdS_3 $ should be dual to a conformal field theory on a two-dimensional cylinder, that is the boundary of $\AdS_3 $ in global coordinates. The continuous isometries of~$\AdS_3 $ form the special orthogonal group $SO(2,2)$, while the $\CFT_2 $ has an \emph{infinite-dimensional} symmetry algebra, given by two copies of the Virasoro algebra. The relation between these two sets of symmetries was elucidated by Brown and Henneaux in ref.~\cite{Brown:1986nw}. The group $SO(2,2)$ is the one generated by the Virasoro elements that can be defined globally. The remaining symmetry generators are only asymptotic symmetries of $\AdS_3 $, and acting with them does not leave the gravity vacuum invariant, as it modifies the stress-energy tensor.

An interesting feature of three-dimensional gravity is that, despite being much simpler than its higher-dimensional counterparts, it admits black-hole solutions, that exist precisely in the case of negative curvature. Such black holes, first found by Ba{\~n}ados, Teitelboim and Zanelli (BTZ)~\cite{Banados:1992gq,Banados:1992wn}, are essentially given by a discrete quotient of $\AdS_3 $ and as such are locally isometric to the maximally symmetric background. Therefore, they have no curvature singularity. However, they do have (inner and outer) horizons and an ergosphere, see ref.~\cite{Carlip:1995qv} for a review. Such solutions exist also in supersymmetric extensions of the gravity theory (supergravities), and preserve supersymmetry as long as they have vanishing temperature~\cite{Coussaert:1993jp,Izquierdo:1994jz,Steif:1995zm}.

Indeed supersymmetry comes naturally into the picture if we want to obtain $\AdS_3 $ (and, as we will see, the dual $\CFT_2 $) from string theory, in the context of the celebrated D1-D5 system of branes that played a pivotal role in the investigation of black-hole microstates~\cite{Strominger:1996sh, Strominger:1997eq}. The AdS/CFT setup for the D1-D5 system was first detailed in the seminal paper by Maldacena~\cite{Maldacena:1997re}, see also ref.~\cite{Aharony:1999ti} for a review of that setup and ref.~\cite{David:2002wn} for a more detailed review of the D1-D5 system and an extensive list of references concerning early investigations thereof. Let us briefly overview some features of the D1-D5 system in AdS/CFT~\cite{Vafa:1995bm, Seiberg:1997zk,Witten:1997yu, Larsen:1999uk,Seiberg:1999xz}.

We start by compactifying four directions in target space on%
\footnote{%
More generally, these could be compactified on a $K3$ manifold, \ie\ essentially on a discrete quotient of~$\T^4 $.
}
 a four-torus~$\T^4 $. Then let us consider~$Q_1 $ D1 branes along a non-compact direction and~$Q_5$ D5 branes that extend along the same non-compact direction and wrap the four compact ones. This configuration is invariant under\footnote{%
We will denote Lie algebras and superalgebras in Gothic letters.
}%
~$\so(1,1)$, the algebra of boosts along the string, and under $\so(4) $ rotations in the space orthogonal to the branes. Such a brane configuration preserves eight complex supersymmetry  generators, that can be decomposed chirally with respect to~$\so(1,1)$, yielding $\mathcal{N}=(4,4) $ supersymmetry.\footnote{For a review on the related supergravity solutions, see ref.~\cite{Youm:1997hw}.}
In the near-horizon limit, this geometry reduces to~$\AdS_3\times\S^3\times\T^4 $, where the curvature radii of the warped spaces are equal,
\begin{equation}
R^{2}_{\AdS_3}=R^{2}_{\S^3}=\sqrt{Q_1\,Q_5}\,,
\end{equation}
while the volume of the~$\T^4 $ is~$Q_1/Q_5$. The picture further simplifies in the 't Hooft (or planar) limit of the duality, whereby the strings propagate freely in the fixed $\AdS_3\times\S^3\times\T^4 $ background. It is worth noticing that the superisometries of this background are given---up to some abelian factors---by
\begin{equation}
\label{eq:psu112:2}
\psu(1,1|2)_{\L}\oplus\psu(1,1|2)_{\R}\,,
\end{equation}
which is the subalgebra of~$\mathcal{N}=(4,4) $ which can be defined globally---in analogy with $\so(2,2) $ and Virasoro. The two copies of $\psu(1,1|2)$ carry labels ``L'' (left) and ``R'' (right) to identify their respective chiralities with respect to~$\so(1,1)$ algebra of boosts along the non-compact D-brane direction---that is, the chirality in the dual CFT. In total, they amount to sixteen real supercharges, as expected. This is half of the maximum possible amount of supersymmetry, which is instead attained in the case of the~$\AdS_5\times\S^5$ background.

If we consider the same brane construction but focus on its low-energy excitations, we will find a supersymmetric Yang-Mills theory (SYM) with~$\su(2)_{\L}\oplus\su(2)_{\R} $ R~symmetry (coming from the aforementioned $\so(4) $ isometries). This theory contains both vector and hypermultiplets, that have different transformation properties under the chiral R~symmetry. 
The presence of matter in both the fundamental and adjoint representations of the gauge group is noteworthy, as it makes this gauge theory less special and more realistic than $\mathcal{N}=4$ SYM. Another very interesting feature of this AdS/CFT construction is that the gauge theory description (being two-dimensional) is not conformal, and has a non-trivial flow. This mean that its low-energy limit is some to-be-determined~$\CFT_2$.

Luckily, the brane construction offers some guidance in characterising this~$\CFT_2$. Let us focus on SYM the D5 branes, and view the D1 branes as $SU(Q_{5})$ instantons, with instanton number $Q_{1}$~\cite{Douglas:1995bn}. The instanton configurations are parametrised by moduli, and fluctuations  around a given configuration can be understood as fluctuations of the moduli along the time direction or along the non-compact direction of the D1-D5 branes. Therefore, the low-energy dynamics of this system is the 1+1~dimensional QFT taking values in the space of  instanton moduli space. Such a space is a deformation of the symmetric product of~$Q_1Q_5$ copies of~$\T^4 $, \ie\  $(\T^4)^{Q_1Q_5}/S_{Q_1Q_5}$ where $S_{N}$ is the symmetric group on~$N$ elements~\cite{Vafa:1995bm,Maldacena:1997re,Witten:1997yu, Larsen:1999uk,Seiberg:1999xz}.
The symmetric-product orbifold description has been validated by a number of comparison with the dual string theory, including comparison of moduli spaces and of the sprectrum of protected operators~\cite{Dijkgraaf:1998gf, Maldacena:1998bw, deBoer:1998ip, Kutasov:1998zh}, as well as more recently of correlation functions~\cite{Gaberdiel:2007vu, Dabholkar:2007ey, Pakman:2007hn, Taylor:2007hs}.

It is also worth mentioning another closely related but somewhat more involved background that preserves the same amount of supersymmetry, \ie\ sixteen real supercharges. This is given by~$\AdS_3\times\S^3\times\S^3\times\S^1 $, provided that the curvature radii of the two spheres spaces satisfy
\begin{equation}
\frac{1}{R^2_{\S^3_{(1)}}}=\frac{\alpha}{R^2_{\AdS_3}}\,,\qquad
\frac{1}{R^2_{\S^3_{(2)}}}=\frac{1-\alpha}{R^2_{\AdS_3}}\,,
\qquad\quad
0<\alpha<1\,.
\end{equation}
The parameter $\alpha $ gives the relative size of the two spheres. In the limits $\alpha\to 0 $ or $\alpha\to 1 $ either sphere becomes flat and, up to compactifying back to a torus, we go back to the~$\AdS_3\times\S^3\times\T^4 $ background. The AdS/CFT correspondence for $\AdS_3\times\S^3\times\S^3\times\S^1 $ backgrounds has also been studied~\cite{Elitzur:1998mm,Boonstra:1998yu,deBoer:1999rh, FigueroaO'Farrill:2000ei, Gukov:2004ym,Tong:2014yna}, but it remains difficult to characterise its dual $\CFT_2 $. It is known that its symmetry algebra should be the \emph{large} $\mathcal{N}=(4,4) $ superconformal algebra~\cite{Sevrin:1988ew,Schoutens:1988ig, Spindel:1988sr, VanProeyen:1989me,Sevrin:1989ce}, which differs from the one of $\AdS_3\times\S^3\times\T^4 $ by the presence of two additional~$\su(2) $ subalgebras. The rigid part of such infinite dimensional symmetry is given by the exceptional Lie superalgebra~\cite{Gauntlett:1998kc}
\begin{equation}
\d21a_{\L}\oplus \d21a_{\R}\,.
\end{equation}
Sending $\alpha\to 0 $ or $\alpha\to 1 $ amounts to a contraction of the Lie superalgebra, which indeed yields~\eqref{eq:psu112:2} up to abelian factors.\footnote{It should be noted that such a limit, as we will see in chapter~\ref{ch:outlook}, requires great care.}

It is very interesting to note that similar constructions can be realised in terms of NS5 branes and fundamental strings, rather than D~branes. In fact, such a setup is S-dual to the D1-D5 system. The advantage in this case is that the near-horizon limit of the NS-brane system is supported only by NSNS fluxes. In this case, worldsheet CFT techniques can be efficiently used to study the propagation of strings there~\cite{Giveon:1998ns, deBoer:1998pp, Kutasov:1999xu, Maldacena:2000hw, Maldacena:2000kv, Maldacena:2001km}.
Let us be slightly more specific. In absence of RR fluxes, the worldsheet theory can be described nicely in the NSR formalism. The bosonic theory on $\AdS_3\times\S^3 $ amounts to a Wess-Zumino-Witten (WZW) model with gauge group $SL(2)\times SU(2) $. The non-compact $SL(2)$ factor presented a major obstacle to the CFT approach, that has however been overcome by Maldacena and Ooguri yielding a solution of that sector of the theory~\cite{Maldacena:2000hw, Maldacena:2000kv, Maldacena:2001km}.  Fermions can also be included in the picture, which leads to  a supersymmetric $SL(2)\times SU(2) $~WZW model.
Its spectrum can also be investigated by considering it as a supergroup coset $\sigma $ model, as it was done in ref.~\cite{Gaberdiel:2011vf} by taking advantage of the hybrid formalism introduced in~\cite{Berkovits:1999im} in the hope to extend the approach to backgrounds with RR fluxes.

The super-coset description is the most interesting for our purposes. Super-cosets are known to be a useful tool for writing down target-space supersymmetric string actions in flat space~\cite{Henneaux:1984mh} as well as in curved AdS backgrounds supported by RR fluxes\footnote{%
Most notably, the $\AdS_5\times\S^5 $ background can be described as the super-coset $PSU(2,2|4)/SO(1,4)\times SO(5)$. As shown by Metsaev and Tseytlin~\cite{Metsaev:1998it}, this can be used to write down the Green-Schwarz string action.}, and to study their classical properties. However, such an approach does not immediately offer a good way to quantise the theory. Quantisation can be done in light-cone gauge, which becomes quickly very cumbersome. In practice, even when restricting to the 't Hooft limit, observables can only be explicitly computed at the first orders of  a perturbative expansion in the string tension.

Therefore, it would appear that there are little chances to study the D1-D5 system without resorting to any (non-perturbative and non-planar) S~duality. Studying a background supported by mixed RR and NSNS fluxes---which can be constructed by considering D- and NS-branes simultaneously---appears even harder~\cite{Berkovits:1999im,Rahmfeld:1998zn,Pesando:1998wm}. It is in this context that the notion of integrability can save the day, and provide an effective tool to study the spectrum of such theories, at least as long as we are in the 't~Hooft limit. Remarkably, this seem to be possible for both the~$\AdS_3\times\S^3\times\T^4 $ and $\AdS_3\times\S^3\times\S^3\times\S^1 $ backgrounds, and even for mixed-flux backgrounds. Before discussing how that happens, let us the introduce the observables that we are interested in computing in the 't~Hooft limit, \ie\ let us present the \emph{the spectral problem} of AdS/CFT.

\section{The spectral problem}
\label{sec:intro:string}
From now on, and in all of this review, let us restrict to AdS/CFT \emph{in the 't~Hooft limit}. We are dealing with free strings, so that the natural observables are the string energy levels. These are the eigenvalues~$\{E^{\text{t.s.}}_{j}\}$ of the generator~$\mathbf{H}^{\text{t.s.}}$ of time-translations in the target space%
\footnote{More precisely, this is true when using global coordinates for~$\AdS $.}.
 For some string states (such as the vacuum) these are protected by supersymmetry, but in general, they are a non-trivial function of the dimensionless string length~$\ell_s/R$, due to the fact that the free strings probe the curved~$\AdS_{n+1}\times\mathcal{M}_{9-n}$ geometry. In the 't~Hooft limit~\cite{'tHooft:1973jz} of the CFT, the leading observables are the two-point functions, whose form is constrained by conformal symmetry---schematically
\begin{equation}
\left\langle
\mathcal{O}_{j}(x)\,\mathcal{O}_{j}(y)
\right\rangle
=
\frac{1}{|x-y|^{2\Delta_{j}}}\,,
\end{equation} 
where $\Delta_j$ is the eigenvalue of the generator of dilatations~$\mathbf{D}$ acting on $\mathcal{O}_j$, and is in general a non-trivial function of the 't Hooft coupling~$\lambda$. The spectrum~$\{\Delta_j\}$ should then be dual to the string energy spectrum. 

Perturbative calculations in the string worldsheet theory will give $E_j^{\text{t.s.}}$ at~$\ell_s/R\ll1$, while in the CFT we would find $\Delta_j$ at $\lambda\ll1$, \ie\ in the opposite regime. The aim of integrability is to give a description valid at any intermediate coupling. This can usually be set-up by considering either side of the AdS/CFT duality,\footnote{%
While the description arising on the CFT side played a very important role in the development of integrability for AdS/CFT, it should be noted that it is only from the worldsheet theory point of view that the so-called wrapping effects can be accounted for.
}
 but in our case it will be more convenient to focus more on the string side of it.

\subsection*{String theory in light-cone gauge}

Let us consider a theory of closed (super)strings only. In absence of string interactions, the worldsheet of the string is a cylinder of circumference~$\ell $, and the classical string theory is defined by an action of the form
\begin{equation}
\label{eq:bosonicstring}
S_{\text{bos}}=-\frac{h}{2}\int\de\tau\int_{-{\ell}/{2}}^{{\ell}/{2}}\de\sigma\, \sqrt{|\gamma|}\,\gamma^{\alpha\beta}\partial_\alpha X^\mu\partial_\beta X^\nu\, G_{\mu\nu}(X)\,,
\end{equation}
for the bosons, which should be supplemented by fermionic terms---here we avoid doing so to keep the discussion simple. Here $h\approx{R^2}/{\ell_{s}^2}$ is a coupling constant, $\gamma_{\alpha\beta}$ is the metric on the worldsheet, $X^\mu$ can be thought of as coordinates in the target space, whose metric is~$G_{\mu\nu}$. For more complicated backgrounds, an antisymmetric $B_{\mu\nu}(X)$ field can also appear, but we will not include it here.

The action~\eqref{eq:bosonicstring} is invariant under reparametrisations of the worldsheet and Weyl rescalings, and should be gauge fixed. To briefly illustrate the strategy, let us assume for the moment that the target space is flat, $G_{\mu\nu}=\eta_{\mu\nu}$.\footnote{We will discuss the case of curved AdS backgrounds at length in the next chapter.} Then, if we introduce light-cone coordinates
\begin{equation}
X_\pm=X^0\pm X^9\,,
\end{equation}
we can both eliminate the world-sheet metric~$\gamma^{\alpha\beta} $  and set
\begin{equation}
X_+(\sigma,\tau)= \tau\,,
\end{equation}
where we ignore the winding terms.
In this way, one can use the Virasoro constraints
\begin{equation}
\frac{\delta\, S}{\delta\,\gamma^{\alpha\beta}}=0\,,
\end{equation}
to solve for $X^-$ in terms of the remaining fields. This still leaves one non-linear constraint, the so-called \emph{level-matching condition}, which enforces periodicity of the strings along $\sigma$ and amounts to the vanishing of the worldsheet momentum $\mathbf{P}$.

This gauge-fixed theory of free strings (or rather, a suitable supersymmetric version on certain curved backgrounds) is what we want to quantise. In the bosonic sector we are left with eight physical fields $X^j$ defined on a cylinder. The physical Hilbert space will consist of the excitation of these eight fields subject to the level-matching condition, which is realised as a projection on the Hilbert space
\begin{equation}
\mathbf{P}\Ket{\{M_1\},\dots,\{M_8\}}_{\text{phys.}}=0\,,
\end{equation}
where $\{M_j\}$ are label the excitations of each field. A preferred basis is the one of eigenstates of the worldsheet Hamiltonian $\Ham$, that is the operator generating time evolution on the worldsheet in the sense of Stone's theorem.
The spectrum we eventually want to compute is the one corresponding to time evolution \emph{in the target space}. However, by light-cone gauge fixing we have related the worldsheet time~$\tau$ to $X_{+}$ and hence to $X^{0}$, that is the target-space time. From this it follows that the eigenvalues of~$\mathbf{H}^{\text{t.s.}}$ are simply related to the ones of~$\mathbf{H}$, and it will be enough to compute the latter.

It is worth recalling that the spectrum of physical states will organise itself into multiplets of the symmetry algebra of the theory. Even if we did not construct the string Hamiltonian, it is immediate to realise that in the flat case the physical fields $X^j$, $j=1,\dots,8$ will appear in~\eqref{eq:bosonicstring} in $SO(8)$-invariant combinations. This is the \emph{manifest} symmetry of the theory in light-cone gauge. However, as there are massive string excitations, whose little group is $SO(9)$, we expect that the theory should enjoy a larger symmetry, and that the $\so(8)$ Lie algebra multiplets should arrange themselves into irreducible representations of~$\so(9)$. This illustrates how in general the manifest symmetries of the light-cone gauge-fixed theory form a subalgebra of the whole symmetry algebra, which in fact for flat space should be the full~$\so(1,9)$ when we also take boosts into account.

In practice, for curved supersymmetric backgrounds there will be several complications: the action will involve fermions and non-linear terms. As we mentioned, an useful approach is to rewrite the action (or part of it) as a coset action of a suitable supergroup~\cite{Metsaev:1998it}. Still, the gauge fixed Hamiltonian will be highly non-linear,  so that only perturbative quantisation will be possible. A way around this complication---valid as long as we can identify asymptotic states---is to find a (hopefully unique) S matrix that preserves the symmetries of the theory, and use that to find the spectrum, as we will detail.

\subsection*{The decompactification limit and the worldsheet S matrix}
In the light-cone gauge, the worldsheet is no longer invariant under rescalings. In fact, a more careful analysis would reveal that the size $\ell$ of the worldsheet is fixed in terms of the momentum conjugated to the light-cone coordinate $X_-$. It is interesting to consider the limit $\ell\to\infty$, whereby the worldsheet cylinder decompactifies to a plane. In this case, we are dealing with a two-dimensional QFT with well-defined asymptotic states. In particular, the spectrum can be described in terms of $M$-particle states on the worldsheet created by raising operators from a vacuum,
\begin{equation}
\Ket{p_1,\dots,p_n}^{(\IN ,\OUT)}_{\a_1,\dots,\a_M}=a^{\dagger\,(\IN ,\OUT)}_{\a_1}(p_1)\cdots a^{\dagger\,(\IN ,\OUT)}_{\a_n}(p_n)\Ket{\vac}\,,
\end{equation}
where the in and out raising operators~$a^{\dagger\,(\IN,\OUT)}$ satisfy canonical commutation relations with the in and out lowering operators~$a^{(\IN,\OUT)}$. If we want to consider physical states, we will have to impose the level matching condition,
\begin{equation}
\label{eq:onshell}
\mathbf{P}\Ket{p_1,\dots,p_n}^{(\IN,\OUT)}_{\a_1,\dots,\a_M}=\left(p_1+\dots+p_n\right)\Ket{p_1,\dots,p_n}^{(\IN,\OUT)}_{\a_1,\dots,\a_M}=0\,.
\end{equation}
We will call a state satisfying eq.~\eqref{eq:onshell} ``on-shell'', as opposed to a generic (off-shell) state.
In either case, the action of the Hamiltonian is very simple
\begin{equation}
\label{eq:dispersionrel}
\begin{aligned}
\Ham^{(\IN,\OUT)}\Ket{p_1,\dots,p_n}^{(\IN,\OUT)}_{\a_1,\dots,\a_M}&=E(\{\a_j,p_j\})\,\Ket{p_1,\dots,p_n}^{(\IN,\OUT)}_{\a_1,\dots,\a_M}\,,\\
E(\{\a_i,p_i\})&=\sum_{j=1}^{M} \omega_{\a_j}(p_j)\,,
\end{aligned}
\end{equation}
where $\omega_\a(p)$ is the dispersion relation, and $m_\alpha $ accounts for the fact that particles of different flavor may have different mass. In the case of a relativistic theory we should find
\begin{equation}
\label{eq:omega-disp-rel}
\omega_{\a}(p)=\sqrt{m^2_{\a}+p^2}\,.
\end{equation}
However, for $\AdS$ background it has been found that light-cone gauge fixing breaks the relativistic invariance on the worldsheet so that the dispersion relation takes a lattice-like form,  \ie\ schematically
\begin{equation}
\label{eq:omega-disp-nonrel}
\omega_{\a}(p)=\sqrt{m^2_{\a}+4\,h^2\sin^2\frac{p}{2}}\,,
\end{equation}
where $h$ is the coupling constant.

Given that the two sets of raising and lowering operators~$\{a^{\dagger\,(\IN )},\,a^{(\IN )}\}$ and~$\{a^{\dagger\,(\OUT )},$ $a^{(\OUT )}\}$ both satisfy canonical commutation relations, by virtue of the Stone-von Neumann theorem they must be related by an unitary operator $\Smat$ satisfying
\begin{equation}
\Smat^{\dagger}\,\Smat = \Smat\ \Smat^{\dagger} = \Id\, ,
\quad \Smat\ket{\vac}=\Ket{\vac}\,,\qquad
a^{\dagger\,(\IN )} = \Smat\ a^{\dagger\,(\OUT )}\Smat^{\dagger} \, ,
\quad a^{(\IN )} = \Smat\ a^{(\OUT )}\Smat^{\dagger}\,,
\end{equation}
from which one immediately finds that $\Smat$ is the familiar S matrix that relates in- and out-states,
\begin{equation}
\label{eq:genericSmat}
\Ket{p_1,\dots,p_n}^{(\IN)}_{\a_1,\dots,\a_M}=\Smat\,
\Ket{\tilde{p}_1,\dots,\tilde{p}_{\tilde{M}}}^{(\OUT)}_{\tilde{\a}_1,\dots,\tilde{\a}_{\tilde{M}}}\,.
\end{equation}

In practice, finding the S~matrix is hard---it can be done perturbatively only at one- or two-loop order for the models of our interest. We will circumvent this problem by dealing with theories whose S~matrix can be determined by the symmetries of the theory---that is, \emph{integrable theories}. 
We will discuss at length what this means further on in this review, see in particular chapter~\ref{ch:smatrix}.

Let us assume that we somehow have found the all-loop complete S~matrix and dispersion relation. Equipped with these, and remembering that we are dealing with a QFT in 1+1 dimensions, we can start looking at some observables. Let us prepare a one-particle state $\Ket{p}^{(\IN )}_\a$ of definite momentum and flavour. In absence of external fields such a stable asymptotic state will satisfy  $\Ket{p}^{(\IN )}_\a=\Ket{p}^{(\OUT )}_\a$ and its energy will be
\begin{equation}
E= \omega_{\a}(p)\,,
\end{equation}
so that the spectrum is continuous. If, however, we take into account the fact that the worldsheet is a cylinder of size $\ell$, we should also impose the spatial periodicity of the wave-function, which amounts to the quantisation condition for the momentum
\begin{equation}
e^{i\,p\,\ell}= 1\,,
\end{equation}
resulting in a \emph{discrete} spectrum. Let us now take a state $\Ket{p,q}_{\a,\a}^{(\IN)}$ consisting of two particles of the same flavour~$\a$, with momenta~$p>q$ so that they are asymptotically well separated. Let us also assume that they scatter elastically without producing any other particle---which is generally not the case in a QFT. Then the corresponding out-state will contain again two particles of the same flavour and momenta $p,q$. If we now impose periodicity, however, we have to account for the fact that each particle underwent a phase shift due to the scattering, so that we have
\begin{equation}
\label{eq:BA-twopartciles}
e^{i\,p\,\ell}\,S(p,q)=1\,,\qquad
e^{i\,q\,\ell}\,S(q,p)=1\,,
\end{equation}
where~$S(p,q)$ is the diagonal matrix element for the flavour $\a$,~$S_{\a\a}^{\a\a}(p,q)$.
Inserting the solutions for~$p,q$ into the dispersion relations~\eqref{eq:dispersionrel} will yield again a discrete energy spectrum.

Since we are interested in the \emph{physical} spectrum, we must impose the level matching condition~\eqref{eq:onshell}, finding that there are no non-trivial one-particle states in the on shell theory, and that for two particles one must have $q=-p$. If one were able to follow a similar recipe for any number of particles of arbitrary flavours, then he would have a description of the string spectrum. It turns out that this is possible provided that $\Smat$  ``factorises'', \ie\ provided that a $M$-body scattering event can be understood as a sequence of two-body ones. Again, this is a typical feature of integrable theories, which amounts to satisfying the celebrated \emph{Yang-Baxter equation}.
Then, the $M$-particle analogue of eq.~\eqref{eq:BA-twopartciles} are the~\emph{Bethe-Yang equations} (BY equations), that are schematically of the form
\begin{equation}
\label{eq:schematic-BA}
e^{i\,p_k\,\ell}\,\prod_{j\neq k}^M S(p_k,p_j)=1\qquad\text{for}\qquad k=1,\dots,M\,,
\end{equation}
where $S(p,q)$ is a suitable S-matrix element. Equations of this type where first found in the context of QFTs by Yang~\cite{Yang:1967bm}, inspired by the ansatz that Bethe proposed in the context of quantum spin chains~\cite{Bethe:1931hc}.

\subsection*{A spin-chain picture}
The appearance of the periodic dispersion relation~\eqref{eq:omega-disp-nonrel} and the fact that the spectrum can be described in terms equations of the Bethe ansatz type strongly hint that the underlying theory may have an alternative description in terms of a discrete model, perhaps of a quantum spin chain.

It has been long known that some NLSMs are equivalent to certain quantum spin chains, see \eg\ chapter 5 in ref.~\cite{Fradkin:1991nr}. For instance, a coset model similar to the one emerging from string theory but with target space $\S^2$ is equivalent to the long wave-lenght limit of the Heisenberg spin chain.
The Heisenberg chain is perhaps the prototype of a quantum spin chain, defined on $\ell$ sites periodically identified and with Hamiltonian
\begin{equation}
\label{eq:Heisenberg}
\mathbf{H}=\mu\, \sum_{j=1}^{\ell} \left(\frac{1}{4}-\vec{S}_{j}\cdot\vec{S}_{j+1}\right)\,,
\end{equation}
where $\vec{S}_{j}$ is a spin at the $j$th site and $\mu$ is a coupling constant.

The sphere~$\S^2$ may be part of the target space for our string NLSM, \eg\ in the case of~$\AdS_3\times\S^3\times\mathcal{M}_4$.  Therefore the spectrum of the Heisenberg chain could somehow be part of the one that we wish to compute. For several integrable strings backgrounds this is actually the case~\cite{Kruczenski:2003gt,Kruczenski:2004kw, Hernandez:2004uw,Stefanski:2004cw, Stefanski:2005tr,Stefanski:2007dp}, and for instance in the case of $\AdS_5/\CFT_4$ the first evidence of integrability was discovered in four-dimensional $\mathcal{N}=4$ SYM theory precisely in form of the Heisenberg spin chain Hamiltonian by Minahan and Zarembo~\cite{Minahan:2002ve}. Let us briefly review that prototypical picture.

The operator dual to the string Hamiltonian is proportional to the generator of dilatations~$\mathbf{D}$ in the $\CFT$. String states are dual to local  operators which in the 't~Hooft limit take the form \eg
\begin{equation}
\mathcal{O}=\text{tr}\big[
Z\,Z\,\partial^{\mu}X\,\partial_{\mu}Z\,\psi\cdots \bar{\psi}\, Z
\big]\,,
\end{equation}
where $X,Z,\psi,\bar{\psi}$ are some of the  (bosonic and fermionic) fundamental fields appearing in the $\CFT_{n}$ Lagrangian, which are all evaluated at the same spacetime point, and the trace ensures gauge invariance. The spectral problem now reads
\begin{equation}
\mathbf{D}\,\mathcal{O}= \Delta\,\mathcal{O}\,.
\end{equation}
Each of the fields that compose $\mathcal{O}$ are identified by their transformation properties under the $n$-dimensional superconformal algebra. In fact, we can think of them as of ``spins'' of that algebra, so that $\mathcal{O}$ becomes a state of the \emph{periodic} (due to the trace) spin chain.

In particular, in the case of $\mathcal{N}=4$ SYM, if we restrict to two $\su(2)$-charged scalar fields $X,Z$, we have that an operator $\mathcal{O}$ is equivalent to a state $\ket{\Psi}$
\begin{equation}
\mathcal{O}=\text{tr}\big[
Z\,Z\,X\,Z\,X\cdots X\, Z
\big]
\qquad
\longleftrightarrow
\qquad
\ket{\Psi}=\ket{
\downarrow\,\downarrow\,\uparrow\,\downarrow\,\uparrow\,\cdots\uparrow\,\downarrow
}\,,
\end{equation}
where the arrows indicate $\su(2)$ spins, $S^{3}\ket{\downarrow}=-\frac{1}{2}\ket{\downarrow}$ and $S^{3}\ket{\uparrow}=+\frac{1}{2}\ket{\uparrow}$. The breakthrough of ref.~\cite{Minahan:2002ve} was realising that at 1-loop in the weakly coupled $\CFT$, the dilatation operator $\mathbf{D}$ coincides with $\mathbf{H}$ of~\eqref{eq:Heisenberg} up to suitably identifying the coupling constants. This is particularly remarkable because $\mathbf{H}$ enjoys a large number of symmetries that make the explicit solution of the spectral problem possible in terms of a Bethe ansatz.

As it turns out, the whole spectral problem of $\mathcal{N}=4$ SYM can be related to a spin chain for the superconformal algebra, and solved by considering an asymptotic \emph{spin-chain S matrix}  
\cite{Beisert:2005fw,Beisert:2006ib}, in a procedure that strongly resembles the one described for the string worldsheet theory. The first step is to consider the limit of a long spin chain, $\ell\to \infty$ and effectively decompactify the chain%
\footnote{%
In simpler cases such as the Heisenberg chain we might not need to first consider the $\ell\to\infty$ limit, and the Bethe ansatz would be exact for any~$\ell$. However, one peculiarity of the spin chains arising from $\AdS/\CFT$ is that the Hamiltonian may couple states of different length---in the case of~$\mathcal{N}=4$ SYM this can be seen from the  presence of a Yukawa interaction in the SYM Lagrangian. Then the Bethe ansatz description is only asymptotic, \ie~valid for~$\ell\to\infty $.
}%
. We then consider collective excitations of definite momentum, called \emph{spin waves} or \emph{magnons} of the form \eg
\begin{equation}
\ket{p}=\sum_{j=1}^{\infty} e^{i\,p\,j}\,
\ket{\downarrow\,\downarrow\,\downarrow\,\downarrow\,\cdots
\downarrow\,\downarrow\,\uparrow\,\downarrow\,\downarrow\,\cdots}\,,
\end{equation}
where the overturned spin sits at the $j$th site. Notice that we have implicitly picked a \emph{vacuum state} of downward spins. This operation is akin to the light-cone gauge fixing and breaks the manifest symmetry of the model. Remarkably, it still may happen that the S~matrix corresponding to the scattering of two magnons is uniquely fixed by the residual symmetries, and the multiparticle scattering ``factorises'' in the sense alluded to earlier. In that case, we can proceed to write down \emph{asymptotic} Bethe ansatz equations of the form~\eqref{eq:schematic-BA} by requiring that $M$ magnons live on a periodic spin chain of finite 	length~$\ell$. Finally, requiring that the states are invariant under cyclic permutations precisely reproduces the level-matching constraint~\eqref{eq:onshell}, as we will see in chapter~\ref{ch:betheansatz} where we will consider in detail a similar construction for~$\AdS_3/\CFT_2 $.

The above picture can be seen to hold for several instances of $\AdS/\CFT$. However, in the cases where a Lagrangian description of the $\CFT$ cannot be efficiently used, as it happens in  $\AdS_3/\CFT_2$, it is less clear how the spin chain emerges and how to relate it to the CFT observables. Still, if such a description exists it may be used as a tool to investigate the CFT side of the duality, as well as to give an alternative albeit similar way to solve the spectral problem.
One may speculate that a spin-chain description may emerge just by considering a discretisation of the string worldsheet, but it is hard to make such a statement rigorous. In fact, very recently for~$\AdS_3/\CFT_2$ there appeared evidence that this may not always be the case, as we will describe in the last chapter of this review.

\subsection*{The spectrum of the finite-size theory}
\begin{figure}
  \centering
  \subfloat[Real process]{%
    \label{fig:Fterm}
  \begin{tikzpicture}
  
    \coordinate (a1) at (-1cm,0.5cm);
    \coordinate (a2) at (0cm,0cm);
    \coordinate (a3) at (1cm,0.5cm);
    \coordinate (a4) at (0cm,1cm);
    
    \coordinate (b1) at (-1cm,2cm);
    \coordinate (b2) at (0cm,1.5cm);
    \coordinate (b3) at (1cm,2cm);
    \coordinate (b4) at (0cm,2.5cm);
    
    \coordinate (c1) at (-1cm,3.5cm);
    \coordinate (c2) at (0cm,3cm);
    \coordinate (c3) at (1cm,3.5cm);
    \coordinate (c4) at (0cm,4cm);
   
    \draw [thick] [out=270,in=180] (a1) to (a2);
    \draw [thick] [out=0,in=270] (a2) to (a3);
    \draw [thick, dotted] [out=90,in=0] (a3) to (a4);
    \draw [thick, dotted] [out=180,in=90] (a4) to (a1);

    \draw [thick,cyan] [out=270,in=180] (b1) to (b2);
    \draw [thick,cyan] [out=0,in=270] (b2) to (b3);
    \draw [thick,cyan,dotted] [out=90,in=0] (b3) to (b4);
    \draw [thick,cyan,dotted] [out=180,in=90] (b4) to (b1);

    \draw [thick,cyan,directed]  (a2) to (b2);
    \draw [thick,cyan,dotted,directed]  (b4) to (c4);
    
    \draw [thick] [out=270,in=180] (c1) to (c2);
    \draw [thick] [out=0,in=270] (c2) to (c3);
    \draw [thick] [out=90,in=0] (c3) to (c4);
    \draw [thick] [out=180,in=90] (c4) to (c1);

    \draw [thick] (a1) to (c1);
    \draw [thick]  (a3) to (c3);
    
    \draw [time] (-2cm, 1cm) to (-2cm, 2.7cm);
    \node at (-1.8cm, 1.9cm) {$t$};
    
  \end{tikzpicture}
}%
   \hspace{2.5cm}
  \subfloat[Virtual process]{%
    \label{fig:muterm}
  \begin{tikzpicture}
  
    \coordinate (a1) at (-1cm,0.5cm);
    \coordinate (a2) at (0cm,0cm);
    \coordinate (a3) at (1cm,0.5cm);
    \coordinate (a4) at (0cm,1cm);
    
    \coordinate (b1) at (-1cm,2cm);
    \coordinate (b2) at (0cm,1.5cm);
    \coordinate (b3) at (1cm,2cm);
    \coordinate (b4) at (0cm,2.5cm);
    
    \coordinate (c1) at (-1cm,3.5cm);
    \coordinate (c2) at (0cm,3cm);
    \coordinate (c3) at (1cm,3.5cm);
    \coordinate (c4) at (0cm,4cm);
   
    \draw [thick] [out=270,in=180] (a1) to (a2);
    \draw [thick] [out=0,in=270] (a2) to (a3);
    \draw [thick, dotted] [out=90,in=0] (a3) to (a4);
    \draw [thick, dotted] [out=180,in=90] (a4) to (a1);

    \draw [thick,cyan] [out=270,in=180] (b1) to (b2);
    \draw [thick,cyan] [out=0,in=270] (b2) to (b3);
    \draw [thick,cyan,dotted] [out=90,in=0] (b3) to (b4);
    \draw [thick,cyan,dotted] [out=180,in=90] (b4) to (b1);

    \draw [thick,cyan,directed]  (a2) to (b2);
    \draw [thick,cyan,directed]  (b2) to (c2);
    
    \draw [thick] [out=270,in=180] (c1) to (c2);
    \draw [thick] [out=0,in=270] (c2) to (c3);
    \draw [thick] [out=90,in=0] (c3) to (c4);
    \draw [thick] [out=180,in=90] (c4) to (c1);

    \draw [thick] (a1) to (c1);
    \draw [thick]  (a3) to (c3);
    
    \draw [time] (-2cm, 1cm) to (-2cm, 2.7cm);
    \node at (-1.8cm, 1.9cm) {$t$};
  
  \end{tikzpicture}
}%
  \caption{%
Two processes involving exchange of real or virtual particles that wrap around the worldsheet cylinder, and therefore are not captured by the Bethe-Yang equations.
  }%
  \label{fig:wrapping}
\end{figure}
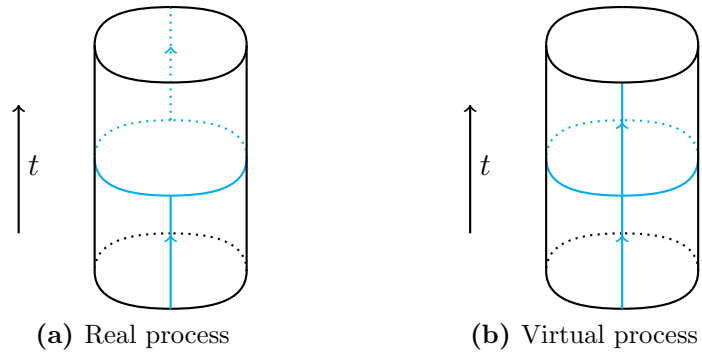

The Bethe-Yang equations~\eqref{eq:schematic-BA} do \emph{not} describe the spectrum of the finite-size $\AdS/\CFT$ duality~\cite{Ambjorn:2005wa}. They rather describe the spectrum of a QFT defined on a plane, after periodic identification of its spatial direction. To appreciate the difference between the two cases, let us consider the processes of figure~\ref{fig:wrapping}. There we depict the propagation of (possibly virtual) particles wrapping the worldsheet cylinder, which cannot be accounted for by the S~matrix derived in the decompactified theory. In terms of the spin chain, a similar effect arises due to the presence of long-range interactions---eventually, for a finite-length chain, such interactions wrap around the chain, invalidating the asymptotic Bethe ansatz approach.

In the QFT context, it is possible to estimate that wrapping effects are exponentially suppressed when~$\ell$ is large, but nonetheless they are to be taken into account. One possibility, pioneered by L\"uscher~\cite{Luscher:1985dn,Luscher:1986pf}, is to treat them as (perturbative) corrections to the spectrum predicted by the BY equations.

One can do even better by fully exploiting the integrability properties of~$\Smat$ and the fact that, while directly dealing with (integrable) S matrices in finite volume is even hardly self-consistent, doing so at finite temperature is quite natural. The thermodynamic Bethe ansatz (TBA) is a tool to compute the free energy of finite-temperature integrable QFTs~\cite{Yang:1968rm}. One describes the thermal bath of particles in the grand canonical ensemble by taking into account all the particles appearing in the BY equations and their bound states, and requiring thermodynamic equilibrium. This means that the  number of excitations is large, and their densities are fixed. Consequently, starting from the logarithm of eq.~\eqref{eq:schematic-BA} one then obtains a set of non-linear coupled integral equations\footnote{%
More precisely, one obtains such an integral equation for each species of particle configurations appearing in the thermodynamic limit. Identifying those usually requires some assumptions that go under the name of string hypothesis.%
}.

Zamolodchikov realised that this can be used to find the ground-state energy of a finite-size theory from the free energy of a finite-temperature one~\cite{Zamolodchikov:1989cf}, provided that these are related by exchanging the role of time and space by two Wick rotations---a ``mirror'' transformation. This procedure does not introduce any additional complication in the case of a relativistic theory, but is not straightforward in non-relativistic models. This can be seen by considering the effects of the mirror transformation 
\begin{equation}
\label{eq:mirrorTrans}
p\mapsto -i\,E = p_{\text{mirror}}\,,\qquad
E\mapsto -i\,p = E_{\text{mirror}}\,,
\end{equation}
on the dispersion relation $E^2=\omega_j(p)^2$, which leaves invariant the relativistic case~\eqref{eq:omega-disp-rel} yields an entirely new relation when one uses eq.~\eqref{eq:omega-disp-nonrel}. In fact, the mirror transformation  produces a novel~\emph{mirror theory}. The thermodynamic properties of such ancillary theory yield the finite-size properties of the original one.

\section{The integrability approach to AdS/CFT}
Let us briefly overview the developments that lead to successfully employing integrability to the spectral problem in AdS/CFT.

\subsection*{The case of $\AdS_5/\CFT_4 $}
The best understood example of integrability in AdS/CFT is the case of type IIB superstrings on $\AdS_5\times\S^5 $ and~$\mathcal{N}=4$ SYM. This is unsurprising since such string theory background preserves as much supersymmetry as possible, and $\N=4$ SYM can be easily studied in perturbation theory. A detailed review of the string side of the story can be found in ref.~\cite{Arutyunov:2009ga}, while a broader account and an extensive list of references can be found in ref.~\cite{Beisert:2010jr}.

The first hints of integrability were found on the gauge theory side. Early on, it was noticed by Lipatov, building up on the existence of integrable structure in Yang-Mills and QCD,\footnote{%
Such structures emerge in particular when considering the reggeised hihg-energy gluon dynamics. This is described by the BFKL Hamiltonian~\cite{Lipatov:1976zz,Kuraev:1976ge, Kuraev:1977fs,Fadin:1975cb,Balitsky:1978ic}
which was found to be integrable and related to a generalisation of the Heisenberg model~\cite{Lipatov:1994xy,Faddeev:1994zg}, see also ref.~\cite{Lipatov:2009nt}.
}
 that that such structures and the extended supersymmetry of gauge theory were closely connected~\cite{Lipatov:1997vu,Lipatov:1998as}. Later on, integrability was rediscovered by Minahan and Zarembo~\cite{Minahan:2002ve}, by explicitly investigating the one-loop spectrum of the dilatation operator. As we mentioned in the previous subsection, the key point was to recognise that such operator could be interpreted as an integrable spin-chain Hamiltonian---again closely related to the Heisenberg one. It was then realised that similar structures persist at higher loop order in the 't~Hooft coupling~\cite{Beisert:2003tq,Beisert:2003jj,Beisert:2004ry}.

Almost in parallel, similar investigations were performed on the string side. As we mentioned, the light-cone gauge Hamiltonian there is highly non-linear~\cite{Metsaev:2000yu,Arutyunov:2004yx,Frolov:2006cc}. However, exploiting the Metsaev-Tseytlin~\cite{Metsaev:1998it} coset description of the Green-Schwarz action~\cite{Green:1983wt,Grisaru:1985fv}, it was shown that the worldsheet theory is integrable as a classical field theory~\cite{Bena:2003wd}. Classical integrability allowed to consider special solutions of the equations of motion~\cite{Arutyunov:2003uj, Tseytlin:2003ii,Arutyunov:2003za}, including ``giant magnons''~\cite{Hofman:2006xt,Chen:2006gea}, and to write down generalised Landau-Lifshitz equations~\cite{Kruczenski:2003gt,Kruczenski:2004kw, Hernandez:2004uw,Stefanski:2004cw, Stefanski:2005tr,Stefanski:2007dp}. These are string solitons that can be thought of as semi-classical limit of spin chain magnons in the dual theory. In particular, they feature the non-linear dispersion relation~\eqref{eq:omega-disp-nonrel}, which was also found in refs.~\cite{Frolov:2006cc,Klose:2007rz}. While classical integrability is in no way guaranteed to carry over to the quantum theory, in the case of~$\AdS_5\times\S^5 $ it was possible to find indications that this is the case~\cite{Kazakov:2004qf,Arutyunov:2004vx}.\footnote{%
Integrability was later identified also by studying string theory in the pure-spinor formulation, see ref.~\cite{Puletti:2010ge} for a review.}

Eventually, integrability was established as an all-loop feature of the $\AdS_5/\CFT_4$ duality, at least up to including wrapping effects. This was first realised on the gauge theory side, where an integrable all-loop S~matrix was proposed~\cite{Beisert:2005tm,Beisert:2005fw,Beisert:2006ez}, and then on the string side as well~\cite{Arutyunov:2006ak}. In fact, the two descriptions can be precisely mapped into one another~\cite{Arutyunov:2006yd}.

As for wrapping effects, it was shown~\cite{Bajnok:2008bm} that the approach of L\"uscher~\cite{Luscher:1985dn,Luscher:1986pf} can be extended to the $\AdS_5\times\S^5$ NLSM, and in fact that the whole mirror TBA approach can be applied, yielding an exact description of the spectrum. Following the reasoning of ref.~\cite{Ambjorn:2005wa}, the mirror model was constructed~\cite{Arutyunov:2007tc}. Since this is related to the original theory by an analytic continuation~\cite{Arutyunov:2009kf}, its all-loop S~matrix is automatically integrable. Then, the mirror TBA equations (or equivalently, the ``Y system'') were worked out~\cite{Arutyunov:2009zu, Gromov:2009tv,Arutyunov:2009ur, Bombardelli:2009ns,Gromov:2009bc, Cavaglia:2010nm,Balog:2011nm}. These can also be simplified to a finite set of non-linear integral equations~\cite{Suzuki:2011dj,Balog:2012zt}, ultimately taking the form of a ``quantum spectral curve''~\cite{Gromov:2013pga, Gromov:2014caa}. All these descriptions by construction yield a spectrum organised in multiplets of the superconformal algebra~\cite{Arutyunov:2011uz}, for which finite-size effects are essential~\cite{Sfondrini:2011rr}.  The study of spectroscopy for $\AdS_5/\CFT_4$ by either analytical~\cite{Arutyunov:2009ax,Arutyunov:2010gb, Balog:2010xa,Arutyunov:2011mk,Arutyunov:2012tx, Leurent:2012ab,Leurent:2013mr}, or entirely numerical methods~\cite{Gromov:2009zb,Frolov:2010wt,Frolov:2012zv} has been initiated and both provided substantial evidence in favour of the holographic duality and demonstrated the power of the integrability approach.

\subsection*{The case of $\AdS_4/\CFT_3 $}
\begin{figure}
  \centering
\begin{tikzpicture}[%
    arrow/.style={-latex}
    ]%
  \begin{scope}[xshift=0cm]

    \node (left) at (-4cm,0cm) [spin] {$\quad U(N)\quad  $};
    \node (right) at (4cm,0cm) [spin] {$\quad \hat{U}(N)\quad  $};
    
    \draw [->,out=0,in=90,thick] ($(right.east)+(0,0.4cm)$) to ($(right.east)+(1cm,0)$) to [out=270,in=0] ($(right.east)-(0,0.4cm)$);
    \draw [->,out=180,in=270,thick] ($(left.west)-(0,0.4cm)$) to ($(left.west)-(1cm,0)$) to [out=90,in=180] ($(left.west)+(0,0.4cm)$);
    
	\node (a1) at (-6.5cm,0) [] {$A_{\mu}$};    
	\node (a1) at (6.5cm,0) [] {$\hat{A}_{\mu}$};

    \draw [->,out=10,in=180,thick] ($(left.east)+(0,0.4cm)$) to ($(0cm,0.7cm)$) to [out=0,in=170] ($(right.west)+(0,0.4cm)$);
    \draw [->,out=190,in=0,thick] ($(right.west)-(0,0.4cm)$) to ($(0cm,-0.7cm)$) to [out=180,in=-10] ($(left.east)-(0,0.4cm)$);

	\node (y1) at (0,1cm) [] {$Y^{A},\,\psi_{A}$};
	\node (y2) at (0,-1cm) [] {$Y^{\dagger}{}_{\!\!A},\,\psi^{\dagger\,A}$};  
  \end{scope}
\end{tikzpicture}

  \caption{The quiver diagram of ABJM theory, where we have drawn an arrow from the fundamental to the anti-fundamental representation of each gauge group. Note that the scalars $Y^{A}$ and the fermions $\psi_{A}$, as well as their conjugates, carry an index of the R-symmetry group $SU(4)$, $A=1,\dots 4$.}
  \label{fig:abjm}
\end{figure}
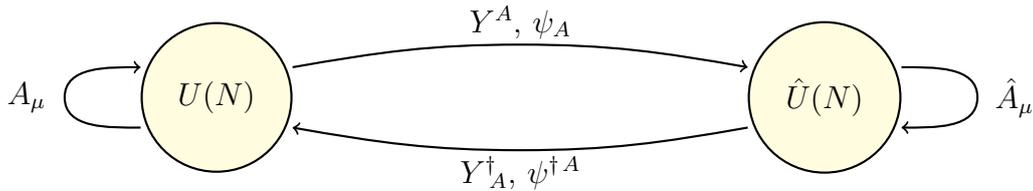

It is worth briefly reviewing another holographic set-up that is amenable to integrability and is in many ways ``intermediate'' between the prototypical one of $\AdS_5/\CFT_4$ and the one of  $\AdS_3/\CFT_2 $ to which this review is devoted. This is the planar limit of the correspondence between type IIA strings on%
\footnote{%
Following standard conventions we denote by $\mathbbm{C}\text{P}^n$ the $n$-dimensional complex projective space, which can be regarded as the real quotient manifold~$\S^{2n+1}/U(1)$.
}
 $\AdS_4\times\mathbbm{C}\text{P}^3$ and the ``ABJM'' theory of Aharony, Bergman, Jafferis and Maldacena~\cite{Aharony:2008ug}.%
\footnote{%
This is really a limit of a more general correspondence between M~theory on $\AdS_4\times\S^7/\mathbbm{Z}_k$ and ABJM, see \eg\ ref.~\cite{Klebanov:2009sg}. However, there is no loss of generality in restricting to type IIA strings if we are only interested in the planar limit of the correspondence.
}
ABJM is a three-dimensional $\mathcal{N}=6$ superconformal Chern-Simons gauge theory with gauge group $U(N)\times \hat{U}(N)$ and Chern-Simons levels $k$ and $-k$, see the quiver diagram in figure~\ref{fig:abjm}. The planar limit consist in sending $k,N\to\infty$ while keeping the 't~Hooft coupling $\lambda=N/k$ fixed.

Besides its dimensionality, the amount of supersymmetry of $\AdS_4/\CFT_3$---24 real supercharges~\cite{Nilsson:1984bj}---is also intermediate  between $\AdS_5/\CFT_4$ and $\AdS_3/\CFT_2 $, that preserve 36 and 18 supersymmetries, respectively. This results in some features that we will also encounter later on for~$\AdS_3/\CFT_2$, such as the fact that fundamental excitations transform in reducible representations of the symmetry algebra, or the existence of a ``slope function''. In fact, while in $\AdS_5/\CFT_4$ S~duality fixes the expression of the coupling constant in terms of the 't~Hooft coupling $h=h(\lambda)= \sqrt{\lambda}/2\pi$~\cite{Berenstein:2009qd},\footnote{%
This was \textit{a posteriori} verified by comparing the results for the exact \textit{Bremsstrahlung} function found by integrability with the one known from supersymmetric localisation, see refs.~\cite{Correa:2012at,Correa:2012hh, Gromov:2012eu}.
} when we write all-loop expressions such as eq.~\eqref{eq:omega-disp-nonrel} in $\AdS_4/\CFT_3$~\cite{Berenstein:2008dc}, $h$ is some to-be-determined function of~$\lambda$.

Integrability manifests itself in the ABJM theory, where an integrable spin-chain Hamiltonian was found first at two~loop in perturbation theory~\cite{Minahan:2008hf,Gaiotto:2008cg}.
In parallel, integrability can be studied for the string NLSM; to this end, it is useful to represent the Green-Schwarz action~\cite{Grisaru:1985fv,Gomis:2008jt} as the supergroup coset action
\begin{equation}
\frac{OSp(6|4)}{SO(1,3)\times U(3)}\,,
\end{equation}
as done in refs.~\cite{Stefanski:2008ik,Arutyunov:2008if}. However, such a coset description is equivalent to the original Green-Schwarz action only in a specific $\kappa$~gauge~\cite{Stefanski:2008ik, Arutyunov:2008if}.%
\footnote{%
Additionally, such a $\kappa$~gauge becomes singular for certain bosonic string configurations; we will see that something similar happens also in $\AdS_3/\CFT_2$.
}
Classical integrability was found for such a coset description~\cite{Stefanski:2008ik, Arutyunov:2008if}, and then generalised to the Green-Schwarz action~\cite{Sorokin:2010wn}. Semi-classical integrability was found in terms of giant magnons~\cite{Gaiotto:2008cg,Grignani:2008is, Ahn:2008hj, Abbott:2009um,Hollowood:2009sc} and of an algebraic-curve description~\cite{Gromov:2008bz}, and an all-loop S-matrix was proposed~\cite{Ahn:2008aa} by postulating the symmetry that then emerged from the off-shell analysis of the gauge-fixed theory~\cite{Bykov:2009jy}. From such an S~matrix, the all-loop Bethe ansatz originally proposed in ref.~\cite{Gromov:2008qe} could be validated.

These investigations culminated in a mirror-TBA and Y-system description for the exact spectrum of the duality, first conjectured in ref.~\cite{Gromov:2009tv} and then derived (with some modifications with respect to the original proposal) in refs.~\cite{Bombardelli:2009xz,Gromov:2009at}. Recently, these were reformulated as a quantum spectral curve~\cite{Cavaglia:2014exa}, which in turn was used to exactly compute the slope function~$h(\lambda)$, see ref.~\cite{Gromov:2014eha}. For a more detailed discussion of this duality we refer the reader to ref.~\cite{Klose:2010ki}.

\subsection*{The case of $\AdS_3/\CFT_2 $}
The first indication of integrability for $\AdS_3$ backgrounds was the presence of giant-magnon solutions to $\AdS_3\times\S^3\times\T^4 $ equations of motion~\cite{David:2008yk,David:2010yg}.This is not entirely surprising since the~$\AdS_5\times\S^5$ solutions are contained in $\mathbbm{R}\times\S^2$, which can be embedded in  $\AdS_3\times\S^3 $.

The confirmation of classical integrability was found in ref.~\cite{Babichenko:2009dk} by relating the Green-Schwarz string action in a specific $\kappa $ gauge to an appropriate supercoset~\cite{Rahmfeld:1998zn,Park:1998un, Metsaev:2000mv}. Up to some $U(1)$ factors, this is
\begin{equation}
\frac{PSU(1,1|2)_{\L}\times PSU(1,1|2)_{\R}}{SO(1,2)\times SO(3)}\,,
\end{equation}
for the (pure Ramond-Ramond) $\AdS_3\times\S^3\times\T^4 $ background and by
\begin{equation}
\frac{D(2,1;\alpha)_{\L}\times D(2,1;\alpha)_{\R}}{SO(1,2)\times SO(3)\times SO(3)}\,,
\end{equation}
for the (pure Ramond-Ramond)  $\AdS_3\times\S^3\times\S^3\times\S^1 $ one. This was then generalised to arbitrary $\kappa $ gauges in ref.~\cite{Sundin:2012gc}. Another indication of integrability emerged in studying the Gubser-Klebanov-Polyakov ``spinning string'' classical solution~\cite{Sundin:2013uca}.

A peculiar feature of these backgrounds is the presence of \emph{massless} excitations in the string spectrum, which could not be straightforwardly included in the integrability machinery.
For this reason, the efforts to determine the integrable S~matrix and Bethe-Yang equations focused on the sub-sector where only massive asymptotic states are involved. After some initial investigations based on semiclassical integrability properties~\cite{OhlssonSax:2011ms,Ahn:2012hw}, these were fixed in refs.~\cite{Borsato:2012ud,Borsato:2012ss} for $\AdS_3\times\S^3\times\S^3\times\S^1 $ and in refs.~\cite{Borsato:2013qpa,Borsato:2013hoa} for~$\AdS_3\times\S^3\times\T^4 $.
As for the massless modes, they were initially considered in a weakly-coupled spin chain description~\cite{Sax:2012jv} and in the semi-classical string integrability picture~\cite{Lloyd:2013wza}. Only very recently an all-loop S~matrix for all fundamental (massive and massless) string modes was proposed~\cite{Borsato:2014exa,Borsato:2014hja}.
These investigations were supplemented by a number of perturbative or semiclassical calculations that, as we will see in chapter~\ref{ch:comparison}, confirmed the integragrability picture~\cite{Zarembo:2010sg,Zarembo:2010yz,Rughoonauth:2012qd, Sundin:2013ypa,Beccaria:2012kb,Beccaria:2012pm, Abbott:2013ixa, Bianchi:2013nra,Engelund:2013fja,Sundin:2014sfa, Bianchi:2014rfa}.
Even if perturbative calculations of non-protected quantities in the symmetric-produt CFT are also possible~\cite{Pakman:2009zz},  it is still not completely clear how integrability would enter the gauge or CFT side of the duality~\cite{Pakman:2009mi}.

It is also very interesting to note that the~$\AdS_3$ backgrounds discussed above remain classically integrable even when supported by a mixture of RR and NSNS fluxes, as it was recently shown by Cagnazzo and Zarembo~\cite{Cagnazzo:2012se}. This lead to very rapid developments in their study~\cite{Hoare:2013pma,Hoare:2013ida, Hoare:2013lja,Babichenko:2014yaa,Lloyd:2014bsa}, to which we will come back in chapter~\ref{ch:outlook}.

\section{Plan of the review}
As we have seen, the recent progress towards integrability for~$\AdS_3/\CFT_2 $ touches upon a number of different models. To keep our discussion simple, we discuss in full detail what is perhaps the simplest instance, \ie\ the massive sector of pure-RR $\AdS_3\times\S^3\times\T^4$ superstrings.
This will be done both from the point of view of the worldsheet theory and of a spin-chain, which as we will show are precisely related to one another.
After detailing that prototypical case, it will be relatively straightforward to describe the more general ones, which indeed share several features with it.  Our presentation will not always follow the one original adopted in the original literature, in an effort to be more pedagogical and self-contained.

In chapter~\ref{ch:sigmamodel} we will discuss the worldsheet theory of free strings in~$\AdS_3\times\S^3\times\T^4$ in terms of a coset NLSM. We will focus on the massive excitations and derive their symmetries. This follows closely ideas that were first developed for~$\AdS_5\times\S^5$~\cite{Frolov:2006cc,Arutyunov:2006ak}, but highlights some novel unexpected features of the~$\AdS_3\times\S^3\times\T^4$ background. Some of the results presented here have also been found, among other things, in refs.~\cite{Borsato:2014exa,Borsato:2014hja} by different techniques. Our presentation here strives to be as pedagogical as possible by drawing a parallel with the case of~$\AdS_5\times\S^5$.

In chapter~\ref{ch:smatrix} we will review the main ideas behind integrability, and in particular factorised scattering approach of Zamolodchikov. We will then use the symmetries found in the previous chapter to find the two-body S~matrix. This can be determined up to two scalar functions---the dressing factors---and satisfies several non-trivial consistency checks, most notably the Yang-Baxter equation, which is a necessary requisite for integrability. The results of this chapter were first found in~\cite{Borsato:2013qpa,Borsato:2012ud}, while the presentation follows~\cite{Arutyunov:2006yd,Arutyunov:2009ga}.

Chapter~\ref{ch:crossing} is devoted to the study of crossing symmetry. Firstly we will discuss crossing invariance for worldsheet excitations, which will be a generalisation of the familiar relativistic one~\cite{Janik:2006dc}. Then we discuss a proposal for crossing-invariant dressing factors, originally put forward in ref.~\cite{Borsato:2013hoa}, and discuss some of their analytic properties, notably their compatibility with the expected massive bound-state spectrum of the theory. This is a relevant check because crossing symmetry alone does not fix the form of the dressing factors completely.

We will switch gears in chapter~\ref{ch:spinchain}, and introduce a spin-chain picture, constructed in such a way as to be dual to the worldsheet theory of the preceding chapters. Again the focus will be on its symmetries, and it will lead us to a two-magnon S~matrix. We will relate it in a precise way to the worldsheet S~matrix, and discuss the  notion analogue to crossing in the spin-chain picture. This chapter follows more closely the original approach~\cite{Borsato:2012ud,Borsato:2013qpa} in which the S~matrix was found.

In chapter~\ref{ch:betheansatz} we will see in some detail how the S~matrix we computed can be used to find the string energy spectrum up to the so-called wrapping corrections. To this end, we introduce in quite some detail the asymptotic ``coordinate'' Bethe ansatz for both the spin-chain (which was originally worked out in refs.~\cite{Borsato:2012ss,Borsato:2013qpa}) and worldsheet pictures. Since we will be dealing with non-diagonal S~matrices, we will need to use the nesting procedure in order to write the Bethe equations. As we will show, the Bethe ansatz in the spin-chain and worldsheet picture describe the same physical spectrum.

In chapter~\ref{ch:comparison} we will discuss how the integrability construction was put to the test. Up to that point we will have assumed the worldsheet theory to be integrable at the quantum level, and derived an S~matrix based on that assumption. We will now check that S~matrix, including the proposed dressing factors, against perturbative and semiclassical calculations in the worldsheet theory~\cite{Babichenko:2009dk, Zarembo:2010sg,Zarembo:2010yz,Rughoonauth:2012qd, Sundin:2013ypa,Hoare:2013pma,Beccaria:2012kb,Beccaria:2012pm, Abbott:2013ixa, Bianchi:2013nra,Engelund:2013fja,Sundin:2014sfa} up to one-loop and including a non-trivial two-loop consistency check, finding complete agreement. Unfortunately it is much harder to perform such comparisons with the perturbative expansion of the dual~$\CFT_{2}$, where therefore some further validation remains  necessary.

In chapter~\ref{ch:outlook}, the final chapter, we will overview several directions in which the topics discussed here can be and are being evolved. The most natural one is the inclusion of the massless fundamental excitations to the integrability picture~\cite{Sax:2012jv,Lloyd:2013wza, Borsato:2014exa,Borsato:2014hja}. Another is the extension of integrability to the other maximally supersymmetric~$\AdS_3$ background, \ie~$\AdS_3\times \S^3\times\S^3\times\S^1 $. For that case, the all-loop massive S~matrix and Bethe ansatz was proposed in~\cite{Borsato:2012ud,Borsato:2012ss} up to the dressing factors, but several open questions remain---notably what happens in the limit where one of the spheres blows up to give back~$\AdS_3\times \S^3\times \T^4$, up to a compactification. Finally, we will discuss the $\AdS_3$ backgrounds supported by a mixture of Ramond-Ramond and Neveu-Schwarz-Neveu-Schwarz fluxes~\cite{Cagnazzo:2012se, Hoare:2013pma,Hoare:2013ida,Hoare:2013lja, Babichenko:2014yaa,Lloyd:2014bsa}, which may be a novel important playground to further our understanding of~$\AdS/\CFT$ integrability.

\subsubsection*{A note on notation}
The bulk of the material presented here has already appeared elsewhere~\cite{Borsato:2012ud, Borsato:2012ss,Borsato:2013qpa, Borsato:2013hoa, Borsato:2014exa,Borsato:2014hja,Hoare:2013pma, Hoare:2013ida,Hoare:2013lja}. To streamline our presentation, our notations here may differ from some of the original ones. Furthermore, for the same reason we sometimes use slightly different convention and normalisations---most notably for the S~matrix, which after all is a matrix and as such depends on the choice of basis.
We also should warn the reader that some quantities  that appear both in the worldsheet picture and in the spin-chain one (such as the S~matrix, the generators of the symmetry algebra, \etc) will be indicated by the same letter, \emph{despite not having the same value in the two cases}. We do so because these objects play the very same role and are \emph{almost} identical in the two pictures, and because we feel that introducing different notations for each of them would be an unnecessary burden. When confusion may arise, we do clarify in the text what quantities we are referring to.

Let us also point out that throughout this review we will not discuss the form of the interpolating function~$h(\lambda)$ for $\AdS_3/\CFT_2$. We will always indicate the coupling constant as~$h$, without specifying its relation with the 't~Hooft coupling~$\lambda$.

\section{Acknowledgements}
I would like to thank Riccardo Borsato, Olof Ohlsson Sax, Bogdan Stefa{\'n}ski and Alessandro Torrielli for the fruitful and enjoyable collaboration whence some of the results discussed here originated.
I am also grateful to Gleb Arutyunov, Diego Bombardelli, Sergey Frolov, Ben Hoare, Thomas Klose, Grisha Korchemsky, Gustavo Lucena~G{\'o}mez, Vladimir Mitev, Matthias Staudacher, Ryo Suzuki, Stijn van Tongeren, Arkady Tseytlin and Kostya Zarembo for many stimulating discussions.
I especially thank Gleb Arutyunov, Riccardo Borsato, Ben Hoare, Thomas Klose, Grisha Korchemsky, Gustavo Lucena~G{\'o}mez, Olof Ohlsson Sax, Bogdan Stefa{\'n}ski and Alessandro Torrielli for reading parts of the manuscript and providing useful comments.

The author's work is funded by the People Programme (Marie Curie Actions) of the European Union's Seventh Framework Programme FP7/2007-2013/ under REA Grant Agreement No.~317089 (GATIS).
The work leading to this review was initiated during the author's employment at the Institute for Theoretical Physics of Utrecht University, supported by the 
Netherlands Organisation for Scientific Research (NWO) under VICI grant No.~680-47-602 and in the framework of  the ERC Advanced grant research programme No.~246974,
 ``Supersymmetry: a window to non-perturbative physics''.

\bigskip
This review is partially based on the author's doctoral dissertation, defended \textit{cum laude} at the University of Utrecht on April 28${}^{\text{th}}$, 2014.

\chapter{The non-linear \texorpdfstring{$\sigma$}{sigma} model and its symmetries}
\label{ch:sigmamodel}
In this chapter we will analyse in more detail some features of the NLSM on~$\AdS_3\times \S^3\times \T^4$. In particular, we will write down its gauge-fixed action, restricting to massive excitations for simplicity. Out of that we will work out the symmetries of the theory, which will be important in the subsequent chapters.

As anticipated in the introduction, we want to study our theory in the light-cone gauge. However, the technique to do so will be different and slightly more involved than the one sketched there. In fact, in $\AdS$ backgrounds it is impossible to impose conformal and light-cone gauge on top of each other~\cite{Metsaev:2000yu,Tseytlin:2000na}. The way around this issue is to use first-order formalism, which has the further advantage naturally producing the Hamiltonian, whose eigenvalues are what we are after.

Firstly, we will demonstrate this approach in the case of a bosonic NLSM on $\AdS_3\times \S^3\times \T^4$. This will allow us to describe the strategy in some detail, without dealing with the complications due to fermions.
In order accommodate these, we could take two routes
\begin{enumerate}
\item Consider the Green-Schwarz (GS) action for the superstring~\cite{Green:1983wt,Grisaru:1985fv}. This is known up to quartic order in the fermions~\cite{Wulff:2013kga}, which would give the complete action for the case of our interest.
\item Following the approach used to study classical integrability in ref.~\cite{Babichenko:2009dk}, write the superstring action as a coset action
\begin{equation}
\label{eq:supercoset}
\frac{PSU(1,1|2)\times PSU(1,1|2)}{SO(1,2)\times SO(3)}\times U(1)^4\,.
\end{equation}
\end{enumerate}
The former method is completely general, but makes it harder to  see how the isometries are realised. This is instead manifest in the coset formulation, which however requires a specific choice of which fermions we consider physical (the $\kappa$-gauge fixing). This choice is not suitable for studying massive and massless excitations at the same time. Since our focus will be on the massive excitations only, we will use the coset action, and work out its symmetries.
As we will discuss in chapter~\ref{ch:outlook}, the GS approach has been used in ref.~\cite{Borsato:2014exa,Borsato:2014hja} precisely to understand the role played by the massless excitations.

\section{Bosonic strings in light-cone gauge}
\label{sec:bosonic1ord}
To exemplify the procedure we will follow later, let us first consider a bosonic NLSM action for closed strings\footnote{%
Integrable open superstrings can also be considered in principle, see ref.~\cite{Zoubos:2010kh} for a review in the case of~$\AdS_5\times\S^5$. While such an investigation has been initiated in the case of $\AdS_3/\CFT_2$~\cite{Mandal:2007ug,Prinsloo:2014dha}, the understanding of the integrability properties in this case is still limited.}, of the form
\begin{equation}
\label{eq:bosonicNLSM}
S=-\frac{h}{2}\int_{-\ell/2}^{\ell/2}\!\de\sigma\,\de\tau\, \gamma^{\alpha\beta} \partial_\alpha X^\mu\partial_\beta X^\nu \,G_{\mu\nu}(X)\,,
\end{equation}
where we consistently set to zero the Fradkin-Tseytlin $R_{(2)}$ term.
Notice further that we replaced the worldsheet metric by the conformally invariant combination ${\gamma}^{\a\b}\sqrt{|{\gamma}|}\to\gamma^{\a\b}$.
Let us introduce the conjugate momenta
\begin{equation}
p_{\mu}=\frac{\delta S}{\delta \dot{X}^\mu}=-h\gamma^{0\beta} \partial_{\beta}X^{\nu} G_{\mu\nu}(X)\,,
\end{equation}
where $\dot{X}^\mu=\partial_{0}X^\mu$. Then the action can be rewritten as~\cite{Metsaev:2000yu,Tseytlin:2000na}
\begin{equation}
\label{eq:firstorder-bos}
S=\int_{-\ell/2}^{\ell/2}\de^2\sigma \left(p_\mu \dot{X}^\mu+\frac{\gamma^{01}}{\gamma^{00}}\mathcal{C}_1+\frac{1}{2h\,\gamma^{00}}\mathcal{C}_2\right)\,,
\end{equation}
where
\begin{equation}
\mathcal{C}_1=p_{\mu} X'{}^{\mu}\,,
\qquad
\mathcal{C}_2=G^{\mu\nu}p_{\mu}p_{\nu}+h^2 G_{\mu\nu}X'{}^{\mu}X'{}^{\nu}\,,
\end{equation}
with $X'{}^{\mu}=\partial_1 X^{\mu}$. The action~\eqref{eq:firstorder-bos} is no longer manifestly covariant on the worldsheet. However, due to the Virasoro constraints we have that it must be
\begin{equation}
\label{eq:virasoro-constr}
\mathcal{C}_1=0\,,
\qquad
\mathcal{C}_2=0\,,
\end{equation}
so that the~$\gamma^{\alpha\beta}$ plays the role of a Langrange multiplier and will not appear in the Hamiltonian after~\eqref{eq:virasoro-constr} is imposed.

Let us specialise to the background $\AdS_3\times \S^3\times \T^4$. Let $t$ be the time-coordinate in $\AdS_3$ and $\phi$ be an angle in $\S^3$. Translations and rotations in these directions are isometries, to which correspond two conserved charges
\begin{equation}
\mathbf{H}^{\text{t.s.}}=-\int_{-\ell/2}^{\ell/2}\de\sigma\,p_{t}\,,
\qquad
\mathbf{J}=\int_{-\ell/2}^{\ell/2}\de\sigma\,p_{\phi}\,,
\end{equation}
where $p_{t},p_{\phi}$ are conjugate momenta. The target space energy~$\mathbf{H}^{\text{t.s.}}$ is what, after quantisation, will give us the energy spectrum we are interested in, and $\mathbf{J}$ is a distinguished angular momentum. Out of these two coordinates, we can construct light-cone ones~\cite{Arutyunov:2005hd}%
\footnote{%
In fact, a more general coordinate choice is possible~\cite{Arutyunov:2004yx,Arutyunov:2006gs}, where $x_{+}=a\phi+(1-a)t$.
}
\begin{equation}
x_{-}=\phi-t\,,
\qquad
x_{+}=\frac{1}{2}\big(\phi+t\big)\,,
\end{equation}
and denote by $x^i$ the remaining coordinates. The light-cone gauge-fixing condition is then
\begin{equation}
x_{+}=\tau\,,
\qquad
p_{+}=1\,,
\end{equation}
where we assumed that there is no winding.%
\footnote{%
One can account for a winding of the form $x_{+}(r)-x_{+}(-r)=2\pi {W}$, with $W\in\mathbbm{Z}$ by allowing for a term linear in $\sigma$ in $x_{+}$. 
}
We will always restrict to this case, which is the only one where the large-tension expansion of the string is well-defined.
Our coordinate choice implies that the light-cone momentum is $p_{+}=\frac{1}{2}(p_{\phi}-p_t)$, so that the corresponding Noether charge is~$\mathbf{P}_{+}=\frac{1}{2}(\mathbf{J}+\mathbf{H}^{\text{t.s.}})$. In addition, the eigenvalue $P_{+} $ of~$\mathbf{P}_{+}$ is
\begin{equation}
P_{+}=\int_{-\ell/2}^{\ell/2}\de\sigma\,p_{+}=\ell
\qquad
\Longrightarrow
\qquad
\ell=\frac{1}{2}\big(J+E^{\text{t.s.}}\big)\,.
\end{equation}
This shows explicitly how in light-cone gauge, invariance under worldsheet rescaling is lost, and in fact the worldsheet radius is fixed in term of physical charges. Moreover, relating $t$ with $\tau$ will allow us to compute the target space energy $E^{\text{t.s.}}$ in terms of the worldsheet Hamiltonian.

One of the advantages of light-cone gauge fixing is that spurious degrees of freedom can be eliminated, which makes quantisation easier. In particular, we can get rid of the metric $\gamma^{\alpha\beta}$ by a suitable gauge fixing, whose precise form is irrelevant here, 
and eliminate the longitudinal modes of the string~$x_{\pm},p_{\pm}$. To this end we solve $\mathcal{C}_1=0$ for $x_{-}'$
\begin{equation}
\label{eq:c1constrainbos}
x_{-}'=-p_{j}x'{}^{k}\,.
\end{equation}
Then, $p_{-}$ can be eliminated by solving the non-linear constraint $\mathcal{C}_2=0$. We have, plugging in the light-cone gauge conditions
\begin{equation}
\label{eq:C2explicit}
\begin{aligned}
\mathcal{C}_2=&\frac{1}{4}(G^{\phi\phi}-G^{tt})(p_{-}^2+4)+
(G^{\phi\phi}+G^{tt})p_{-}
+\frac{h^2}{4}(G_{\phi\phi}-G_{tt})x'_{-}{}^2\\
&+G^{jk}p_jp_k+h^2G_{jk}x'{}^{j}x'{}^{k}\,,
\end{aligned}
\end{equation}
where $G_{tt},G_{\phi\phi},G_{jk}$ are metric elements and $x_{-}'$ is a function of the transverse fields as in eq.~\eqref{eq:c1constrainbos}. Formally inverting the constraint equation gives
\begin{equation}
p_{-}=p_{-}\big(x^j,x'{}^{j},p_{j}\big)\,.
\end{equation}
Plugging these expressions into the action~\eqref{eq:firstorder-bos} and dropping the total time derivative $\dot{x}_{-}$, we find
\begin{equation}
S=\int_{-\ell/2}^{\ell/2}\de^2\sigma\,\left(p_j\dot{x}^j+p_{-}\big(x^j,x'{}^{j},p_{j}\big)\right)\,,
\end{equation}
whence we can immediately identify the worldsheet Hamiltonian density
\begin{equation}
\mathcal{H}=-p_{-}\big(x^j,x'{}^{j},p_{j}\big)\,,
\end{equation}
and the Poisson structure
\begin{equation}
\big\{x^k(\sigma),\,p_j(\tilde{\sigma})\big\}=\delta^{k}_{\ j}\,\delta(\sigma-\tilde{\sigma})\,.
\end{equation}
For~$\mathcal{H}$ to be positive, $p_{-}$ should be taken to be the \emph{negative} root of~\eqref{eq:C2explicit}. We also see that the worldsheet and target-space energies are related by
\begin{equation}
\label{eq:HEJrel}
\mathbf{H}=\mathbf{H}^{\text{t.s.}}-\mathbf{J}\,,
\end{equation}
where $\mathbf{H}$ is the worldsheet Hamiltonian.
When eliminating the longitudinal degrees of freedom we did not solve for~$x_{-}$, but only for its derivative~$x_{-}'$. For consistency, we have to impose that $x_{-}$ is periodic,
\begin{equation}
0=\int_{-\ell/2}^{\ell/2}\de\sigma\,x_{-}'=-\int_{-\ell/2}^{\ell/2}\de\sigma\,p_{j}x'{}^{j}\,.
\end{equation}
This last condition, the \emph{level-matching constraint}, is difficult to impose before quantisation. Instead, we will impose it on the Hilbert space of the quantum theory. To this end, notice that the worldsheet momentum, \ie\ the Noether charge corresponding to $\sigma$-translations, is given precisely by
\begin{equation}
\label{eq:p-ws-bos}
\mathbf{P}=\int_{-\ell/2}^{\ell/2}\de\sigma\,p_{j}x'{}^{j}\,.
\end{equation}
This should not be confused with the Noether charge~$\mathbf{P}_{+}$, which is constant and equal to~$\ell$. 
Therefore, the physical states in the quantum theory will be the ones annihilated by the (quantum) worldsheet momentum.

\subsection*{Perturbative expansion of the action}
It is still not straightforward to quantise the Hamiltonian $\mathbf{H}=\int \de \sigma \,\mathcal{H}$, which comes from inverting the non-linear constraint $\mathcal{C}_2$ and is highly interacting. In order to proceed, we need to systematically expand it. A way of doing so is to suitably redefine fields and coordinates so that the Hamiltonian density can be written as
\begin{equation}
\mathcal{H}=\mathcal{H}_2+\frac{1}{h}\mathcal{H}_4+\frac{1}{h^2}\mathcal{H}_6+\dots\,,
\end{equation}
where $\mathcal{H}_2$ is quadratic in the fields, $\mathcal{H}_4$ quartic, \etc, and take a large-$h$ limit. One such procedure is the Berenstein-Maldacena-Nastase (BMN) limit~\cite{Berenstein:2002jq}, whereby
\begin{equation}
P_{+}\to\infty\,,
\qquad
h\to\infty\,,
\qquad
\frac{P_{+}}{h}\ \text{fixed.}
\end{equation}
The action then admits a perturbative expansion in~$1/P_{+}$.

Alternatively, one can first consider a \emph{decompactification limit}
\begin{equation}
P_{+}\to\infty\,,
\qquad
h\ \text{fixed},
\end{equation} 
which is enough to identify asymptotic states and define an S matrix, since~$\ell=P_{+}$. Then, to perform perturbative calculations we can take a large-tensions expansion in $1/h$. Building on that, one can go on and construct a perturbative Hamiltonian and S matrix, see \eg\ ref.~\cite{Arutyunov:2009ga} for more details. The advantage is this approach is to distinguish the decompactification of the worldsheet from the expansions of the Hamiltonian. In what follows, we will adopt this latter procedure.

\subsubsection*{Evaluation of the quadratic Hamiltonian}
The line element of this geometry reads
\begin{equation}
\de s^2=-G_{tt}\,\de t^2+G_{\phi\phi}\,\de\phi^2+ f_{ii}\,\de z_{i}+g_{ii}\,\de y_{i}+\de X_{i}^2\,,
\end{equation}
where we denoted the transverse coordinates by $z_1,z_2$ for $\AdS_3$, by $y_1,y_2$ for $\S^3$ and by $X_1,\dots,X_4$ for the torus. The metric elements are
\begin{equation}
G_{tt}=\left(\frac{4+|z|^2}{4-|z|^2}\right)^2\!\!,
\ 
G_{\phi\phi}=\left(\frac{4-|y|^2}{4+|y|^2}\right)^2\!\!,
\ 
f_{ii}=\left(\frac{4}{4-|z|^2}\right)^2 \!\!,
\ 
g_{ii}=\left(\frac{4}{4+|y|^2}\right)^2\!\!.
\end{equation}
From~\eqref{eq:C2explicit}, we see that it is convenient to perform the rescaling
\begin{equation}
\label{eq:bosonicrescaling}
\sigma\to h\,\sigma\,,
\qquad
x^j\to h^{-1/2} x^j
\,,
\qquad
p_j\to h^{-1/2} p_j\,.
\end{equation}
The first replacement suitably rescales the $\sigma$-derivatives, and the other two implement a field expansion.
In~\eqref{eq:C2explicit}, keeping track of all the sub-leading terms for now, and picking the appropriate solution have
\begin{equation}
p_{-}=2\frac{G^{tt}+G^{\phi\phi}}{G^{tt}-G^{\phi\phi}}+2\frac{\sqrt{4G^{\phi\phi}G^{tt}+ (G^{tt}-G^{\phi\phi})\mathcal{H}^{\perp}+\tfrac{h^2}{4}(G^{tt}G_{\phi\phi}+G^{\phi\phi}G_{tt}-2)x_-^2}}{G^{\phi\phi}-G^{tt}}\,,
\end{equation}
with
\begin{equation}
\mathcal{H}^{\perp}=\frac{1}{h}\big(\delta^{jk}p_{j}p_{k}+\delta_{jk}x'{}^{j}x'{}^{k}\big)+O\big(\frac{1}{h^2}\big)\,.
\end{equation}
Expanding to leading order gives
\begin{equation}
\label{eq:bosonicH2}
\mathcal{H}_2=\frac{1}{2}\left(|p_y|^2+|y'|^2+|y|^2\right)+
\frac{1}{2}\left(|p_z|^2+|z'|^2+|z|^2\right)+\frac{1}{2}\left(|p_X|^2+|X'|^2\right)\,.
\end{equation}
This is the (relativistic) Hamiltonian of four massive excitations (the four transverse coordinates in~$\AdS_3\times\S^3$) and four massless ones (the tours coordinates). By supersymmetry, we expect the superstring action at quadratic order to be given by the above~$\mathcal{H}_2$ plus $4+4$ free fermions of the same masses.

At this order~$\mathbf{H}$ is free and can be quantised in a standard way in terms of raising and lowering operators~$a_{j}^\dagger(p),\,a^{j}(p)$ satisfying canonical commutation relations
\begin{equation}
\big[a^k(p),\,a^\dagger_j(\tilde{p})\big]=\delta^{k}_{\ j}\,\delta(p-\tilde{p})\,,
\end{equation}
yielding%
\footnote{%
Owing to the decompactification limit, the integration ranges from $-\infty$ to $+\infty$, and we omit to indicate the limits in the integral.
}
\begin{equation}
\mathbf{H}=\int \de\sigma\,\mathcal{H}_2=
\int \de p\,\sum_{j=1}^8 \omega_j(p)\,a^j(p)\,a^\dagger_j(p)\,,
\qquad
\omega_j=\sqrt{m_j^2+p^2}\,,
\end{equation}
with $m_j=0$ for the four number operators of the torus and $m_j=1$ for $\AdS_3\times\S^3$.
The worldsheet momentum~\eqref{eq:p-ws-bos} is then
\begin{equation}
\mathbf{P}=\int \de p\,\sum_{j=1}^8 p\,a^j(p)\,a^\dagger_j(p)\,,
\end{equation}
so that a multiparticle state $|p_1,\dots p_n\rangle$ is physical if and only if $p_1+\cdots +p_n=0$.

\section{The \texorpdfstring{$\AdS_3\times\S^3\times\T^4$}{AdS3xS3xT4} supercoset}
\label{sec:supercoset}
In this section we will write down the coset action for the superstring, following~\cite{Babichenko:2009dk}. Even if this construction is inspired by the one orginally performed in $\AdS_5\times\S^5$~\cite{Metsaev:1998it}, there are some remarkable differences due to the presence of the~$\T^4$ flat directions. In that case, the coset action coincided with the GS action \emph{before} fixing the $\kappa$-gauge. Here, the correspondence holds only when $\kappa$-gauge is \emph{completely fixed}. This will lead to some complications that fortunately are irrelevant as long as we focus on the massive modes alone.
We will describe $\AdS_3\times\S^3\times\T^4$ as the coset~\eqref{eq:supercoset} which can be essentially constructed out of \emph{two copies} of the superalgebra~$\psu(1,1|2)$.

\subsection*{The superalgebra $\psu(1,1|2)$}
The superalgebra $\psu(1,1|2)$ consists of an even (``bosonic'') part given by a non-compact $\su(1,1)$ or $\sl(2)$, and a compact $\su(2)$ algebra. We can think of the former as being some of the isometries from $\AdS_3$ and the latter as coming  from $\S^3$. These are supplemented by eight (``fermionic'') supercharges.

Let us denote the generators of the non-compact bosonic algebra as $\mathbf{L}_{i}$, the ones from $\su(2)$ as $\mathbf{J}_{i}$, and the supercharges as $\mathbf{Q}_{a\kappa\iota}$. In terms of raising and lowering operators the even part of the algebra, in a suitable real form, is given by
\begin{equation}
  \begin{aligned}
    \comm{\mathbf{L}_3}{\mathbf{L}_\pm} &= \mp i\, \mathbf{L}_\pm , &
    \qquad
    \comm{\mathbf{L}_+}{\mathbf{L}_-} &= +2i\, \mathbf{L}_3 , \\
    \comm{\mathbf{J}_3}{\mathbf{J}_\pm} &= \mp i\, \mathbf{J}_\pm , &
    \qquad
    \comm{\mathbf{J}_+}{\mathbf{J}_-} &= -2i\, \mathbf{J}_3 , \\
  \end{aligned}
\end{equation}
and the supercharges are charged under the bosonic subalgebra
\begin{equation}
  \begin{aligned}
    \comm{\mathbf{L}_3}{\mathbf{Q}_{\pm\kappa\iota}} &= \pm\frac{i}{2} \mathbf{Q}_{\pm\kappa\iota} , &
    \qquad
    \comm{\mathbf{L}_\pm}{\mathbf{Q}_{\pm\kappa\iota}} &= +\ \mathbf{Q}_{\mp\kappa\iota} , \\
    \comm{\mathbf{J}_3}{\mathbf{Q}_{a\pm\iota}} &= \mp\frac{i}{2} \mathbf{Q}_{a\pm\iota} , &
    \qquad
    \comm{\mathbf{J}_\pm}{\mathbf{Q}_{a\mp\iota}} &=-i\, \mathbf{Q}_{a\pm\iota} , \\
  \end{aligned}
\end{equation}
with $\kappa=\pm,\iota=\pm$ and $a=\pm $.
Finally, the supercharges' anticommutators read
\begin{equation}
  \begin{aligned}
    \acomm{\mathbf{Q}_{\pm++}}{\mathbf{Q}_{\pm--}} &= \pm\  \mathbf{L}_{\mp} , \! &
    \acomm{\mathbf{Q}_{\pm+-}}{\mathbf{Q}_{\pm-+}} &= \mp\  \mathbf{L}_{\mp}  , \\
    \acomm{\mathbf{Q}_{+\pm+}}{\mathbf{Q}_{-\pm-}} &= \mp i\, \mathbf{J}_{\pm} , \! &
    \acomm{\mathbf{Q}_{+\pm-}}{\mathbf{Q}_{-\pm+}} &= \pm i\, \mathbf{J}_{\pm}  ,\\
    \acomm{\mathbf{Q}_{+\pm\pm}}{\mathbf{Q}_{-\mp\mp}} &= i\,(+ \mathbf{L}_3 \pm \mathbf{J}_3) , \!
    &\qquad
    \acomm{\mathbf{Q}_{+\pm\mp}}{\mathbf{Q}_{-\mp\pm}} &= i\,(- \mathbf{L}_3 \mp \mathbf{J}_3) .
  \end{aligned}
\end{equation}

In what follows, we will need to identify two copies of this algebra with part of the superisometries of $\AdS_3\times \S^3\times\T^4$, and use an explicit representation for it. In order to do so, we will consider a realisation of the $\su(1,1|2)$ in terms of supermatrices, and find $\psu(1,1|2)$ as a quotient subalgebra.

\subsubsection*{Supermatrix realisation}
Let us consider $\mathbbm{Z}_2$-graded vector space~$\mathbbm{C}^{2|2}$. The set of its linear endomorphisms form the superalgebra~$\gl(2|2)$, that can be represented in terms of $4\times4$ supermatrices
\begin{equation}
\label{eq:supermatrix}
\mathcal{M}=\left(
\begin{array}{cc}
A & \Theta\\
\Xi & B
\end{array}\,
\right)\,,
\end{equation}
where the $2\times2$ blocks $A,B$ are even and $\Theta,\,\Xi$ are odd%
\footnote{%
Note that the odd elements of the supermatrix are nonetheless commuting (non-Grassmann) scalars.
}. The subalgebra $\u(1,1|2)$ can be singled out by imposing a suitable hermiticity condition
\begin{equation}
\label{eq:algebrareality}
\mathcal{M}^{\dagger}+H^{-1}\,\mathcal{M}\,H=0\,,
\qquad
H=H^{-1}=\text{diag}(1,-1,1,1)\,,
\end{equation}
where $H$ implements the non-euclidean signature. There are 8 odd and 8 even independent solutions to such condition. Among the latter, there are two central elements
\begin{equation}
\mathbf{I}=\text{diag}(1,1,1,1)\,,
\qquad
\mathbf{I}^{s}=\text{diag}(1,1,-1,-1)\,.
\end{equation}
The supertrace
\begin{equation}
\str\,\mathcal{M}=\tr\,A -\tr\,B\,,
\end{equation}
is an invariant of~$\gl(2|2)$ and~$\u(1,1|2)$ which can be used to impose the condition
$\str \mathcal{M} = 0$,
that mods out~$\mathbf{I}^{s}$ and defines the matrix algebra~$\su(1,1|2)$. However, we cannot get rid of~$\mathbf{I}$ by consistently imposing~$\tr \mathcal{M}=0$ because for a generic odd (and therefore traceless) element of~$\u(1,1|2)$ we have
\begin{equation}
\tr \big(\big\{\mathcal{M}_{\text{odd}},\mathcal{M}_{\text{odd}}\big\}\big)=-2\,\tr \big(\mathcal{M}_{\text{odd}} H^{-1}\mathcal{M}_{\text{odd}}^{\dagger}H\big) <0\,.
\end{equation}
Therefore $\psu(1,1|2)$ does not admits a matrix realisation. With a small abuse of language we will refer to the quotient $\su(1,1|2)/\mathbf{I}$ as the ``matrix realisation'' of~$\psu(1,1|2)$ and, when writing anticommutators, understand equalities modulo a multiple of the identity. In appendix~\ref{app:psu112} we give the explicit form of the generators in terms of $4\times4$~supermatrices, as well as some additional properties of the superalgebra.

\subsection*{The full algebra and $\mathbbm{Z}_4$-automorphism}
The algebra generating the supercoset~\eqref{eq:supercoset} is given by two copies of $\psu(1,1|2)$, which we will denote by ``L'' or ``left'' and ``R'' or ``right'' and four copies of $\u(1)$. As it is always the case when considering a superalgebra that is direct sum of two subalgebras, $\psu(1,1|2)_{\L}\oplus\psu(1,1|2)_{\R}$ enjoys a $\mathbbm{Z}_4$-automorphism~$\Omega$, on top of the natural $\mathbbm{Z}_2$-grading due to the superalgebra structure. This can be defined, for instance, by a permutation of the two copies together with multiplication by the fermion sign of one of them.

There are several inequivalent ways to define a realisation of $\psu(1,1|2)_{\L}\oplus\psu(1,1|2)_{\R}$ in terms of $8\times8$ supermatrices~$\mathcal{M}$. A possibility would be to consider two identical realisations of~$\psu(1,1|2)$ and take their direct sum. However, this is not the most suitable choice for the coset construction that we will carry out in the next section.%
\footnote{%
In fact, it is easy to check that carrying out the construction with the more na\"ive representation would lead to an ill-defined Poisson structure in the resulting coset action.
}
 The elements of~$\psu(1,1|2)_{\L}\oplus\psu(1,1|2)_{\R}$ can be split in two,
\begin{equation}
\label{eq:matrixrepr2copies}
\mathcal{M}_{\L}=\mathcal{M}\oplus \mathbf{0}\,,
\qquad
\mathcal{M}_{\R}=\mathbf{0}\oplus \widetilde{\mathcal{M}}\,,
\end{equation} 
where $\mathbf{0}$ is the zero matrix, and $\mathcal{M}$ and $\widetilde{\mathcal{M}}$ are in two matrix representations whose bosonic subalgebras have opposite notions of what the highest weight vectors are. In appendix~\ref{app:psu112} we give the explicit form of these matrices and comment more on this choice.
There are also several inequivalent ways of defining the $\mathbbm{Z}_4$ automorphism~$\Omega$. We will rely on an exchange of the two $\psu(1,1|2)$ copies supplemented by a ``fermionic'' operation. In particular, $\Omega$ will take the form
\begin{equation}
\label{eq:automorphism}
\Omega(\mathcal{M})=\mathcal{K}^{-1}\mathcal{M}\,\mathcal{K}\,,
\qquad
\Omega^4(\mathcal{M})=\mathcal{M}\,,
\end{equation}
where $\mathcal{K}$  permutes left and right and takes into account the fermionic signs, so that in terms of the (Hermitian) Pauli matrices we have
\begin{equation}
\mathcal{K}=\mathcal{F}_{\L}\,\mathcal{P}\,,
\qquad
\mathcal{F}_{\L}=\mathbf{I}_2\otimes \big(\sigma_3\oplus\mathbf{I}_2\big),
\quad
\mathcal{P}=\sigma_1\otimes
\big(\mathbf{I}_2\otimes\sigma_1\big).
\end{equation}
In this way, $\Omega$ maps the left and right bosonic subalgebras into each other and accounts for their different matrix representations, see appendix~\ref{app:psu112}.

The advantage of such a construction is that it gives a natural way to decompose the symmetry algebra into the direct sum of four eigenspaces relative to the eigenvalues~$i^{k}$ with $k=0,\dots 3$. Using the matrix representation~(\ref{eq:supermatrix}--\ref{eq:matrixrepr2copies}) it is easy to see that the eigenspaces relative to $\pm1$ consist of even elements. In particular, the eigenspace given by matrices such that
$\Omega(\mathcal{M})=\mathcal{M}$
consists of $\su(1,1)\oplus\su(2)$. The corresponding group is precisely the quotient part of the coset.
%
It only remains to extend the automorphism to~$\u(1)^4$. A simple way of doing so is declaring that all of the $\u(1)$ generators have eigenvalue~$-1$ under~$\Omega$. A more ``symmetric'' choice is to introduce a $\mathbbm{Z}_2$-grading on~$\u(1)^8$, which amounts to realising~$U(1)^4$ as a coset $(U(1)^2/U(1))^4$. Since the latter is trivial, the two descriptions are equivalent.

Therefore, we have a vector-space decomposition of
\begin{equation}
\mathcal{A}=\psu(1,1|2)\oplus\psu(1,1|2)\oplus\u(1)^4\,,
\end{equation}
as
\begin{equation}
\label{eq:z4decomp}
\mathcal{A}=\bigoplus_{k=0}^3\mathcal{A}^{(k)}\,,
\qquad
\Omega\left(\mathcal{A}^{(k)}\right)=i^k\,\mathcal{A}^{(k)}\,.
\end{equation}
Using the fact that $\Omega$ is realised linearly and that
\begin{equation}
\Omega(\mathcal{M}_1\,\mathcal{M}_2)=\Omega(\mathcal{M}_1)\,\Omega(\mathcal{M}_2)\,,
\end{equation}
we have that the decomposition turns~$\mathcal{A}$ into a $\mathbbm{Z}_4$-graded Lie superalgebra, \ie
\begin{equation}
\big[\mathcal{A}^{(j)},\mathcal{A}^{(k)}\big]\subset \mathcal{A}^{(j+k\,\text{mod}4)}\,.
\end{equation}

This structure has two important consequences:
\begin{itemize}
\item It provides a simple way to realise the supergroup coset starting from the superalgebra elements and to write down the string action and its symmetries, as we will do in the rest of this chapter.
\item It ensures that the string equations of motions can be written as the flatness condition for a suitable (Lax) connection, a feature that guarantees classical integrability, see section~\ref{sec:integrability}.
\end{itemize}

\subsection*{Parametrisation of group elements}
It is convenient to parametrise a generic element $g\in PSU(1,1|2)^2\times U(1)^4$ in terms of the ``exponential'' of the algebra elements. Several choices are possible. Here we will set
\begin{equation}
\label{eq:gparam}
g(t,\phi,\chi,x,X)=\Lambda(t,\phi)\,g(\Psi_m)\,g(\Psi_l)\,g(x)\,g(X)\,,
\end{equation}
where
\begin{itemize}
\item $t,\phi$ are the time coordinate in $\AdS_3 $ and an angle in $\S^3$, respectively. We will later use them as light-cone coordinates.
\item $\Psi_{m,l}$ are all the fermions, which we will split into massive fermions~$\theta^{i}_{\smallL,\smallR}$ contained in~$\Psi_{m}$, and massless fermions~$\eta^{i}_{\smallL,\smallR}$ contained in~$\Psi_{l}$.
\item $x=(z_{i},y_{i})$ are the remaining four bosonic coordinates on $\AdS_3\times\S^3$.
\item $X=(X_1,\dots X_{4})$ are the four bosons from~$\T^4$, so that $g(X)$ commutes with all the other parameters. 
\end{itemize}
In particular, let us set
\begin{equation}
\label{eq:Lambdaparam}
\Lambda(t,\phi)=e^{\frac{t}{2}(\mathbf{L}_3^{\smallL}-\mathbf{L}_3^{\smallR}) +\frac{\phi}{2}(\mathbf{J}_3^{\smallL}-\mathbf{J}_3^{\smallR})}=
e^{i\,x_{+}\Sigma_{+}-\frac{i}{2}x_{-}\Sigma_{-}}\,,
\end{equation}
where we introduced the light-cone matrices in $\psu(1,1|2)^2$
\begin{equation}
i\,\Sigma_{\pm}=\big(\mathbf{L}_3^{\smallL}-\mathbf{L}_3^{\smallR}\big) \pm\big(\mathbf{J}_3^{\smallL}-\mathbf{J}_3^{\smallR}\big)\,.
\end{equation}
The remaining four fields~$\{x_k\}=(z_1,z_2,y_1,y_2)$ appear through~$g(x)$ which is written in terms of the remaining transverse angular momenta. For their form as well as for the forms of the fermion parametrisation we refer the reader to appendix~\ref{app:coset}.

\subsection*{Lagrangian and Noether current}
Let $g\in PSU(1,1|2)^2\times U(1)^4$. The one-form
\begin{equation}
A=-g^{-1}\de g\,,
\end{equation}
takes values in~$\mathcal{A}$, and therefore can be decomposed according to~\eqref{eq:z4decomp} as
\begin{equation}
A=A^{(1)}+A^{(2)}+A^{(3)}+A^{(4)}\,.
\end{equation}
Inspired by the case of $\AdS_5\times\S^5$, let us define the Lagrangian
\begin{equation}
\label{eq:secondorderLagr}
\mathcal{L}=-\frac{h}{2}\left(\gamma^{\a\b}\str\big(A^{(2)}_{\a}A^{(2)}_{\b}\big) +\eps^{\a\b}\str\big(A^{(1)}_{\a}A^{(3)}_{\b}\big)\right)\,,
\end{equation}
where definiteness we fixed the coefficient of the Wess-Zumino term~$\str A^{(1)}\wedge A^{(3)}$.
This Lagrangian and the resulting action have several good properties. Firstly, as it is easy to check they are manifestly real. Furthermore, $A$ and therefore $\mathcal{L}$ is invariant under left multiplication of the group element~$g$ by any constant element~$g_0$---the theory is manifestly invariant under \emph{global} (super)isometries.

One may worry that the action depends on $g$, rather than on the corresponding coset representative. However, it is easy to see that under a transformation of the $SU(1,1)\times SU(2)$ from~$\mathcal{A}^{(0)}$ all the current components appearing in the action $A^{(k)}$, $k=1,\dots3$ transform covariantly, so that~$
\mathcal{L}$ is left unchanged. Moreover, if we take~$g$ to be parametrised by~\eqref{eq:gparam} and set the coefficients of the odd algebra elements to zero, we have that the Lagrangian~$\mathcal{L}$ reduces to the one of the bosonic NLSM on $\AdS_3\times\S^3\times\T^4$ given by eq.~\eqref{eq:bosonicNLSM}.

An important feature of~\eqref{eq:secondorderLagr} is that the resulting equations of motion can be written as a flatness condition for a Lax connection. This fact, familiar from $\AdS_5\times\S^5$~\cite{Bena:2003wd} was pointed out in ref.~\cite{Babichenko:2009dk} for the $\AdS_3\times\S^3\times\S^3\times\S^1$ background of which  $\AdS_3\times\S^3\times\T^4$ can be seen as a limit,
at least in absence of winding. In ref.~\cite{Babichenko:2009dk} it was also argued that the coset action coincides with the GS superstring action in a suitable $\kappa$-gauge, namely one where all the massless fermions are eliminated by the gauge fixing. This was checked up to quadratic order in the fermions.

Finally, owing to the global $\mathcal{A}$-symmetry, there exists a conserved Noether current, which takes a simple form in the coset formulation~\cite{Babichenko:2009dk}
\begin{equation}
\label{eq:noethercurrent}
\mathcal{J}^{\alpha}=g\left(\gamma^{\a\b}A^{(2)}_{\b}-\frac{1}{2}\eps^{\a\b}\big(A^{(1)}_{\b}-A^{(3)}_{\b}\big)\right)g^{-1}\,.
\end{equation}

\section{Massive modes in light-cone gauge}
\label{sec:lightconegauge}
Following what we did in section~\ref{sec:bosonic1ord}, let us rewrite the Lagrangian~\eqref{eq:secondorderLagr} in the first-order formalism,
\begin{equation}
\mathcal{L}=-\left(\str\big(\varpi A^{(2)}_{0}\big)
+\frac{h}{2}\eps^{\a\b}\str\big(A^{(1)}_{\a}A^{(3)}_{\b}\big)
+\frac{\gamma^{01}}{\gamma^{00}}\mathcal{C}_1
-\frac{1}{2h\,\gamma^{00}}\mathcal{C}_{2}
\right)\,,
\end{equation}
where we introduced the auxiliary field~$\varpi$ that without loss of generality we take to be equal to its component~$\varpi^{(2)}$. The constraints are $\mathcal{C}_1=\mathcal{C}_2=0$ where now
\begin{equation}
\mathcal{C}_1=\str\big(\varpi A^{(2)}_{1}\big),
\qquad
\mathcal{C}_2=\str\big(\varpi^2+h (A^{(2)}_{1})^2\big).
\end{equation}

To preserve as much supersymmetry as possible, we want to fix light-cone gauge in terms of the coordinates $x_{\pm}$ constructed in section~\ref{sec:supercoset}. Such a gauge fixing is incompatible with the $\kappa$-symmetry fixing that was necessary to assume to have a coset description~\cite{Babichenko:2009dk}, and should we proceed in this way we would find that the massless fermions lack a good kinetic term (\ie~quadratic in a field expansion).
In what follows, we will restrict to considering \emph{only massive excitations}, \ie\ truncate the coset to~$\AdS_3\times\S^3$. This will be enough to elucidate at least some of the symmetries of the theory, which is what we will later need to find the worldsheet S~matrix for massive particles. Let us therefore set, from now on,
\begin{equation}
\Psi_l=0\,,
\qquad
X_1=\cdots =X_4=0\,.
\end{equation}

\subsection*{Evaluation of the action}
It is useful to split the current into an even and odd part under the $\mathbbm{Z}_4$ decomposition, $A_{\text{even}}$ and~$A_{\text{odd}}$ respectively.
We also single out the part~$A^{\perp}_{\text{even}}$ of~$A_{\text{even}}$ which depends only on the transverse bosonic coordinates.
In terms of these quantities (whose explicit expression we give in appendix~\ref{app:coset}) and up to the imposing the Virasoro constrains, the Lagrangian can be rewritten as
\begin{equation}
\label{eq:lagrangianevenodd}
\mathcal{L}=p_{+}\dot{x}_{-}+\mathbf{p}_{-}\dot{x}_{+}-\str\big(\varpi A^{\perp}_{\text{even}}\big)
-i\frac{h}{4}\eps^{\a\b}\str\big( A^{\text{odd}}_{\a}\Omega(A^{\text{odd}}_{\b})\big)
\,,
\end{equation}
where we also made use of the $\mathbbm{Z}_4$ automorphism. Note that $p_{+}$ is the momentum conjugate to $x_{-}$, while $\mathbf{p}_{-}$ differs from the conjugate momentum to $x_{+}$ due to the contribution of the Wess-Zumino term. To see this, let us define the decomposition
\begin{equation}
\varpi=\frac{i}{2}\varpi_{+}\Sigma_{+}+\frac{i}{4}\varpi_{-}\Sigma_{-}-\frac{1}{2}\varpi_{k}\Sigma_{k}\,,
\end{equation}
valid up to trace contributions, where $\Sigma_k$ corresponds to the bosonic generators of the transverse directions, see eq.~\eqref{eq:Sigmai-def}. Then, using the explicit expressions from appendix~\ref{app:coset}, we get indeed
\begin{equation}
p_{+}=\varpi_{+}G_{+}-\frac{1}{2}\varpi_{-}G_{-}\,,
\qquad
\mathbf{p}_{-}=\frac{i}{2}\str\left(\varpi\Sigma_{+}g_{x} \big(1+2\Psi_m^2\big)g_{x}\right)\,.
\end{equation}
Note that we introduced the short-hand notation $g_{x}=g(x)$ for the transverse bosonic coordinates and the metric in the light-cone directions,
$G_{\pm}=\tfrac{1}{2}\big(G_{tt}\pm G_{\phi\phi}\big)$.

\subsubsection*{Gauge fixing}
Let us now fix light-cone gauge (with zero winding)
\begin{equation}
x_{+}=\tau\,,
\qquad
p_{+}=1\,.
\end{equation}
From the latter equation we can immediately find the value of 
\begin{equation}
\varpi_{+}=\frac{1}{G_{+}}\big(1+\frac{1}{2}\varpi_{-}G_{-}\big)\,.
\end{equation}
Substituting $\varpi_{+}$ in the $\mathcal{C}_1=0$ constraint we find as expected
\begin{equation}
\begin{aligned}
-x'_{-}=p_{k}x'_{k}+\text{fermions}\,,
\end{aligned}
\end{equation}
where the complete expression of the fermion contributions are given in eq.~\eqref{eq:c1fermions}. Recall that it is this expression that appears in the level matching constrain, which as in the bosonic case amounts to vanishing of the worldsheet momentum
\begin{equation}
\label{eq:levelmatching}
0=\int_{-\ell /2}^{\ell/2}\de\sigma\, x_{-}'=\mathbf{P}\,.
\end{equation}
The last longitudinal component of the auxiliary field~$\varpi_{-}$ can be found from the quadratic constraint $\mathcal{C}_2=0$, with
\begin{equation}
\mathcal{C}_{2}=\varpi_{+}\varpi_{-}+ \varpi_{k}^2+h\,\str\left((A^{(2)}_\sigma)^2\right).
\end{equation}
The expansion of and solution to this constraint is given in appendix~\ref{app:coset}, where we also give the form of the auxiliary fields $\varpi_{k}$ in terms of the conjugate momenta $p_k$ as well as some other useful formulae. Using all this, in the next section we will write down explicit expressions to the leading order in a field expansion.

\subsection*{Noether charge}
We conclude this section by writing down the conserved charges corresponding to the Noether current~\eqref{eq:noethercurrent}. By making use of the equations of motion for the auxiliary field~$\varpi$
\begin{equation}
\varpi=h\, \gamma^{0\beta}A^{(2)}_{\beta}\,,
\end{equation}
we can write down the charges in a form that is independent of the worldsheet metric,
\begin{equation}
\mathcal{Q}=\int\limits_{-\ell /2}^{\ell /2} \de\sigma\,g(x_{\pm},x,\Psi_m)\,\left(\varpi-\frac{1}{2}h\big(A^{(1)}_{1}-A^{(3)}_{1}\big)\right)g^{-1}(x_{\pm},x,\Psi_m)\,,
\end{equation}
which by using the $\mathbbm{Z}_4$-automorphism~$\Omega$ can be recast in the form
\begin{equation}
\label{eq:noethercharge}
\mathcal{Q}=\int\limits_{-\ell /2}^{\ell /2}  \de\sigma\, \Lambda\, g_{\Psi_{m}}g_x \left(\varpi+\frac{i}{2}h\,g_{x}\,\Omega\big(\Psi_{m}'\big) g_{x}^{-1}\right)g_x^{-1}g^{-1}_{\Psi_{m}}\Lambda^{-1},
\end{equation}
where we also expressed $g(x_{\pm},x,\Psi_m)$ by means of~\eqref{eq:gparam}. The Noether charge $\mathcal{Q}\in\psu(1,1|2)^2$ is written as a matrix, and its independent components may be projected by defining
\begin{equation}
\label{eq:chargemap}
\mathbf{Q}_{\mathcal{M}}=\str\big(\mathcal{M}\,\mathcal{Q}\big)\,,
\qquad
\mathcal{M}\in\psu(1,1|2)^2\,,
\end{equation}
which relates the superalgebra structure of $\psu(1,1|2)^2$ to the one induced in the phase space by the Poisson brackets%
\footnote{%
This pairing can be understood in terms of the moment map~\cite{Arutyunov:2006ak,Souriau}.
}. 
By construction $\mathcal{Q}_{\mathcal{M}}$ will be conserved,
\begin{equation}
0=\frac{\de\mathbf{Q}_{\mathcal{M}}}{\de\tau}=\frac{\partial\mathbf{Q}_{\mathcal{M}}}{\partial\tau}+\big\{\mathbf{H},\mathbf{Q}_{\mathcal{M}}\big\}\,,
\end{equation}
where in the last equality we have expressed the conservation law in terms of the Poisson structure. This highlights the fact that \emph{some charges will not commute with the Hamiltonian}, namely the ones that depend explicitly on $\tau=x_{+}$. Only the remaining charges will constitute the manifest symmetry algebra of the theory, with~$\mathbf{H}$ as central element. This is similar to what we discussed in the introduction for strings in flat space, where the $\so(1,9)$~symmetry is broken down to~$\so(8)$.

The $x_{+}$ dependence enters~\eqref{eq:noethercharge} in a simple way, \ie\ only trough $\Lambda=\Lambda(x_{+},x_{-})$. This makes it easy to identify the elements of $\mathcal{M}$ that give rise to charges commuting with $\mathbf{H}$. These are depicted in figure~\ref{fig:Mmatrix}, and consist of the bosonic charges lying on the diagonal of $\mathcal{M}$ and eight supercharges. Similarly, it is also  easy to see that all of these supercharges carry a dependence on the unphysical (and highly non-local) field $x_{-}$, a fact that will be important later.

\begin{figure}
  \centering
\[
\mathcal{M}=\left(\begin{array}{cccc}
\KinCC	&	\DynNC	&	\DynCC	&	\KinNC	\\
\DynNC	&	\KinCC	&	\KinNC	&	\DynCC	\\
\DynCC	&	\KinNC	&	\KinCC	&	\DynNC	\\
\KinNC	&	\DynCC	&	\DynNC	&	\KinCC	
\end{array}\right)
\oplus
\left(\begin{array}{cccc}
\KinCC	&	\DynNC	&	\DynCC	&	\KinNC	\\
\DynNC	&	\KinCC	&	\KinNC	&	\DynCC	\\
\DynCC	&	\KinNC	&	\KinCC	&	\DynNC	\\
\KinNC	&	\DynCC	&	\DynNC	&	\KinCC	
\end{array}\right)
\]
  \caption{The elements of the $\psu(1,1|2)^2$ matrix $\mathcal{M}$ of~\eqref{eq:chargemap}, distinguished by the dependence on $x_{\pm}$ in the resulting charge~$\mathbf{Q}_{\mathcal{M}}$. Elements on a white background yield an $x_{+}$-dependent charge (that does not commute with $\mathbf{H}$), while the one highlighted in yellow $\KinCC,\DynCC$ yield conserved charges. We further distinguish between kinematical~($\mathtt{K}$), \ie\ $x_{-}$-independent charges, and dynamical ones~($\mathtt{D}$).}
  \label{fig:Mmatrix}
\end{figure}

\subsection*{Perturbative evaluation at leading order}
We now want to find explicit expressions for the action and its symmetries perturbatively in a field expansion. To this end, as discussed in section~\ref{sec:bosonic1ord} we will take the decompactification limit and perform a field expansion. We are interested in the quadratic Hamiltonian, that comes at leading order in $1/h$, and should be a suitable supersymmetric completion of~\eqref{eq:bosonicH2}, together with its symmetry algebra.

\subsubsection*{Quadratic expressions in the fields}
Using the expansion of the auxiliary fields provided in appendix~\ref{app:coset} we find that the Lagrangian can be written, at leading order, as
\begin{equation}
\mathcal{L}_2=p_k\dot{x}_k
-\frac{i}{2}\str\big(\Sigma_{+}\Psi_{m}\,\dot{\Psi}_{m}\big)
-\mathcal{H}_2\,,
\end{equation}
where the quadratic Hamiltonian is
\begin{equation}
\begin{aligned}
\label{eq:quadrH-matrix}
\mathcal{H}_2=&\frac{1}{2}|p_k|^2+\frac{1}{2}|x'_k|^2+\frac{1}{2}|x_k|^2
-\frac{1}{2}\str\big(\Sigma_{+}\Psi_{m} \,\Omega(\Psi_{m}')\big)+
\frac{1}{2}\str\big(\Psi_{m}^2\big)\,.
\end{aligned}
\end{equation}
As expected, this is a supersymmetric relativistic extension of what computed in the purely bosonic NLSM.
In order to write down more explicitly these results, let us use the fermion parametrisation~\eqref{eq:Psim-param}
%
so that the Lagrangian becomes
\begin{equation}
\mathcal{L}_2=p_k\dot{x}_k
+i\,\bar{\theta}_{j\,\smallL}\dot{\theta}^{j\,\smallL}
+i\,\bar{\theta}_{j\,\smallR}\dot{\theta}^{j\,\smallR}
-\mathcal{H}_2\,,
\end{equation}
where the fermion fields are conjugate to each other,~$\bar{\theta}=\theta^{\dagger}$, and indices are raised and lowered with~$\delta^{jk},\delta_{jk}$. The quadratic Hamiltonian is
\begin{equation}
\label{eq:quadraticH}
\begin{aligned}
\mathcal{H}_2=
&
\frac{1}{2}|p_k|^2+\frac{1}{2}|x'_k|^2+\frac{1}{2}|x_k|^2
+\theta^{j}_{\smallL}\theta_{j\smallR}'
-\bar{\theta}^{j}_{\smallL}\bar{\theta}_{j\smallR}'
+\bar{\theta}_{j\,\smallL}{\theta}^{j}_{\smallL}
+\bar{\theta}_{j\,\smallR}{\theta}^{j}_{\smallR}\,.
\end{aligned}
\end{equation}
In particular, this implies that upon quantisation we find the the non-vanishing canonical commutation relations
\begin{equation}
\label{eq:PB-boson}
\big[x_{j}(\sigma),p_{k}(\tilde{\sigma})\big]=
i\,\delta_{jk}\,\delta(\sigma-\tilde{\sigma})\,,
\end{equation}
and anti-commutation relations
\begin{equation}
\label{eq:PB-fermion}
\big\{\theta^{j}_{\L}(\sigma),\bar{\theta}_{k\,\L}(\tilde{\sigma})\big\}=
\delta^{j}_{k}\,\delta(\sigma-\tilde{\sigma})\,,
\qquad
\big\{\theta^{j}_{\R}(\sigma),\bar{\theta}_{k\,\R}(\tilde{\sigma})\big\}=
\delta^{j}_{k}\,\delta(\sigma-\tilde{\sigma})\,.
\end{equation}

Let us now work out the charges that commute with the Hamiltonian.
We start from the bosonic charges. From figure~\ref{fig:Mmatrix} we see that they can be found from contracting $\mathcal{Q}$ with a diagonal matrix $\mathcal{M}$. Taking into account that $\mathcal{M}\in\psu(1,1|2)^2$, we are left with four independent choices of $\mathcal{M}$, yielding as many central charges. Two of these are well familiar:
\begin{equation}
\frac{1}{2i}\str\big(\Sigma_{+}\,\mathcal{Q}\big)=\mathbf{H}\,,
\qquad
\frac{1}{2i}\str\big(\Sigma_{-}\,\mathcal{Q}\big)=-2\,\mathbf{P}_{+}=-2\,P_{+}\,.
\end{equation}
The remaining two come from
\begin{equation}
\mathcal{M}=\mathbf{L}^3_{\L}+\mathbf{L}^3_{\R} +\mathbf{J}^3_{\L}+\mathbf{J}^3_{\R}\,,
\qquad
\mathcal{N}=\mathbf{L}^3_{\L}+\mathbf{L}^3_{\R}- \mathbf{J}^3_{\L}-\mathbf{J}^3_{\R}\,,
\end{equation}
and are given in appendix~\ref{app:quadraticcharges}.
We define the supercharges that commute with the Hamiltonian by contracting $\mathcal{Q}$ with odd supermatrices with a single non-vanishing entry (akin to raising and lowering operators). As expected we find eight supercharges, that we label
\begin{equation}
\mathbf{Q}^{1\,\L,\R},\qquad
\mathbf{Q}^{2\,\L,\R},\qquad
\overline{\mathbf{Q}}{}^{1\,\L,\R},\qquad
\overline{\mathbf{Q}}{}^{2\,\L,\R},
\end{equation}
and which are Hermitian conjugate in pairs, $(\mathbf{Q}^{j\,\L,\R})^{\dagger}=\overline{\mathbf{Q}}{}^{j\,\L,\R}$.
The general form of the charges in terms of fields is
\begin{equation}
\label{eq:genericcharge}
\begin{aligned}
\mathbf{Q} \approx	&
\int \de\sigma\,e^{+ \frac{i}{2}x_{-}}
\left(p\,\theta+x\,\bar{\theta}{\,}'+x\,\theta \right)\,,\\
\overline{\mathbf{Q}} \approx &
	\int \de\sigma\,e^{- \frac{i}{2}x_{-}}
\left(p\,\theta+x\,\bar{\theta}{\,}'+x\,\theta \right)\,,
\end{aligned}
\end{equation} 
where we considered only the leading order contribution inside the brackets but kept track of the $e^{\pm \frac{i}{2}x_{-}}$ factor coming from $\Lambda(x_{\pm})$. The precise form of all charges is given in appendix~\ref{app:quadraticcharges}. It is however more  convenient to take a Fourier transform and introduce a basis of raising and lowering operators, thereby making the action of the $\sigma$-derivatives more apparent, and diagonalizing some of the charges. In particular, we will take all of the central elements to be proportional to the number operator.

\subsection*{Momentum-space representation}
Let us introduce the bosonic creation and annihilation operators $a_{\smallL,\smallR\,\pm\pm}^\dagger$ and $a^{\pm\pm}_{\smallL,\smallR}$, as defined in appendix~\ref{app:quadraticcharges}. An annihilation operator takes the schematic form
\begin{equation}
\label{eq:generic-annihilation}
a_{\pm\pm}(q)\approx\frac{1}{\sqrt{2\pi}}
\int \frac{\de\sigma}{\sqrt{\omega_q}}
\left(\omega_q\,x(\sigma)+i\,p(\sigma)\right)e^{-iq\sigma}
\end{equation}
where
\begin{equation}
\label{eq:bosonicparam}
\omega_p = \sqrt{m^2+p^2}\,.
\end{equation}
The respective creation operator is the complex conjugate of~\eqref{eq:generic-annihilation}. The definitions in appendix~\ref{app:quadraticcharges} are given in such a way that the operators have canonical commutation relations
\begin{equation}
\big[ a^{\kappa\iota}_{\smallL}(p),
a_{\smallL\,\tilde{\kappa}\tilde{\iota}}^\dagger(\tilde{p})\big]
=\delta^{\kappa}_{\tilde{\kappa}}\,\delta^{\iota}_{\tilde{\iota}} \,\delta(p-\tilde{p})\,,
\qquad
\big[a^{\kappa\iota}_{\smallR}(p),
a_{\smallR\,\tilde{\kappa}\tilde{\iota}}^\dagger(\tilde{p})\big]
=\delta^{\kappa}_{\tilde{\kappa}}\,\delta^{\iota}_{\tilde{\iota}} \,\delta(p-\tilde{p})\,.
\end{equation}
Similarly, we introduce the fermionic operators $a_{\smallL,\smallR\,\pm\mp}^\dagger$ and $a^{\pm\mp}_{\smallL,\smallR}$, also defined in appendix~\ref{app:quadraticcharges}, that have the general form \eg
\begin{equation}
a_{\pm\mp}(q)\approx\frac{1}{\sqrt{2\pi}}
\int \frac{\de\sigma}{\sqrt{\omega_q}}
\left(f_q\, \bar{\theta}(\sigma)+
 g_q\, \theta(\sigma)\right)e^{-iq\sigma}\,,
\end{equation}
where we used the fermion wave-function parameters
\begin{equation}
\label{eq:fermionicparam}
f_p = \sqrt{\frac{\omega_p+m}{2}}\,,
\quad
g_pf_p =
-\frac{p}{2}\,,
\quad
f_p^2-g_p^2=m\,,
\quad f_p^2+g_p^2=\omega_p \,.
\end{equation}
The fermionic operators are defined to satisfy
\begin{equation}
\big\{ a^{\kappa\iota}_{\smallL}(p),
a_{\smallL\,\tilde{\kappa}\tilde{\iota}}^\dagger(\tilde{p})\big\}
=\delta^{\kappa}_{\tilde{\kappa}}\,\delta^{\iota}_{\tilde{\iota}} \,\delta(p-\tilde{p})\,,
\qquad
\big\{a^{\kappa\iota}_{\smallR}(p),
a_{\smallR\,\tilde{\kappa}\tilde{\iota}}^\dagger(\tilde{p})\big\}
=\delta^{\kappa}_{\tilde{\kappa}}\,\delta^{\iota}_{\tilde{\iota}} \,\delta(p-\tilde{p})\,.
\end{equation}
Since the Hamiltonian~\eqref{eq:quadraticH} contains only particles of unit mass, from now on we set $m=1$.
At leading order, the Hamiltonian and the other central charges take the form
\begin{equation}
\label{eq:bosonic-charges1}
\begin{aligned}
\mathbf{H}=&\int \de p\sum_{\kappa,\iota=\pm}  \left(a^\dagger_{\smallL\,\kappa\iota}a_{\smallL}^{\kappa\iota}+a^\dagger_{\smallR\,\kappa\iota}a_{\smallR}^{\kappa\iota}\right)\omega_p\,,
\\
\mathbf{M}=&\int \de p\sum_{\kappa,\iota=\pm} \left(a^\dagger_{\smallL\,\kappa\iota}a_{\smallL}^{\kappa\iota}-a^\dagger_{\smallR\,\kappa\iota}a_{\smallR}^{\kappa\iota}\right),
\\
\mathbf{N}=&\int \de p \sum_{\kappa=\pm}\kappa\,\left(a^\dagger_{\smallL\,\kappa\kappa}a_{\smallL}^{\kappa\kappa}-a^\dagger_{\smallR\,\kappa\kappa}a_{\smallR}^{\kappa\kappa}\right).
\end{aligned}
\end{equation}
In particular, we see that L-excitations have charge $+1$ under the angular momentum  $\mathbf{M}$, while R-excitations have charge $-1$. We will therefore refer to these excitations as ``left'' and ``right''. However, \emph{these do not correspond to left- and right-moving excitations on the worldsheet}. Since the angular momentum~$\mathbf{M}$ has a component on~$\AdS_3$,  these excitations can instead be thought of as left- and right-movers in the dual $\CFT_2$.

The supercharges are given by
\begin{equation}
\label{eq:supercharges1}
\begin{aligned}
\mathbf{Q}^{1\L}=\int\! \de p\sum_{\iota=\pm}\left(
f_p \,a_{\smallL-\iota}^\dagger a_{\smallL}^{+\iota} +
g_p\, a_{\smallR+\iota}^\dagger a_{\smallR}^{-\iota} \right),
\\
\mathbf{Q}^{1\R}=\int\! \de p\sum_{\iota=\pm}\left(
f_p \,a_{\smallR-\iota}^\dagger a_{\smallR}^{+\iota} +
g_p\, a_{\smallL+\iota}^\dagger a_{\smallL}^{-\iota} \right),
\end{aligned}
\end{equation}
and
\begin{equation}
\label{eq:supercharges2}
\begin{aligned}
\mathbf{Q}^{2\L}=\int\!\! \de p\!\sum_{\kappa=\pm}\left(
f_p \,a_{\smallL\kappa-}^\dagger a_{\smallL}^{\kappa+} +
g_p\, a_{\smallR\kappa+}^\dagger a_{\smallR}^{\kappa-} \right)\kappa,
\\
\mathbf{Q}^{2\R}=\int\!\! \de p\!\sum_{\kappa=\pm}\left(
f_p \,a_{\smallR\kappa-}^\dagger a_{\smallR}^{\kappa+} +
g_p\, a_{\smallL\kappa+}^\dagger a_{\smallL}^{\kappa-} \right)\kappa,
\end{aligned}
\end{equation}
together with their Hermitian conjugates $\overline{\mathbf{Q}}{}^{j\L,\R}=(\mathbf{Q}^{j\L,\R})^{\dagger}$. Notice how the form of the supercharges is completely symmetric under exchange of the labels $\bigL\leftrightarrow\bigR$, a fact that is true also when we write them in terms of the field, see eq.~\eqref{eq:charges-fields}.

Finally, let us recall that the worldsheet momentum is given by
\begin{equation}
\mathbf{P}=\int \de p\sum_{\kappa,\iota=\pm}  \left(a^\dagger_{\smallL\,\kappa\iota}a_{\smallL}^{\kappa\iota}+a^\dagger_{\smallR\,\kappa\iota}a_{\smallR}^{\kappa\iota}\right)p\,.
\end{equation}

\section{Symmetry algebra and its representation}
\label{sec:stringsymm}
We can now use the perturbative information we just obtained to study the symmetry algebra and its representations.

\subsection*{Commutation relations}
From the explicit results of the previous section it is easy to read off the anticommutation relations. The non-vanishing ones are
\begin{equation}
\label{eq:anticommHM}
\begin{aligned}
\big\{\mathbf{Q}^{1\L},\overline{\mathbf{Q}}{}^{1\L}\big\}=
\big\{\mathbf{Q}^{2\L},\overline{\mathbf{Q}}{}^{2\L}\big\}=
\frac{1}{2}\mathbf{H}+\frac{1}{2}\mathbf{M}\,,\\
\big\{\mathbf{Q}^{1\R},\overline{\mathbf{Q}}{}^{1\R}\big\}=
\big\{\mathbf{Q}^{2\R},\overline{\mathbf{Q}}{}^{2\R}\big\}=
\frac{1}{2}\mathbf{H}-\frac{1}{2}\mathbf{M}\,,
\end{aligned}
\end{equation}
and
\begin{equation}
\label{eq:anticommP}
\begin{aligned}
\big\{\mathbf{Q}^{1\L},{\mathbf{Q}}^{1\R}\big\}=
\big\{\mathbf{Q}^{2\L},{\mathbf{Q}}^{2\R}\big\}=
-\frac{1}{2}\mathbf{P}\,,\\
\big\{\overline{\mathbf{Q}}{}^{1\L},\overline{\mathbf{Q}}{}^{1\R}\big\}=
\big\{\overline{\mathbf{Q}}{}^{2\L},\overline{\mathbf{Q}}{}^{2\R}\big\}=
-\frac{1}{2}\mathbf{P}\,.
\end{aligned}
\end{equation}
From~\eqref{eq:anticommHM} we recognise two copies of $\su(1|1)^2$, and from~\eqref{eq:anticommP} we see that each of them gets centrally extended by two $\u(1)$ charges, which at leading order coincide and are proportional to the worldsheet momentum%
\footnote{%
From this leading-order analysis we cannot establish whether the two right-hand-sides of~\eqref{eq:anticommP} also coincide at higher orders, but only that such charges should be Hermitian conjugate to each other. We will later see they do not coincide, by more carefully investigating the structure of the supercharges.
}. We conclude that the symmetry algebra~is
\begin{equation}
\label{eq:offshellalgebra}
\left[\big(\su(1|1)_{\L}\oplus\su(1|1)_{\R}\big)\oplus \big(\u(1)\oplus\u(1)\big)\right]^2/\,\u(1)^4\,,
\end{equation}
where \eg\ $\mathbf{Q}^{1\L},\overline{\mathbf{Q}}{}^{1\L}$ and their central charge form one of the four $\su(1|1)$, and $\cdots\oplus \u(1)^2$ denotes the central extension by two copies of $\mathbf{P}$. The quotient ensures that the central charges of commutators differing by $1\leftrightarrow2$ coincide. Since this algebra can be found from centrally extending~$\psu(1|1)^4$, we indicate it as~$\psu(1|1)^4_{\text{c.e.}}$.

It is quite remarkable that the algebra has been extended by two elements proportional to the worldsheet momentum~$\mathbf{P}$, which was not part of the original superisometries. Such an extended algebra should not be a symmetry of the physical states, which is guaranteed by the fact that the level matching constrain~\eqref{eq:levelmatching} imposes 
\begin{equation}
\mathbf{P}\,|\text{physical state}\rangle=0\,.
\end{equation}
Therefore the algebra~\eqref{eq:offshellalgebra} is valid \emph{off-shell}, whereas \emph{the on-shell algebra} is just
\begin{equation}
\label{eq:onshellalgebra}
\left(\su(1|1)_{\L}\oplus\su(1|1)_{\R}\right)^2/\,\u(1)^2\,.
\end{equation}

\subsection*{Representations on one-excitation states}
Let us consider excitations of a defined momentum $p$, which can be created by acting on a vacuum as
\begin{equation}
\label{eq:wsexcitation}
|\Phi_{\pm\pm}^{\smallL,\smallR}(p)\rangle=
 a^{\dagger}_{\pm\pm\smallL,\smallR}(p)\,|0\rangle\,,
\qquad
|\Phi_{\pm\mp}^{\smallL,\smallR}(p)\rangle=
 a^{\dagger}_{\pm\mp\smallL,\smallR}(p)\,|0\rangle\,,
\end{equation}
where the former and latter expressions correspond to bosonic and fermionic excitations, respectively.
We should distinguish between representation of the full, off-shell algebra and representations of the on-shell one. The former is realised for general $p$, while the latter when we impose $p=0$. Let us start from this simpler case.

\subsubsection*{On-shell representations}
\begin{figure}
  \centering
  \begin{tikzpicture}[%
    box/.style={outer sep=1pt},
    Q node/.style={inner sep=1pt,outer sep=0pt},
    arrow/.style={-latex}
    ]%

    \node [box] (PhiM) at ( 0  , 1.5cm) {\small $\ket{\Phi^{\smallL}_{++}}$};
    \node [box] (PsiP) at (-1.5cm, 0cm) {\small $\ket{\Phi^{\smallL}_{-+}}$};
    \node [box] (PsiM) at (+1.5cm, 0cm) {\small $\ket{\Phi^{\smallL}_{+-}}$};
    \node [box] (PhiP) at ( 0  ,-1.5cm) {\small $\ket{\Phi^{\smallL}_{--}}$};

    \newcommand{\horshift}{0.09cm,0cm}
    \newcommand{\vershift}{0cm,0.10cm}
 
    \draw [arrow] ($(PhiM.west) +(\vershift)$) -- ($(PsiP.north)-(\horshift)$) node [pos=0.5,anchor=south east,Q node] {\scriptsize $\mathbf{Q}^{1\L}$};
    \draw [arrow] ($(PsiP.north)+(\horshift)$) -- ($(PhiM.west) -(\vershift)$) node [pos=0.5,anchor=north west,Q node] {\scriptsize $\overline{\mathbf{Q}}{}^{1\L}$};

    \draw [arrow] ($(PsiM.south)-(\horshift)$) -- ($(PhiP.east) +(\vershift)$) node [pos=0.5,anchor=south east,Q node] {\scriptsize $\mathbf{Q}^{1\L}$};
    \draw [arrow] ($(PhiP.east) -(\vershift)$) -- ($(PsiM.south)+(\horshift)$) node [pos=0.5,anchor=north west,Q node] {\scriptsize $\overline{\mathbf{Q}}{}^{1\L}$};

    \draw [arrow] ($(PhiM.east) -(\vershift)$) -- ($(PsiM.north)-(\horshift)$) node [pos=0.5,anchor=north east,Q node] {\scriptsize $\mathbf{Q}{}^{2\L}$};
    \draw [arrow] ($(PsiM.north)+(\horshift)$) -- ($(PhiM.east) +(\vershift)$) node [pos=0.5,anchor=south west,Q node] {\scriptsize $\overline{\mathbf{Q}}{}^{2\L}$};

    \draw [arrow] ($(PsiP.south)-(\horshift)$) -- ($(PhiP.west) -(\vershift)$) node [pos=0.5,anchor=north east,Q node] {\scriptsize $-\mathbf{Q}^{2\L}$};
    \draw [arrow] ($(PhiP.west) +(\vershift)$) -- ($(PsiP.south)+(\horshift)$) node [pos=0.5,anchor=south west,Q node] {\scriptsize $-\overline{\mathbf{Q}}{}^{2\L}$};
  \end{tikzpicture}
\hspace{1cm}
  \begin{tikzpicture}[%
    box/.style={outer sep=1pt},
    Q node/.style={inner sep=1pt,outer sep=0pt},
    arrow/.style={-latex}
    ]%

    \node [box] (PhiM) at ( 0  , 1.5cm) {\small $\ket{\Phi^{\smallR}_{--}}$};
    \node [box] (PsiP) at (-1.5cm, 0cm) {\small $\ket{\Phi^{\smallR}_{+-}}$};
    \node [box] (PsiM) at (+1.5cm, 0cm) {\small $\ket{\Phi^{\smallR}_{-+}}$};
    \node [box] (PhiP) at ( 0  ,-1.5cm) {\small $\ket{\Phi^{\smallR}_{++}}$};

    \newcommand{\horshift}{0.09cm,0cm}
    \newcommand{\vershift}{0cm,0.10cm}
 
    \draw [arrow] ($(PhiM.west) +(\vershift)$) -- ($(PsiP.north)-(\horshift)$) node [pos=0.5,anchor=south east,Q node] {\scriptsize $\overline{\mathbf{Q}}{}^{1\R}$};
    \draw [arrow] ($(PsiP.north)+(\horshift)$) -- ($(PhiM.west) -(\vershift)$) node [pos=0.5,anchor=north west,Q node] {\scriptsize ${\mathbf{Q}}^{1\R}$};

    \draw [arrow] ($(PsiM.south)-(\horshift)$) -- ($(PhiP.east) +(\vershift)$) node [pos=0.5,anchor=south east,Q node] {\scriptsize $\overline{\mathbf{Q}}{}^{1\R}$};
    \draw [arrow] ($(PhiP.east) -(\vershift)$) -- ($(PsiM.south)+(\horshift)$) node [pos=0.5,anchor=north west,Q node] {\scriptsize ${\mathbf{Q}}^{1\R}$};

    \draw [arrow] ($(PhiM.east) -(\vershift)$) -- ($(PsiM.north)-(\horshift)$) node [pos=0.5,anchor=north east,Q node] {\scriptsize $\overline{\mathbf{Q}}{}^{2\R}$};
    \draw [arrow] ($(PsiM.north)+(\horshift)$) -- ($(PhiM.east) +(\vershift)$) node [pos=0.5,anchor=south west,Q node] {\scriptsize 
    ${\mathbf{Q}}^{2\R}$};

    \draw [arrow] ($(PsiP.south)-(\horshift)$) -- ($(PhiP.west) -(\vershift)$) node [pos=0.5,anchor=north east,Q node] {\scriptsize $-\overline{\mathbf{Q}}{}^{2\R}$};
    \draw [arrow] ($(PhiP.west) +(\vershift)$) -- ($(PsiP.south)+(\horshift)$) node [pos=0.5,anchor=south west,Q node] {\scriptsize $-{\mathbf{Q}}^{2\R}$};
  \end{tikzpicture}
  \caption{The action of the supercharges on a zero-momentum (on-shell) worldsheet  excitation~\eqref{eq:wsexcitation}. In this case an L-excitation is charged only under $\mathbf{Q}^{j\L}$ and $\overline{\mathbf{Q}}{}^{j\L}$, and similarly for R-excitations. As explained in the text, the raising operators are~$\overline{\mathbf{Q}}{}^{j\L}$ and~${\mathbf{Q}}^{j\R}$.}
  \label{fig:representation}
\end{figure}
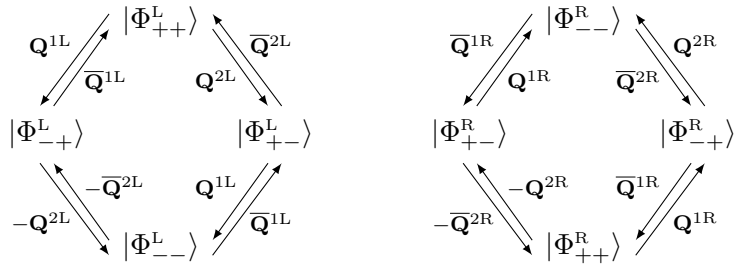

On-shell states are annihilated by the total momentum operator~$\mathbf{P}$, so that one-particle on-shell states are just given by particles with momentum~$p=0$. As a consequence,
the on-shell action of the supercharges~(\ref{eq:supercharges1}--\ref{eq:supercharges2}) and their conjugates is easily found, by using that when $p=0$, we have $f_{p=0}=\omega_{p=0}=1$ and $g_{p=0}=0$. We see that L- and R-excitations form two distinct representations that are charged under $\mathbf{Q}^{\L}$'s and  $\mathbf{Q}^{\R}$'s only, respectively. In figure~\ref{fig:representation} we draw the action of the supercharges. We see that the left excitations transform into a bifundamental of a~$\psu(1|1)^2_{\L}$ consisting of the four L-supercharges. The right excitations are in a bifundamental representation of~$\psu(1|1)^2_{\R}$.

In figure~\ref{fig:representation} we have depicted the left and right representation slightly differently. This is because we cannot  take the same raising operators to be the same in the L- and R-algebra if we want this representation to smoothly extend to the off-shell one. If we took the raising operators to be, \eg
\begin{equation}
\overline{\mathbf{Q}}{}^{1\L},
\qquad
\overline{\mathbf{Q}}{}^{2\L},
\qquad
\overline{\mathbf{Q}}{}^{1\R},
\qquad
\overline{\mathbf{Q}}{}^{2\R},
\end{equation}
it would be impossible for the anticommutator~$\{\overline{\mathbf{Q}}{}^{j\L},\overline{\mathbf{Q}}{}^{k\L}\}$ to be non-vanishing on an highest weight state, which should happen if we deform $p$ to be different from zero. Therefore, we will take the raising operators to be 
\begin{equation}
\overline{\mathbf{Q}}{}^{1\L},
\qquad
\overline{\mathbf{Q}}{}^{2\L},
\qquad
{\mathbf{Q}}^{1\R},
\qquad
{\mathbf{Q}}^{2\R}.
\end{equation}
If we want to think of this algebra as embedded into~$\psu(1,1|2)_{\L}\oplus\psu(1,1|2)_{\R}$, it then follows that the two copies of~$\psu(1,1|2)$ are in \emph{different gradings} with respect to each other. As we discuss in appendix~\ref{app:psu112}, this nicely ties to our choice of different supermatrix representatives for the left and right copies of~$\psu(1,1|2)$.

\subsubsection*{Off-shell representations}
Let us now consider an excitation of arbitrary momentum~$p$, \eg\ a left excitation. Now this is charged under all of the supercharges---the left ones act proportionally to $f_p$, and the right ones to~$g_p$. However, no supercharge can transform it into a right excitation, since nowhere in~(\ref{eq:supercharges1}--\ref{eq:supercharges2}) appears an operator of the form $a^{\dagger}_{\smallR}a_{\smallL}$. The left representation is then an irreducible representation of the whole~$\psu(1|1)^4_{\text{c.e.}}$.

Clearly, the same reasoning can be applied to any right-moving excitation. Therefore, fundamental massive particles of the superstring transform into two irreducible (L and R) representations of the off-shell symmetry algebra, as it can also be seen by the fact that they have different eigenvalue under~$\mathbf{M} $.

\subsubsection*{Tensor-product structure}
From the form of the on-shell representation depicted in figure~\ref{fig:representation} it is easy to see that we can describe the symmetries by means of a tensor product structure, whose factors are related to the indices $1$ and $2$  of the supercharges.  We then can obtain the bifundamental representation of \eg\ $\psu(1|1)^2_{\L}$ from the tensor product of two fundamental representations of a $\psu(1|1)_{\L}$.

Moreover, a tensor product structure exists in the off-shell algebra too. To see this let us introduce bosonic and fermionic operators $a^{\dagger}_{\smallL,\smallR},a_{\smallL,\smallR}$ and $d^{\dagger}_{\smallL,\smallR},d_{\smallL,\smallR}$, respectively. They obey canonical (anti)commutation relations
\begin{equation}
\big[a_{\smallL}(p),a^{\dagger}_{\smallL}(\tilde{p})\big]=\delta(p-\tilde{p})\,,
\qquad
\big\{d_{\smallL}(p),d^{\dagger}_{\smallL}(\tilde{p})\big\}=\delta(p-\tilde{p})\,,
\end{equation}
and similarly for $\bigL\leftrightarrow\bigR$. By means of them we can define the supercharges
\begin{equation}
\label{eq:smallalgebra}
\mathbf{q}^{\smallL}=\int \de p \big(f_p d^{\dagger}_{\smallL}a_{\smallL} + g_{p}a^{\dagger}_{\smallR}d_{\smallR}\big)\,,
\qquad
\mathbf{q}^{\smallR}=\int \de  p \big(f_p d^{\dagger}_{\smallR}a_{\smallR} + g_{p}a^{\dagger}_{\smallL}d_{\smallL}\big)\,,
\end{equation}
and their Hermitian conjugates~$\bar{\mathbf{q}}^{\smallL}$ and~$\bar{\mathbf{q}}^{\smallR}$. It is easy to verify that these satisfy a centrally extended $\psu(1|1)^2$ algebra given by
\begin{equation}
\label{eq:smallrepr-ferm}
\begin{aligned}
&\big\{\mathbf{q}^{\smallL},\bar{\mathbf{q}}^{\smallL}\big\}=\frac{1}{2}\mathbf{h}+\frac{1}{2}\mathbf{m}\,,
&\qquad &
\big\{\mathbf{q}^{\smallR},\bar{\mathbf{q}}^{\smallR}\big\}=\frac{1}{2}\mathbf{h}-\frac{1}{2}\mathbf{m}\,,\\
&\big\{\mathbf{q}^{\smallL},{\mathbf{q}}^{\smallR}\big\}=-\frac{1}{2}\mathbf{p}\,,
&\qquad &
\big\{\bar{\mathbf{q}}^{\smallL},\bar{\mathbf{q}}^{\smallR}\big\}=-\frac{1}{2}\mathbf{p}\,,\\
\end{aligned}
\end{equation}
where the central charges are
\begin{equation}
\label{eq:smallrepr-bos}
\begin{aligned}
\mathbf{h}=&\int \de p\,\left(a^\dagger_{\smallL}a_{\smallL}+ d^\dagger_{\smallL}d_{\smallL}+
a^\dagger_{\smallR}a_{\smallR}+ d^\dagger_{\smallR}d_{\smallR}\right)\omega_p\,,\\
\mathbf{m}=&\int \de p\,\left(a^\dagger_{\smallL}a_{\smallL}+ d^\dagger_{\smallL}d_{\smallL} 
-a^\dagger_{\smallR}a_{\smallR}-d^\dagger_{\smallR}d_{\smallR}\right)\,,\\
\mathbf{p}=&\int \de p\,\left(a^\dagger_{\smallL}a_{\smallL}+ d^\dagger_{\smallL}d_{\smallL}+
a^\dagger_{\smallR}a_{\smallR}+ d^\dagger_{\smallR}d_{\smallR}\right)p\,.
\end{aligned}
\end{equation}

Let us introduce excitations
\begin{equation}
\label{eq:smallexcitation}
\ket{\phi^{\smallL,\smallR}(p)}=a^{\dagger}_{\smallL,\smallR}(p)\ket{0}\,,
\qquad
\ket{\psi^{\smallL,\smallR}(p)}=d^{\dagger}_{\smallL,\smallR}(p)\ket{0}\,.
\end{equation}
On shell we have $f_{p=0}=1$ and $g_{p=0}=0$, so that the excitations indeed transform in two fundamental representations of $\su(1|1)_{\L}$ and $\su(1|1)_{\R}$, respectively. Off-shell, these two representations get deformed as depicted in figure~\ref{fig:offshellrepr}.

\begin{figure}
  \centering
  \begin{tikzpicture}[%
    box/.style={outer sep=1pt},
    Q node/.style={inner sep=1pt,outer sep=0pt},
    arrow/.style={-latex}
    ]%

    \node [box] (PhiM) at ( 0.5cm  , 1.5cm) {\small $\ket{\phi^{\smallL}(p)}$};
    \node [box] (PsiP) at (-1.5cm, -0.5cm) {\small $\ket{\psi^{\smallL}(p)}$};

    \newcommand{\horshift}{0.09cm,0cm}
    \newcommand{\vershift}{0cm,0.10cm}
 
    \draw [arrow] ($(PhiM.west) +(\vershift)$) -- ($(PsiP.north)-(\horshift)$) node [pos=0.5,anchor=south east,Q node] { $\mathbf{q}^{\smallL},\bar{\mathbf{q}}^{\smallR}$};
    \draw [arrow] ($(PsiP.north)+(\horshift)$) -- ($(PhiM.west) -(\vershift)$) node [pos=0.5,anchor=north west,Q node] { $\bar{\mathbf{q}}^{\smallL},\mathbf{q}^{\smallR}$};

  \end{tikzpicture}
\hspace*{2cm}
  \begin{tikzpicture}[%
    box/.style={outer sep=1pt},
    Q node/.style={inner sep=1pt,outer sep=0pt},
    arrow/.style={-latex}
    ]%

    \node [box] (PhiM) at ( 0.5cm  , 1.5cm) {\small $\ket{\psi^{\smallR}(p)}$};
    \node [box] (PsiP) at (-1.5cm, -0.5cm) {\small $\ket{\phi^{\smallR}(p)}$};

    \newcommand{\horshift}{0.09cm,0cm}
    \newcommand{\vershift}{0cm,0.10cm}
 
    \draw [arrow] ($(PhiM.west) +(\vershift)$) -- ($(PsiP.north)-(\horshift)$) node [pos=0.5,anchor=south east,Q node] { $\bar{\mathbf{q}}^{\smallR},\mathbf{q}^{\smallL}$};
    \draw [arrow] ($(PsiP.north)+(\horshift)$) -- ($(PhiM.west) -(\vershift)$) node [pos=0.5,anchor=north west,Q node] { $\mathbf{q}^{\smallR},\bar{\mathbf{q}}^{\smallL}$};

  \end{tikzpicture}
  \caption{The action of the supercharges $\mathbf{q}^{\smallL,\smallR}$ and $\bar{\mathbf{q}}^{\smallL,\smallR}$ of $\psu(1|1)_{\L}\oplus\psu(1|1)_{\R}$ centrally extended on an arbitrary momentum (off-shell) excitation~\eqref{eq:smallexcitation}.}
  \label{fig:offshellrepr}
\end{figure}

The symmetry algebra of our worldsheet theory is in fact just given by the tensor product of two copies of~\eqref{eq:smallalgebra} together with the relative central charges. To see this, note that the annihilation operators are
\begin{equation}
\label{eq:tensorfileds}
a^{++}_{\smallL}=a_{\smallL}\otimes a_{\smallL}\,,
\quad
a^{--}_{\smallL}=d_{\smallL}\otimes d_{\smallL}\,,
\qquad
a^{+-}_{\smallL}=a_{\smallL}\otimes d_{\smallL}\,,
\quad
a^{-+}_{\smallL}=d_{\smallL}\otimes a_{\smallL}\,,
\end{equation}
and similarly for the creation operators and for $\bigL\leftrightarrow\bigR$. When we want to emphasise that the excitations $\phi^{\smallL,\smallR}$ and $\psi^{\smallL,\smallR}$ are to be taken as elements of a tensor product we will instead denote them as
\begin{equation}
\label{eq:tensor-excit-repr}
\ket{\Phi_{+}^{\smallL,\smallR}}=\ket{\phi^{\smallL,\smallR}}\,,
\qquad
\ket{\Phi_{-}^{\smallL,\smallR}}=\ket{\psi^{\smallL,\smallR}}\,,
\end{equation}
so that $\Phi_{\kappa\iota}^{\smallL}= \Phi_{\kappa}^{\smallL}\otimes\Phi_{\iota}^{\smallL}$, and $\Phi_{\kappa\iota}^{\smallR}= \Phi_{\kappa}^{\smallR}\otimes\Phi_{\iota}^{\smallR}$.
For the charges, we have
\begin{equation}
\label{eq:tensorcharges}
\mathbf{Q}^{1\L}=\mathbf{q}^{\smallL}\otimes \mathbf{I}\,,
\quad
\mathbf{Q}^{2\L}=\mathbf{I}\otimes\mathbf{q}^{\smallL}\,,
\qquad
\mathbf{Q}^{1\R}=\mathbf{q}^{\smallR}\otimes \mathbf{I}\,,
\quad
\mathbf{Q}^{2\R}=\mathbf{I}\otimes\mathbf{q}^{\smallR}\,,
\end{equation}
and similarly  for the $\overline{\mathbf{Q}}$'s. The tensor product here should respect the $\mathbbm{Z}_2$ grading, and yield a minus sign when two fermionic operators are swapped. By taking this into account we see that the charges defined in this way precisely agree with~(\ref{eq:supercharges1}--\ref{eq:supercharges2}).

This scenario is quite similar to  what happens for the $\AdS_5\times \S^5$ superstring. In that case, the off-shell algebra is given by two copies of $\su(2|2)$ centrally extended, and the excitations transform in a representation that is the tensor product of two fundamental representations of that algebra. In our case, $\su(1|1)_{\L}\oplus\su(1|1)_{\R}$ plays the role of~$\su(2|2)$. Excitations now transform in \emph{two distinct representations} that are once again of the tensor product form. This will have important consequences for the form of the S matrix that we will compute in the next section. Before moving to that, let us further investigate the central charges appearing in the off-shell algebra.

\subsection*{The central charges}
Studying the symmetry algebra at quadratic order in the fields we have seen that it has the form of a subalgebra of the original isometries, supplemented by two central charges that we will denote by $\mathbf{C}$ and $\overline{\mathbf{C}}$ and that vanish on-shell. At quadratic order it so happens that these two central charges are equal and proportional to the worldsheet momentum $\mathbf{P}$. In particular, in terms of the fields we had \eg
\begin{equation}
\big\{\mathbf{Q}^{j\L},{\mathbf{Q}}^{k\R}\big\}=
-\frac{1}{2}\,\delta^{jk} \,\int \de \sigma\,x_{-}'\,,
\end{equation}
at leading order.

While---as the coset construction highlighted---the rest of the algebra is fixed by the embedding into $\psu(1,1|2)^2$, these additional central charges are not, and it is important to understand how they are modified if we consider higher order terms in our field expansion. In particular, we would like to understand
\begin{itemize}
\item whether there is any additional central charge that vanishes on-shell and appearing at higher order in the expansion,
\item and what is the form of $\mathbf{C},\overline{\mathbf{C}}$ when we account for higher order terms. 
\end{itemize}

The former point is negatively answered by observing that the one we are considering is already the \emph{maximal} non-trivial central extension of $\psu(1,1)^4$. As for the latter point, the scenario we have here is very similar to the one found in $\AdS_5\times\S^5$, and in fact find an answer by repeating an argument employed there, which spares us the---hardly feasible---computation of the charges at very high orders in the field expansion.

The key observation is that in the coset construction the unphysical coordinate $x_{-}$ is neatly packaged in $\Lambda(x_{\pm})$ so that the supercharges can be cast in the general form 
\begin{equation}
\mathbf{Q}=\int \de \sigma\,e^{+ \frac{i}{2}x_{-}}
\left(\theta\cdot\big(
\mathcal{G}^{(1)}(p,x)+\frac{1}{h}\mathcal{G}^{(3)}(p,x)+\dots\big)
+\mathcal{O}\big(\theta^3\big)\,
\right),
\end{equation}
see also appendix~\ref{app:quadraticcharges}. 
Here we have expanded the charge density first order by order in the fermions, and then we have expanded the coefficient of each such term in $1/h$. Therefore, $\mathcal{G}^{(1)}(p,x)$ is linear in the bosons, $\mathcal{G}^{(3)}(p,x)$ is cubic, and so on. Since the central charges should vanish on-shell, they should be a function of $\mathbf{P}$ which vanishes at zero. In particular, if we find the functional dependence of~$\mathbf{C}$ and~$\overline{\mathbf{C}}$ on the bosonic fields alone, we will be able to unambiguously fix their full form~\cite{Arutyunov:2006ak}. The bosonic part of the Poisson bracket of two supercharges is then \eg
\begin{equation}
\big\{\mathbf{Q}^{j\L},{\mathbf{Q}}^{k\R}\big\}\approx 
\delta^{jk}\int \de \sigma
e^{i x_{-}}
 \Big(\mathcal{G}_{j\L}^{(1)}\,{\mathcal{G}}_{k\R}^{(1)}+\frac{1}{h}\big(\mathcal{G}_{j\L}^{(1)}\,{\mathcal{G}}_{k\R}^{(3)}+\mathcal{G}_{j\L}^{(3)}\,{\mathcal{G}}_{k\R}^{(1)}\big)+\dots\Big)\,.
\end{equation}
In fact, the leading order term is precisely what we have already calculated, that is $x_{-}'$ restricted to the bosonic fields only. Therefore, at the leading order of this hybrid expansion we find
\begin{equation}
\label{eq:centralcahrgeexp}
\big\{\mathbf{Q}^{j\L},{\mathbf{Q}}^{k\R}\big\} =
-\frac{h}{2}\delta^{jk}\int \de \sigma
e^{i x_{-}}x_{-}'= i\frac{h}{2}\big(e^{i x_{-}(+\infty)}-e^{i x_{-}(-\infty)}\big)\delta_{jk}\,.
\end{equation}
Repeating this calculation to include higher orders of the bosonic expansion, one can check that these do not spoil the form of eq.~\eqref{eq:centralcahrgeexp}. It is actually convenient to rewrite this in a way that makes the dependence on the worldsheet momentum manifest,%
\footnote{%
Eq.~\eqref{eq:anticommP} can be recovered by performing the rescaling~$\mathbf{P}\to\mathbf{P}/h$ and expanding in~$1/h$.
}
\begin{equation}
\label{eq:zeta-def}
\big\{\mathbf{Q}^{j\L},{\mathbf{Q}}^{k\R}\big\}=\delta^{jk}\,\mathbf{C}=
i\,\delta^{jk}\,\zeta\,\frac{h}{2}\big(e^{i\mathbf{P}}-1\big)\,,
\end{equation}
where we isolated the boundary condition $\zeta=e^{i x_{-}(-\infty)}$. In fact, on the one-particle representation we can consistently set \eg\ $x_{-}(-\infty)=0$ so that $\zeta=1$. A similar calculation shows that, as required by hermiticity
\begin{equation}
\big\{\overline{\mathbf{Q}}^{j\L},\overline{\mathbf{Q}}^{k\R}\big\}=\delta_{jk}\,\overline{\mathbf{C}}=
-i\,\delta^{jk}\,\bar{\zeta}\,\frac{h}{2}\big(e^{-i\mathbf{P}}-1\big)\,.
\end{equation}
Of course the non-linearity of $\mathbf{C}, \overline{\mathbf{C}}$ in the off-shell algebra means that the same must be true in each copy of the tensor product for~$\mathbf{c}$ and~$\bar{\mathbf{c}}$, while the last line of~\eqref{eq:smallrepr-ferm} holds only at leading order in~$1/h$.

\section{Chapter summary}
The main result of this chapter is that we derived the off-shell symmetries of the massive excitations of $\AdS_3\times\S^3\times\T^4$ from a perturbative analysis of the NLSM action. The symmetry algebra is given by two copies of a central extension of $\su(1|1)_{\L}\oplus\su(1|1)_{\R}$, which has anticommutation relations
\begin{equation}
\label{eq:smallalgebra-nonlin}
\begin{aligned}
&\big\{\mathbf{q}^{\smallL},\bar{\mathbf{q}}^{\smallL}\big\}=\frac{1}{2}\mathbf{h}+\frac{1}{2}\mathbf{m}\,,
&\qquad &
\big\{\mathbf{q}^{\smallR},\bar{\mathbf{q}}^{\smallR}\big\}=\frac{1}{2}\mathbf{h}-\frac{1}{2}\mathbf{m}\,,\\
&\big\{\mathbf{q}^{\smallL},{\mathbf{q}}^{\smallR}\big\}=\mathbf{c}\,,
&\qquad &
\big\{\bar{\mathbf{q}}^{\smallL},\bar{\mathbf{q}}^{\smallR}\big\}=\bar{\mathbf{c}}\,.\\
\end{aligned}
\end{equation}
The central charges $\mathbf{c},\bar{\mathbf{c}}$ vanish on-shell, and otherwise are non-linear functions of the worldsheet momentum $\mathbf{p}$, which for an appropriate choice of boundary conditions can be written as
\begin{equation}
{\mathbf{c}}=+i\frac{h}{2}\big(e^{+i\, \mathbf{p}}-1\big)\,,
\qquad
\bar{\mathbf{c}}=-i\frac{h}{2}\big(e^{-i\, \mathbf{p}}-1\big)\,,
\end{equation}
where we stress once more that we are not investigating the relation of the parameter~$h$ with the 't~Hooft coupling~$\lambda$.
The role of the central extensions for $\su(1|1)^2$ was originally discussed in ref.~\cite{Beisert:2007sk}, in the context of $\AdS_5/\CFT_4$ duality, and in the case of $\AdS_3/\CFT_2$ in refs.~\cite{David:2008yk,Babichenko:2009dk} and more recently in ref.~\cite{Hoare:2013pma}.%
\footnote{The same symmetry algebra was also found in the analysis of the Pohlmeyer reduced sigma model of the $\AdS_3 \times \S^3$ in ref.~\cite{Hoare:2011fj}.}
We found two one-particle representation of the algebra~\eqref{eq:smallalgebra-nonlin}, which are deformations of the fundamental representation of $\psu(1|1)$: the ``left'' representation (L) has a bosonic highest weight state, whereas the ``right'' representation (R) has a fermionic one. Massive particles of the worldsheet theory are in two tensor product representations, $\bigL\otimes\bigL$ and $\bigR\otimes\bigR$. This accounts for all of the $8=4_{\L}+4_{\R}$ massive excitations.

Some remarks are in order. Firstly, owing to the coset construction, even if our approach is perturbative we were able to capture the general form of the algebra including the non-linear central extension. We should still bear in mind that our calculation was entirely classical, and in principle some of these result might still be spoiled by quantum anomalies.

To exploit the coset formulation we ``froze'' all the massless excitations. Could taking them into account modify the symmetries we identified? We know that the $\AdS_3\times\S^3\times\T^4$ backgrounds has $16$ supersymmetries, and that the choice of light-cone gauge will break at least half of those. Therefore, we cannot expect any new supercharge to appear from including the massless modes. In principle, we may be overlooking some bosonic symmetries that should supplement the algebra we found. Still, we know that  the off-shell theory has \emph{at least} the symmetries given by two copies of~\eqref{eq:smallalgebra-nonlin}. Of course this does not say anything on what are the symmetries of the \emph{massless} excitations. We do know that since $\mathbf{H}$ (and therefore the mass) appears in the  symmetry algebra, the massless excitations cannot transform in the same representation as the massive ones. We will come back to this in chapter~\ref{ch:outlook}.

Let us finally stress that even if physical excitations transform under the on-shell symmetry algebra, our interest lies mainly in the off-shell symmetries, which are the ones that constrain the S matrix.

\chapter{The all-loop integrable S matrix}
\label{ch:smatrix}
In this chapter we will conjecture the scattering matrix for fundamental massive excitations of the $\AdS_3\times\S^3\times\T^4$ superstring. To do so, we will rely on two results:
\begin{itemize}
\item the off-shell symmetry algebra found in the previous chapter,
\item and the fact that the underlying classical field theory enjoys a large amount of symmetries that make it ``integrable''~\cite{Babichenko:2009dk, Sundin:2012gc,Cagnazzo:2012se}.%
\footnote{Classical integrability is also seen by studying the  Gubser-Klebanov-Polyakov ``spinning string''~\cite{Sundin:2013uca}.}
\end{itemize}
We will supplement these fact by two assumptions, namely
\begin{itemize}
\item that the off-shell symmetries which we found persist in the full quantum theory,
\item and that the integrable structure also extends to the quantum theory.
\end{itemize}
In this way, we will be able to derive an essentially unique S matrix~$\mathbf{S}$, up to some prefactors---the so-called dressing factors.

We begin by exploring the idea of integrability in  classical and quantum theories, and derive restrictions on the resulting S~matrix. Then we will formulate the off-shell symmetries at the level of the S matrix. This will be sufficient to find an~$\mathbf{S}$ which will have several desirable physical properties, guaranteeing the self-consistency of our procedure, as we will see.
Ultimately, the validity of our assumptions will have to be tested by comparing the S~matrix against perturbative calculations. We will come back to this in chapter~\ref{ch:comparison}.

\section{Classical and quantum integrability}
\label{sec:integrability}
Integrability is a broad concept that first emerged in the context of Hamiltonian systems and later was extended to classical and quantum field theories. There is no universal mathematical definition of what an integrable theory is. Colloquially, an integrable theory is one that enjoys so much symmetry so that its dynamics is completely constrained and can be solved ``exactly''. Here we will try to make this idea more precise.
Note that our discussion of this general framework will be quite essential. We refer the reader to \eg~refs.\cite{Babelon,faddeev2007} for further details on classical integrability and the inverse scattering method and to refs.~\cite{Gaudin,Zamolodchikov:1978xm,Faddeev:1996iy} for scattering factorisation and quantum integrable theories.

\subsection*{Classical integrable theories}
\begin{figure}
  \centering
    \includegraphics[width=60mm]{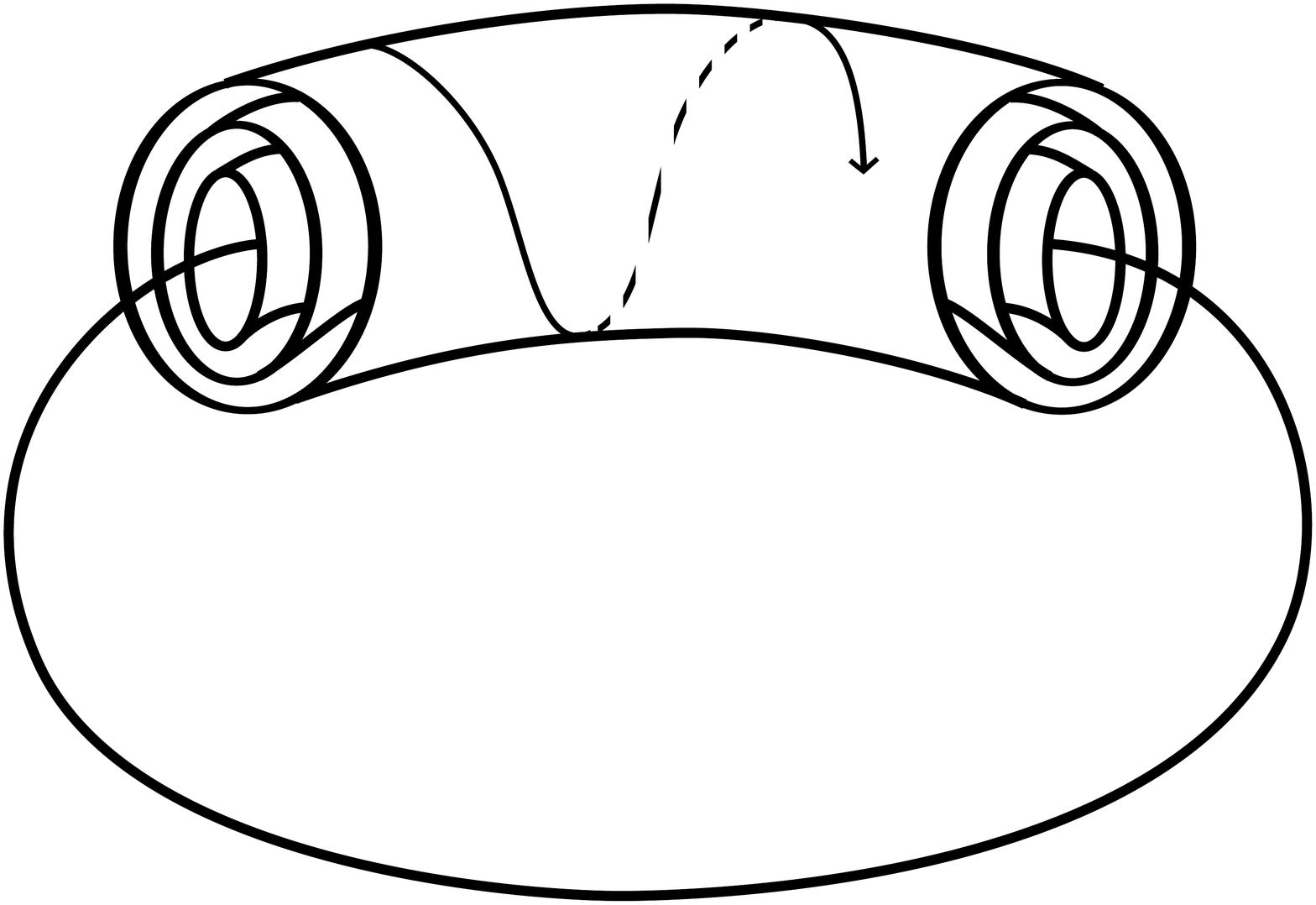}
  \caption{%
    In classical mechanics, the Liouville-Arnol'd theorem establishes that, when enough conserved quantities in involution exist, the phase space is foliated in tori where the dynamics is supported. Here we depict such a foliation, with the arrow representing the motion of a~particle.
  }%
  \label{fig:torifoliation}
\end{figure}

In the case of an Hamiltonian system with $M$ degrees of freedom $(p_1,q_1,\dots, p_{M},q_{M})$, integrability can be defined through the Liouville-Arnol'd theorem. This states that if there exists $M$ Poisson-commuting independent quantities $\mathbf{F}_1,\dots\mathbf{F}_M$, then there exists a foliation of the phase space into invariant tori%
\footnote{If the orbits are compact. The generalisation to non-compact orbits is straightforward.} on which the motion is supported, as depicted in figure~\ref{fig:torifoliation}. Furthermore there exists a canonical transformation to action-angle variables %
$({K}_1,\varphi_1,\dots, {K}_{M},\varphi_{M})$ that takes the equations of motion to the form
\begin{equation}
\dot{{K}}_j=0\,,
\qquad
\dot{\varphi}_{j}=1\,,
\qquad\qquad
j=1,\dots M\,.
\end{equation}
Typically, we know only one of the $\mathbf{F}_j$'s, which is the Hamiltonian. The other charges can sometimes be found from the geometry of the problem. 
When that is possible---which happens quite exceptionally---even if the Hamiltonian appears to be highly non-linear, the theorem guarantees that we are in fact dealing with a simple system ``in disguise''---such as the Kepler problem or the Euler top.

The idea of having enough commuting conserved quantities is at the hearth of integrability, but it is not the most convenient approach to extend it to classical field theory. In that case there are infinitely many degrees of freedom, so that we would need to exhibit infinitely many quantities in involution. This may be possible, but how many should we actually produce? Two infinite sets of charges may have the same cardinality while one is strictly included in the other. A way around this complication is given by the \emph{Lax formalism}%
\footnote{%
Such a formalism can also be very useful when dealing with finite dimensional integrable systems, but this is beyond the purpose of our discussion.
}%
. Consider a two-dimensional classical field theory, and assume that the resulting equations of motion can be cast in the form
\begin{equation}
\label{eq:Lax-equations}
\partial_{\tau}U(\tau,\sigma,x)-\partial_{\sigma}V(\tau,\sigma,x)-\big[V(\tau,\sigma,x),U(\tau,\sigma,x)\big]=0\,,
\end{equation}
where $U,V$ are matrices depending on the fields and on the complex parameter~$x$. Let us define the \emph{monodromy matrix} $T(\tau,x)$ by the path-ordered exponential
\begin{equation}
T(\tau,x)=\;\stackrel{\longleftarrow}{\exp}\left(\int_{-\ell/2}^{\ell/2}\de \sigma\,U(\tau,\sigma,x)\right)\,.
\end{equation}
An explicit computation shows that this obeys the evolution equation
\begin{equation}
\partial_{\tau}T(\tau,x)=\left[V(\tau,-\ell/2,x),\,T(\tau,x)\right]\,.
\end{equation}
As a consequence, $\partial_{\tau}\text{tr}[T(\tau,x)^j]=0$ and all of the eigenvalues of the monodromy matrix are conserved by the time evolution. In fact, an object that nicely encodes all of these conserved quantities is the complex curve defined by the eigenvalue equation
\begin{equation}
\label{eq:spectralcurve}
\Gamma(x,\mu(x))=0\,,
\qquad
\Gamma(x,\mu)=\text{det}\big(T(\tau,x)-\mu\,1\big)\,,
\end{equation}
where $\mu(x)$ is such as to solve the leftmost equation. The resulting~$\Gamma(x)$ is called the \emph{spectral curve}.

On top of this set of conserved quantities, the Lax formulation guarantees more.  Without loss of generality, we may think of the equations of motion~\eqref{eq:Lax-equations} as arising from a linear system%
\footnote{%
In fact, the monodromy matrix takes its name from representing the monodromy of a solution of this linear system, $\Psi(\tau,r,x)=T(\tau,x)\Psi(\tau,-r,x)$.
}
\begin{equation}
\begin{aligned}
\partial_{\sigma}\Psi(\tau,\sigma,x)=\,&U(\tau,\sigma,x)\,\Psi(\tau,\sigma,x)\,,\\
\partial_{\tau}\Psi(\tau,\sigma,x)=\,&V(\tau,\sigma,x)\,\Psi(\tau,\sigma,x)\,,
\end{aligned}
\end{equation}
by requiring the compatibility condition $(\partial_{\sigma}\partial_{\tau} -\partial_{\tau}\partial_{\sigma})\Psi=0$. This gives us the means of solving the non-linear equations~\eqref{eq:Lax-equations} in terms of linear ones---a procedure sometimes called \emph{the inverse scattering method}~\cite{Gardner:1967,Lax:1968,Zakharov:1971,Shabat:1972}.
The matrices $U,V$ are functions of the fields, and in particular $U(0,\sigma,x)$ is completely determined once we fix the initial conditions for the fields at $\tau=0$. From this, finding $\Psi(0,\sigma,x)$ amounts to solving a linear equation.  So does finding the time evolution for~$\Psi(\tau,\sigma,x)$. Once this is determined, the only missing step is to reconstruct $U(\tau,\sigma,x)$ (and therefore the value of the fields at any given $\tau$) from~$\Psi(\tau,\sigma,x)$. This can be done by solving an integral equation, the Gel'fand-Levitan-Marchenko equation~\cite{gelfand1955,marchenko}, which again is linear.

In conclusion, the mere fact that the equations of motion for a two-dimensional classical field theory can be cast in the form~\eqref{eq:Lax-equations} implies that the system has infinitely many conserved charges and that its equations of motions are relatively simple to solve. Moreover, for such systems it is generally possible to relate the matrix structure with the Poisson structure of the original Hamiltonian description in such a way as to guarantee that the conserved quantities Poisson-commute.

\subsection*{Integrability for two-dimensional QFTs}
Let us assume that we can quantise an integrable theory as the one described before, and that we can do so without spoiling its infinitely many symmetries.
Let
\begin{equation}
\mathbf{F}_1,\dots\mathbf{F}_n,\dots\,,
\qquad
[\mathbf{F}_j,\,\mathbf{F}_k]=0
\end{equation}
be the set of commuting conserved charges. These charges can be simultaneously diagonalised. For definiteness, let us suppose that this happens in a basis where Hilbert space states are identified by their momentum and a flavour label~$\a$. Then we will have
\begin{equation}
\mathbf{F}_n\,\Ket{p}^{(\IN ,\OUT)}_{\alpha}=
F_n(p;\alpha)\Ket{p}^{(\IN ,\OUT)}_{\alpha}\,.
\end{equation}
For instance, in the particular case of the sine-Gordon model where there is only one scalar excitation of mass~$m$, one would have
\begin{equation}
F_{2n+1}(p)=p^{2n+1}\,,
\qquad
F_{2n}(p)=p^{2n}\,\sqrt{p^2+m^2}\,,
\end{equation}
which give \emph{higher charges} generalizing momentum and energy.
In a more general theory, the higher charges will feature some invariant tensor constructed out of the labels $\alpha$ and infinitely many functions of the particle's momentum.

Let us consider the action of such a charge on an $M$-particle state. Considering \eg\ an in-state, we have
\begin{equation}
\mathbf{F}_n\,
\Ket{p_1,\dots p_M}^{(\IN)}_{\alpha_1,\dots \alpha_M}=
\big(
F_n(p_1;\alpha_1)+
\cdots +
F_n(p_M;\alpha_M)
\big)
\Ket{p_1,\dots p_M}^{(\IN)}_{\alpha_1,\dots \alpha_M}\,.
\end{equation}
After we evolve the state $\Ket{p_1,\dots p_M}^{(\IN)}_{\alpha_1,\dots \alpha_M}$ we obtain some corresponding out-state, which we denote by $\Ket{\tilde{p}_1,\dots \tilde{p}_{\tilde{M}}}^{(\OUT)}_{\tilde{\alpha}_1,\dots \tilde{\alpha}_{\tilde{M}}}$.
Since the charges are conserved, it must be that
\begin{equation}
\label{eq:ZFconservationlaw}
\sum_{j=1}^{M}
F_n(p_j;\alpha_{j})
=
\sum_{{k}=1}^{\tilde{M}}
F_n(\tilde{p}_k;\tilde{\alpha}_{k})\,.
\end{equation}
The only way for these sums to be equal~\emph{for all of the} $F_{n}(p)$ is if the set of ``in''~momenta appearing $\{p_j\}$  corresponds to the set of ``out''~momenta $\{\tilde{p}_k\}$. In particular, it must be~$M=\tilde{M}$.

This scenario, together with the peculiar topology of two-dimensional QFTs, has deep implications for a scattering event. Let us start by considering a two-particle scattering event. The in-state is $\Ket{p_1, p_2}^{(\IN)}_{\alpha_1,\alpha_2}$. Since this state is defined at time $\tau=-\infty$, momenta are ordered so that $p_1>p_2$. As time evolves, the particles move on a line until at some point they come together and scatter. After the scattering, the products move away from each other and can again be considered as two real particles. Because of the conservation law~\eqref{eq:ZFconservationlaw}, the resulting state is proportional to $\Ket{p_2, p_1}_{\tilde{\alpha}_2,\tilde{\alpha}_1}$, where the momenta are exactly the same and the labels $\tilde{\alpha}_2,\tilde{\alpha}_1$ may have changed.

This can be readily extended to a $M$-particle in-state $\Ket{p_1,\dots p_M}^{(\IN)}_{\alpha_1,\dots \alpha_M}$ with $p_1>p_2>\dots>p_M$. At some point two of the particles undergo a $2\to2$ scattering like the one described above. After that, the we are left with $M$ real (as opposed to virtual) particles propagating on a line. After a sequence of $\tfrac{1}{2}M(M-1)$ scattering events, the particles are spatially ordered as in the out-state, having momenta~$p_{M},p_{M-1},\dots p_{1}$ from left to right.
Therefore, an $M$-particle scattering event \emph{factorises} into a sequence of two-particle events, and was first put forward by the Zamolodchikov brothers in the seminal paper~\cite{Zamolodchikov:1978xm}.
%
%
 This is the special property of the multiparticle S matrix that we alluded to in the introduction. The object that we need to determine is only the two-particle S~matrix~$\mathbf{S}(p_1,p_2)$.

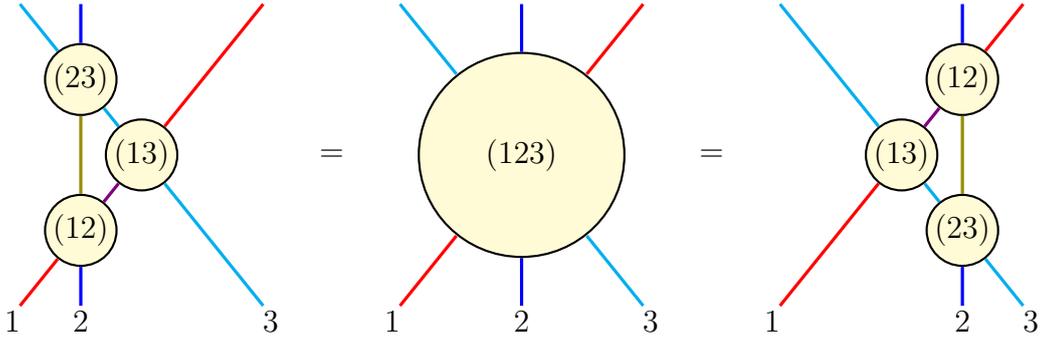
\begin{figure}
  \centering
\begin{tikzpicture}
  \begin{scope}[xshift=-1cm]
    \coordinate (i1) at (-1.6cm,0);
    \coordinate (i2) at (-0.8,    0);
    \coordinate (i3) at (+1.6cm,0);

    \coordinate (o1) at (-1.6cm,4cm);
    \coordinate (o2) at (-0.8, 4cm);
    \coordinate (o3) at (+1.6cm,4cm);

    \node (v1) at (-0.8cm,1cm) [S-mat] {${(12)}$};
    \node (v2) at ( 0cm, 2cm)    [S-mat] {${(13)}$};
    \node (v3) at (-0.8cm,3cm) [S-mat] {${(23)}$};

    \draw [very thick,red]    (i1) to (v1);
    \draw [very thick,blue]   (i2) to (v1);
    \draw [very thick,cyan] (i3) to (v2);
    
    \draw [very thick,violet]    (v1) to (v2);
    \draw [very thick,olive]   (v1) to (v3);
    \draw [very thick,cyan] (v2) to (v3);
    
    \draw [very thick,red]    (v2) to (o3);
    \draw [very thick,blue]   (v3) to (o2);
    \draw [very thick,cyan] (v3) to (o1);
    
\node at (-1.7cm,-0.2cm) {$1$};
\node at (-0.8cm,-0.2cm) {$2$};
\node at (+1.7cm,-0.2cm) {$3$};

  \end{scope}
  \node at (1.5,2cm) {$=$};
  \begin{scope}[xshift=4cm]
    \coordinate (i1) at (-1.6cm,0);
    \coordinate (i2) at (0,    0);
    \coordinate (i3) at (+1.6cm,0);

    \coordinate (o1) at (-1.6cm,4cm);
    \coordinate (o2) at (0, 4cm);
    \coordinate (o3) at (+1.6cm,4cm);

    \node (v0) at (0,2cm) [S-mat] {${\qquad(123)\qquad}$};

    \draw [very thick,red]    (i1) to (v0);
    \draw [very thick,blue]   (i2) to (v0);
    \draw [very thick,cyan] (i3) to (v0);
    
    \draw [very thick,red]    (v0) to (o3);
    \draw [very thick,blue]   (v0) to (o2);
    \draw [very thick,cyan] (v0) to (o1);

\node at (-1.7cm,-0.2cm) {$1$};
\node at (0cm,-0.2cm) {$2$};
\node at (+1.7cm,-0.2cm) {$3$};

  \end{scope}
  \node at (6.5cm,2cm) {$=$};
  \begin{scope}[xshift=9cm]
    \coordinate (i1) at (-1.6cm,0);
    \coordinate (i2) at (0.8,    0);
    \coordinate (i3) at (+1.6cm,0);

    \coordinate (o1) at (-1.6cm,4cm);
    \coordinate (o2) at (0.8, 4cm);
    \coordinate (o3) at (+1.6cm,4cm);

    \node (v1) at (0.8cm,1cm) [S-mat] {${(23)}$};
    \node (v2) at ( 0cm, 2cm)    [S-mat] {${(13)}$};
    \node (v3) at (0.8cm,3cm) [S-mat] {${(12)}$};

    \draw [very thick,red]    (i1) to (v2);
    \draw [very thick,blue]   (i2) to (v1);
    \draw [very thick,cyan] (i3) to (v1);
    
    \draw [very thick,violet]    (v2) to (v3);
    \draw [very thick,olive]   (v1) to (v3);
    \draw [very thick,cyan] (v1) to (v2);
    
    \draw [very thick,red]    (v3) to (o3);
    \draw [very thick,blue]   (v3) to (o2);
    \draw [very thick,cyan] (v2) to (o1);
    
\node at (-1.7cm,-0.2cm) {$1$};
\node at (0.8cm,-0.2cm) {$2$};
\node at (+1.7cm,-0.2cm) {$3$};

  \end{scope}
\end{tikzpicture}%

  \caption{Different ways of  resolving the $3\to3 $ particle scattering (central panel). The lines are colored in such a way as to allow for a non-trivial flavor structure, which may be affected by the scattering. Taking time to flow upwards, we have in the left panel the sequence of scattering~(12)-(13)-(23), while in the right panel we have~(23)-(13)-(12).
  The Yang-Baxter equation imposes the equality of the two resolutions.}
  \label{fig:YBE-Rmat}
\end{figure}

For internal consistency, we have to require some conditions on the S matrix. The most obvious one is that there is no $2\to M$ scattering unless $M=2$, in sharp contrast with what happens \eg\ in a typical particle collider experiment. Another obstacle is that there are several apparently inequivalent ways of resolving an $M\to M$ scattering in terms of a sequence of $2\to 2$ ones. In figure~\ref{fig:YBE-Rmat} we depict a generic $3\to3$ particle scattering, and the two ways of resolving it. We must require that these two sequences of scattering events yield the same result. Each of them gives a cubic expression in the S~matrix, and their equality results in a non-linear matrix equation---the \emph{Yang-Baxter equation}. The physical reason for  the equivalence of the pictures in fig.~\ref{fig:YBE-Rmat} can be traced back to the existence of the higher conserved charges $\mathbf{F}_n$. These generate unitary transformations that shift the momentum~$p_j$ of a particle by an amount of order~$(p_j)^{n-1}$, and effectively transform the leftmost panel of fig.~\ref{fig:YBE-Rmat} into the rightmost one.

To better investigate the constraints that~$\mathbf{S}(p_{1},p_{2})$ should satisfy it is convenient to introduce a more formal algebraic framework.

\subsection*{The Zamolodchikov-Faddeev algebra}
The Zamolodchikov-Faddeev (ZF) algebra~\cite{Zamolodchikov:1978xm,Faddeev:1980zy} is a tool to encode the integrability properties of a two-dimensional QFT. It is defined in terms of abstract raising and lowering operators $A_{\a}^{\dagger}(p)$ and $A^{\a}(p)$ which create or destroy a particle of definite momentum and flavour. Their action on the vacuum is
\begin{equation}
\begin{aligned}
A^{\a}(p)\,\Ket{0}=&0\,,\\
A_{\a}^{\dagger}(p)\,\Ket{0}=&\Ket{p}^{(\IN)}_\a=\Ket{p}^{(\OUT)}_\a\,,\\
\end{aligned}
\end{equation}
where we used the fact that in an integrable theory like the one we want to describe a one-particle state undergoes a trivial time evolution. Following the discussion of the previous section, it is natural to relate the two-particle in- and out-states as
\begin{equation}
\label{eq:2part-ZF}
\begin{aligned}
\Ket{p_1,p_2}^{(\IN)}_{\a_1,\a_2}=&
A_{\a_1}^{\dagger}(p_1)\,A_{\a_2}^{\dagger}(p_2)\Ket{0}\,,\\
\Ket{p_1,p_2}^{(\OUT)}_{\a_3,\a_4}=&
(-1)^{\eps(\a_{3})\eps(\a_{4})}
A_{\a_3}^{\dagger}(p_2)\,A_{\a_4}^{\dagger}(p_1)\Ket{0}\,,\\
\end{aligned}
\qquad
\text{with }p_1 > p_2\,.
\end{equation}
Notice that we used that the set of momenta is conserved and we ordered the action of the creation and annihilation operators according to the ordering of the particles in the final states. We have also written explicitly the sign arising from the exchange of two fermions, where $\eps(\a)=1$ for fermionic particles and zero otherwise.

The two states in~\eqref{eq:2part-ZF} are precisely the ones related by the S~matrix from~\eqref{eq:genericSmat} in the $2\to2$ case
\begin{equation}
\Ket{p_1,p_2}^{(\IN)}_{\a_1,\a_2}=
{S}_{\a_{1},\a_{2}}^{\a_{3},\a_{4}}(p_1,p_2)\,
\Ket{p_1,p_2}^{(\OUT)}_{\a_3,\a_4}\,,
\end{equation}
where we indicated by~${S}_{j_{1},j_{2}}^{j_{3},j_{4}}$ the matrix element of $\mathbf{S}$. This equation, together with~\eqref{eq:2part-ZF}, yields the commutation relations for the ZF operators
\begin{equation}
\label{eq:ZF-fundamental-1}
A_{\a_1}^{\dagger}\,(p_1)A_{\a_2}^{\dagger}(p_2)=
(-1)^{\eps(\a_{3})\eps(\alpha_{4})}{S}_{\a_{1},\a_{2}}^{\a_{3},\a_{4}}(p_1,p_2)\,
A_{\a_3}^{\dagger}\,(p_2)A_{\a_4}^{\dagger}(p_1)\,,
\end{equation}
which makes it evident that $A^\dagger,A$ satisfy \emph{a different algebra} than the one of the canonical raising and lowering operators~$a^\dagger,a$.

To avoid carrying around too many indices it is useful to introduce a matrix basis. Let $E_{j}$ and $E^{j}$ be rows and column vectors with a single non-vanishing unit entry in the $j$th place. These gives bases of the two dual vector spaces $\mathscr{V}^*$ and $\mathscr{V}$. A basis of the space of matrices on~$\mathscr{V}$ is given by $E_{j}^{\ k}=E_j\otimes E^k$, so that by indices are naturally contracted to give
\begin{equation}
E^l E_{k}^{\ j}=\delta^{l}_{k}\,E^{j}\,,
\qquad
E_{l} E_{j}^{\ k}=\delta^{k}_{k}\,E_{j}\,,
\qquad
E_{k}^{\ l}E_{i}^{\ j}=\delta^{l}_{i}\,E^{\ j}_{k}\,.
\end{equation}
We can then introduce vector and row operators of the ZF algebra
\begin{equation}
\mathbf{A}^\dagger=\mathbf{A}^\dagger(p)= A^\dagger_\a(p)\,E^\a\,,
\qquad
\mathbf{A}=\mathbf{A}(p)=A^\a(p)\,E_\a\,.
\end{equation}
It is also useful to reabsorb a permutation and the fermionic signs into the~$\mathbf{R}$ matrix
\begin{equation}
\mathbf{R}(p,q)=
{R}_{\a_{1},\a_{2}}^{\a_{3},\a_{4}}(p,q)\,E_{\a_3}^{\ \a_1}\otimes E_{\a_4}^{\ \a_2}\,,\qquad
{R}_{\a_{1},\a_{2}}^{\a_{4},\a_{3}}(p,q)=
(-1)^{\eps(\a_{3})\eps(\a_{4})}{S}_{\a_{1},\a_{2}}^{\a_{3},\a_{4}}(p,q)\,,
\end{equation}
so that $\mathbf{R}$ differs from $\mathbf{S}$ by a graded permutation~$\Pi^g$, $\mathbf{R}=\Pi^g\,\mathbf{S}$. In these terms the commutation relations~\eqref{eq:ZF-fundamental-1} take the form
\begin{equation}
\mathbf{A}^\dagger_{(1)}
\mathbf{A}^\dagger_{(2)}=
\mathbf{A}^\dagger_{(2)}
\mathbf{A}^\dagger_{(1)}
\mathbf{R}_{(12)}\,,
\end{equation}
where we added subscript indices specifying on which factors of the tensor product~$\mathscr{V}\otimes\mathscr{V}$ the operators act.

In the same way as we have obtained~\eqref{eq:ZF-fundamental-1} we can now derive similar relations involving the annihilation operators, thus completing the ZF algebra
\begin{equation}
\label{eq:ZF-algebra-full}
\begin{aligned}
\mathbf{A}^\dagger_{(1)}
\mathbf{A}^\dagger_{(2)}=\,
\mathbf{A}^\dagger_{(2)}
\mathbf{A}^\dagger_{(1)}
\mathbf{R}_{(12)}\,,
& \qquad\qquad\qquad
\mathbf{A}_{(1)}
\mathbf{A}_{(2)}=\,
\mathbf{R}_{(12)}
\mathbf{A}_{(2)}
\mathbf{A}_{(1)}\,,
\\
\mathbf{A}_{(1)}
\mathbf{A}^\dagger_{(2)}=&\,
\mathbf{A}_{(2)}
\mathbf{R}_{(21)}
\mathbf{A}^\dagger_{(1)}+\delta(p_1-p_2)\,\mathbf{I}\,.
\end{aligned}
\end{equation}

\subsubsection*{Consistency conditions}
The algebra~\eqref{eq:ZF-algebra-full} in principle consitutes a tool to express an in-particle state in terms of the out-particle basis and \textit{vice versa}, \ie\ a way of computing S-matrix elements.
For this to be true, however, we have to impose some further consistency conditions. We immediately find that
\begin{equation}
\mathbf{A}^\dagger_{(1)}
\mathbf{A}^\dagger_{(2)}=\,
\mathbf{A}^\dagger_{(2)}
\mathbf{A}^\dagger_{(1)}
\mathbf{R}_{(12)}=
\mathbf{A}^\dagger_{(1)}
\mathbf{A}^\dagger_{(2)}
\mathbf{R}_{(21)}\mathbf{R}_{(12)}\,,
\end{equation}
so that it must be
\begin{equation}
\label{eq:braidingunitarity}
\mathbf{R}_{(21)}\mathbf{R}_{(12)}=
\mathbf{R}_{(12)}\mathbf{R}_{(21)}=
\mathbf{I}\,.
\end{equation}
This condition, called \emph{braiding unitarity}, supplements the usual \emph{physical unitarity} condition, which states that~$\mathbf{S}$ and therefore~$\mathbf{R}$ should be unitary as a matrix:
\begin{equation}
\label{eq:physicalunitarity}
\mathbf{R}_{(12)}\,\mathbf{R}_{(12)}^\dagger=
\mathbf{R}_{(12)}^\dagger\,\mathbf{R}_{(12)}=
\mathbf{I}\,.
\end{equation}

So far we have used the ZF algebra only on two-particle states. We now want to extend it to arbitrary multiparticle states in such a way as to implement the factorisation of the scattering. To do so,
we extend the definition~\eqref{eq:2part-ZF} by
\begin{equation}
\begin{aligned}
\Ket{p_1,\dots p_M}^{(\IN)}_{\a_1,\dots \a_M}=&
A_{\a_1}^{\dagger}(p_1)\cdots A_{\a_M}^{\dagger}(p_M)\Ket{0}\,,\\
\Ket{p_1,\dots p_M}^{(\OUT)}_{{\a}_1,\dots {\a}_M}=&
(-1)^{\sum_{k<l}\eps(\a_{k})\eps(\a_{l})}
A_{\a_1}^{\dagger}(p_M)\cdots A_{\a_M}^{\dagger}(p_1)\Ket{0}\,,\\
\end{aligned}
\end{equation}
again with $p_1>p_2>\cdots>p_M$. If we want to use~\eqref{eq:ZF-algebra-full} to express in-states in terms of out-states, we are faced with an ambiguity. Consider, for instance, the combination $\mathbf{A}^{\dagger}_{(1)} \mathbf{A}^{\dagger}_{(2)} \mathbf{A}^{\dagger}_{(3)}$ acting on $\mathscr{V}\otimes\mathscr{V}\otimes\mathscr{V}$. We can rewrite it in two different ways:
\begin{equation}
\begin{aligned}
\mathbf{A}^{\dagger}_{(1)} \mathbf{A}^{\dagger}_{(2)} \mathbf{A}^{\dagger}_{(3)} = &\, 
\mathbf{A}^{\dagger}_{(3)} \mathbf{A}^{\dagger}_{(2)} \mathbf{A}^{\dagger}_{(1)}
\mathbf{R}_{(12)}
\mathbf{R}_{(13)}
\mathbf{R}_{(23)}\,,
\\
\mathbf{A}^{\dagger}_{(1)} \mathbf{A}^{\dagger}_{(2)} \mathbf{A}^{\dagger}_{(3)} = &\, 
\mathbf{A}^{\dagger}_{(3)} \mathbf{A}^{\dagger}_{(2)} \mathbf{A}^{\dagger}_{(1)}
\mathbf{R}_{(23)}
\mathbf{R}_{(13)}
\mathbf{R}_{(12)}\,.
\end{aligned}
\end{equation}
These coincide only provided that
\begin{equation}
\label{eq:YBE-Rmatrix}
\mathbf{R}_{(12)}
\mathbf{R}_{(13)}
\mathbf{R}_{(23)}
=
\mathbf{R}_{(23)}
\mathbf{R}_{(13)}
\mathbf{R}_{(12)}\,.
\end{equation}
This is the \emph{Yang-Baxter equation} that we described in the previous subsection. In fact, comparing with figure~\ref{fig:YBE-Rmat} one can see that the product of the matrices appearing in the equation precisely corresponds to the sequences of scattering events depicted there.
It is then straightforward to see that, by repeatedly using the Yang-Baxter equation, it is possible to rearrange a string of ZF operators of arbitrary length, which ensures factorisation of any multiparticle scattering.

\subsubsection*{Symmetries}
In our subsequent study, it will be important to make use of the transformation properties of S~matrix under the off-shell symmetry algebra.

The simplest case is given by the central charges, which make up the whole bosonic part of the off-shell symmetries. Let us consider a charge that is proportional to the number operator
\begin{equation}
\mathbf{X}_{f}=\int \de p\,f(p)\,\mathbf{A}^\dagger(p)\,\mathbf{A}(p)\,,
\end{equation}
where $f(p)$ is an arbitrary function of the momentum. Using braiding unitarity it is easy to prove~\cite{Arutyunov:2009ga} that $\mathbf{X}_{f}$ must satisfy the relations
\begin{equation}
\label{eq:abelian-ZF}
\mathbf{X}_{f}\,\mathbf{A}^\dagger(p)=
\mathbf{A}^\dagger(p)\,
\big(f(p)+\mathbf{X}_{f}\big)\,,
\qquad
\mathbf{X}_{f}\,\mathbf{A}(p)=
\mathbf{A}(p)\,
\big(-f(p)+\mathbf{X}_{f}\big)\,.
\end{equation}
Therefore, these functions must form an abelian subalgebra of the ZF  algebra. In particular, we expect the worldsheet momentum, the Hamiltonian and the two central charges found in the previous section to be part of it.

Let us now consider a more general non-abelian (super)algebra~$\mathcal{G}$, and let us assume that the one-particle vector space~$\mathscr{V}$ carries one or more irreducible representations of~$\mathcal{G}$. Eventually we will identify~$\mathcal{G}$ with the full off-shell algebra.
It is then natural to take~$\mathcal{G}$ to commute with the worldsheet momentum and the particle number%
\footnote{%
Since we are dealing with a supersymmetric theory, we also require that~$(-1)^{\mathcal{F}}$, where~$\mathcal{F}$ is the fermion number, to be conserved. Note that~$\mathcal{F}$ itself is generally not conserved, resulting in processes such as fermion-fermion~$\to$~boson-boson.%
}
 as well as with all the higher conserved charges---compatibly with our assumption of integrability. Then we can write the linear action of a symmetry generator~$\mathbf{Q}\in\mathcal{G}$ on the zero-, one-, two-particle Hilbert spaces, \etc\ as
\begin{equation}
\begin{aligned}
&\mathbf{Q}\cdot\Ket{0}=\,
0\,,\\
&\mathbf{Q}\cdot A^{\dagger}_{\a}(p)\Ket{0}=\,
{Q}_{\a}^{\b}(p)\,A^{\dagger}_{\b}(p)\Ket{0}\,,\\
&\mathbf{Q}\cdot A^{\dagger}_{\a_1}(p)\,A^{\dagger}_{\a_2}(q)\Ket{\Omega}=\,
{Q}_{\a_1,\a_2}^{\b_1,\b_2}(p,q)\,A^{\dagger}_{\b_1}(p)\,A^{\dagger}_{\b_2}(q)\Ket{0}\,,
\end{aligned}
\end{equation}
and so on.

Since $\mathbf{Q}_{\alpha}$ is a symmetry of the theory, we can simultaneously transform the in- and out-states without affecting the S-matrix elements. In particular using~\eqref{eq:2part-ZF} in
\begin{equation}
\mathbf{Q}\cdot\Ket{p_1,p_2}^{(\IN)}_{\a_1,\a_2}=
\mathbf{S}\cdot\mathbf{Q}\cdot
\Ket{p_1,p_2}^{(\OUT)}_{\a_3,\a_4}\,,
\end{equation}
we get the invariance condition
\begin{equation}
\begin{aligned}
S_{\b_3,\b_4}^{\a_3,\a_4}(p_1,p_2)\,
Q_{\a_1,\a_2}^{\b_3,\b_4}(p_1,p_2)=&
(-1)^{\eps(\a_3)\eps(\a_4)+\eps(\b_3)\eps(\b_4)}\\
&\qquad
Q_{\b_3,\b_4}^{\a_3,\a_4}(p_2,p_1)\,
S^{\b_3,\b_4}_{\a_1,\a_2}(p_1,p_2)\,,
\end{aligned}
\end{equation}
which can be more compactly expressed in the matrix notation in terms of~$\mathbf{R}$:
\begin{equation}
\label{eq:R-matrix-comm-rel}
\mathbf{R}_{(12)}(p_1,p_2)\,\mathbf{Q}_{(12)}(p_1,p_2)=
\mathbf{Q}_{(21)}(p_2,p_1)\,\mathbf{R}_{(12)}(p_1,p_2)\,.
\end{equation}

Let us further investigate the form of the structure constants. Since our symmetry algebra~$\mathcal{G}$ has a non-trivial centre, any of its irreducible representations is labelled by the value of the momentum and the other central charges~$\{c\}$. Therefore we in principle we have to allow for%
\footnote{%
This notation is a bit heavy, and we will keep the dependence on the central charges implicit where no confusion may arise.
}
\begin{equation}
\begin{aligned}
&{Q}_{\a}^{\b}(p)
&\to &
\qquad
{Q}_{\a}^{\b}(p;\{c\})\,,
\\
&Q_{\a_1,\a_2}^{\b_1,\b_2}(p_1,p_2)
&\to  &
\qquad
Q_{\a_1,\a_2}^{\b_1,\b_2}(p_1,p_2;\{c_1,c_2\})\,.
\end{aligned}
\end{equation}
In a matrix notation, we can decompose the action of $\mathbf{Q}_{\a\,(12)}$ on the factors of the tensor product as
\begin{equation}
\label{eq:generic-2part-charge}
\begin{aligned}
\mathbf{Q}_{(12)}(p_1,p_2;\{c_1,c_2\})=
\mathbf{Q}(p_1,\{c_1\})\otimes \mathbf{I}+
\mathbf{I}^{g}\big(
\mathbf{I}\otimes\mathbf{Q}(p_2,\{c_2\})
\big)\mathbf{I}^{g},
\end{aligned}
\end{equation}
where $\mathbf{I}^{g}$ is the graded identity matrix.

In order to further specify the form of this representation we will need to input some information on the symmetry algebra. We will do so in the next section.

\section{Representations of \texorpdfstring{$\su(1|1)^2$}{su(1|1)**2} centrally extended}
\label{sec:fullrepr}
Rather than focusing on the whole off-shell symmetry algebra~$\psu(1|1)^4_{\text{c.e.}}$, let us focus on a single $\su(1|1)^2_{\text{c.e.}}$. The commutation relations are given by~\eqref{eq:smallalgebra-nonlin}, and we rewrite them here for convenience
\begin{equation}
\label{eq:smallalgebra-nonlin2}
\begin{aligned}
&\big\{\mathbf{q}^{\smallL},\bar{\mathbf{q}}^{\smallL}\big\}=\mathbf{h}^{\smallL}\,,
&\qquad &
\big\{\mathbf{q}^{\smallR},\bar{\mathbf{q}}^{\smallR}\big\}=\mathbf{h}^{\smallR}\,,\\
&\big\{\mathbf{q}^{\smallL},{\mathbf{q}}^{\smallR}\big\}=\mathbf{c}\,,
&\qquad &
\big\{\bar{\mathbf{q}}^{\smallL},\bar{\mathbf{q}}^{\smallR}\big\}=\bar{\mathbf{c}}\,,\\
\end{aligned}
\end{equation}
where we introduce a left- and right-``Hamiltonian'' $\mathbf{h}^{\smallL,\smallR}$ so that
\begin{equation}
\mathbf{h}=\mathbf{h}^{\smallL}+\mathbf{h}^{\smallR}\,,
\qquad
\mathbf{m}=\mathbf{h}^{\smallL}-\mathbf{h}^{\smallR}\,.
\end{equation}
We already know that how to represent this algebra on the space of left- and right-excitations $\phi^{\smallL,\smallR},\psi^{\smallL,\smallR}$ when $\mathbf{c}=\bar{\mathbf{c}}=-\frac{1}{2}\mathbf{p}$. In terms of oscillators, that representation is given by~\eqref{eq:smallrepr-ferm} and~\eqref{eq:smallrepr-bos}. Here we want to deform that representation to allow for
\begin{equation}
\label{eq:cbarc-p}
{\mathbf{c}}=+i\zeta\,\frac{h}{2}\big(e^{+i\mathbf{p}}-1\big)\,,
\qquad
\bar{\mathbf{c}}=-i\bar{\zeta}\,\frac{h}{2}\big(e^{-i\mathbf{p}}-1\big)\,.
\end{equation}
This will amount to suitably deforming the parameters~$f_p,\,g_p$ and $\omega_p$ appearing in the representation. The deformed parameters will depend on the momentum~$p$, the coupling constant~$h$ and the phase~$\zeta$ that we found in~\eqref{eq:zeta-def}.

The phase~$\zeta$ has the meaning of a boundary condition on the unphysical field $x_{-}$ in the coset model, $\zeta=e^{i x_{-}(-\infty)}$. When we want the representation of~$\su(1|1)^2_{\text{c.e.}}$ to describe excitations of the superstring, we will set $\zeta=1$. Then
\begin{equation}
\label{eq:cbarc-p-nozeta}
{\mathbf{c}}=+i\,\frac{h}{2}\big(e^{+i\mathbf{p}}-1\big)\,,
\qquad
\bar{\mathbf{c}}=-i\,\frac{h}{2}\big(e^{-i\mathbf{p}}-1\big)\,.
\end{equation}

\subsection*{Shortening condition}
In the previous chapter, we have found two irreducible representations of dimension two. These are both short~(or atypical) representations of $\su(1|1)^2_{\text{c.e.}}$. In fact, if we consider \eg\ the left representation, we have that $\ket{\phi^{\smallL}}$ is the highest weight state, annihilated by the two raising operators
\begin{equation}
\bar{\mathbf{q}}^{\smallL}\,\ket{\phi^{\smallL}}=0\,,
\qquad
{\mathbf{q}}^{\smallR}\,\ket{\phi^{\smallL}}=0\,.
\end{equation}
However, the highest weight is also annihilated by a combination of the lowering operators
\begin{equation}
\label{eq:shortening1}
\big(\mathbf{h}^{\smallR}\,\mathbf{q}^{\smallL}-\mathbf{c}\,\bar{\mathbf{q}}^{\smallR}\big)\,\ket{\phi^{\smallL}}=0\,.
\end{equation}
The vanishing of this combination of charges is the \emph{shortening condition}. It also implies the vanishing of the anticommutator
\begin{equation}
\label{eq:shortening2}
0=\big\{\bar{\mathbf{q}}^{\smallL},\,\mathbf{h}^{\smallR}\,\mathbf{q}^{\smallL}-\mathbf{c}\,\bar{\mathbf{q}}^{\smallR}\big\}=
\mathbf{h}^{\smallL}\,\mathbf{h}^{\smallR}-\bar{\mathbf{c}}\,\mathbf{c}
\,,
\end{equation}
which we will also refer to as shortening condition.

Had we considered the right-representation, we would have found a fermionic highest weight state
\begin{equation}
\bar{\mathbf{q}}^{\smallL}\,\ket{\psi^{\smallR}}=0\,,
\qquad
{\mathbf{q}}^{\smallR}\,\ket{\psi^{\smallR}}=0\,.
\end{equation}
The same combination appearing in~\eqref{eq:shortening1} annihilates~$\ket{\psi^{\smallR}}$. In fact, we can take~\eqref{eq:shortening2} as shortening condition of both the left- and right-representations.

\subsection*{One-particle representation}
Since the excitations span a space~$\mathscr{V}$ of dimension only four, it is quite handy to introduce a $4\times4$~matrix representation for the (super)charges. Let us pick a basis
\begin{equation}
\label{eq:B-basis}
\mathscr{B}=\left(\phi^{\smallL},\,\psi^{\smallL},\,\phi^{\smallR},\,\psi^{\smallR}\right)\,.
\end{equation}
We can then make an ansatz for the supercharges in the one-particle representation,
\begin{equation}
\label{eq:supercharges-1part-matrix}
\begin{aligned}
&\mathbf{q}^{\smallL}=
\left(
\begin{array}{cc|cc}
0	&	0	&	0	&	0	\\
a^{\smallL}&	0	&	0	&	0	\\
\hline
0	&	0	&	0	&	\bar{b}^{\smallR}\\
0	&	0	&	0	&	0	
\end{array}
\right),
\qquad
&{\mathbf{q}}^{\smallR}=
\left(
\begin{array}{cc|cc}
0	&	{b}^{\smallL}	&	0	&	0	\\
0	&	0	&	0	&	0	\\
\hline
0	&	0	&	0	&	0	\\
0	&	0	&{a}^{\smallR}&	0	
\end{array}
\right),\\
&\bar{\mathbf{q}}^{\smallL}=
\left(
\begin{array}{cc|cc}
0	&	\bar{a}^{\smallL}	&	0	&	0	\\
0	&	0	&	0	&	0	\\
\hline
0	&	0	&	0	&	0	\\
0	&	0	&	\bar{b}^{\smallR}	&	0	
\end{array}
\right),
\qquad
&\bar{\mathbf{q}}^{\smallR}=
\left(
\begin{array}{cc|cc}
0	&	0	&	0	&	0	\\
\bar{b}^{\smallL}	&	0	&	0	&	0	\\
\hline
0	&	0	&	0	&	\bar{a}^{\smallR}	\\
0	&	0	&	0	&	0	
\end{array}
\right),
\end{aligned}
\end{equation}
where we made it manifest that~$(\mathbf{q}^{\smallL,\smallR})^\dagger=\bar{\mathbf{q}}^{\smallL,\smallR}$. The parameters $a^{\smallL,\smallR}, b^{\smallL,\smallR}$ and their conjugates characterise the L- and R-representations, which sit in the diagonal matrix blocks. We can think of these matrices as explicit realisations of the tensors~${Q}_{j}^{k}(p;\{c\})$ introduced in the previous section. The set of parameters~$\{c\}$ consists of $h$ and $\zeta$, on which $a^{\smallL,\smallR}, b^{\smallL,\smallR}$ and their conjugates depend.

From the anticommutators, we find immediately the central charges
\begin{equation}
\begin{aligned}
&\mathbf{h}^{\smallL}
=
\left(
\begin{array}{c|c}
\phantom{\Big[}|a^{\smallL}|^2\, \mathbf{I}\phantom{\Big]}&		0		\\
\hline
0		&	\phantom{\Big[}|b^{\smallR}|^2\, \mathbf{I}\phantom{\Big]}
\end{array}
\right),
\quad
&\mathbf{h}^{\smallR}
=
\left(
\begin{array}{c|c}
\phantom{\Big[}|b^{\smallL}|^2\, \mathbf{I}\phantom{\Big]}&		0		\\
\hline
0		&	\phantom{\Big[}|a^{\smallR}|^2\, \mathbf{I}\phantom{\Big]}
\end{array}
\right),\\
&\mathbf{c}^{\phantom{\smallL}}
=
\left(
\begin{array}{c|c}
\phantom{\Big[}{a}^{\smallL}b^{\smallL}\, \mathbf{I}\phantom{\Big]}&		0		\\
\hline
0		&	\phantom{\Big[}{a}^{\smallR}b^{\smallR}\, \mathbf{I}\phantom{\Big]}
\end{array}
\right),
\quad
&\bar{\mathbf{c}}^{\phantom{\smallR}}
=
\left(
\begin{array}{c|c}
\phantom{\Big[}\bar{a}^{\smallL}\bar{b}^{\smallL}\, \mathbf{I}\phantom{\Big]}&		0		\\
\hline
0		&	\phantom{\Big[}\bar{a}^{\smallR}\bar{b}^{\smallR}\, \mathbf{I}\phantom{\Big]}
\end{array}
\right).
\end{aligned}
\end{equation}

It can be explicitly checked that the shortening condition holds. Furthermore, can use the fact that the angular momentum
\begin{equation}
\mathbf{m}=\mathbf{h}^{\smallL}-\mathbf{h}^{\smallR}=\text{diag}(+1,+1,-1,-1),
\end{equation}
is quantised
and should not receive corrections, together with the explicit form~\eqref{eq:cbarc-p} of $\mathbf{c},\bar{\mathbf{c}}$ in terms of the momentum to solve for the representation parameters. To do so, let us introduce the \emph{Zhukovski paramterisation} of the momentum in terms of variables~$x^{\pm}_{p}$ satisfying
\begin{equation}
\label{eq:zhukovski-def}
\frac{x^+_{p}}{x^-_{p}}=e^{ip}\,,
\qquad
\left(x^{+}_{p}+\frac{1}{x^+_{p}}\right)
-\left(x^{-}_{p}+\frac{1}{x^-_{p}}\right)
=\frac{2i}{h}\,.
\end{equation}
We find
\begin{equation}
\label{eq:repr-coeffs}
\begin{aligned}
&a^{\smallL}_p=a^{\smallR}_p=
\sqrt{\zeta}\,e^{\frac{1}{4}i\,p}\eta_{p}\,,
\qquad
&\bar{a}^{\smallL}_p=\bar{a}^{\smallR}_p=
\frac{1}{\sqrt{\zeta}}\,e^{-\frac{3}{4}i\,p}\,\eta_p\,,\\
&b^{\smallL}_p=b^{\smallR}_p=-
\sqrt{\zeta}\,\frac{e^{-\frac{3}{4}i\,p}}{x^{-}_p}\eta_{p}\,,
\qquad
&\bar{b}^{\smallL}_p=\bar{b}^{\smallR}_p=-
\frac{1}{\sqrt{\zeta}}\,\frac{e^{\frac{1}{4}i\,p}}{x^{+}_p}\,\eta_p\,,
\end{aligned}
\end{equation}
where we explicitly indicated the momentum dependence on the representation parameters and introduced the function
\begin{equation}
\label{eq:eta-def}
\eta_{p}=e^{ip/4}\sqrt{i\,\frac{h}{2}\left(x^{-}_p-x^{+}_p\right)}\,.
\end{equation}
The definition~\eqref{eq:eta-def} is given in such a way that~$\eta_p$ will have nice analyticity properties that we will use in the next chapter. 
Note that if we want to identify the vectors in~$\mathscr{V}$ with states of the superstring, we should restrict to representations having~$\zeta=1$ to reproduce~\eqref{eq:cbarc-p-nozeta}.

There is some arbitrariness the choice of the representation coefficients, \eg\ corresponding to a change of normalisation of the basis vectors in~$\mathscr{B}$.  We fix some of this freedom by requiring that we are truly deforming the relativistic representation found perturbatively in the previous chapter. In this way, if we rescale $p\to p/h$ and expand the coefficients in $1/h$, we find as expected
\begin{equation}
\label{eq:ab-expansion}
a^{\smallL,\smallR}_p=
f_p+O\big(1/h\big)
=\bar{a}^{\smallL,\smallR}_p\,,
\qquad\quad 
b^{\smallL,\smallR}_p=
g_p+O\big(1/h\big)
=\bar{b}^{\smallL,\smallR}_p
\,.
\end{equation}

\subsubsection*{Dispersion relation}
By explicitly evaluating $\mathbf{h}=\mathbf{h}^{\smallL}+\mathbf{h}^{\smallR}$ we find 
\begin{equation}
\label{eq:dispersion-xpm}
\mathbf{h}=\frac{h}{2i}
\left(x^{+}_p-\frac{1}{x^{+}_p}-x^{-}_p+\frac{1}{x^{-}_p}\right)\,\mathbf{I}=\sqrt{1+4h^2\sin^2\left(\frac{p}{2}\right)}\,\mathbf{I}\,.
\end{equation}
We anticipated the presence of such a non-relativistic dispersion relation in the introduction. In fact, it is an immediate consequence of the shortening condition~\eqref{eq:shortening2} and the non-linear form of~$\mathbf{c}$ and~$\bar{\mathbf{c}}$. Let us rewrite~\eqref{eq:shortening2} as
\begin{equation}
0=\big(\mathbf{h}+\mathbf{m}\big)\,\big(\mathbf{h}-\mathbf{m}\big)-4\,\bar{\mathbf{c}}\,\mathbf{c}\,,
\end{equation}
Using the fact that the eigenvalues of $\mathbf{m}$ are~$\pm1$ so that $\mathbf{m}^2=1$, we have
\begin{equation}
\label{eq:dispersion-charges}
\mathbf{h}^2=1+4\,\bar{\mathbf{c}}\,\mathbf{c}=1+4h^2\sin^2\left(\frac{\mathbf{p}}{2}\right)\,,
\end{equation}
\ie\ precisely the dispersion relation~\eqref{eq:dispersion-xpm}.

\subsubsection*{Left-right symmetry}
The identity between left- and right-representation coefficients in~\eqref{eq:repr-coeffs} suggests the presence of a discrete symmetry relating them. This is not surprising when we think of the original coset model, where L and R where introduced as arbitrary labels for the two copies of~$\psu(1,1|2)$. Furthermore, as discussed in the previous chapter (see also appendix~\ref{app:quadraticcharges}), the supercharges in the symmetry algebra are naturally split in two sets differing only by the relabelling $\bigL\leftrightarrow\bigR$, while the energy and the central charges take have the same form on both representations. As a consequence, up to exchanging raising and lowering operators, we can map the left representation into the right one and \textit{vice versa}.

This results in a discrete~$\mathbbm{Z}_2$-symmetry, which we will call left-right (LR) symmetry. When extended to multiparticle states, this will tell us that \eg\ any configuration involving only left excitations has an equivalent realisation in terms of right excitations only, \etc, which  will yield restrictions on the S-matrix elements.

\subsection*{Two-particle representation}
The two-particle representation is the tensor product of two one-particle representations like the one we just constructed. Its structure constants will be of the form $Q_{\a_1,\a_2}^{\b_1,\b_2}(p_1,p_2;\{c_1,c_2\})$ where $p_1,p_2$ and ${c_1,c_2}$ are the momenta and the central charges of the one-particle representations. Even if the two-particle representation will have central charges of the form~\eqref{eq:cbarc-p-nozeta}, \ie\ with $\zeta=1$, this may not necessarily be the case for the two factors. To see that, let us evaluate the structure constant~$Q_{\a_1,\a_2}^{\b_1,\b_2}(p_1,p_2;\{c_1,c_2\})$ for the central charge $\mathbf{c}$ of~\eqref{eq:cbarc-p-nozeta}. Since this acts diagonally, we have
\begin{equation}
\mathbf{c}\cdot A^{\dagger}_{\a_1}(p_1) A^{\dagger}_{\a_2}(p_2)\Ket{0}
=
i\frac{h}{2}\left(e^{i(p_1+p_2)}-1\right)\,\delta^{\b_1}_{\a_1}\delta^{\b_2}_{\a_2}
\, A^{\dagger}_{\a_1}(p_1) A^{\dagger}_{\a_2}(p_2)\Ket{0}\,,
\end{equation}
where we used~\eqref{eq:abelian-ZF} which implies that the worldsheet momentum of this state is $p_1+p_2$ as we expect. From this, we can read off
\begin{equation}
\label{eq:cc1stexpr}
C_{\a_1,\a_2}^{\b_1,\b_2}(p_1,p_2;\{c_1,c_2\})
=
i\frac{h}{2}\left(e^{i(p_1+p_2)}-1\right)\,\delta^{\b_1}_{\a_1}\delta^{\b_2}_{\a_2}\,,
\end{equation}

On the other hand, we can commute~$\mathbf{c}$ with one creation operator at the time, finding instead
\begin{equation}
\label{eq:cc2ndexpr}
C_{\a_1,\a_2}^{\b_1,\b_2}(p_1,p_2;\{c_1,c_2\})
=
i\frac{h}{2}\left(\zeta_1(e^{ip_1}-1)+\zeta_2(e^{ip_2}-1)\right)\,\delta^{\b_1}_{\a_1}\delta^{\b_2}_{\a_2}\,,
\end{equation}
Clearly, \eqref{eq:cc1stexpr} and \eqref{eq:cc2ndexpr} should match, which cannot happen if~$\zeta_1=\zeta_2=1$. We conclude that the two-particle representation is the tensor product of two one-particle representations \emph{with non-trivial central-charge values $\zeta_{1,2}$}. It is easy to check that the two expressions we found match if and only if
\begin{equation}
\label{eq:coproductchoice}
\text{(I):}\quad
\zeta_1=1\,,
\ 
\zeta_2=e^{ip_1}
\qquad
\text{or}
\qquad
\text{(II):}\quad
\zeta_1=e^{ip_2}\,,
\ 
\zeta_2=1\,.
\end{equation}
As discussed, for consistency with left-right symmetry it natural to take the same choice of $\zeta_{1,2}$ for both the L~and R~irreducible representations. Therefore, we can label the whole representation space~$\mathscr{V}$ by the choice of the central charge,~$\mathscr{V}=\mathscr{V}(p_i,\zeta_i)$. Then the action of the S matrix on the two-particle (reducible) representation is
\begin{equation}
\mathbf{S}:
\quad
\mathscr{V}(p_1,\zeta_1)\otimes
\mathscr{V}(p_2,\zeta_2)
\quad\to\quad
\mathscr{V}(p_2,\zeta_2)\otimes
\mathscr{V}(p_1,\zeta_1)\,.
\end{equation}
If we assign the value of $\zeta_{1,2}$ on the intial states according to either choice in~\eqref{eq:coproductchoice}, we find that~$\mathbf{S}$ exchanges (I) with (II) or \textit{vice versa}. In what follows we will consider the choice~(I) for the initial states, which as we will see will reproduce the perturbative results.

We can now rewrite more explicitly~\eqref{eq:generic-2part-charge} for a supercharge as
\begin{equation}
\label{eq:16x16charges}
\begin{aligned}
&\mathbf{Q}_{(12)}(p_1,p_2)=
\mathbf{Q}(p_1,\zeta_1=1)\otimes \mathbf{I}+
\Sigma\otimes\mathbf{Q}(p_2,\zeta_2=e^{ip_1}),
\end{aligned}
\end{equation}
where we used that~$\mathbf{Q}$ is odd in order to rewrite the action of the graded identity in terms of the fermion sign matrix, which in the basis~$\mathscr{B}$ is
\begin{equation}
\label{eq:Sigma-def}
\Sigma=\text{diag}(+1,-1,+1,-1)\,.
\end{equation}
The $16\times16$ matrix representation is then found by taking the one-particle charge~$\mathbf{Q}$ to be equal to each of the matrices~\eqref{eq:supercharges-1part-matrix}.

The choice of a non-trivial~$\zeta_1$ gives the general form of the two-particle representation
\begin{equation}
\label{eq:coproduct-deformed}
\mathbf{Q}_{(12)}= \mathbf{Q}(p_1)\otimes \mathbf{I}
+e^{\pm \frac{i}{2} p_1}\,
\Sigma\otimes \mathbf{Q}(p_2)\,,
\end{equation}
where the sign in the exponent depends on which supercharge we consider---positive for~$\mathbf{Q}$ and negative for~$\overline{\mathbf{Q}}$. This can be compared to the more usual relation
\begin{equation}
\mathbf{Q}_{(12)}= \mathbf{Q}(p_1)\otimes \mathbf{I}+\Sigma\otimes \mathbf{Q}(p_2)\,.
\end{equation}
In algebraic terms, the latter is a trivial (graded) \emph{coproduct}, \ie\ the most natural way of extending the action of an operator to a tensor product space. We can then say that the symmetries of the superstring have a non-trivial coproduct. The mathematical object to deal with such structures is an \emph{Hopf albegra}%
\footnote{%
More precisely, an Hopf algebra is also equipped with an ``antipode'' involution, which has the physical interpretation of a particle-to-antiparticle transformation.
}. The relevance of Hopf algebras in $\AdS/\CFT$ is well known~\cite{Plefka:2006ze}, see also refs.~\cite{Torrielli:2010kq,Torrielli:2011gg} for a review. In fact, it is possible to use such an algebraic approach to find~$\mathbf{R}$ in the case at hand. Here we will follow an equivalent and more direct route, and refer the reader to ref.~\cite{Borsato:2013qpa} for the Hopf-algebraic derivation of the S~matrix.

In principle we could go on and construct the representation of the symmetries on higher multiparticle states. This can be done either explicitly~\cite{Arutyunov:2009ga} or by subsequent applications of the Hopf-algebra coproduct. However, since our focus is on the two-particle S~matrix, we will not need to work out the resulting expressions.

\section{Finding the S matrix}
\label{sec:stringsmat}
In this section we want to find the S~matrix of massive fundamental excitations of the~$\AdS_3\times\S^3\times\T^4$ superstring. This can be done working with~$\mathbf{R}$ or~$\mathbf{S}$, but it will be useful for us to introduce the graded matrix $\check{\mathbf{S}}(p,q)$ satisfying
\begin{equation}
\check{\mathbf{S}}(p,q)=\mathbf{I}^g\,\mathbf{S}(p,q)=\Pi\,\mathbf{R}(p,q)\,,
\end{equation}
where $\Pi$ is the permutation and $\mathbf{I}^g$ is the graded identity. This will make some of our expressions easier to manipulate, because all permutations of tensor product factors will be automatically accounted for, and match with refs.~\cite{Borsato:2012ud,Borsato:2013qpa}.

 Our strategy will be to write down the most general operator
\begin{equation}
\check{\mathbf{S}}:
\quad
\mathscr{V}(p,1)\otimes
\mathscr{V}(q,e^{ip})
\quad\to\quad
\mathscr{V}(q,e^{ip})\otimes
\mathscr{V}(p,1)\,,
\end{equation}
as~$16\times16$ matrix, and require that it satisfies suitable physical properties, which we now list.

\subsection*{Off-shell symmetries}
The S~matrix should commute with the whole off-shell symmetry algebra. Since the latter is generated by the supercharges, it is enough to impose that the S~matrix commutes with $\mathbf{q}^{\smallL,\smallR}$ and  $\bar{\mathbf{q}}^{\smallL,\smallR}$ in the $16\times16$ matrix representation given by~\eqref{eq:16x16charges}.

Explicitly, in terms of~$\check{\mathbf{S}}$, we have
\begin{equation}
\label{eq:Smat-commutation}
\check{\mathbf{S}}_{(12)}(p,q)\,
{\mathbf{Q}}_{(12)}(p,q)
=
{\mathbf{Q}}_{(12)}(q,p)\,
\check{\mathbf{S}}_{(12)}(p,q)\,,
\end{equation}
where~$\mathbf{Q}_{(12)}(p,q)$ is defined by~\eqref{eq:16x16charges}. This should be imposed for all four supercharges. Notice that, unlike~\eqref{eq:R-matrix-comm-rel}, this equations does not feature~$\mathbf{Q}_{(21)}$.

Since~\eqref{eq:Smat-commutation}  is a linear equation, any solution can be multiplied by a prefactor. In our case, the charges~\eqref{eq:supercharges-1part-matrix} have a block-diagonal structure due to the presence of two (L and~R) irreducible representations in~$\mathscr{V}$. Therefore, we expect~\eqref{eq:Smat-commutation} to determine~$\check{\mathbf{S}}$ at best up to four scalar factors for the LL, LR, RL and RR blocks. We will refer to the part of~$\mathbf{S}$ determined independently from these factors (\ie, suitable ratios of the matrix elements) as its ``matrix part''.

\subsubsection*{Left-right symmetry}
An additional constraint is the discrete left-right symmetry, which amounts to imposing that scattering processes differing only by relabelling~$\bigL\leftrightarrow\bigR$ should be indistinguishable. In particular, this removes part of the ambiguity due to scalar factors, relating the LL~block to the RR~one, and the LR~one to the RL~one.

\subsection*{Braiding and physical unitarity}
In section~\ref{sec:integrability} we have established that the $\mathbf{R}$~matrix must satisfy braiding unitarity as well as be unitary as a matrix. In terms of~$\check{\mathbf{S}}$ this gives
\begin{equation}
\label{eq:braidingunitarity-Scheck}
\check{\mathbf{S}}_{(12)}(q,p)\,
\check{\mathbf{S}}_{(12)}(p,q)=\mathbf{I}\,,
\qquad
\big(\check{\mathbf{S}}_{(12)}(p,q)\big)^{\dagger}\,
\check{\mathbf{S}}_{(12)}(p,q)=\mathbf{I}\,.
\end{equation} 
One of these two equations quadratic constrains can be eliminated in terms of the linear one
\begin{equation}
\check{\mathbf{S}}_{(12)}(q,p)=\big(\check{\mathbf{S}}_{(12)}(p,q)\big)^{\dagger}\,.
\end{equation}
While this linear equation will put a restriction on the reality properties of the scalar factors, it is easy to see that the matrix part should automatically satisfy it. In fact, taking the conjugate of~\eqref{eq:Smat-commutation}, we find
\begin{equation}
\label{eq:linear-unitarty}
\overline{\mathbf{Q}}_{(12)}(p,q)\,
\big(\check{\mathbf{S}}_{(12)}(p,q)\big)^{\dagger}
=
\big(\check{\mathbf{S}}_{(12)}(p,q)\big)^{\dagger}\,
\overline{\mathbf{Q}}_{(12)}(q,p)\,.
\end{equation}
Since for any charge~${\mathbf{Q}}_{(12)}$ its conjugate~$\overline{\mathbf{Q}}_{(12)}$ is also part of the algebra, we have that $\check{\mathbf{S}}_{(12)}(p,q)^\dagger$ satisfies the same invariance condition that define~$\check{\mathbf{S}}_{(12)}(q,p)$. Therefore the part of each of the two S~matrices that is completely determined by symmetries must coincide.

\subsection*{The reflectionless~$\su(1|1)^2_{\text{c.e.}}$ S matrix}
It turns out that imposing all the symmetries together with the unitarity requirements gives two physically distinct solutions for $\check{\mathbf{S}}_{pq}=\check{\mathbf{S}}_{(12)}(p,q)$. These coincide on the LL  and RR~sectors, but are different on the mixed sectors. One solution gives
\begin{equation}
\text{(T):}\qquad
\bra{\mathcal{X}^{\smallL}\mathcal{Y}^{\smallR}}
\check{\mathbf{S}}_{pq}
\ket{\mathcal{X}^{\smallL}\mathcal{Y}^{\smallR}}=0
\quad\text{and}\quad
\bra{\mathcal{X}^{\smallR}\mathcal{Y}^{\smallL}}
\check{\mathbf{S}}_{pq}
\ket{\mathcal{X}^{\smallL}\mathcal{Y}^{\smallR}}\neq0\,,
\end{equation}
and similarly for~$\bigL\leftrightarrow\bigR$, whereas the other gives
\begin{equation}
\text{(R):}\qquad
\bra{\mathcal{X}^{\smallL}\mathcal{Y}^{\smallR}}
\check{\mathbf{S}}_{pq}
\ket{\mathcal{X}^{\smallL}\mathcal{Y}^{\smallR}}\neq0
\quad\text{and}\quad
\bra{\mathcal{X}^{\smallR}\mathcal{Y}^{\smallL}}
\check{\mathbf{S}}_{pq}
\ket{\mathcal{X}^{\smallL}\mathcal{Y}^{\smallR}}=0\,,
\end{equation}
where~$\mathcal{X},\mathcal{Y}$ are two generic excitations.
Keeping into account that the matrix~$\check{\mathbf{S}}$ permutes the final states, the case we indicated by (T) corresponds to \emph{pure transmission} of the target-space chirality, while (R) corresponds to \emph{pure reflection}. From the symmetry properties, there is no reason to choose one over the other. However, by a perturbative calculation it is easy to check which are the non-vanishing matrix elements. In fact, a tree-level calculation suffices~\cite{Rughoonauth:2012qd}, and shows that the case~(R) cannot reproduce the worldsheet superstring S~matrix. From now on, we will restrict our  considerations to the case (T) of a \emph{pure-transmission} (or \emph{reflectionless}) S~matrix. 

Constructing the tensor-product basis out of~$\mathscr{B}$, we can explicitly represent $\check{\mathbf{S}}_{pq}$ as a~$16\times16$ matrix which naturally splits in four blocks depending on the target-space chiralities of the particle scattered
\begin{equation}
\check{\mathbf{S}}_{pq} = 
\left(
\begin{array}{c|c}
\phantom{\Big[}\check{\mathbf{S}}^{\smallLL}_{pq}\phantom{\Big]} &
\phantom{\Big[}\check{\mathbf{S}}^{\smallRL}_{pq}\phantom{\Big]}\\
\hline
\phantom{\Big[}\check{\mathbf{S}}^{\smallLR}_{pq}\phantom{\Big]} &
\phantom{\Big[}\check{\mathbf{S}}^{\smallRR}_{pq}\phantom{\Big]}
\end{array}
  \right) .
\end{equation}
The full form of the S~matrix is given in figure~\ref{fig:Smatrix}. Let us further investigate its structure. Due to LR symmetry, it is natural to distinguish the scattering of particles of the same or opposite chirality.

\begin{figure}
  \centering
\begin{equation*}
  \newcommand{\0}{\color{black!40}0}
  \renewcommand{\arraystretch}{1.1}
  \setlength{\arraycolsep}{3pt}
  \left(\!
    \mbox{\footnotesize$
      \begin{array}{cccccccc|cccccccc}
        \Ael^{\smallLL}_{pq} & \0 & \0 & \0 & \0 & \0 & \0 & \0 & \0 & \0 & \0 & \0 & \0 & \0 & \0 & \0 \\
        \0 & \Cel^{\smallLL}_{pq} & \0 & \0 &\Del^{\smallLL}_{pq} & \0 & \0 & \0 & \0 & \0 & \0 & \0 & \0 & \0 & \0 & \0 \\
        \0 & \0 & \0 & \0 & \0 & \0 & \0 & \0 & \Ael^{\smallRL}_{pq}  & \0 & \0 & \0 & \0 & \Fel^{\smallRL}_{pq} & \0 & \0 \\
        \0 & \0 & \0 & \0 & \0 & \0 & \0 & \0 & \0 & \0 & \0 & \0 & \Del^{\smallRL}_{pq} & \0 & \0 & \0 \\
        \0 & \Bel^{\smallLL}_{pq} & \0 & \0 & \Eel^{\smallLL}_{pq} & \0 & \0 & \0 & \0 & \0 & \0 & \0 & \0 & \0 & \0 & \0 \\
        \0 & \0 & \0 & \0 & \0 & \Fel^{\smallLL}_{pq} & \0 & \0 & \0 & \0 & \0 & \0 & \0 & \0 & \0 & \0 \\
        \0 & \0 & \0 & \0 & \0 & \0 & \0 & \0 & \0 & \Cel^{\smallRL}_{pq} & \0 & \0 & \0 & \0 & \0 & \0 \\
        \0 & \0 & \0 & \0 & \0 & \0 & \0 & \0 & \Bel^{\smallRL}_{pq} & \0 & \0 & \0 & \0 &\Eel^{\smallRL}_{pq} & \0 & \0 \rule[-1.4ex]{0pt}{0pt} \\
        \hline
        \0 & \0 & \Ael^{\smallLR}_{pq} & \0 & \0 & \0 & \0 & \Fel^{\smallLR}_{pq}  & \0 & \0 & \0 & \0 & \0 & \0 & \0 & \0 \\
        \0 & \0 & \0 & \0 & \0 & \0 & \Del^{\smallLR}_{pq}  & \0 & \0 & \0 & \0 & \0 & \0 & \0 & \0 & \0 \\
        \0 & \0 & \0 & \0 & \0 & \0 & \0 & \0 & \0 & \0 & \Ael^{\smallRR}_{pq} & \0 & \0 & \0 & \0 & \0 \\
        \0 & \0 & \0 & \0 & \0 & \0 & \0 & \0 & \0 & \0 & \0 &\Cel^{\smallRR}_{pq} & \0 & \0 & \Del^{\smallRR}_{pq} & \0 \\
        \0 & \0 & \0 & \Cel^{\smallLR}_{pq}  & \0 & \0 & \0 & \0 & \0 & \0 & \0 & \0 & \0 & \0 & \0 & \0 \\
        \0 & \0 & \Bel^{\smallLR}_{pq}  & \0 & \0 & \0 & \0 & \Eel^{\smallLR}_{pq}  & \0 & \0 & \0 & \0 & \0 & \0 & \0 & \0 \\
        \0 & \0 & \0 & \0 & \0 & \0 & \0 & \0 & \0 & \0 & \0 & \Bel^{\smallRR}_{pq} & \0 & \0 &\Eel^{\smallRR}_{pq} & \0 \\
        \0 & \0 & \0 & \0 & \0 & \0 & \0 & \0 & \0 & \0 & \0 & \0 & \0 & \0 & \0 & \Fel^{\smallRR}_{pq}
      \end{array}$}\!
  \right)
\end{equation*}
 \caption{The matrix representation of~$\check{\mathbf{S}}_{(12)}(p,q)$ in the basis constructed out of the tensor product of~$\mathscr{B}$.
Notice how each block is given by a $4\times4$ S~matrix suitably embedded in a tensor product structure. 
 }
  \label{fig:Smatrix}
\end{figure}

\subsubsection*{Same-chirality scattering}
Let us consider particles of the same, let us say~LL, chirality. Then the non-vanishing scattering processes are
\begin{equation}
  \begin{aligned}
    \check{\mathbf{S}} \ket{\fixedspaceL{\psi_p^{\smallL} \psi_q^{\smallL}}{\phi_p^{\smallL} \phi_q^{\smallL}}} 
    &= \fixedspaceR{\Del^{\smallLL}_{pq}}{\Ael^{\smallLL}_{pq}} \ket{\fixedspaceL{\psi_p^{\smallL} \psi_q^{\smallL}}{\phi_q^{\smallL} \phi_p^{\smallL}}} , \qquad &
    \check{\mathbf{S}} \ket{\fixedspaceL{\psi_p^{\smallL} \psi_q^{\smallL}}{\phi_p^{\smallL} \psi_q^{\smallL}}} 
    &= \fixedspaceR{\Del^{\smallLL}_{pq}}{\Bel^{\smallLL}_{pq}} \ket{\fixedspaceL{\psi_p^{\smallL} \psi_q^{\smallL}}{\psi_q^{\smallL} \phi_p^{\smallL}}} + \Cel^{\smallLL}_{pq} \ket{\fixedspaceL{\psi_p^{\smallL} \psi_q^{\smallL}}{\phi_q^{\smallL} \psi_p^{\smallL}}}, \\
    \check{\mathbf{S}} \ket{\fixedspaceL{\psi_p^{\smallL} \psi_q^{\smallL}}{\psi_p^{\smallL} \psi_q^{\smallL}}} &= \fixedspaceR{\Del^{\smallLL}_{pq}}{\Fel^{\smallLL}_{pq}} \ket{\fixedspaceL{\psi_p^{\smallL} \psi_q^{\smallL}}{\psi_q^{\smallL} \psi_p^{\smallL}}} , \qquad &
    \check{\mathbf{S}} \ket{\fixedspaceL{\psi_p^{\smallL} \psi_q^{\smallL}}{\psi_p^{\smallL} \phi_q^{\smallL}}} 
    &= \fixedspaceR{\Del^{\smallLL}_{pq}}{\Del^{\smallLL}_{pq}} \ket{\fixedspaceL{\psi_p^{\smallL} \psi_q^{\smallL}}{\phi_q^{\smallL} \psi_p^{\smallL}}} + \Eel^{\smallLL}_{pq} \ket{\fixedspaceL{\psi_p^{\smallL} \psi_q^{\smallL}}{\psi_q^{\smallL} \phi_p^{\smallL}}}. \\
  \end{aligned}
\end{equation}
These are determined up to a single, overall scalar factor~$\mathscr{S}^{\smallLL}_{pq}$, and they read
\begin{equation}
\label{eq:stringSmatLL}
\begin{aligned}
&\Ael^{\smallLL}_{pq}=
\mathscr{S}^{\smallLL}_{pq} e^{\frac{i}{2}(p-q)}
\frac{x^{-}_p-x^{+}_q}{x^{+}_p-x^{-}_q}\,,
\qquad 
&&\Bel^{\smallLL}_{pq}=
\mathscr{S}^{\smallLL}_{pq} e^{-\frac{i}{2}q}
\frac{x^{+}_p-x^{+}_q}{x^{+}_p-x^{-}_q}\,,
\\
&\Cel^{\smallLL}_{pq}=
\mathscr{S}^{\smallLL}_{pq} e^{\frac{i}{4}(p-3q)}
\frac{-\tfrac{2i}{h}\,\eta_p\,\eta_q}{x^{+}_p-x^{-}_q}
\,,
\qquad 
&&\Del^{\smallLL}_{pq}=
\mathscr{S}^{\smallLL}_{pq} e^{\frac{i}{2}p}
\frac{x^{-}_p-x^{-}_q}{x^{+}_p-x^{-}_q}\,,
\\
&\Eel^{\smallLL}_{pq}=
\mathscr{S}^{\smallLL}_{pq} e^{-\frac{i}{4}(p+q)}
\frac{-\tfrac{2i}{h}\,\eta_p\,\eta_q}{x^{+}_p-x^{-}_q}
\,,
\qquad 
&&\Fel^{\smallLL}_{pq}=
-\mathscr{S}^{\smallLL}_{pq}\,.
\end{aligned}
\end{equation}

In the RR sector we find exactly the same formulae, in terms of a scalar factor~$\mathscr{S}^{\smallRR}_{pq}$. Requiring LR-symmetry implies, for instance,
\begin{equation}
\bra{\phi_q^{\smallL} \phi_p^{\smallL}}    \check{\mathbf{S}} \ket{\phi_p^{\smallL} \phi_q^{\smallL}}
=
\bra{\phi_q^{\smallR} \phi_p^{\smallR}}    \check{\mathbf{S}} \ket{\phi_p^{\smallR} \phi_q^{\smallR}}\,,
\end{equation}
so that it must be
\begin{equation}
\mathscr{S}^{\smallLL}_{pq}
=
\mathscr{S}^{\smallRR}_{pq}
=
\mathscr{S}_{pq}\,.
\end{equation}

\subsubsection*{Opposite-chirality scattering}
If we now consider processes of LR~chirality we find
\begin{equation}
  \begin{aligned}
    \check{\mathbf{S}} \ket{\fixedspaceL{\psi_p^{\smallL}\psi_q^{\smallR}}{\phi_p^{\smallL} \phi_q^{\smallR}}} 
    &= \fixedspaceR{\Ael^{\smallLR}_{pq}}{\Ael^{\smallLR}_{pq}} \ket{\fixedspaceL{\psi_q^{\smallR}\psi_p^{\smallL}}{\phi_q^{\smallR}\phi_p^{\smallL}}} + \fixedspaceR{\Ael^{\smallLR}_{pq}}{\Bel^{\smallLR}_{pq}} \ket{\psi_q^{\smallR} \psi_p^{\smallL}}, \qquad &
    \check{\mathbf{S}} \ket{\fixedspaceL{\psi_p^{\smallL}\psi_q^{\smallR}}{\phi_p^{\smallL}\psi_q^{\smallR}}} 
    &= \fixedspaceR{\Ael^{\smallLR}_{pq}}{\Cel^{\smallLR}_{pq}} \ket{\fixedspaceL{\psi_q^{\smallR}\psi_p^{\smallL}}{\psi_q^{\smallR}\phi_p^{\smallL}}}, \\
    \check{\mathbf{S}} \ket{\fixedspaceL{\psi_p^{\smallL}\psi_q^{\smallR}}{\psi_p^{\smallL}\psi_q^{\smallR}}} 
    &= \fixedspaceR{\Ael^{\smallLR}_{pq}}{\Eel^{\smallLR}_{pq}} \ket{\fixedspaceL{\psi_q^{\smallR}\psi_p^{\smallL}}{\psi_q^{\smallR}\psi_p^{\smallL}}} + \fixedspaceR{\Ael^{\smallLR}_{pq}}{\Fel^{\smallLR}_{pq}} \ket{\phi_q^{\smallR}\phi_p^{\smallL}}, \qquad &
    \check{\mathbf{S}} \ket{\fixedspaceL{\psi_p^{\smallL}\psi_q^{\smallR}}{\psi_p^{\smallL}\phi_q^{\smallR}}} 
    &= \fixedspaceR{\Ael^{\smallLR}_{pq}}{\Del^{\smallLR}_{pq}} \ket{\fixedspaceL{\psi_q^{\smallR}\psi_p^{\smallL}}{\phi_q^{\smallR}\psi_p^{\smallL}}},
\end{aligned}
\end{equation}
while for RL
\begin{equation}
\begin{aligned}
        \check{\mathbf{S}} \ket{\fixedspaceL{\psi_p^{\smallR}\psi_q^{\smallL}}{\phi_p^{\smallR} \phi_q^{\smallL}}} 
    &= \fixedspaceR{\Ael^{\smallRL}_{pq}}{\Ael^{\smallRL}_{pq}} \ket{\fixedspaceL{\psi_q^{\smallL}\psi_p^{\smallR}}{\phi_q^{\smallL}\phi_p^{\smallR}}} + \fixedspaceR{\Ael^{\smallRL}_{pq}}{\Bel^{\smallRL}_{pq}} \ket{\psi_q^{\smallL} \psi_p^{\smallR}}, \qquad &
    \check{\mathbf{S}} \ket{\fixedspaceL{\psi_p^{\smallR}\psi_q^{\smallL}}{\phi_p^{\smallR}\psi_q^{\smallL}}} 
    &= \fixedspaceR{\Ael^{\smallRL}_{pq}}{\Cel^{\smallRL}_{pq}} \ket{\fixedspaceL{\psi_q^{\smallL}\psi_p^{\smallR}}{\psi_q^{\smallL}\phi_p^{\smallR}}}, \\
    \check{\mathbf{S}} \ket{\fixedspaceL{\psi_p^{\smallR}\psi_q^{\smallL}}{\psi_p^{\smallR}\psi_q^{\smallL}}} 
    &= \fixedspaceR{\Ael^{\smallRL}_{pq}}{\Eel^{\smallRL}_{pq}} \ket{\fixedspaceL{\psi_q^{\smallL}\psi_p^{\smallR}}{\psi_q^{\smallL}\psi_p^{\smallR}}} + \fixedspaceR{\Ael^{\smallRL}_{pq}}{\Fel^{\smallRL}_{pq}} \ket{\phi_q^{\smallL}\phi_p^{\smallR}}, \qquad &
    \check{\mathbf{S}} \ket{\fixedspaceL{\psi_p^{\smallR}\psi_q^{\smallL}}{\psi_p^{\smallR}\phi_q^{\smallL}}} 
    &= \fixedspaceR{\Ael^{\smallRL}_{pq}}{\Del^{\smallRL}_{pq}} \ket{\fixedspaceL{\psi_q^{\smallL}\psi_p^{\smallR}}{\phi_q^{\smallL}\psi_p^{\smallR}}}.\\
  \end{aligned}
\end{equation}
The elements read
\begin{equation}
\label{eq:stringSmatLR}
\begin{aligned}
&\Ael^{\smallLR}_{pq}=
\mathscr{S}^{\smallLR}_{pq} e^{\frac{i}{2}(p+q)}
\frac{1-x^{+}_px^{-}_q}{1-x^{+}_px^{+}_q}\,,
\quad 
&&\Bel^{\smallLR}_{pq}=
\mathscr{S}^{\smallLR}_{pq}e^{\frac{i}{4}(3p-q)}
\frac{\tfrac{2i}{h}\,\eta_p\,\eta_q}{1-x^{+}_px^{+}_q}
\,,
\\
&\Cel^{\smallLR}_{pq}=
\mathscr{S}^{\smallLR}_{pq} e^{\frac{i}{2}(2p+q)}
\frac{1-x^{-}_px^{-}_q}{1-x^{+}_px^{+}_q}\,,
\quad 
&&\Del^{\smallLR}_{pq}=
\mathscr{S}^{\smallLR}_{pq} e^{\frac{i}{2}p}\,,
\\
&\Eel^{\smallLR}_{pq}=
-\mathscr{S}^{\smallLR}_{pq} e^{ip}
\frac{1-x^{-}_px^{+}_q}{1-x^{+}_px^{+}_q}\,,
\quad 
&&\Fel^{\smallLR}_{pq}=
\mathscr{S}^{\smallLR}_{pq}e^{\frac{i}{4}(p-q)}
\frac{-\tfrac{2i}{h}\,\eta_p\,\eta_q}{1-x^{+}_px^{+}_q}\,,
\end{aligned}
\end{equation}
and
\begin{equation}
\begin{aligned}
&\Ael^{\smallRL}_{pq}=
\mathscr{S}^{\smallRL}_{pq} e^{-\frac{i}{2}(p+q)}
\frac{1-x^{+}_px^{-}_q}{1-x^{-}_px^{-}_q}\,,
\quad 
&&\Bel^{\smallRL}_{pq}=
\mathscr{S}^{\smallRL}_{pq}e^{-\frac{i}{4}(p+3q)}
\frac{\tfrac{2i}{h}\,\eta_p\,\eta_q}{1-x^{-}_px^{-}_q}
\,,
\\
&\Cel^{\smallRL}_{pq}=
\mathscr{S}^{\smallRL}_{pq} e^{-\frac{i}{2}q}\,,
\quad 
&&\Del^{\smallRL}_{pq}=
\mathscr{S}^{\smallRL}_{pq} e^{-\frac{i}{2}(p+2q)}
\frac{1-x^{+}_px^{+}_q}{1-x^{-}_px^{-}_q}\,,
\\
&\Eel^{\smallRL}_{pq}=
-\mathscr{S}^{\smallRL}_{pq} e^{-iq}
\frac{1-x^{-}_px^{+}_q}{1-x^{-}_px^{-}_q}\,,
\quad 
&&\Fel^{\smallRL}_{pq}=
\mathscr{S}^{\smallRL}_{pq}e^{-\frac{i}{4}(3p+5q)}
\frac{-\tfrac{2i}{h}\,\eta_p\,\eta_q}{1-x^{-}_px^{-}_q}\,.
\end{aligned}
\end{equation}
Now LR-symmetry implies \eg
\begin{equation}
\bra{\psi_q^{\smallR}\phi_p^{\smallL}}
\check{\mathbf{S}}
\ket{\phi_p^{\smallL}\psi_q^{\smallR}}
=\bra{\psi_q^{\smallL}\phi_p^{\smallR}}
\check{\mathbf{S}}
\ket{\phi_p^{\smallR}\psi_q^{\smallL}}\,.
\end{equation}
 This can be solved in terms of a single scalar factor~$\widetilde{\mathscr{S}}_{pq}$
\begin{equation}
\begin{aligned}
\mathscr{S}^{\smallLR}_{pq}=
&\,
\widetilde{\mathscr{S}}_{pq}
e^{-\frac{i}{2}(p+q)}\left(\frac{1-x^{+}_px^{+}_q}{1-x^{-}_px^{-}_q}\right)^{+1/2},\\
\qquad
\mathscr{S}^{\smallRL}_{pq}=
&\,
\widetilde{\mathscr{S}}_{pq}
e^{+\frac{i}{2}(p+q)}\left(\frac{1-x^{+}_px^{+}_q}{1-x^{-}_px^{-}_q}\right)^{-1/2}.
\end{aligned}
\end{equation}
By this choice, we find not only $\Cel^{\smallRL}_{pq}=\Cel^{\smallLR}_{pq}$, but also $\Ael^{\smallRL}_{pq}=\Ael^{\smallLR}_{pq}$, $\Bel^{\smallRL}_{pq}=\Bel^{\smallLR}_{pq}$, and so on.

\subsubsection*{Scalar factors}
The requirements of braiding unitarity and physical unitarity pose constrains on the scalar factors~${\mathscr{S}}_{pq}$ and~$\widetilde{\mathscr{S}}_{pq}$. In particular, they must satisfy
\begin{equation}
\begin{aligned}
{\mathscr{S}}(p,q)\,
{\mathscr{S}}(q,p)=1\,,
\qquad
\big|{\mathscr{S}}(p,q)\big|^2=1\,,\\
\widetilde{\mathscr{S}}(p,q)\,
\widetilde{\mathscr{S}}(q,p)=1\,,
\qquad
\big|\widetilde{\mathscr{S}}(p,q)\big|^2=1\,.
\end{aligned}
\end{equation}
\ie\ they are given by two antisymmetric phases.

There is an additional condition that we can require on the S~matrix, that is \emph{crossing symmetry}. This is an extension of the well-known relativistic covariance under particle-antiparticle transformation to our non-relativistic S~matrix. Crossing invariance will put stringent requirements on the analytic structure of the dressing factors, which we will analyse at length in the next chapter.

\subsection*{The full S~matrix}
The full S~matrix that we are eventually interested in is not the~$\su(1|1)^2_{\text{c.e.}}$-invariant one. Instead, it is invariant under two copies of such an algebra, with the excitations transforming in a tensor product representation~(\ref{eq:tensorfileds}--\ref{eq:tensorcharges}). One way to find it is to construct a representation of the full~$\psu(1|1)^4_{\text{c.e.}}$ similar to~\eqref{eq:16x16charges}, in terms of $64\times64$~matrices. The S~matrix one would find then by repeating the procedure of this section is precisely the tensor product of two copies of the~$16\times16$ matrix we just found,%
\footnote{%
The tensor product yields a~$256\times256$ matrix with many vanishing elements, rather than~$64\times64$~one. This is because in~\eqref{eq:tensorfileds} we do not consider any state of the form~\eg~$\phi^{\smallL}\otimes\phi^{\smallR}$, that do exist in the tensor product space. 
}
\begin{equation}
\label{eq:S-matrix-tensorpr}
\check{\mathbf{S}}_{\psu(1|1)^4_{\text{c.e.}}}=
\check{\mathbf{S}}_{\su(1|1)^2_{\text{c.e.}}}
\otimes\,
\check{\mathbf{S}}_{\su(1|1)^2_{\text{c.e.}}}\,.
\end{equation}
The tensor product should take into account the signs arising from permuting the fer\-mions. In components this gives
\begin{equation}
\begin{aligned}
&\big(\check{S}_{\psu(1|1)^4_{\text{c.e.}}}\big)_{\a_1,\a_2}^{\a_3,\a_4}
=
\big(\check{S}_{\psu(1|1)^4_{\text{c.e.}}}\big)_{\kappa_1\iota_1, \kappa_2\iota_2}^{\kappa_3\iota_3, \kappa_4\iota_4}=\\
&\qquad\qquad\qquad\qquad\qquad
(-1)^{\eps(\iota_1)\eps(\kappa_2)+ \eps(\iota_3)\eps(\kappa_4)}
\big(\check{S}_{\su(1|1)^2_{\text{c.e.}}}\big)_{\kappa_1,\kappa_2}^{\kappa_3,\kappa_4}
\,
\big(\check{S}_{\su(1|1)^2_{\text{c.e.}}}\big)_{\iota_1,\iota_2}^{\iota_3,\iota_4}\,,
\end{aligned}
\end{equation}
where we used the notation introduced in~\eqref{eq:tensor-excit-repr}, with $\a=\kappa\iota$.
From this, all of the scattering elements can be readily calculated, and again we have that the scattering can be decomposed based on the LR chirality, with the matrix elements being the ``square'' of what found in the previous section. To illustrate this, let us work out a few processes in the LL sector
\begin{equation}
  \begin{aligned}
\check{\mathbf{S}}_{\psu(1|1)^4_{\text{c.e.}}}
 \ket{\Phi_{+{+}\,p}^{\smallL} \Phi_{+{+}\,q}^{\smallL}} =& \Ael^{\smallLL}_{pq} \Ael^{\smallLL}_{pq} \ket{\Phi_{+{+}\,q}^{\smallL} \Phi_{+{+}\,p}^{\smallL}} , \\
\check{\mathbf{S}}_{\psu(1|1)^4_{\text{c.e.}}}
 \ket{\Phi_{+{+}\,p}^{\smallL} \Phi_{-{-}\,q}^{\smallL}} =& 
    \Bel^{\smallLL}_{pq} \Bel^{\smallLL}_{pq} \ket{\Phi_{-{-}\,q}^{\smallL} \Phi_{+{+}\,p}^{\smallL}} + \Cel^{\smallLL}_{pq} \Cel^{\smallLL}_{pq} \ket{\Phi_{+{+}\,q}^{\smallL} \Phi_{-{-}\,p}^{\smallL}}\\
    &\ + \Bel^{\smallLL}_{pq} \Cel^{\smallLL}_{pq}\left(\ket{\Phi_{-{+}\,q}^{\smallL} \Phi_{+{-}\,p}^{\smallL}} - \ket{\Phi_{+{-}\,q}^{\smallL} \Phi_{-{+}\,p}^{\smallL}}\right), 
  \end{aligned}
\end{equation}
where we used the short-hand notation~$\ket{\Phi_{\pm{\pm}\,p}^{\smallL}} =\ket{\Phi_{\pm{\pm}}^{\smallL}(p)}$ and so on.
Similarly, in the LR sector we have for instance
\begin{equation}
\label{eq:LRscattering}
  \begin{aligned}
\check{\mathbf{S}}_{\psu(1|1)^4_{\text{c.e.}}}
\ket{\Phi_{+{+}\,p}^{\smallL} {\Phi}_{+{+}\,q}^{\smallR}} =& 
    \Ael^{\smallLR}_{pq} \Ael^{\smallLR}_{pq} \ket{{\Phi}_{+{+}\,q}^{\smallR} \Phi_{+{+}\,p}^{\smallL}} - \Bel^{\smallLR}_{pq} \Bel^{\smallLR}_{pq} \ket{{\Phi}_{-{-}\,q}^{\smallR} \Phi_{-{-}\,p}^{\smallL}}  \\
   &\ + \Ael^{\smallLR}_{pq} \Bel^{\smallLR}_{pq}\left( \ket{{\Phi}_{+{-}\,q}^{\smallR} \Phi_{+{-}\,p}^{\smallL}} +  \ket{{\Phi}_{-{+}\,q}^{\smallR} \Phi_{-{+}\,p}^{\smallL}}\right) , \\
\check{\mathbf{S}}_{\psu(1|1)^4_{\text{c.e.}}}
\ket{\Phi_{+{+}\,p}^{\smallL} {\Phi}_{-{-}\,q}^{\smallR}} =& \Cel^{\smallLR}_{pq} \Cel^{\smallLR}_{pq} \ket{{\Phi}_{-{-}\,q}^{\smallR} \Phi_{+{+}\,p}^{\smallL}} .
  \end{aligned}
\end{equation}
In particular, in~$\check{\mathbf{S}}_{\psu(1|1)^4_{\text{c.e.}}}$ there are still only~\emph{two} undetermined scalar factors, \ie\ $\mathscr{S}^2$ and~$\widetilde{\mathscr{S}}^2$.
It is also  interesting to note that the tensor-product structure yields symmetric and anti-symmetric combinations of S-matrix elements, in a way reminiscent of~$\su(2)$ invariance. As we will see in chapter~\ref{ch:outlook}, this is no accident: there is indeed an hidden~$\su(2)$ structure in this theory, which we will make manifest when we will address the massless sector.

In what follows, unless confusion may arise, we will drop the subscript invariant algebra from~$\check{\mathbf{S}}$.

\subsection*{Yang-Baxter equation and integrability}
Once we have successfully derived the $2\to2$ S~matrix by maxing use of the off-shell symmetry, the question remains as to whether such an S~matrix can be used to define an arbitrary $M\to M$ scattering. The condition for this is expressed by Yang-Baxter equation~\eqref{eq:YBE-Rmatrix}, which in terms of~$\check{\mathbf{S}}_{pq}=\check{\mathbf{S}}(p,q)$ reads
\begin{equation}
\label{eq:YB-scheck}
\check{\mathbf{S}}_{qr}\otimes\mathbf{I}
\,\cdot\,
\mathbf{I}\otimes\check{\mathbf{S}}_{pr}
\,\cdot\,
\check{\mathbf{S}}_{pq}\otimes\mathbf{I}
=
\mathbf{I}\otimes\check{\mathbf{S}}_{pq}
\,\cdot\,
\check{\mathbf{S}}_{pr}\otimes\mathbf{I}
\,\cdot\,
\mathbf{I}\otimes\check{\mathbf{S}}_{qr}\,,
\end{equation}
where we made the three-particle tensor product explicit.

It is easy to check that the Yang-Baxter equation is non-trivially%
\footnote{
In the case of $\check{\mathbf{S}}_{\su(1|1)^2_{\text{c.e.}}}$, once the YB equation is spelled out in components, one finds 112~equations involving the non-zero matrix elements, which vanish by using their explicit form.
} satisfied by the $\su(1|1)^2_{\text{c.e.}}$ S~matrix that we found, as well as by the~$\psu(1|1)^4_{\text{c.e.}}$ one---a fact following from the tensor-product structure. Therefore, they can be both used to define an integrable two-dimensional QFT.

\subsection*{Comparison with the $\su(2|2)$-symmetric S~matrix}
The symmetry algebra $\su(1|1)_{\L}\oplus\su(1|1)_{\R}$~centrally extended can be embedded into $\su(2|2)_{\text{c.e.}}$. This last algebra is also related to an integrable S~matrix. The S~matrix of fundamental string excitations in $\AdS_5\times\S^5$ is in fact invariant under two copies of~$\su(2|2)_{\text{c.e.}}$, with the excitations transforming in a tensor product representation akin to the one we encountered here, see \eg~\cite{Arutyunov:2009ga}.

Since $\su(1|1)^{2}_{\text{c.e.}}\subset \su(2|2)_{\text{c.e.}}$ we could expect that the $\su(2|2)_{\text{c.e.}}$-invariant S~matrix should arise of a special limit of our~$\check{\mathbf{S}}$. Differently from~$\su(1|1)^{2}$, in~$\su(2|2)_{\text{c.e.}}$ there exists an additional~$\su(2)\oplus\su(2)$ symmetry. The excitations $\phi^{\smallL,\smallR}$ then would transform as a doublet under the former~$\su(2)$, while~$\psi^{\smallL,\smallR}$ would be a doublet under the latter. This would require the scattering to take the form
\begin{equation}
  \begin{aligned}
    \check{\Smat}_{pq} \left(\ket{\phi^{\smallL}_p {\phi}^{\smallR}_q} + \ket{{\phi}^{\smallR}_p \phi^{\smallL}_q}\right) &= \# \left(\ket{\phi^{\smallL}_p {\phi}^{\smallR}_q} + \ket{{\phi}^{\smallR}_p \phi^{\smallL}_q}\right) , \\
    \check{\Smat}_{pq} \left(\ket{\phi^{\smallL}_p {\phi}^{\smallR}_q} - \ket{{\phi}^{\smallR}_p \phi^{\smallL}_q}\right) 
    &=  \# \left(\ket{\phi^{\smallL}_p {\phi}^{\smallR}_q} - \ket{{\phi}^{\smallR}_p \phi^{\smallL}_q}\right) + \# \left(\ket{\psi^{\smallL}_p {\psi}^{\smallR}_q} - \ket{{\psi}^{\smallR}_p \psi^{\smallL}_q}\right) .
  \end{aligned}
\end{equation}
We immediately see that we cannot obtain this general form from the $\su(1|1)^{2}_{\text{c.e.}}$ S~matrix that we have constructed. The reason is that we have imposed on the S~matrix the discrete LR symmetry, which is incompatible with the additional~$\su(2)\oplus\su(2)$ invariance. It is this symmetry that distinguishes the massive sector of~$\AdS_3\times\S^3\times\T^4$ from a truncation of the~$\AdS_5\times\S^5$ superstring at all~loops. Notice also that in the latter case the excitations form a single irreducible representation of the symmetry algebra, and hence a single scalar factor is left undetermined.

\section{Chapter summary}
The main result of this chapter is the derivation of the two-particle S~matrix~$\check{\mathbf{S}}$ (which is related to~${\mathbf{S}}$ by some fermion signs) for the scattering of fundamental massive excitations of the~$\AdS_3\times\S^3\times\T^4$ superstring, and of the non-relativistic dispersion relation~\eqref{eq:dispersion-charges}.
The matrix~$\check{\mathbf{S}}$ satisfies the Yang-Baxter equation as well as braiding unitarity and physical unitarity, so that we can use it to unambiguously define any multiparticle scattering. $\check{\mathbf{S}}$~is defined up to two antisymmetric unit-norm functions $\mathscr{S}_{pq}, \widetilde{\mathscr{S}}_{pq}$---the \emph{dressing factors}. 

It may appear strange that we could determine~$\check{\mathbf{S}}$ for the massive excitations while completely ignoring the massless ones. Even if we expect those to transform in a different irreducible representation of the symmetry algebra, it is easy to see that perturbatively there exist quartic massive-massless interactions in the Lagrangian---the two sectors \emph{are} coupled. Therefore, the effective vertices for massive particles  will contain massless excitations  running in the loops, even if they cannot appear in the final states.
We can draw a parallel of sorts with $SU(N)$ Yang-Mills theory, with the de~Wit-Faddeev-Popov ghosts playing the role of the massless modes in the massive sector. Even if we ignore the ghosts, we can still correctly predict that the S-matrix elements will be $SU(N)$~invariant. However, to compute their value we need to take the ghosts into account---not everything is fixed by the $SU(N)$ symmetry. Here $SU(N)$ is replaced by  a much larger symmetry, involving the off-shell algebra as well as the higher conserved charges, and the only undertermined elements are the dressing factors. They will contain the dynamical information about the theory, featuring poles and cuts in such a way as to also account for processes involving virtual massless particles. We will return to this topic in the next chapter.

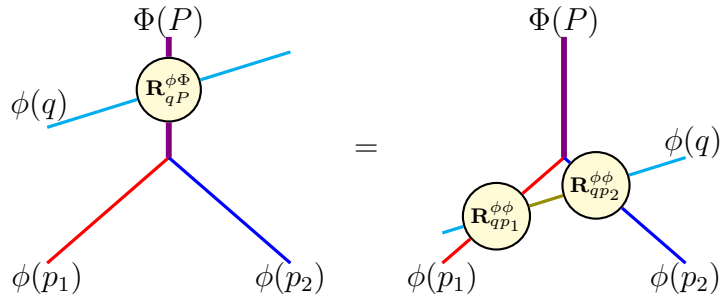
\begin{figure}
  \centering
\begin{tikzpicture}
  \begin{scope}[xshift=-2.6cm]
    \coordinate (i1) at (-1.6cm,0);
    \coordinate (i2) at (+1.6cm,0);
    
    \coordinate (p1) at (-1.6cm,1.8cm);
    \coordinate (p2) at (+1.6cm,2.8cm);

    \coordinate (c) at (0cm,1.4cm);

    \coordinate (o) at (0cm,3cm);
    
    \node (s0) at (0,2.3cm)[S-mat] {${\scriptstyle\mathbf{R}^{\phi\Phi}_{qP}}$};

    \draw [very thick,red]    (i1) to (c);
    \draw [very thick,blue]   (i2) to (c);
    \draw [very thick,cyan]   (p1) to (s0);
    \draw [very thick,cyan]   (s0) to (p2);
    
    \draw [line width=2.2pt,violet] (c) to (s0);
    \draw [line width=2.2pt,violet] (s0) to (o);
    
\node at (-1.6cm,-0.2cm) {$\phi(p_1)$};
\node at (1.6cm,-0.2cm) {$\phi(p_2)$};
\node at (-1.7cm, 2.1cm) {$\phi(q)$};
\node at (0cm,3.2cm) {$\Phi(P)$};

  \end{scope}
  \node at (0,1.5cm) {$=$};
  \begin{scope}[xshift=2.6cm]
    \coordinate (i1) at (-1.6cm,0);
    \coordinate (i2) at (+1.6cm,0);
    
    \coordinate (p1) at (-1.6cm,0.4cm);
    \coordinate (p2) at (+1.6cm,1.4cm);

    \coordinate (c) at (0cm,1.4cm);

    \coordinate (o) at (0cm,3cm);
    
    \node (s1) at (-0.889cm,0.622cm)[S-mat] {${\scriptstyle\mathbf{R}^{\phi\phi}_{qp_1}}$};
    \node (s2) at (0.421cm,1.031cm)[S-mat] {${\scriptstyle\mathbf{R}^{\phi\phi}_{qp_2}}$};

    \draw [very thick,red]    (i1) to (s1);
    \draw [very thick,red]    (s1) to (c);
    \draw [very thick,blue]   (i2) to (s2);
    \draw [very thick,blue]   (s2) to (c);
    \draw [very thick,cyan]   (p1) to (s1);
    \draw [very thick,olive]   (s1) to (s2);
    \draw [very thick,cyan]   (s2) to (p2);
    
    \draw [line width=2.2pt,violet] (c) to (o);
    
\node at (-1.6cm,-0.2cm) {$\phi(p_1)$};
\node at (1.6cm,-0.2cm) {$\phi(p_2)$};
\node at (1.7cm, 1.6cm) {$\phi(q)$};
\node at (0cm,3.2cm) {$\Phi(P)$};

  \end{scope}
\end{tikzpicture}%

  \caption{The bootstrap condition for the bound-state S~matrix. We consider two particles~$\phi$ with complex momenta $p_1=\tfrac{1}{2}P+iv $ and~$p_2=\tfrac{1}{2}P-iv $ that form a bound state~$\Phi$ of momentum~$P$ (thicker violet line). The scattering of a third particle~$\phi(q)$ with~$\Phi(P)$ can be resolved in terms of the scattering with its constituents. Note that the point in the diagram at which the two particle fuse to give a bound state should not be interpreted as a scattering event.}
  \label{fig:boundstates}
\end{figure}
We should also stress that we only computed the S~matrix of \emph{fundamental} massive excitations. If the theory admits bound states, the S~matrix that scatters those should also be determined to have a complete handle on the spectrum, in particular to compute the wrapping effects~\cite{Arutyunov:2007tc,Arutyunov:2009zu}. Typically, this can be done again by means of integrability. The scattering of a particle with a bound state can be defined as illustrated in figure~\ref{fig:boundstates}, \ie\ in term of the scattering with its constituents---a procedure called~\emph{bootstrap}. This is particularly simple  when one takes advantage of the Hopf~algebra structure, see \eg\ refs.~\cite{Delius:1995he,Torrielli:2010kq,Torrielli:2011gg}. In the case of~$\AdS_3\times\S^3\times\T^4$, such a calculation has not been performed so far.

\chapter{Crossing symmetry and dressing factors}
\label{ch:crossing}
In this chapter we will see that crossing symmetry puts stringent requirements on the analytic structure of the dressing factors.

\begin{figure}
  \centering
\begin{tikzpicture}
  \begin{scope}[xshift=-2.6cm]
    \coordinate (i1) at (-1.8cm,0);
    \coordinate (i2) at (+1.8cm,0);

    \coordinate (o2) at (-1.8cm,3cm);
    \coordinate (o1) at (+1.8cm,3cm);   

    \coordinate (c) at (0cm,1.5cm);
    
    \node (s0) at (c)[S-mat] {${\,\scriptstyle\fixedspaceL{\mathbf{R}(\vartheta^{\mathsf{c}})}{\mathbf{R}(\vartheta)}}$};

    \draw [partic1]    (i1) to (s0);
    \draw [partic2]   (i2) to (s0);
    \draw [partic1]    (s0) to (o1);
    \draw [partic2]   (s0) to (o2);
    
    \draw [time]   (-2.2cm,0.8cm) to (-2.2cm,2.2cm);
    
    \draw (-0.6cm,1cm) arc (-140.2:-40.2:0.8);
    
\node at (-1.9cm,-0.2cm){$1$};
\node at (1.9cm,-0.2cm) {$2$};
\node at (-1.9cm,3.2cm) {$2$};
\node at (1.9cm,3.2cm)  {$1$};

\node at (0cm,0.5cm)  {$\vartheta$};

\node at (-2.4cm,1.5cm)  {$t$};

  \end{scope}
  \node at (0,1.5cm) {$=$};
  \begin{scope}[xshift=2.6cm]
    \coordinate (i1) at (-1.8cm,0);
    \coordinate (i2) at (+1.8cm,0);

    \coordinate (o2) at (-1.8cm,3cm);
    \coordinate (o1) at (+1.8cm,3cm);   

    \coordinate (c) at (0cm,1.5cm);
    
    \node (s0) at (c)[S-mat] {${\,\scriptstyle\mathbf{R}(\vartheta^{\mathsf{c}})\,}$};

    \draw [partic1]    (i1) to (s0);
    \draw [partic2]   (i2) to (s0);
    \draw [partic1]    (s0) to (o1);
    \draw [partic2]   (s0) to (o2);
    
    \draw [time]   (-0.8cm,3.4cm) to (0.8cm,3.4cm);
    
    \draw (-0.6cm,1cm) arc (-130:-230:0.65);
    
\node at (-1.9cm,-0.2cm){$1$};
\node at (1.9cm,-0.2cm) {$2$};
\node at (-1.9cm,3.2cm) {$2$};
\node at (1.9cm,3.2cm)  {$1$};

\node at (-1.5cm,1.5cm)  {$\vartheta-\pi$};

\node at (0,3.6cm)  {$t$};

  \end{scope}
\end{tikzpicture}%

  \caption{In a relativistic theory, crossing invariance can be understood by looking at a scattering process in two different ways. In the left panel, time flows upwards, and the scattering of particles~$1$ and~$2$ happens with rapidity~$\vartheta=\vartheta_2-\vartheta_1$.  In the right panel, time flows from left to right, and the same scattering involves the antiparticle of~$2$ (moving backwards in time) and~$1$. The scattering happens with rapidity~$\vartheta^{\mathsf{c}}=\vartheta+i\pi $, which can be understood as a Lorentzian~angle.
  }
  \label{fig:relativcrossing}
\end{figure}
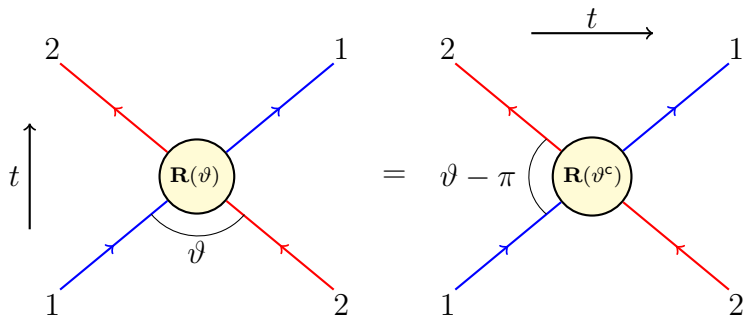

In a relativistic theory, crossing symmetry is a consequence of the fact that fields are constructed out of particle \emph{and} antiparticle creation and annihilation operators. One can then show that any two elements of the S~matrix related by a particle-to-antiparticle transformation are equivalent up to performing an analytic continuation, which exchanges the branches of the dispersion relation. This invariance is illustrated pictorially in figure~\ref{fig:relativcrossing}. There we have introduced the relativistic rapidity~$\vartheta$ satisfying
\begin{equation}
E=m\,\cosh\vartheta\,,\qquad
p=m\,\sinh\vartheta\,,
\end{equation}
which parametrises the positive branch of the dispersion relation~$E^2=m^2+p^2$. In a theory where all particles coincide with their antiparticles (\eg\ photons), each of them can be mapped into each of the other by shifting
\begin{equation}
\label{eq:relativistic-crossing}
\vartheta\to\vartheta+i\pi\,.
\end{equation}
This flips both the sign of~$E$ and~$p$. If particles and antiparticles do not coincide, the transformation~\eqref{eq:relativistic-crossing}  should be supplemented by a linear map sending the particle-representation into the antiparticle one---\eg~in the case of quarks and antiquarks, this would be a map between the fundamental and anti-funtamental representation of~$\su(3)$.

These concept can be extended to the non-relativistic S~matrix that we computed in the previous chapter. In order to do so, it is first convenient to introduce a~\emph{rapidity} variable $z$ akin to~$\vartheta$. Then, we will study the charge-conjugation properties of the $\psu(1|1)^4_{\text{c.e.}}$ or equivalently~$\su(1|1)^2_{\text{c.e.}}$ algebras that we constructed in the previous chapter. This will lead us to formulate the~\emph{crossing equations} constraining the dressing factors, for which we will find a solution in section~\ref{sec:dressingfact}.

\section{Crossing symmetry}
\label{sec:crossing}
Let us start by defining the rapidity~$z$ and its domain.

\subsection*{Uniformising the dispersion relation}
The all-loop dispersion relation for massive excitations on $\AdS_3\times \S^3\times \T^4$ reads
\begin{equation}
E^2(p)=1+4h^2\sin^2\frac{p}{2}\,.
\end{equation}
This is the same functional form appearing in~$\AdS_5\times\S^5$. To uniformise the dispersion relation we can introduce a rapidity variable following ref.~\cite{Janik:2006dc}.
Let~$z$ satisfy
\begin{equation}
\label{eq:elliptic-E-p}
p(z)=2\,\am z\,,\qquad \sin\frac{p(z)}{2}=\sn(z,\kappa)\,,\qquad E(z)=\dn(z,\kappa)\,,
\end{equation}
in terms of Jacobi's elliptic functions, where the elliptic modulus is $\kappa=-4h^2$. This defines a torus with a real period $2\omega_1$ and an imaginary period $2\omega_2$ that depend on $h$ through
\begin{equation}
\omega_1=2\,\K(\kappa)\,,\qquad
\omega_2=2i\,\K(1-\kappa)-2\,\K(\kappa)\,,
\end{equation}
where $\K$ is the complete elliptic integral of the first kind. In this parameterisation the real $z$-axis corresponds to real momentum and positive energy. In order to describe bound states, it is useful to consider complex momenta, and therefore complex rapidities~$z$ in the torus.
One can check that all the definitions we have given are also $\omega_1$-periodic, so that we can always restrict to $|\Re (z)|\leq \omega_1/2$, corresponding to $-\pi < p \leq \pi$.

The Zhukovski variables~$x^\pm(z)$ are meromorphic functions on the torus
\begin{equation}
x^\pm(z)=\frac{1}{2h}\left(\frac{\cn(z,\kappa)}{\sn(z,\kappa)}\pm i\right)\left(1+\dn(z,\kappa)\right)\,,
\end{equation}
and satisfy the relations~\eqref{eq:zhukovski-def}. Furthermore we can resolve the square root in~$\eta_p$ by
\begin{equation}\label{eq:eta-xpm}
\eta(z) =
\frac{\dn\frac{z}{2}\,\big(\cn\frac{z}{2}+i\,\sn\frac{z}{2}\,\dn\frac{z}{2}\big)}{1+4h^2\,\sn^{4}\frac{z}{2}}\,.
\end{equation}
The S-matrix elements that we computed in the previous chapter can now be expressed purely in terms of~$z$. Remarkably, they are all~\emph{rational functions} on two copies%
\footnote{%
Recall that $\mathbf{S}(p_1,p_2)$ or more precisely~$\mathbf{S}(z_1,z_2)$ depends on $z_1$ and $z_2$ separately, in contrast with the relativistic case where it would be 
$\mathbf{S}(\vartheta_1,\vartheta_2)=\mathbf{S}(\vartheta_1-\vartheta_2)$.
}
 of the elliptic torus, up to the dressing factors.

\begin{figure}
  \centering
  \subfloat[Torus with $|x^\pm|=1$ curves]{%
    \label{fig:torusabs}
    \includegraphics[width=54mm]{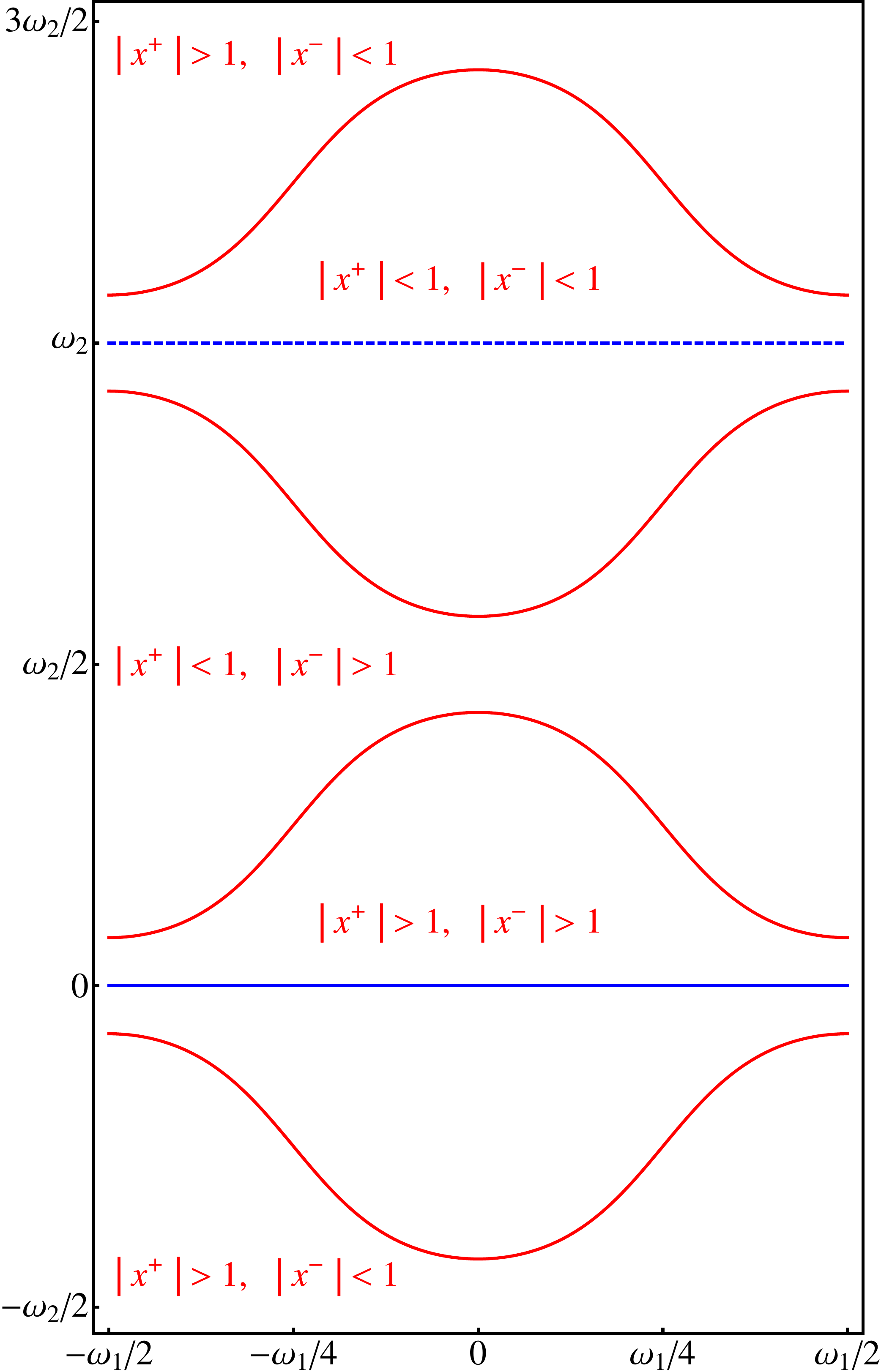}}
  \subfloat[Torus with $\Im(x^\pm)=0$ curves]{%
    \label{fig:torusim}
    \includegraphics[width=54mm]{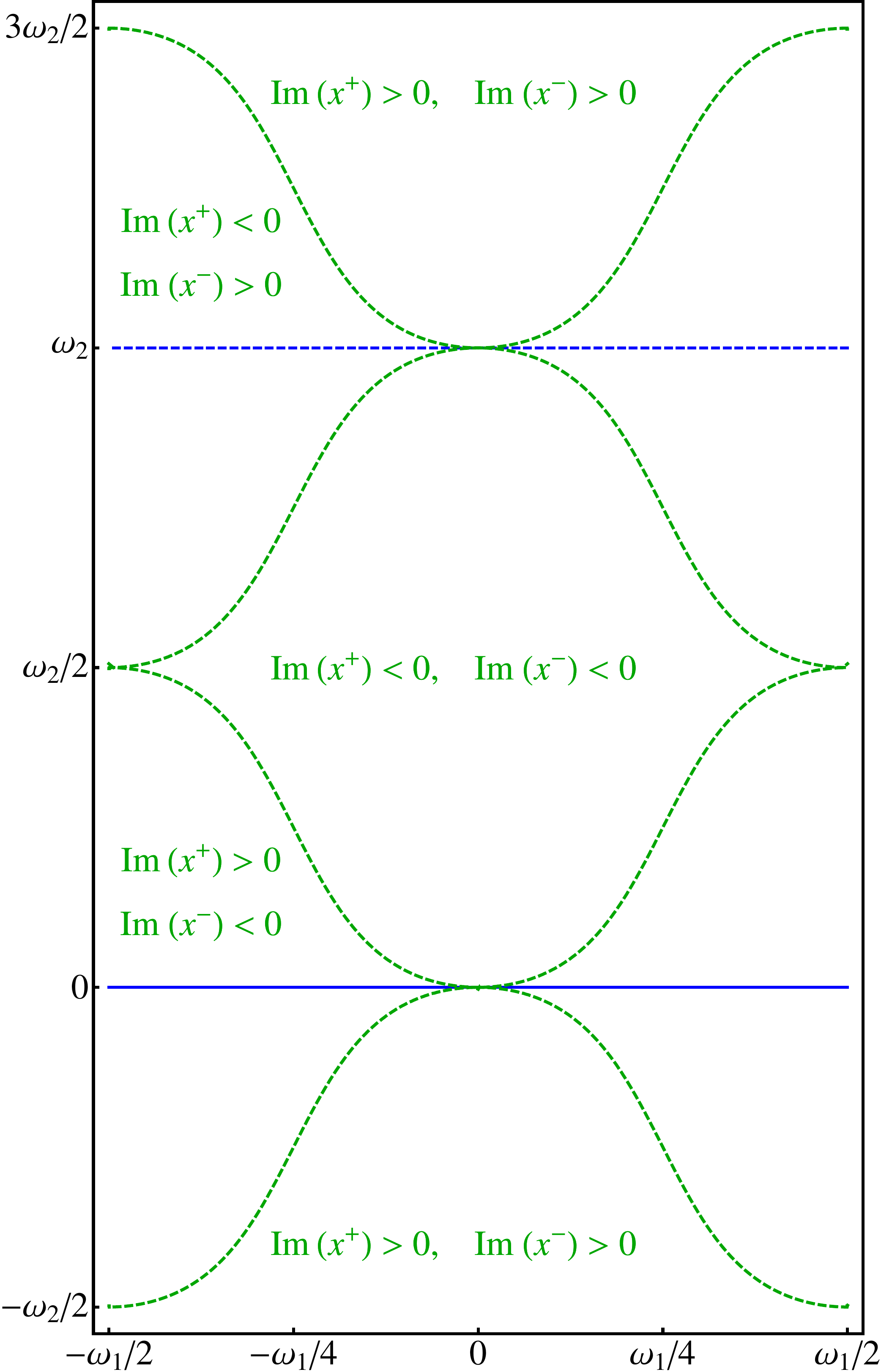}}
  \subfloat[Torus with both curves]{%
    \label{fig:torusabsim}
    \includegraphics[width=54mm]{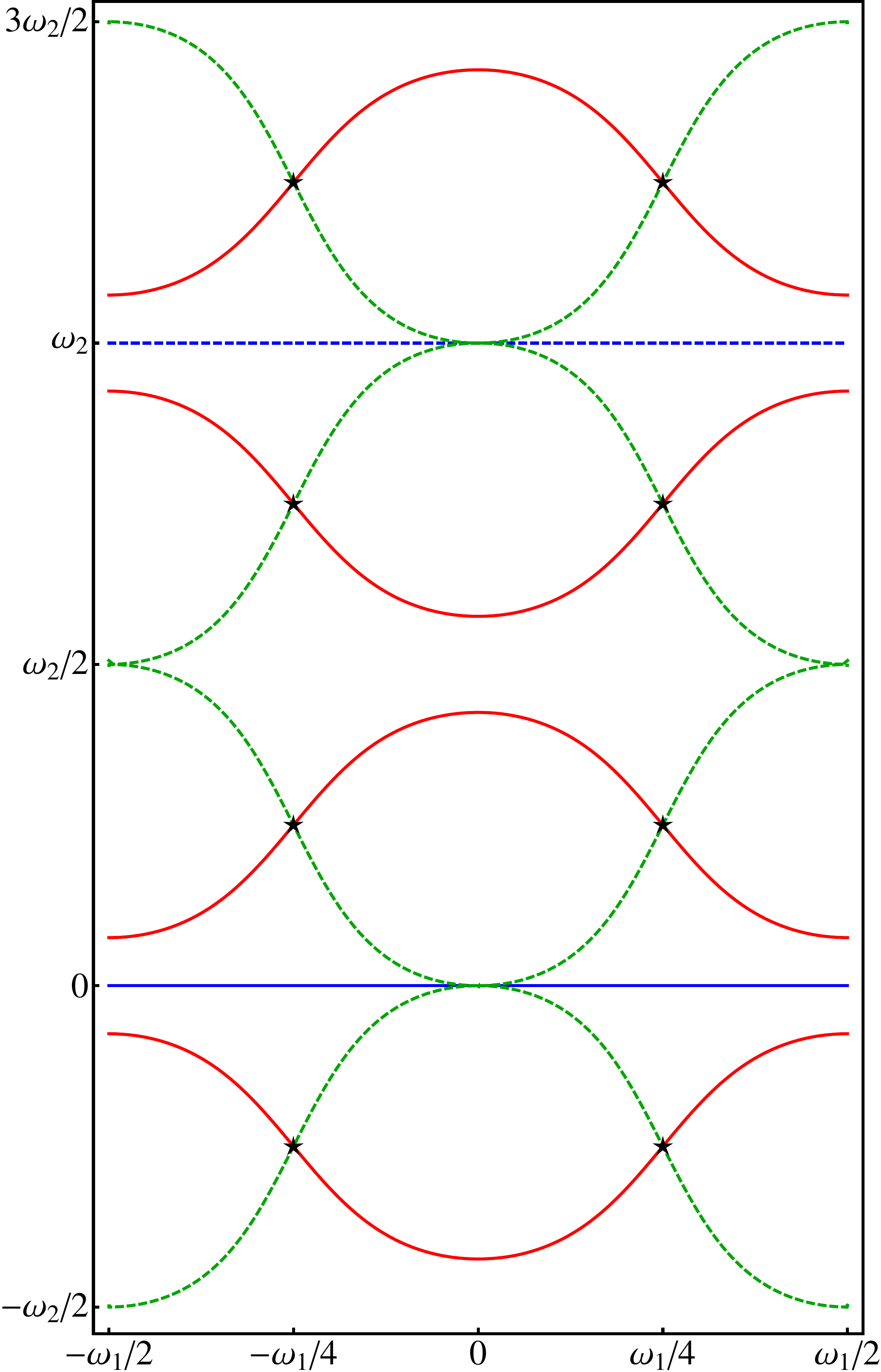}}
  \caption{%
    The rapidity torus with several significant curves. The solid blue line is the real $z$-axis (physical region),  the dashed blue line is the $z=\omega_2$ axis (``crossed'' region). In the leftmost figure the torus is divided in four regions by  $|x^\pm|=1$ and  in the central figure it is divided by $\Im(x^\pm)=0$. The rightmost picture depicts both sets of curves, which intersect in eight points with real part $\pm\omega_1/4$, denoted by stars.
  }%
  \label{fig:torus}
\end{figure}

In figure~\ref{fig:torus} we depict the rapidity torus, highlighting several significant curves. The red ones correspond to~$|x^{\pm}(z)|=1$. In analogy with~$\AdS_5\times\S^5$, we will call the region they delimit which contains the real $z$-axis \emph{the physical region}---this is where the complex momenta of physical bound state are expected to be found~\cite{Arutyunov:2007tc}. It can be identified by requiring $|x^{\pm}(z)|>1$. In what follows, we will consider all S-matrix elements corresponding to physical processes to be evaluated at rapidities lying in the physical region.

A shift $z\to z+\omega_2$ flips the sign of energy and momentum~\eqref{eq:elliptic-E-p}. The region containing the line $\Im(z)=\omega_2$ corresponds to real crossed variables, having real momentum and real negative energy. In terms of the Zukhovski variables we have that
\begin{equation}
x^+(z\pm\omega_2)=\frac{1}{x^{+}(z)}\,,
\quad
x^-(z\pm\omega_2)=\frac{1}{x^{-}(z)}\,,
\quad
\eta(z\pm\omega_2)=\frac{\pm i}{x^{+}(z)}\eta(z)\,,
\end{equation}
so that the shift maps the physical region $|x^{\pm}|>1$ to $|x^{\pm}|<1$.
The scattering  of antiparticles, which take the negative branch of the dispersion relation, will give S-matrix elements evaluated for~$z$ in the crossed region, \ie\ for $|x^{\pm}(z)|<1$.

\subsection*{Charge conjugation}
Let us go back to the representations of~$\psu(1|1)^4_{\text{c.e.}}$, or equivalently of~$\su(1|1)^2_{\text{c.e.}}$. Even better, since we know how to obtain multiparticle representations in terms of a non-trivial coproduct, we can even restrict to the one-particle representation given by the~$4\times4$ supermatrices~\eqref{eq:supercharges-1part-matrix} and by the relative central charges. We can label this pair of representations as
\begin{equation}
\mathscr{V}_{\smallLR}(e,c,\bar{c})=
\mathscr{V}_{\smallL}(e,c,\bar{c})
\oplus
\mathscr{V}_{\smallR}(e,c,\bar{c})\,,
\end{equation}
where $e$ labels the energy and $c,\bar{c}$ the central charges eigenvalues.

By transforming all of the charges and supercharges $\mathbf{x}$ by means of a supertransposition
\begin{equation}
\mathbf{x}\to -\mathbf{x}^{\text{st}}\,,
\end{equation}
we find another representation of~$\su(1|1)^2_{\text{c.e.}}$. Notice that the minus sign is needed to ensure that the anticommutation relations are preserved. This clearly flips the sign of the central charges. In particular, we now have that $E<0$ and that left and right excitation have swapped their charge under~$\mathbf{m}$.

Combining $\mathbf{x}\to -\mathbf{x}^{\text{st}}$ with the crossing transformation $z\to z+\omega_2$, we can \emph{almost} obtain a pair of representations with the same central charges as the one that we started from. If that were the case, that would mean that there exists a (anti)unitary transformation relating $-\mathbf{x}^{\text{st}}(z+\omega_2)$ with~$\mathbf{x}(z)$---that is, a charge conjugation matrix.
There are two obstructions for this to be the case. Firstly, for the eigenvalues of~$\mathbf{m}$ to be swapped, charge conjugation must exchange the left and right irreducible representations.
Secondly, the crossing transformation does flip the sign of energy, but acts on $C(z)$ as
\begin{equation}
c(z)=i\zeta\,\frac{h}{2}\left(e^{ip(z)}-1\right)
\quad\to\quad
c(z+\omega_2)=
-e^{-ip(z)}\,c(z)\,,
\end{equation}
and similarly for~$\bar{c}(z)$. These additional phase factors can be accounted for if we additionally transform the supercharges~$\mathbf{q}=\mathbf{q}^{\smallL,\smallR}$ by a phase
\begin{equation}
\begin{aligned}
&\big(-\mathbf{q}^{\text{st}}(z+\omega_2)\big)\to e^{+\frac{i}{2} p} \, \big(-\mathbf{q}^{\text{st}}(z+\omega_2)\big)\,,\\
&\big(-\bar{\mathbf{q}}^{\text{st}}(z+\omega_2)\big)\to e^{-\frac{i}{2} p}\, \big(-\bar{\mathbf{q}}^{\text{st}}(z+\omega_2)\big)\,.
\end{aligned}
\end{equation}
This transformation is an $U(1)$ automorphism of~$\su(1|1)_{\text{c.e.}}^2$. Using it, we can finally conclude that, for the supercharges it must be
\begin{equation}
\label{eq:chargetransformation}
\begin{aligned}
-e^{+\frac{i}{2} p}\,\mathbf{q}^{\text{st}}(z+\omega_2)
= \mathscr{C}\,\mathbf{q}(z)\,\mathscr{C}^{-1}\,,\\
-e^{-\frac{i}{2} p}\,\bar{\mathbf{q}}^{\text{st}}(z+\omega_2)
= \mathscr{C}\,\bar{\mathbf{q}}(z)\,\mathscr{C}^{-1}\,,
\end{aligned}
\end{equation}
where~$\mathscr{C}$ is the charge-conjugation matrix, which exchanges the left and right representations. It can be written in the basis $\mathscr{B}$ as
\begin{equation}
\mathscr{C}=
\left(
\begin{array}{cc|cc}
0		&		0		&	\xi^{\smallRL}	&		0	\\
0		&		0		&		0		& -i\,\xi^{\smallRL}	\\
\hline
\xi^{\smallLR}	&		0		&		0		&		0	\\
0		&	-i\,\xi^{\smallLR}	&		0		&		0
\end{array}	
\right)\,.
\end{equation}
The two normalisation constants $\xi^{\smallLR}$ and $\xi^{\smallRL}$, one per representation, can be set without loss of generality to \eg
\begin{equation}
\xi^{\smallLR}=\xi^{\smallRL}=1\,.
\end{equation}
In this way $\mathscr{C}^2=\Sigma$ and $\mathscr{C}^{-1}=\mathscr{C}^\dagger$.

By using the fact supertransposition acts on the supercharges as  $\mathbf{q}^{\text{st}}= \mathbf{q}^{\text{t}}\,\Sigma$, we can rewrite
eq.~\eqref{eq:chargetransformation} as
\begin{equation}
\label{eq:chargetransformation2}
\begin{aligned}
&\mathbf{q}^{\text{t}}(z\pm\omega_2)
= \mp e^{-\frac{i}{2}p}\,\Sigma\, \mathscr{C}^{\dagger}\,\mathbf{q}(z)\,\mathscr{C}\,,\\
&\bar{\mathbf{q}}^{\text{t}}(z\pm \omega_2)
= \mp e^{+\frac{i}{2}p}\,\Sigma\, \mathscr{C}^{\dagger}\,\bar{\mathbf{q}}(z)\,\mathscr{C}\,,
\end{aligned}
\end{equation}
where we also used the fact that $\Sigma\,\mathscr{C}=\mathscr{C}\,\Sigma$ and $\Sigma\,\mathbf{q}=-\mathbf{q}\,\Sigma$. Note how these equations are related by Hermitian conjugation.

\subsection*{Crossing equations}
The fact that the charges enjoy the invariance~\eqref{eq:chargetransformation} together with the fundamental invariance property of the S~matrix
\begin{equation}
\mathbf{R}_{(12)}\,\mathbf{Q}_{(12)}
=
\mathbf{Q}_{(21)}\,\mathbf{R}_{(12)}\,,
\end{equation}
allows us to derive an additional property of~$\mathbf{R}_{12}$.%
\footnote{%
We choose to work with~$\mathbf{R}$ here rather than~$\mathbf{S}$ or~$\check{\mathbf{S}}$ to obtain a more compact and familiar-looking final expression the crossing equations.
}
To this end, let us focus on the case where $\mathbf{Q}$ is one of the supercharges $\mathbf{q}^{\smallL,\smallR}$ or of their conjugates---the invariance under the central charges  follows trivially.
Furthermore, let us recall that~$\mathbf{Q}_{(12)}$ is given by the coproduct~\eqref{eq:16x16charges}, and similarly~$\mathbf{Q}_{(21)}$ is given by
\begin{equation}
\mathbf{Q}_{(21)}(p,q)=
\mathbf{Q}(p,\zeta_1=e^{iq})\otimes \Sigma+
\mathbf{I}\otimes\mathbf{Q}(q,\zeta_2=1).
\end{equation}
It is then easy to write down the invariance of~$\mathbf{R}$ under~$\mathbf{q}=\mathbf{q}^{\smallL,\smallR}$
\begin{equation}
\label{eq:R-matrix-invariance}
\begin{aligned}
&\mathbf{R}_{(12)}(z_1,z_2)
\left(
\mathbf{q}(z_1)\otimes\mathbf{I} +
e^{\frac{i}{2}p_1}\Sigma\otimes\mathbf{q}(z_2)
\right)\\
&\qquad\qquad\qquad\qquad\qquad
=
\left(
e^{\frac{i}{2}p_2}\mathbf{q}(z_1)\otimes \Sigma+
\mathbf{I}\otimes\mathbf{q}(z_2)
\right)
\,\mathbf{R}_{(12)}(z_1,z_2)\,,
\end{aligned}
\end{equation}
where $p_j=p(z_j)$. If we take the transpose of this equation with respect to \eg\ the first space, we find
\begin{equation}
\begin{aligned}
&
\left(-
\Sigma\,\mathbf{q}^{\text{t}}(z_1)\otimes\mathbf{I} +
\Sigma\otimes\mathbf{q}(z_2)
\right)\mathbf{R}_{(12)}^{\text{t}_1}(z_1,z_2)\\
&\qquad\qquad\qquad\qquad\qquad
=
\mathbf{R}^{\text{t}_1}_{(12)}(z_1,z_2)\left(-
e^{\frac{i}{2}p_2}\Sigma\,\mathbf{q}^{\text{t}}(z_1)\otimes \Sigma
+
e^{\frac{i}{2}p_1} \mathbf{I}\otimes\mathbf{q}(z_2)
\right)
.
\end{aligned}
\end{equation}
where we also multiplied by $-\Sigma$ from the left.
This equation holds for any $z_1,z_2$. If we now perform a shift $z_1\to z_1+\omega_2$ and use~\eqref{eq:chargetransformation2}, we can recast the equation in the form
\begin{equation}
\begin{aligned}
&
\left(
\mathbf{q}(z_1)\otimes\mathbf{I} +
e^{\frac{i}{2}p_1}\Sigma\otimes\mathbf{q}(z_2)
\right)
\mathscr{C}_{(1)}\,\mathbf{R}_{(12)}^{\text{t}_1} (z_1+\omega_2,z_2)\mathscr{C}^{\dagger}_{(1)}
\\
&\qquad\qquad\qquad\qquad
=
\mathscr{C}_{(1)}\,\mathbf{R}_{(12)}^{\text{t}_1} (z_1+\omega_2,z_2)\mathscr{C}^{\dagger}_{(1)}
\left(
e^{\frac{i}{2}p_2}\mathbf{q}(z_1)\otimes \Sigma+
\mathbf{I}\otimes\mathbf{q}(z_2)
\right),
\end{aligned}
\end{equation}
where $\mathscr{C}_{(1)}=\mathscr{C}\otimes\mathbf{I}$.
Comparing this expression with~\eqref{eq:R-matrix-invariance}, we see that the combination $\mathscr{C}_{(1)}\,\mathbf{R}_{(12)}^{\text{t}_1} (z_1+\omega_2,z_2)\mathscr{C}^{\dagger}_{(1)}$ has the same invariance property with respect to~$\mathbf{q}^{\smallL}$ as the inverse of $\mathbf{R}_{(12)}(z_1,z_2)$. In fact, it is not hard to see that the same calculation applies to all of the supercharges.\footnote{%
Note that in the case of~$\bar{\mathbf{q}}^{\smallL,\smallR}$ the sign in the phase~shifts $e^{\pm \frac{i}{2}p}$ is everywhere opposite to the one we used for~${\mathbf{q}}^{\smallL,\smallR}$.
} We can therefore require that it is
\begin{equation}
\label{eq:crossing1var}
\mathscr{C}_{(1)}\,\mathbf{R}_{(12)}^{\text{t}_1} (z_1+\omega_2,z_2)\mathscr{C}^{\dagger}_{(1)}=\big(\mathbf{R}_{(12)}(z_1,z_2)\big)^{-1}.
\end{equation}
This is the \emph{crossing equation}. If we had repeated a similar calculation taking the transpose in the second space, we would have found
\begin{equation}
\label{eq:crossing2var}
\mathscr{C}_{(2)}\,\mathbf{R}_{(12)}^{\text{t}_2} (z_1,z_2-\omega_2)\mathscr{C}^{\dagger}_{(2)}=\big(\mathbf{R}_{(12)}(z_1,z_2)\big)^{-1}.
\end{equation}
In fact, the two crossing equations are related by braiding unitarity, as we will see explicitly in the next subsection.

What we have just found with regard to crossing symmetry is somewhat similar to what happened for the unitarity condition~\eqref{eq:linear-unitarty}. The invariance property we have found is seemigly new, but it can be found from manipulating eq.~\eqref{eq:R-matrix-invariance}. Since the matrix part of~$\mathbf{R}$ has been found imposing~\eqref{eq:R-matrix-invariance}, it will \emph{automatically} satisfy crossing invariance.
However, this requirement will mean that the scalar factors~$\mathscr{S}_{pq}$ and~$\widetilde{\mathscr{S}}_{pq}$ may not be arbitrary, but satisfy the analyticity requirements coming from~\eqref{eq:crossing1var} or equivalently~\eqref{eq:crossing2var}.

Here we have derived the crossing symmetry requirements for the~$\su(1|1)^2_{\text{c.e.}}$ S~matrix. The ones for the bigger~$\psu(1|1)^4_{\text{c.e.}}$ S~matrix follow immediately by requiring that it is given by the tensor product~\eqref{eq:S-matrix-tensorpr} of two \emph{crossing invariant} $\su(1|1)^2_{\text{c.e.}}$ S~matrices.
Equivalently, it is easy to repeat the construction of the charge-conjugation matrix for the charges in~$\psu(1|1)^4_{\text{c.e.}}$ and obtain the crossing equations in that way.

\subsection*{Constraints on the scalar factors}
Let us write the requirements imposed by crossing symmetry on the scalar factors of the $\psu(1|1)^4_{\text{c.e.}}$ S~matrix. For later convenience, let us normalise the scalar factors by introducing two functions
$\sigma(p_1,p_2)$ and $\widetilde{\sigma}(p_1,p_2)$ satisfying
\begin{equation}
\label{eq:sigmaSrel}
\begin{aligned}
&\mathscr{S}^2(p_1,p_2)=
\sigma^{-2}(p_1,p_2)\,\frac{x^+_1-x^-_2}{x^-_1-x^+_2}
\frac{1-\frac{1}{x^-_1 x^+_2}}{1-\frac{1}{x^+_1 x^-_2}}\,,\\
&\widetilde{\mathscr{S}}^2(p_1,p_2)=
\widetilde{\sigma}^{-2}(p_1,p_2)\,
\frac{1-\frac{1}{x^-_1 x^+_2}}{1-\frac{1}{x^+_1 x^-_2}}\,,
\end{aligned}
\end{equation}
In this way the phase for the same-chirality diagonal processes features the Beisert-Dippel-Staudacher~\cite{Beisert:2004hm} matrix element. The scalar factors have the form
\begin{equation}
\label{eq:twoads3phases}
\sigma(p_1,p_2)=e^{i\,\theta(p_1,p_2)},\qquad
\widetilde{\sigma}(p_1,p_2)=e^{i\,\tilde{\theta}(p_1,p_2)},
\end{equation} 
where $\theta(p_1,p_2)$ and $\widetilde{\theta}(p_1,p_2)$ are antisymmetric real analytic functions for real $p_1,\,p_2$---a requirement that follows from unitarity.

The crossing equations that we have found then imply
\begin{equation}
\begin{aligned}
\sigma({p}^{\mathsf{c}}_1,p_2)^2\,\widetilde{\sigma}(p_1,p_2)^2= g(p_1,p_2)\,,
&\quad &
\sigma(p_1,p_2)^2\,\widetilde{\sigma}({p}^{\mathsf{c}}_1,p_2)^2= \widetilde{g}(p_1,p_2)\,,\\
\sigma(p_1,{p}^{\mathsf{c}}_2)^2\,\widetilde{\sigma}(p_1,p_2)^2= \frac{1}{\widetilde{g}({p}^{\mathsf{c}}_2,p_1)}\,,
&\quad &
\sigma(p_1,p_2)^2\,\widetilde{\sigma}(p_1,{p}^{\mathsf{c}}_2)^2= \frac{1}{g({p}^{\mathsf{c}}_2,p_1)}\,,
\end{aligned}
\label{eq:crossing12}
\end{equation}
where
\begin{equation}
  \begin{aligned}
    g(p_1,p_2) &= \left(\frac{x_2^-}{x_2^+}\right)^2\frac{\left(1-\frac{1}{x_1^+x_2^+}\right)\left(1-\frac{1}{x_1^-x_2^-}\right)}{\left(1-\frac{1}{x_1^+x_2^-}\right)^2}\frac{x_1^- - x_2^+}{x_1^+ - x_2^-}\,,\\
    \widetilde{g}(p_1,p_2) &= \left(\frac{x_2^-}{x_2^+}\right)^2\frac{\left(x_1^- - x_2^+\right)^2}{\left(x_1^+ - x_2^+\right)\left(x_1^- - x_2^-\right)}\frac{1-\frac{1}{x_1^-x_2^+}}{1-\frac{1}{x_1^+x_2^-}}\,.
\end{aligned}
\label{eq:crossingF}
\end{equation}
and the superscript~${\mathsf{c}}$ indicates crossing. More precisely, we have
\begin{equation}
{p}^{\mathsf{c}}_1= p(z_1+\omega_2)\,,\qquad
{p}^{\mathsf{c}}_2= p(z_2-\omega_2)\,.
\end{equation}
It is then easy to check that the four equations~\eqref{eq:crossing12} are related by the antisymmetry of the scalar factors, \ie\ owing to unitarity. Therefore, it will be enough to restrict ourselves to \eg\ 
\begin{equation}
\begin{aligned}
\sigma({p}^{\mathsf{c}}_1,p_2)^2\,\widetilde{\sigma}(p_1,p_2)^2= g(p_1,p_2)\,,
&\quad &
\sigma(p_1,p_2)^2\,\widetilde{\sigma}({p}^{\mathsf{c}}_1,p_2)^2= \widetilde{g}(p_1,p_2)\,,
\end{aligned}
\label{eq:crossing1}
\end{equation}
which are the ones due to~\eqref{eq:crossing1var}.

We  have remarked how the matrix part of~$\mathbf{R}$ is a meromorphic function on the rapidity torus. However, by iterating the crossing transformation twice we find that the dressing factors are not $2\omega_2$-periodic:
\begin{equation}
\begin{aligned}
\frac{\sigma(z_1+2\omega_2,z_2)^2}{\sigma(z_1,z_2)^2}=\frac{g(z_1+\omega_2,z_2)}{\widetilde{g}(z_1,z_2)} &= \left(\frac{x_1^+ - x_2^+}{x_1^+ - x_2^-}\frac{x_1^- - x_2^-}{x_1^- - x_2^+}\right)^2\,,\\
\qquad
\frac{\widetilde{\sigma}(z_1+2\omega_2,z_2)^2}{\widetilde{\sigma}(z_1,z_2)^2}=\frac{\widetilde{g}(z_1+\omega_2,z_2)}{g(z_1,z_2)} &= \left(\frac{1-\frac{1}{x_1^+x_2^-}}{1-\frac{1}{x_1^+x_2^+}}\frac{1-\frac{1}{x_1^-x_2^+}}{1-\frac{1}{x_1^-x_2^-}}\right)^2\,.
\end{aligned}
\end{equation}
Therefore, they must have cuts on the rapidity torus, so that the whole crossing-invariant S~matrix is defined on some more complicated surface. In fact, even if we considered shifts by $4\omega_2,\,6\omega_2,$ \etc\ we would still find no periodicity, meaning that the dressing phases should live on an infinite cover of the $z$-torus.

\begin{figure}
  \centering
  \includegraphics[width=75mm]{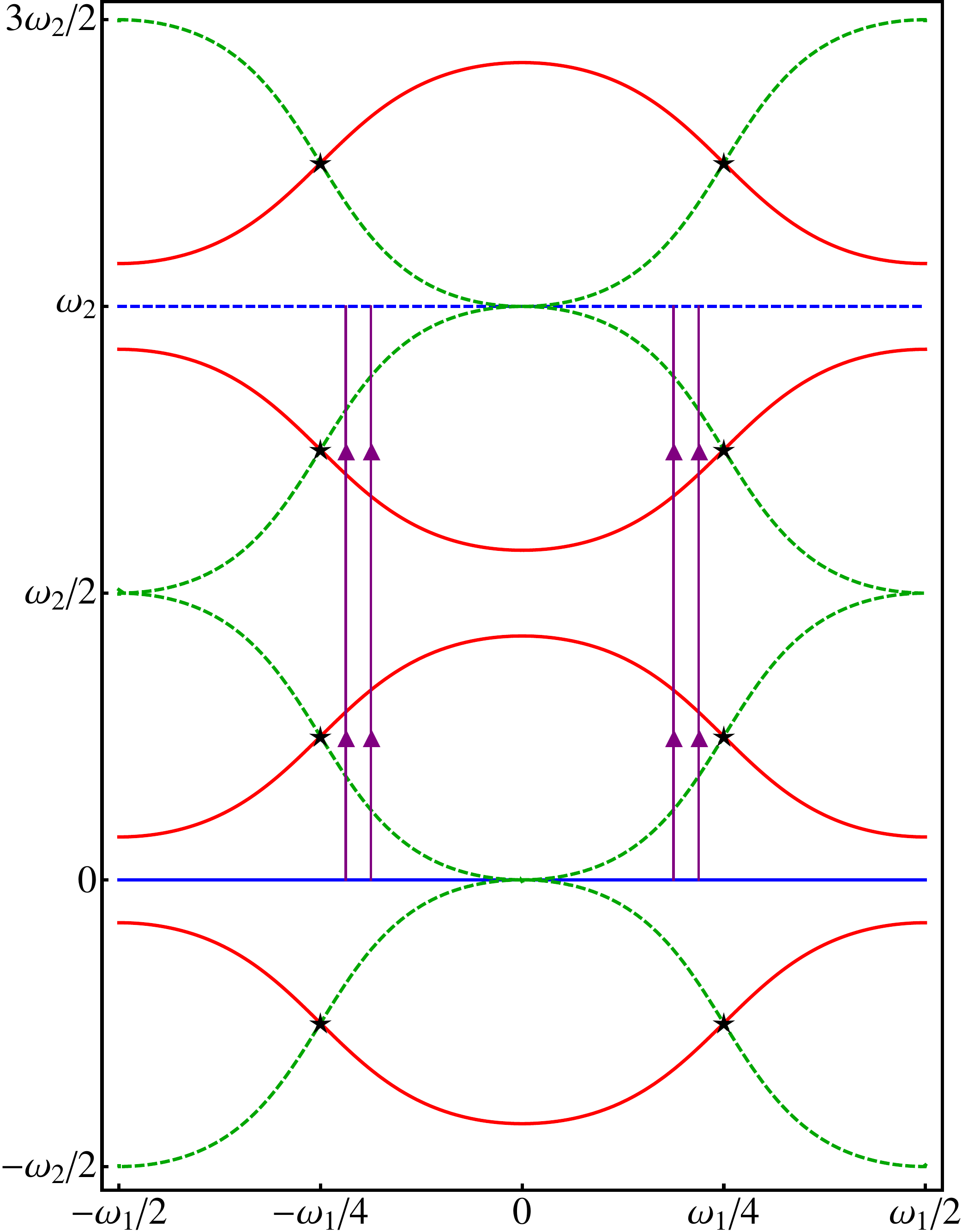}
  \caption{%
    The paths used for analytic continuation from $z$ to $z+\omega_2$ (in purple) are vertical segments that lie close to the boundary of $|\Re(\gamma)|<\omega_1/4$. They cross the red lines $|x^\pm|=1$ when $\Im(x^\pm)<0$.
  }%
  \label{fig:toruspaths}
\end{figure}

Even in the better-understood case of~$\AdS_5\times\S^5$ it is unknown how to define a generalised rapidity that resolves the additional cuts and describes that cover. Therefore, in what follows we will continue using $z$ (or $x^{\pm}_z$), keeping in mind that when describing paths on the torus additional care should be used in the case where a cut is crossed.

\section{Solving the crossing equations}
\label{sec:dressingfact}
A solution for the crossing equations is given by two antisymmetric phases satisfying eq.~\eqref{eq:crossing1}.
It is convenient to rewrite this in terms of crossing equations for the sum and the difference of the two phases  $\theta(p_1,p_2)$ and~$\widetilde{\theta}(p_1,p_2)$.
Let us denote the product  and the ratio of the dressing factors by
\begin{equation}
\prodsigma(p_1,p_2)= \sigma(p_1,p_2)\, \widetilde{\sigma}(p_1,p_2)\,,
\qquad
\ratiosigma(p_1,p_2)= \frac{\sigma(p_1,p_2)}{\widetilde{\sigma}(p_1,p_2)}\,,
\end{equation}
and corresponding phases by~$\sumtheta(p_1,p_2)$ and~$\difftheta(p_1,p_2)$. By analogy with the case of $\AdS_5\times\S^5$~\cite{Arutyunov:2004vx}, it is also useful to rewrite each phase as
\begin{equation}
\label{eq:thchi}
\theta(p_1,p_2) = \chi(x_1^+,x_2^+) +\chi(x_1^-,x_2^-) -\chi(x_1^+,x_2^-) -\chi(x_1^-,x_2^+)\,,
\end{equation}
where $\chi$ is an antisymmetric function. Similar expressions can be introduced for~$\widetilde{\theta}(p_1,p_2)$, $\sumtheta(p_1,p_2)$ and~$\difftheta(p_1,p_2)$.

\subsection*{Dressing phases in $\AdS_5\times\S^5$}
Due to the many similarities of our S~matrix with the $\AdS_5\times\S^5$, it is worth briefly describing the solution of the crossing equation in that case.
In the~$\AdS_5\times\S^5$ S~matrix, a single dressing phase appears. An all-loop solution to its crossing equation~\cite{Janik:2006dc} was found by Beisert, Eden and Staudacher (BES)~\cite{Beisert:2006ez}
\begin{equation}\label{eq:cr-BES}
\begin{gathered}
\sigma^\BES(z_1,z_2) \sigma^\BES(z_1+\omega_2,z_2) = h(x_1^\pm,x_2^\pm), \\
h(x_1^\pm,x_2^\pm) =\frac{x_2^-}{x_2^+}\frac{x_1^- - x_2^+}{x_1^- - x_2^-} \frac{1-\frac{1}{x_1^+x_2^+}}{1-\frac{1}{x_1^+x_2^-}}.
\end{gathered}
\end{equation}
A particularly useful representation of this phase was given by Dorey, Hofman and Maldacena (DHM)~\cite{Dorey:2007xn}
\begin{equation}
\chi^\BES(x,y)= i \ointc \frac{\de w}{2 \pi i} \ointc \frac{\de \tilde{w}}{2 \pi i} \, \frac{1}{x-w}\frac{1}{y-\tilde{w}} \log{\frac{\Gamma[1+\frac{i}{2} h(w+1/w-\tilde{w}-1/\tilde{w})]}{\Gamma[1-\frac{i}{2} h(w+1/w-\tilde{w}-1/\tilde{w})]}}.
\label{eq:besdhmrep}
\end{equation}
This is valid in the physical region $|x|<1$, $|y|<1$. On the boundary of this region, the integral representations has cuts so that a path should be chosen to analytically continue the expression to the crossed region.

As it is discussed in detail in~\cite{Arutyunov:2009kf}, for the $\AdS_5\times\S^5$ crossing equations to be solved we can chose the path depicted in figure~\ref{fig:toruspaths}. These are curves $\gamma(z)$ that go from $z$ to $z+\omega_2$ with constant $\Re (\gamma)$, and which lie close to the boundaries of the region $|\Re(\gamma)|<\omega_1/4$ and crossing the lines $|x^\pm|=1$ in the region $\Im (x^\pm) <0$. Then, the precise statement of the crossing equation~\eqref{eq:cr-BES} is that the BES dressing factor evaluated for $z$ in the physical region as defined by the DHM double integral, times itself analytically continued through the cuts of~\eqref{eq:besdhmrep} along~$\gamma(z)$, equals the rational function on the right hand side.%
 \footnote{%
Actually, we are restricting to a subset of all the allowed paths used in the case of $\AdS_5\times \S^5$, see section 4 in~\cite{Arutyunov:2009kf}.
This more special choice will be the one suitable for our crossing equations.
}%

Our crossing equations~\eqref{eq:crossing1} will be interpreted in a similar sense. In fact, we will from now consider the crossing transformation as a continuation along the paths~$\gamma(z)$ described in figure~\ref{fig:toruspaths}. This will be understood every time we write $p^{\mathsf{c}}$, $z^{\mathsf{c}}$  and so on.

At leading order $O(h^{-1})$ in a large-$h$ expansion,%
\footnote{%
  To find the asymptotic expansion of the BES phase at strong coupling one can expand the integrand using that $i\log\frac{\Gamma(1+ix)}{\Gamma(1-ix)}=-x\log\frac{x^2}{e^2}-\frac{\pi}{2} \sign(\Re x)-2\sum_{n=0}^\infty \frac{\zeta(-2n-1)}{2n+1}\frac{(-1)^n}{x^{2n+1}}$ for $\Re x\neq0$. This expression corrects some typos in the expansion given in~\cite{Vieira:2010kb}.%
} %
 the BES phase reduces to the Arutyunov-Frolov-Staudacher (AFS) phase~\cite{Arutyunov:2004vx}. This was found based on rather general considerations on the large-$h$ behaviour of the energy of string states, and for this reason it is expected to be common to several~$\AdS/\CFT$ duals. It can be written as
\begin{equation}
\label{eq:AFS-xpxm}
\sigma^\AFS(x_1,x_2) = \left(  \frac{1-\frac{1}{x_1^-x_2^+}}{1-\frac{1}{x_1^+x_2^-}} \right) \left(  \frac{1-\frac{1}{x_1^+x_2^-}}{1-\frac{1}{x_1^+x_2^+}}  \frac{1-\frac{1}{x_1^-x_2^+}}{1-\frac{1}{x_1^-x_2^-}} \right)^{\frac{i}{2} h (x_1+1/x_1-x_2-1/x_2)}\,,
\end{equation}

The next-to-leading-order term in strong-coupling expansion is the Hern\'andez-L\'opez (HL) phase~\cite{Hernandez:2006tk}, 
\begin{equation}\label{eq:DHM-HL}
\chi^\HL(x,y)= \frac{\pi}{2} \ointc \frac{\de w}{2 \pi i} \ointc \frac{\de \tilde{w}}{2 \pi i} \, \frac{1}{x-w}\frac{1}{y-\tilde{w}} \, \text{sign}(\tilde{w}+1/\tilde{w}-w-1/w)\,.
\end{equation}
The HL phase appears at order~$O(1)$, and indeed its expression is $h$-independent.
For later convenience, let us perform one of the two integrals in (\ref{eq:DHM-HL}) and obtain the representation
\begin{equation}\label{eq:intu-intd-HL}
\chi^\HL(x,y)=\left( \inturl - \intdlr \right)\frac{\de w}{4\pi} \frac{1}{x-w} \left(\log{(y-w)}-\log{\left(y-1/w\right)}\right),
\end{equation}
where the two integrals are performed in the upper and lower unit semi-circle respectively, counterclockwise in both cases.
The HL phase solves the ``odd'' part of the $\AdS_5$ crossing equation~\cite{Beisert:2006ib}
\begin{equation}\label{eq:cr-HL}
\sigma^\HL(z_1,z_2) \sigma^\HL(z_1+\omega_2,z_2) = \sqrt{ \frac{h_{12}}{h_{\bar{1}2}} }=\sqrt{h_{12}\,(h_{12})^*}\,,
\end{equation}
where
\begin{equation}
h_{12}\,(h_{12})^* =\frac{\ell^\HL(x_1^+,x_2^-)\,\ell^\HL(x_1^-,x_2^+)}{\ell^\HL(x_1^+,x_2^+)\,\ell^\HL(x_1^-,x_2^-)}\,, \qquad \ell^\HL(x,y)=\frac{x-y}{1-xy}\,,
\end{equation}
where complex conjugation amounts to sending $x_k^\pm \rightarrow x_k^\mp$.\footnote{One can check that in order for~\eqref{eq:intu-intd-HL} to solve~\eqref{eq:cr-HL} it is necessary to choose the path of analytic continuation as in figure~\ref{fig:toruspaths}. This is discussed in more detail in ref.~\cite{Borsato:2013hoa}.}

\subsection*{Solution for the sum of the phases}\label{sec:sol-diff}
Taking the product of the two crossing equations~(\ref{eq:crossing1}), we find an equation for $\prodsigma$
\begin{equation}
{\prodsigma(z_1,z_2)}^2 \, {\prodsigma(z_1+\omega_2,z_2)}^2 = g_{12} \, \widetilde{g}_{12}\,.
\label{eq:crosssum}
\end{equation}
We observe that the r.h.s of this equation can be written in terms of the function $h_{12}$ appearing on the r.h.s of the $\AdS_5$ crossing equation~(\ref{eq:cr-BES}) 
\begin{equation}
g_{12} \, \widetilde{g}_{12}=\frac{(h_{12})^3}{(h_{12})^*}\,,
\end{equation}
where we used the constrain~\eqref{eq:zhukovski-def}. The above relation allows us to solve the crossing equation~(\ref{eq:crosssum}) using parts of the $\AdS_5$ dressing phase
\begin{equation}
\label{eq:sumsolution}
\prodsigma_{12}=\frac{(\sigma^{\BES}_{12})^2}{\sigma^{\HL}_{12}}\,, \qquad\text{\ie}\qquad \sumtheta_{12}=2\theta^{\BES}_{12}-\theta^{\HL}_{12}
\,.
\end{equation}
To show that such a $\prodsigma_{12}$ satisfies equation~(\ref{eq:crosssum}) one need only use equations~\eqref{eq:cr-BES} and~\eqref{eq:cr-HL}. It is convenient to express $\prodsigma_{12}$ in terms of a DHM-like double-integral representation, by defining $\chi^+(x,y)$ as
\begin{equation}
\label{eq:chi+}
\begin{aligned}
\chi^+(x,y)=&\,2\,\chi^{\BES}(x,y)-\chi^{\HL}(x,y)\\
 =&\ointc \frac{\de w}{2 \pi i} \ointc \frac{\de \tilde{w}}{2 \pi i} \, \frac{1}{x-w}\frac{1}{y-\tilde{w}} \Bigg( 2i \log{\frac{\Gamma[1+i h(w+1/w-\tilde{w}-1/\tilde{w})]}{\Gamma[1-i h(w+1/w-\tilde{w}-1/\tilde{w})]}} \\
&\hspace{5.5cm}- \frac{\pi}{2} \text{sign}(\tilde{w}+1/\tilde{w}-w-1/w) \Bigg),
\end{aligned}
\end{equation}
in the physical region. Notice that the above expression is exact to all orders in the coupling $h$. We postpone the perturbative expansion of this and the following expressions to chapter~\ref{ch:comparison}, where we will also compare them with independent results.

\subsection*{Solution for the difference of the phases}
Taking the ratio of the two crossing equations~(\ref{eq:crossing1}), we get
\begin{equation}\label{eq:cr-ratio}
\frac{{\ratiosigma(z_1,z_2)}^2}{{\ratiosigma(z_1+\omega_2,z_2)}^2}=\frac{\widetilde{g}_{12}}{g_{12}}\,,
\end{equation}
where
\begin{equation}\label{eq:cr-ratio-expl}
\frac{\widetilde{g}_{12}}{g_{12}}=\frac{\ell^-(x_1^+,x_2^-)\ell^-(x_1^-,x_2^+)}{\ell^-(x_1^+,x_2^+)\ell^-(x_1^-,x_2^-)}, \qquad \ell^-(x,y)\equiv(x-y)\left(1-\frac{1}{xy}\right).
\end{equation}
Notice that this equation involves the ratio rather than the product of the dressing factor with its analytic continuation.
As it can be explicitly verified\footnote{See ref.~\cite{Borsato:2013hoa} for the details of such calculation.}, defining $\chi^-(x,y)$ in the physical region ($|x|,|y|>1$) by the integral
\begin{equation}\label{eq:chi-}
\begin{aligned}
\chi^-(x,y) &=\ointc \, \frac{\de w}{8\pi} \frac{1}{x-w} \log{\left[ (y-w)\left(1-\frac{1}{yw}\right)\right]} \, \text{sign}((w-1/w)/i) \ - x \leftrightarrow y \\
&=\left( \inturl - \intdlr \right)\frac{\de w}{8\pi} \frac{1}{x-w} 
\log{\left[ (y-w)\left(1-\frac{1}{yw}\right)\right]} \ - x \leftrightarrow y\,,
\end{aligned}
\end{equation}
solves the crossing equation~\eqref{eq:cr-ratio}. By construction, $\chi^-$ is antisymmetric. Note that the integrand of $\chi^-$ does not depend explicitly on the coupling~$h$, in contrast to the solution of the crossing equation~(\ref{eq:crosssum}) which is solved by an integrand with an infinite series expansion in~$h$.  This is because equation~\eqref{eq:cr-ratio} is  ``odd'' in the sense of ref.~\cite{Beisert:2006ib}.

The all-loop expressions for $\chi$ and $\tilde{\chi}$ are then given by
\begin{equation}
\label{eq:solution}
\begin{aligned}
  \chi(x,y) &= \chi^{\text{BES}}(x,y)+\frac{1}{2}\left(-\chi^{\text{HL}}(x,y)+\chi^{-}(x,y)\right) \,, \\
  \widetilde{\chi}(x,y) &= \chi^{\text{BES}}(x,y)+\frac{1}{2}\left(-\chi^{\text{HL}}(x,y)-\chi^{-}(x,y)\right) \,.
\end{aligned}
\end{equation}
These solutions are expressed in terms of the non-perturbative BES phase plus terms at the HL order. These latter contributions to $\chi$ and $\tilde{\chi}$ are independent of $h$. As such, they can be added to the DHM representation of the BES  phase without affecting the $h$-resummation.

\section{Poles of the S~matrix}
Now that we have found a solution to the crossing equations, it is natural to ask whether this is \emph{the} solution appearing in the $\AdS_3\times\S^3\times\T^4$ S~matrix. It is clear that our phases could be multiplied by any ``CDD factor''~\cite{Castillejo:1955ed}, that is, any solution of the homogeneous crossing equations
\begin{equation}
\sigma^{\CDD}_{p\,q}\,\widetilde{\sigma}^{\CDD}_{p^{\mathsf{c}}\,q}=1\,,\qquad
\sigma^{\CDD}_{p^{\mathsf{c}}\,q}\,\widetilde{\sigma}^{\CDD}_{p\,q}=1\,.
\end{equation}
Such solutions exist. The simplest ones are meromorphic functions on the torus, which can be defined in terms of 
\begin{equation}
\label{eq:CDD-rational}
\chi^{\CDD}_{pq}=\frac{i}{2}\log\frac{(x-y)^{n_1}}{(1-xy)^{n_2}}\,,\qquad
\widetilde{\chi}^{\CDD}_{pq}=\frac{i}{2}\log\frac{(x-y)^{n_2}}{(1-xy)^{n_1}}\,,
\end{equation}
with $n_1,\,n_2$ integer constants. Such factors will modify the pole structure of the S~matrix.
However, there is a close connection between simple poles in the physical region of the S~matrix and the bound states of the model. Therefore, by exploring the expected bound-state spectrum of the model we will be able to put stringent restriction on the CDD factors.

\subsection*{Bound states and short representations}
Excitations transform in representations of $\psu(1|1)^4_{\text{c.e.}}$. Furthermore, the representations satisfy a shortening condition which immediately follows from the one of~$\su(1|1)^2_{\text{c.e.}}$, see eq.~\eqref{eq:shortening2}
\begin{equation}
\label{eq:su11-shortening}
\mathbf{H}^2 =\mathbf{M}^2+\overline{\mathbf{C}}\,\mathbf{C} \,.
\end{equation}
Bound states preserving some supersymmetry also transform in a short representation of the symmetry algebra. Let us consider a two-particle state containing two left-moving particles, \eg~$\ket{\Phi_{+{+}}^{\smallL} \Phi_{+{+}}^{\smallL}}$. 
For generic values of the momenta $p$ and $q$ the tensor product of two fundamental left representations is an irreducible long representation. However, at special points the tensor product becomes reducible. In particular, we find that the shortening condition~\eqref{eq:su11-shortening} is satisfied for 
\begin{equation}
\label{eq:boundstateLL}
x_p^+ = x_q^-
\qquad
\text{or}
\qquad
x_p^- = x_q^+\,.
\end{equation}
Only at these points it is possible to construct short sub-representations. Therefore any pole in the S-matrix corresponding to a supersymmetric bound state will have to satisfy one of these conditions.

An interesting feature of the $\psu(1|1)^4_{\text{c.e.}}$ algebra is that all short irreducible representations are two-dimensional while all long irreducible representations have dimension four. A two-particle bound state will therefore transform in a representation which has the same form as the fundamental representation, differing only in the values of the central charges. This should be contrasted with the centrally extended $\psu(2|2)$ algebra appearing in $\AdS_5 \times \S^5$~\cite{Beisert:2004ry,Beisert:2005tm,Arutyunov:2006ak}, where the fundamental representation has dimension four while the $M$-particle bound state has dimension $4M$~\cite{Chen:2006gp}.

At the points where the tensor product becomes reducible some of the elements of the S~matrix become zero or develop poles.
These singularities will appear both in the dressing factors, as we will investigate later, and in the ratio of S~matrix elements. In fact, with an appropriate normalisation we will have
\begin{equation}
\mathbf{R}_{pq}\,\mathscr{V}_0 = 0\,,
\qquad
\mathscr{V}_0 \subset \mathscr{V}_{\smallL}(p)\otimes\mathscr{V}_{\smallL}(q)\,,
\end{equation}
when $p$ and $q$ satisfy~\eqref{eq:boundstateLL}, while~$\mathbf{R}_{pq}$ is regular (finite) on the complement of~$\mathscr{V}_0$. The bound state representation is then the factor representation on the quotient space~$\mathscr{V}_{\smallL}(p)\otimes\mathscr{V}_{\smallL}(q)/\mathscr{V}_0$~\cite{Arutyunov:2008zt}. Using the explicit form of the S~matrix, for the point $x_p^+ = x_q^-$ we find that the state $\ket{\Phi_{+{+}}^{\smallL} \Phi_{+{+}}^{\smallL}}$ belongs to the short representation, and we will therefore refer to it as a $\su(2)$ bound state. In the case $x_p^- = x_q^+$ the short representation includes the state $\ket{\Phi^{\smallL}_{-{-}} \Phi^{\smallL}_{-{-}}}$, and is a potential $\sl(2)$ bound state.\footnote{%
  In $\AdS_5 \times \S^5$ the physical bound states correspond to ``$\su(2)$ bound states''. The ``$\sl(2)$ bound states'' appear as  bound states of the mirror theory that we alluded to in the introduction~\cite{Arutyunov:2007tc}.%
}%

To decide which bound state belongs to the physical spectrum we need to impose additional constraints on the momenta of the fundamental excitations. 
In the region $s_1 \ll s_2$ the wave~function of a scattering state takes the general form\footnote{%
  In order to avoid confusion with the dressing phase we denote the world-sheet coordinate by $s$.%
}%
\begin{equation}
\label{eq:2part-wavefunction-QFT}
  \Psi(s_1,s_2) \approx e^{i(p s_1 + q s_2)} + S(p,q) e^{i(p s_2 + q s_1)} ,
\end{equation}
where the first term describes the incoming wave and the second term the outgoing wave, or equivalently
\begin{equation}
  \Psi(s_1,s_2) \approx \frac{1}{S(p,q)}e^{i(p s_1 + q s_2)} + e^{i(p s_2 + q s_1)} .
\end{equation}
To find a bound state we analytically continue the wave~function to complex values of the momenta
\begin{equation}
  p = \frac{\tilde p}{2} + iv , \qquad
  q = \frac{\tilde p}{2} - iv.
\end{equation}
The wave~function then behaves as
\begin{equation}
\Psi(s_1,s_2) \approx \frac{1}{S(p,q)} e^{v(s_2 - s_1)} + e^{-v(s_2 - s_1)}\,,
\qquad s_1 \ll s_2\,.
\end{equation}
If there are bond states in the spectrum, we expect $S(p,q)$ to have a pole, as it can be understood in a diagrammatic expansion in terms of a propagator that goes on the mass shell. Additionally, for the bound-state wave~function to be normalizable, the outgoing-wave exponential should be decaying. Hence we are interested in the solution where the momentum of the first particle has a positive imaginary part,~$v>0$.
By imposing the condition~\eqref{eq:zhukovski-def} for $x_p^\pm$ and $x_q^\pm$ in the physical region, we find that for $x_p^+ = x_q^-$ the momentum $p$ has a positive imaginary part, while $x_p^- = x_q^+$ leads to the imaginary part being negative. We hence conclude that only the $\su(2)$ bound state can appear in the physical spectrum. Of course this will have to be confirmed by the presence of suitable poles in the S~matrix, which will pose a constraint on the dressing factors

So far we have only considered bound states in the LL-sector. If we start with two right-moving excitations we again find an $\su(2)$ bound state at $x_p^+ = x_q^-$, simply by left-right symmetry. It is interesting to consider a state consisting of one left- and one right-moving excitations such as $\ket{\Phi^{\smallL}_{+{+}} {\Phi}^{\smallR}_{-{-}}}$ . In this case the shortening condition~\eqref{eq:su11-shortening} is satisfied for $x_p^+ = 1/x_q^+$ and $x_p^- = 1/x_q^-$. Neither of these solutions lie in the physical region $|x_p^\pm|>1$, $|x_q^\pm|>1$ and hence \emph{there are no supersymmetric bound states in the LR~sector}.

In summary we find that, based on the shortening condition and the matrix form of~$\mathbf{R}_{pq}$, physical two-particle  $\su(2)$ bound states exist in the LL and RR~sectors. The LR~sector, on the other hand, does not contain any bound states.

\subsection*{Semiclassical bound states from giant magnons}

Owing to its integrability, the $\AdS_5\times\S^5$ NLSM admits classical off-shell soliton solutions called \emph{giant magnons}~\cite{Hofman:2006xt}. These can be thought of as a coherent superposition of several many one-particle excitations (magnons in the dual spin chain). The simplest giant magnon is a classical string solution living in a $\mathbbm{R} \times \S^2$ subspace of $\AdS_5 \times \S^5$, and having a definite momentum~$p$. It can be extended to a solution in $\mathbbm{R} \times \S^3$, the \emph{dyonic} giant magnon~\cite{Chen:2006gea}, which carries angular momentum $M$ along the additional angle. This solution corresponds to the semiclassical limit of a bound state of $|M|$ fundamental magnons~\cite{Dorey:2006dq}.  

Since both the fundamental giant magnon and the dyonic extension live in $\mathbbm{R} \times \S^3$ they can be directly embedded in $\AdS_3 \times \S^3$~\cite{Abbott:2012dd}. How does our discussion on the allowed bound states fit together with the giant magnon picture?

An important difference between our case and $\AdS_5 \times \S^5$ is that in the latter space a dyonic giant magnon with positive $\mathbf{M}$-charge $+M$ can be \emph{continuously} rotated to the corresponding magnon with negative charge~$-M$, due to the presence of an additional $\su(2)$ symmetry. However, in the case of $\AdS_3 \times \S^3$ such a rotation is not possible since the intermediate states would not sit inside $\S^3$. Therefore, while in $\AdS_5 \times \S^5$ there is no notion of left  and right giant magnon, in our case the two states with charges $+M$ and~$-M$ are independent and can be distinguished by the sing of their eigenvalue under~$\mathbf{M}$---their target-space chirality. Only configurations of the same chirality can be used to build a dyonic giant magnon, so that only the corresponding microscopic excitations will have bound states.
Since the notion of left and right excitations was defined precisely in terms of their charge under $\mathbf{M}$, we must expect to have LL and RR $\su(2)$-bound states resembling the ones of $\AdS_5\times\S^5$, but no LR or RL bound states. This is precisely the result of our representation-theoretical analysis of the previous subsection.

\subsection*{Simple poles of the S~matrix}

\begin{figure}%
  \centering%
  \subfloat[$s$~channel, LL sector]{%
    \label{fig:s-channel-LL}
    \begin{tikzpicture}[
      thick,
      level/.style={level distance=1.15cm},
      level 2/.style={sibling distance=3.5cm},
      ]
      \coordinate
      child[grow=up]{
        edge from parent [draw=black,line width=1.6]
        child {
          node{$\Phi_{+{+}}^{\smallL}(p)$}
          edge from parent [particle]
        }
        child {
          node{$\Phi_{+{+}}^{\smallL}(q)$}
          edge from parent [particle]
        }
        node [above=3pt]{}
      }
      child[grow=down, level distance=0pt] {
        child {
          node{$\Phi_{+{+}}^{\smallL}(p)$}
          edge from parent [antiparticle]
        }
        child { 
          node{$\Phi_{+{+}}^{\smallL}(q)$}
          edge from parent [antiparticle]               
        }
      };
    \end{tikzpicture}%
  }%
  \hspace{2cm}%
  \subfloat[$t$~channel from crossing]{%
    \label{fig:t-channel-LR}%
    \begin{tikzpicture}[
      thick,
      level/.style={level distance=1.5cm},
      level 2/.style={sibling distance=3.5cm},
      ]
      \coordinate
      child[grow=left]{
        edge from parent [draw=black,line width=1.6]
        child {
          node{${\Phi}_{+{+}}^{\smallR}(p^{\mathsf{c}})$}
          edge from parent [particlecross]
        }
        child {
          node{${\Phi}_{+{+}}^{\smallL}(q)$}
          edge from parent [antiparticle]
        }
        node [above=3pt]{}
      }
      child[grow=right, level distance=0pt] {
        child {
          node{${\Phi}_{+{+}}^{\smallR}( p^{\mathsf{c}})$}
          edge from parent [antiparticlecross]
        }
        child { 
          node{${\Phi}_{+{+}}^{\smallL}(q)$}
          edge from parent [particle]               
        }
      };
    \end{tikzpicture}}%
  \caption{%
    On the left two particles in the same sector form an~$\su(2)$ bound state in the $s$~channel. Applying the crossing transformation to~$\Phi^{\smallL}_{+{+}}(p)$ yields the $t$-channel diagram on the right, where on particle has unphysical momentum~${p}^{\mathsf{c}}$ (red dashed lines).
  }%
  \label{fig:LLpoles}%
\end{figure}
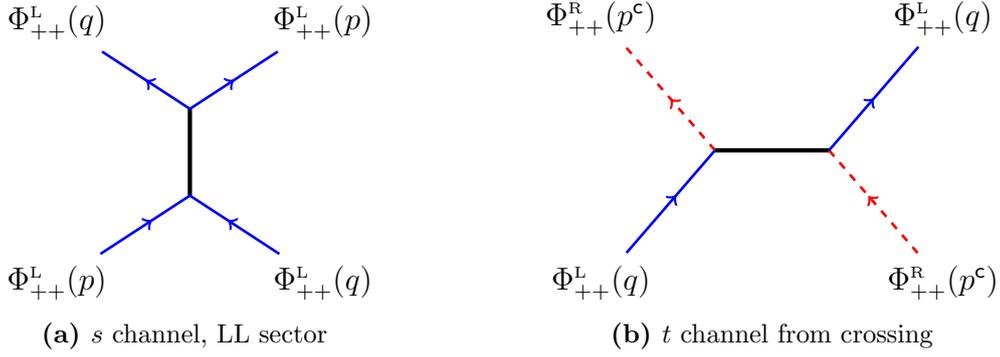

Scattering processes involving formation or exchange of bound states give rise to single poles in the S~matrix for physical values of the spectral parameters~\cite{eden2002}. Let us consider the $s$-channel diagram in figure~\ref{fig:s-channel-LL}. The process involves two fundamental particles from the same sector, \eg\ two left-movers, in the physical region $|x_i|>1$, $i=p,q$, which form an on-shell boundstate and then split up again. 
Similarly to the case of the $\su(2)$ sector in $\AdS_5$~\cite{Dorey:2006dq,Dorey:2007xn}, this should lead to a pole in the corresponding S-matrix element at $x_p^+=x_q^-$.
The relevant element is, up to inessential~$e^{ip}$ prefactors which we will always drop here,
\begin{equation}
\mathcal{A}_{pq}=\bra{\Phi^{\smallL}_{+{+}}\, \Phi^{\smallL}_{+{+}} }\check{\Smat}_{pq} \ket{\Phi^{\smallL}_{+{+}}\, \Phi^{\smallL}_{+{+}}}
  =
   \frac{x_p^- - x_q^+}{x_p^+ - x_q^-} \frac{1-\frac{1}{x_p^- x_q^+}}{1-\frac{1}{x_p^+ x_q^-}} \sigma_{pq}^{-2}.
\end{equation}
As it can be directly checked~\cite{Borsato:2013hoa}, the dressing factor is regular at $x_p^+=x_q^-$, so that $\mathcal{A}_{pq}$ has a simple pole there.

This $s$~channel process is related through crossing symmetry to the exchange of a bound state in the $t$~channel, depicted in figure~\ref{fig:t-channel-LR}. There the particle of momentum~$p$ has been crossed so that $x_{{p}^{\mathsf{c}}}^\pm=1/x_p^\pm$ are not in the physical region. Since the two processes are related by crossing symmetry, the poles in the $s$~channel automatically fix the singularities in the $t$~channel. In fact crossing symmetry implies~\cite{Borsato:2013qpa}
\begin{equation}
\mathcal{A}_{pq} \widetilde{\mathcal{A}}_{{p}^{\mathsf{c}}q}=1, \qquad \text{  where}\quad
 \widetilde{\mathcal{A}}_{pq}=\bra{\Phi_{+ {+}}^{\smallL}\, {\Phi}_{+{+}}^{\smallR} }
\check{\Smat}_{pq}
\ket{{\Phi}_{+{+}}^{\smallR}\, \Phi_{+{+}}^{\smallL}}\,,
\end{equation}
so that a pole of $\mathcal{A}_{pq}$ corresponds to a pole of ${\widetilde{\mathcal{A}}_{{{p}^{\mathsf{c}}}q}}^{-1}$. We can check this explicitly by considering 
\begin{equation}
\widetilde{\mathcal{A}}_{{p}^{\mathsf{c}}q} = \frac{1-\frac{1}{x_{{p}^{\mathsf{c}}}^- x_q^+}}{1-\frac{1}{x_{{p}^{\mathsf{c}}}^- x_q^-}} \,  \frac{1-\frac{1}{x_{{p}^{\mathsf{c}}}^+ x_q^-}}{1-\frac{1}{x_{{p}^{\mathsf{c}}}^+ x_q^+}} \tilde{\sigma}_{{p}^{\mathsf{c}}q}^{-2}.
\end{equation}
Since $\widetilde{\sigma}$ is regular when continued inside the unit circle~\cite{Borsato:2013hoa}, $\widetilde{\mathcal{A}}_{{p}^{\mathsf{c}}q}$ has a zero at $x_{{p}^{\mathsf{c}}}^+=1/x_q^-$, as expected.

If we consider S-matrix elements involving one left- and one right-moving particle we expect no poles, since there are no corresponding bound states. Therefore a process such as the one depicted in figure~\ref{fig:s-channel-LR} should not happen. Indeed, the S-matrix element
\begin{figure}
  \centering%
  \subfloat[RL $s$~channel (forbidden)]{%
    \label{fig:s-channel-LR}%
    \begin{tikzpicture}[
      thick,
      level/.style={level distance=1.15cm},
      level 2/.style={sibling distance=3.5cm},
      ]
      \coordinate
      child[grow=up]{
        edge from parent [draw=black,line width=1.6]
        child {
          node{${\Phi}_{-{-}}^{\smallR}(p)$}
          edge from parent [particle]
        }
        child {
          node{$\Phi_{+{+}}^{\smallL}(q)$}
          edge from parent [particle]
        }
        node [above=3pt]{}
      }
      child[grow=down, level distance=0pt] {
        child {
          node{${\Phi}_{-{-}}^{\smallR}(p)$}
          edge from parent [antiparticle]
        }
        child { 
          node{$\Phi_{+{+}}^{\smallL}(q)$}
          edge from parent [antiparticle]               
        }
      };
    \end{tikzpicture}%
  }%
  \hspace{2cm}%
  \subfloat[crossed LL $t$~channel (forbidden)]{%
    \begin{tikzpicture}[
      thick,
      level/.style={level distance=1.5cm},
      level 2/.style={sibling distance=3.5cm},
      ]
      \coordinate
      child[grow=left]{
        edge from parent [draw=black,line width=1.6]
        child {
          node{${\Phi}_{-{-}}^{\smallL}({p}^{\mathsf{c}})$}
          edge from parent [particlecross]
        }
        child {
          node{$\Phi_{+{+}}^{\smallL}(q)$}
          edge from parent [antiparticle]
        }
        node [above=3pt]{}
      }
      child[grow=right, level distance=0pt] {
        child {
          node{${\Phi}_{-{-}}^{\smallL}({p}^{\mathsf{c}})$}
          edge from parent [antiparticlecross]
        }
        child { 
          node{$\Phi_{+{+}}^{\smallL}(q)$}
          edge from parent [particle]               
        }
      };
    \end{tikzpicture}
    \label{fig:t-channel-LL}}
  \caption{%
    On the left the would-be Landau diagram for one left- and one right-moving particle is depicted. This process should be absent. Similarly, the crossed process on the right should be absent, and the corresponding S-matrix element have no pole. 
  }%
  \label{fig:LRpoles}%
\end{figure}
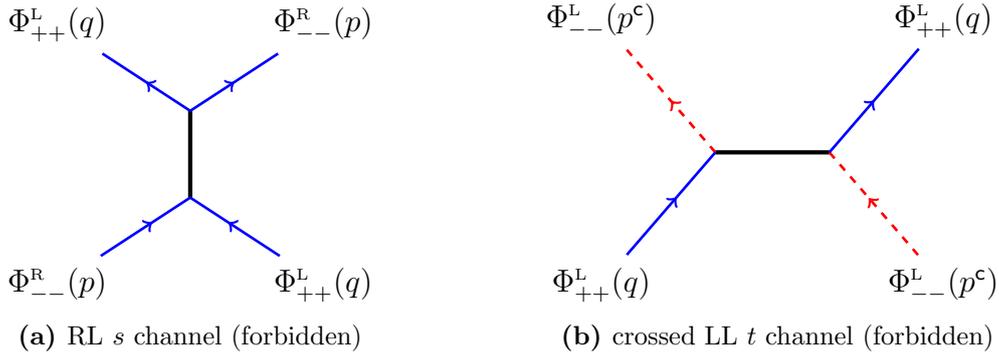%
\begin{equation}
\widetilde{\mathcal{B}}_{pq}=\bra{\Phi_{+ {+}}^{\smallL}\, {\Phi}_{-{-}}^{\smallR} }
\check{\Smat}_{pq}
 \ket{{\Phi}_{-{-}}^{\smallR}\, \Phi_{+{+}}^{\smallL}}  =  \frac{1-\frac{1}{x_p^- x_q^+}}{1-\frac{1}{x_p^+ x_q^-}} \,  \frac{1-\frac{1}{x_p^+ x_q^+}}{1-\frac{1}{x_p^- x_q^-}} \tilde{\sigma}_{pq}^{-2},
\end{equation}
is regular in the physical region, and in particular has no pole at $x_p^+=x_q^-$.
It is an interesting check that the same holds in the crossed channel, whose exchange diagram would be as in figure~\ref{fig:t-channel-LL}, should it exist. 
Again crossing symmetry relates the two processes in a simple way (because we are scattering highest weight states), \ie
\begin{equation}
\widetilde{\mathcal{B}}_{pq} \mathcal{B}_{{p}^{\mathsf{c}}q}=1, \qquad \text{where}\quad \mathcal{B}_{pq}=\bra{\Phi_{+{+}}^{\smallL}\, \Phi_{-{-}}^{\smallL} }
\check{\Smat}_{pq}
\ket{\Phi_{-{-}}^{\smallL}\, \Phi_{+{+}}^{\smallL}}\,,
\end{equation}
which implies the first crossing equation in~\eqref{eq:crossing12}.
Since $\widetilde{\mathcal{B}}_{pq}$ has no singularity at $x_p^+=x_q^-$ we expect $\mathcal{B}_{{p}^{\mathsf{c}}q}$ to have no singularity at $x_{{p}^{\mathsf{c}}}^+=1/x_q^-$. Explicitly we have 
\begin{equation}
\mathcal{B}_{{p}^{\mathsf{c}}q} = \frac{(x_{{p}^{\mathsf{c}}}^- - x_q^-)^2}{(x_{{p}^{\mathsf{c}}}^- - x_q^+)(x_{{p}^{\mathsf{c}}}^+ - x_q^-)} \, \frac{1-\frac{1}{x_{{p}^{\mathsf{c}}}^- x_q^+}}{1-\frac{1}{x_{{p}^{\mathsf{c}}}^+ x_q^-}}  \sigma_{{p}^{\mathsf{c}}q}^{-2}.
\end{equation}
The rational terms have a pole at $x_{{p}^{\mathsf{c}}}^+=1/x_q^-$, but  once the dressing factor is continued to the crossed region, this is canceled by a zero of $\sigma^{-2}$, so that the result is non-singular.

\subsection*{Conditions on CDD factors}

The fact that we correctly match the structure of single poles in the physical region tells us that for any CDD factors of the form~\eqref{eq:CDD-rational} we must set $n_1=n_2=0$. However, in principle we could still allow for different solutions of the homogeneous crossing equations. If these are defined on a cover of the rapidity torus and feature no poles in the physical region, for instance, they would not be ruled out by our bound-state analysis. Since the dressing phases we propose live on such a cover, such solutions cannot be ruled out.

\section{Chapter summary}
In this chapter we have formalised the particle-to-antiparticle symmetry of the non-relativistic worldsheet theory in terms of a set of crossing equations, in the spirit of what was done in $\AdS_5\times\S^5$ by Janik~\cite{Janik:2006dc}. In our case, we find that the particle-to-antiparticle transformation exchanges the left and right representations. This is a qualitatively new feature, which tells us that for consistency of our 2-dimensional worldsheet theory we cannot restrict ourselves to particles of a single target-space chirality, even if they form an irreducible representation of the off-shell symmetry algebra.

The crossing equation couples the LL and LR  dressing factors factors~$\mathscr{S}_{pq}$ and~$\widetilde{\mathscr{S}}_{pq}$ or equivalently~$\sigma_{pq}$ and~$\widetilde{\sigma}_{pq}$. Specifically, it relates one scalar factor evaluated in the physical region with another analytically continued to the crossed region. We see that, were it not for the dressing factors, the S~matrix would be a rational function on a double complex torus---instead, it is defined on an infinite cover of it.

The main result of the chapter is an all-loop solution to the crossing equations. This features the well known Beisert-Eden-Staudacher~\cite{Beisert:2006ez} and Hern\'andez-L\'opez~\cite{Hernandez:2006tk} dressing factors that originally emerged in~$\AdS_5\times\S^5$, together with a novel factor defined by~\eqref{eq:chi-}. This also shows that the BES phase, which essentially constitutes the dressing factors of integrable~$\AdS_5/\CFT_4$  and~$\AdS_4/\CFT_3$, is not universal to all $\AdS/\CFT$ duals.

Besides being crossing-symmetric, the dressing factors we found reproduce the bound-state spectrum we expect to find in the physical region. Still, it is unclear whether they are \emph{the} physical solution of the crossing equations, or should be modified by multiplying them by a solution of the homogeneous equations, which could be highly non-trivial. To check whether this is the case, we can compare the large-$h$ expansion of the dressing factors to independent perturbative calculations. We will successfully do so in chapter~\ref{ch:comparison}. Additionally, we could try to investigate further the analytic properties of the S~matrix, especially beyond the physical region. For instance, its properties in the region where a mirror theory should be defined%
\footnote{%
This is the region in the middle of the green curves in figure~\ref{fig:toruspaths}, corresponding to a shift of~$\tfrac{1}{2}\omega_2$, which in fact reproduces the mirror transformation~\eqref{eq:mirrorTrans}.
} will be important to formulate the mirror thermodynamical Bethe ansatz description that we mentioned in the introduction. So far, such an investigation has not been performed.

Finally, we should remark that in principle the dressing factors of massive particles should also contain some information due to the presence of massless virtual particles. This could appear through a solution of the homogeneous crossing equations, too. At the moment, however, it is not clear what its contribution should be, or whether somehow it has already been accounted for.

\chapter{The \texorpdfstring{$\psu(1,1|2)^2$}{psu(1,1|2)**2} spin chain}
\label{ch:spinchain}
In this chapter we will see how the all-loop S~matrix can be found in a \emph{spin-chain picture}. Inspired from what happens in the case $\AdS_5\times\S^5$, which we briefly discussed in the introduction, it is rather natural to assume that strings on~$\AdS_3\times\S^3\times\T^4$ can be described in terms of a spin chain. If we restrict to the massive sector, the ``spins'' should be in modules of $\psu(1,1|2)^2$. By picking a ground state for the chain, and considering fluctuations around it, we will find an off-shell symmetry algebra. Out of this we will able to fix a two-body S~matrix. While the S~matrix that we found out of the Zamolodchikov-Faddeev algebra in chapter~\ref{ch:smatrix} describes the reordering of  two ZF creation operators, the one we study in this chapter acts on pairs of spin-chain sites.
Consistently extending it to an~$M$-sites S~matrix will again require a sort of Yang-Baxter equation to hold.
Equivalently, we can describe the S~matrix as acting on plane-wave excitations (magnons), which will allow us to show that it is in fact equivalent to the worldsheet S~matrix, in a sense that we will specify.

This spin-chain picture was the way in which the S~matrix was originally derived~\cite{Borsato:2012ud,Borsato:2013qpa}. To keep our presentation as homogeneous as possible, here we use slightly different conventions than in refs.~\cite{Borsato:2012ud,Borsato:2013qpa}.

\section{The weakly-coupled spin chain}
\label{eq:weaklycoupledSC}

We want to describe the all-loop $\psu(1,1|2)^2$  spin-chain dual to free $\AdS_3 \times \S^3$ strings.
At weak coupling---that is, for~$h\ll1 $---the spin chain was originally constructed in~\cite{OhlssonSax:2011ms}. There one finds two copies of the superalgebra $\psu(1,1|2)$, describing the left- and right-moving sectors of the dual CFT. At leading order in a $h\to0$ expansion, left- and right-movers decouple. The spectrum is then described by~\emph{two} homogeneous spin-chains with the sites of each transforming in the representation%
\footnote{%
To make contact with the coset construction, we take the~$\su(1,1) $ spin to be positive. Still, the $\tfrac{1}{2} $ representation of~$\su(1,1) $ is infinite-dimensional, while the $\tfrac{1}{2} $ representation of~$\su(2) $ is finite-dimensional. This is in contrast with ref.~\cite{Borsato:2013qpa} where we considered the $(-\tfrac{1}{2};\tfrac{1}{2})$ representation of $\psu(1,1|2)$.
} $(\tfrac{1}{2};\tfrac{1}{2})$ of $\psu(1,1|2)$---one for the left and one for the right sector. At higher orders in~$h$ the two sectors couple to each other through local interactions. We will be able to account for all of these interactions  from symmetry arguments.

\subsection*{The spin-chain algebra and representation}
\label{sec:spinchainrepr}

The sites of each $\psu(1,1|2)$ spin-chain transform in the infinite-dimensional representation $(\tfrac{1}{2};\tfrac{1}{2})$, consisting of the bosonic $\su(2)$ doublets $\phi^{(n)}_{\pm}$ and the two sets of fermionic $\su(2)$ singlets $\psi^{(n)}_{\pm}$, where the index $n$ indicates the $\sl(2)$ quantum number. In the spin-chain picture it is convenient to consider a real form of~$\psu(1,1|2)$ which differs from the one we saw in section~\ref{sec:supercoset}.%
\footnote{%
This difference in the real form is familiar from~$\AdS_5/\CFT_4$, see \eg~ref.~\cite{Beisert:2010kp}.
}
 Let us consider
\begin{equation}
  \begin{aligned}
    \comm{\mathbf{L}_3}{\mathbf{L}_\pm} &= \pm \mathbf{L}_\pm , &
    \qquad
    \comm{\mathbf{L}_+}{\mathbf{L}_-} &= 2\, \mathbf{L}_3 , \\
    \comm{\mathbf{J}_3}{\mathbf{J}_\pm} &= \pm \mathbf{J}_\pm , &
    \qquad
    \comm{\mathbf{J}_+}{\mathbf{J}_-} &= 2\, \mathbf{J}_3 , \\
  \end{aligned}
\end{equation}
for the bosonic charges and
\begin{equation}
  \begin{aligned}
    \comm{\mathbf{L}_3}{\mathbf{Q}_{\pm\kappa\iota}} &= \mp\frac{1}{2} \mathbf{Q}_{\pm\kappa\iota} , &
    \qquad
    \comm{\mathbf{L}_\pm}{\mathbf{Q}_{\pm\kappa\iota}} &= \mathbf{Q}_{\mp\kappa\iota} , \\
    \comm{\mathbf{J}_3}{\mathbf{Q}_{a\pm\iota}} &= \pm\frac{1}{2} \mathbf{Q}_{a\pm\iota} , &
    \qquad
    \comm{\mathbf{J}_\pm}{\mathbf{Q}_{a\mp\iota}} &= \mathbf{Q}_{a\pm\iota} , \\
  \end{aligned}
\end{equation}
with $\kappa=\pm,\iota=\pm$ and $a=\pm $. The anticommutators then are
\begin{equation}
  \begin{aligned}
    \acomm{\mathbf{Q}_{\pm++}}{\mathbf{Q}_{\pm--}} &= \pm \mathbf{L}_{\mp} , \! &
    \acomm{\mathbf{Q}_{\pm+-}}{\mathbf{Q}_{\pm-+}} &= \mp \mathbf{L}_{\mp}  , \\
    \acomm{\mathbf{Q}_{+\pm+}}{\mathbf{Q}_{-\pm-}} &= \mp \mathbf{J}_{\pm} , \! &
    \acomm{\mathbf{Q}_{+\pm-}}{\mathbf{Q}_{-\pm+}} &= \pm \mathbf{J}_{\pm}  ,\\
    \acomm{\mathbf{Q}_{+\pm\pm}}{\mathbf{Q}_{-\mp\mp}} &= + \mathbf{L}_3 \pm \mathbf{J}_3 , \!
    &\qquad
    \acomm{\mathbf{Q}_{+\pm\mp}}{\mathbf{Q}_{-\mp\pm}} &= - \mathbf{L}_3 \mp \mathbf{J}_3 .
  \end{aligned}
\end{equation}
Further  properties of~$\psu(1,1|2) $ are given in appendix~\ref{app:psu112}.

\begin{figure}
  \centering
  \begin{tikzpicture}
    \node (phi0) at (-\hordist,0) {$ \Ket{\phi_{\kappa}^{(0)}} $};
    \node (phi1) at (-\hordist,-2\vertdist) {$ \Ket{\phi_{\kappa}^{(1)}} $};
    \node (phi2) at (-\hordist,-4\vertdist) {$ \Ket{\phi_{\kappa}^{(2)}} $};

    \node (psi0) at (\hordist,-\vertdist) {$ \Ket{\psi_{{\iota}}^{(0)}} $};
    \node (psi1) at (\hordist,-3\vertdist) {$ \Ket{\psi_{{\iota}}^{(1)}} $};

    \coordinate (q1) at (0,-0.5\vertdist);
    \coordinate (q2) at (0,-1.5\vertdist);
    \coordinate (q3) at (0,-2.5\vertdist);
    \coordinate (q4) at (0,-3.5\vertdist);
    \coordinate (q5) at ($(0,-4.25\vertdist)+(0,-0.4ex)$);

    \coordinate (s1) at ($(-\hordist,-\vertdist)+(0,-0.4ex)$);
    \coordinate (s2) at ($( \hordist,-2\vertdist)+(0,-0.4ex)$);
    \coordinate (s3) at ($(-\hordist,-3\vertdist)+(0,-0.4ex)$);
    \coordinate (s4) at ($( \hordist,-3.75\vertdist)+(0,-0.4ex)$);
    \coordinate (s5) at ($(-\hordist,-4.75\vertdist)+(0,-0.4ex)$);

    \coordinate (L1) at (-1.6\hordist,0);
    \coordinate (L2) at (-1.6\hordist,-2\vertdist);
    \coordinate (L3) at (-1.6\hordist,-4\vertdist);

    \coordinate (R1) at (1.6\hordist,-\vertdist);
    \coordinate (R2) at (1.6\hordist,-3\vertdist);

    \draw [->] ($(phi0.east)-(0,0.30cm)$) -- ($(psi0.west)+(0,0.15cm)$);
    \node at ($(q1)+(-0.2cm,0cm)$) [anchor=south west] {$\textstyle \epsilon^{\iota\b}\mathbf{Q}_{+\kappa\b}$};
    
    \draw [<-] ($(phi0.east)-(0,0.15cm)$) -- ($(psi0.west)+(0,0.30cm)$);
    \node at ($(q1)+(0.2cm,-0cm)$) [anchor=north east] {$ \epsilon^{\kappa\a}\mathbf{Q}_{-\a\iota}$};

    \draw [->] ($(phi1.east)+(0,0.30cm)$) -- ($(psi0.west)-(0,0.15cm)$);
    \node at ($(q2)+(0.2cm,0.1cm)$) [anchor=south east] {$ \epsilon^{\kappa\a}\mathbf{Q}_{+\a\iota}$};

    \draw [<-] ($(phi1.east)+(0,0.15cm)$) -- ($(psi0.west)-(0,0.30cm)$);
    \node at ($(q2)+(-0.2cm,0cm)$) [anchor=north west] {$ \epsilon^{\iota\b}\mathbf{Q}_{-\kappa\b}$};

    \draw [->] ($(phi1.east)-(0,0.30cm)$) -- ($(psi1.west)+(0,0.15cm)$);
    \node at ($(q3)+(-0.2cm,0cm)$) [anchor=south west] {$ \epsilon^{\iota\b}\mathbf{Q}_{+\kappa\b}$};

    \draw [<-] ($(phi1.east)-(0,0.15cm)$) -- ($(psi1.west)+(0,0.30cm)$);
    \node at ($(q3)+(0.2cm,-0cm)$) [anchor=north east] {$ \epsilon^{\kappa\a}\mathbf{Q}_{-\a\iota}$};

    \draw [->] ($(phi2.east)+(0,0.30cm)$) -- ($(psi1.west)-(0,0.15cm)$);
    \node at ($(q4)+(0.2cm,0cm)$) [anchor=south east] {$ \epsilon^{\kappa\a}\mathbf{Q}_{+\a\iota}$};

    \draw [<-] ($(phi2.east)+(0,0.15cm)$) -- ($(psi1.west)-(0,0.30cm)$);
    \node at ($(q4)+(0cm,-0.1cm)$) [anchor=north west] {$ \epsilon^{\iota\b}\mathbf{Q}_{-\kappa\b}$};

    \draw [->] ($(phi0.south)+(-0.08cm,0)$) -- ($(phi1.north)+(-0.08cm,0)$);
    \draw [<-] ($(phi0.south)+(+0.08cm,0)$) -- ($(phi1.north)+(+0.08cm,0)$);
    \node at (s1) [anchor=west] {$\,\mathbf{L}_-$};
    \node at (s1) [anchor=east] {$\mathbf{L}_+$};

    \draw [->] ($(psi0.south)+(-0.08cm,0)$) -- ($(psi1.north)+(-0.08cm,0)$);
    \draw [<-] ($(psi0.south)+(+0.08cm,0)$) -- ($(psi1.north)+(+0.08cm,0)$);
    \node at (s2) [anchor=west] {$\,\mathbf{L}_-$};
    \node at (s2) [anchor=east] {$\mathbf{L}_+$};

    \draw [->] ($(phi1.south)+(-0.08cm,0)$) -- ($(phi2.north)+(-0.08cm,0)$);
    \draw [<-] ($(phi1.south)+(+0.08cm,0)$) -- ($(phi2.north)+(+0.08cm,0)$);
    \node at (s3) [anchor=west] {$\,\mathbf{L}_-$};
    \node at (s3) [anchor=east] {$\mathbf{L}_+$};

    \draw [->,out=170,in=90] ($(phi0.west)+(0,0.2cm)$) to (L1) to [out=270,in=190] ($(phi0.west)-(0,0.2cm)$);
    \node at (L1) [anchor=east] {$\mathbf{J}_{\a}$};

    \draw [->,out=170,in=90] ($(phi1.west)+(0,0.2cm)$) to (L2) to [out=270,in=190] ($(phi1.west)-(0,0.2cm)$);
    \node at (L2) [anchor=east] {$\mathbf{J}_{\a}$};

    \draw [->,out=170,in=90] ($(phi2.west)+(0,0.2cm)$) to (L3) to [out=270,in=190] ($(phi2.west)-(0,0.2cm)$);
    \node at (L3) [anchor=east] {$\mathbf{J}_{\a}$};



    \node at ($(s4)+(0,0.2cm)$) {$\vdots$};
    \node at ($(s5)+(0,0.4cm)$) {$\vdots$};
    \node at ($(q5)+(0,0)$) {$\vdots$};

  \end{tikzpicture}

  \caption{An illustration of the short
    $\left(\tfrac{1}{2};\tfrac{1}{2}\right)$ 
    module, where the action of the (super)charges  is represented up to the numerical coefficients. These are given in eqs.~(\ref{eq:su112-representation}--\ref{eq:su112-representation2}).}
  \label{fig:short-psu112}
\end{figure}
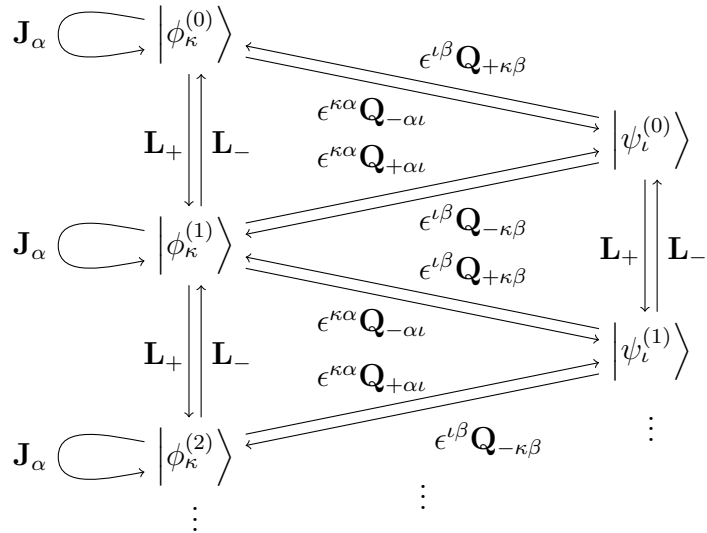

Let us focus our attention on the action of the generators on a single site, which may be in any of the states of the $(\tfrac{1}{2};\tfrac{1}{2})$ module. We have that bosonic states are charged under~$\su(2)$
\begin{equation}\label{eq:su112-representation}
      \mathbf{J}_3 \ket{\phi_{\pm}^{(n)}} = \pm \frac{1}{2} \ket{\phi_{\pm}^{(n)}} , \qquad
      \mathbf{J}_+ \ket{\phi_{-}^{(n)}} = \ket{\phi_{+}^{(n)}} , \qquad
      \mathbf{J}_- \ket{\phi_{+}^{(n)}} = \ket{\phi_-^{(n)}} ,
\end{equation}
and  all states are charged under $\sl(2)$
\begin{equation}
    \begin{aligned}
      \mathbf{L}_3 \ket{\phi_{\kappa}^{(n)}} &=  \left( \tfrac{1}{2} + n \right) \ket{\phi_{\kappa}^{(n)}} , &\qquad
      \mathbf{L}_3 \ket{\psi_{\iota}^{(n)}} &=  \left( 1 + n \right) \ket{\psi_{\iota}^{(n)}} , \\
      \mathbf{L}_- \ket{\phi_{\kappa}^{(n)}} &= +n \ket{\phi_{\kappa}^{(n-1)}} , &
      \mathbf{L}_- \ket{\psi_{\iota}^{(n)}} &= +\sqrt{(n + 1)n} \ket{\psi_{\iota}^{(n-1)}} , \\
      \mathbf{L}_+ \ket{\phi_{\kappa}^{(n)}} &= -(n+1)\ket{\phi_{\kappa}^{(n+1)}} , &
      \mathbf{L}_+ \ket{\psi_{\iota}^{(n)}} &= -\sqrt{(n + 2) (n + 1)} \ket{\psi_{\iota}^{(n+1)}} .
    \end{aligned}
\end{equation}
Finally, the supercharges act as
\begin{equation}
\label{eq:su112-representation2}
    \begin{aligned}
      \mathbf{Q}_{-\pm\iota} \ket{\phi_{\mp}^{(n)}} &= \pm \sqrt{n+1} \ket{\psi_{\iota}^{(n)}} , &
      \mathbf{Q}_{+\pm\iota} \ket{\phi_{\mp}^{(n)}} &= \pm \sqrt{n} \ket{\psi_{\iota}^{(n-1)}} , \\
      \mathbf{Q}_{-\kappa\pm} \ket{\psi_{\mp}^{(n)}} &= \mp \sqrt{n+1} \ket{\phi_{\kappa}^{(n+1)}} ,\qquad &
      \mathbf{Q}_{+\kappa\pm} \ket{\psi_{\mp}^{(n)}} &= \mp \sqrt{n+1} \ket{\phi_{\kappa}^{(n)}} .
    \end{aligned}
\end{equation}
Therefore, the highest weight state $\ket{\phi^{(0)}_+}$ is annihilated by the $\su(2)$ grading raising operators $\mathbf{Q}_{+\pm\pm}$, and by the two generators $\mathbf{Q}_{-+\pm}$. Hence, the representation $(\tfrac{1}{2};\tfrac{1}{2})$ is a short representation, satisfying the shortening conditions
\begin{equation}
  \acomm{\mathbf{Q}_{+-\mp}}{\mathbf{Q}_{-+\pm}} \ket{\phi^{(0)}_+} = \pm (\mathbf{L}_3 - \mathbf{J}_3 ) \ket{\phi^{(0)}_+} = 0.
\end{equation}
The action of the generators on part of the module is depicted in figure~\ref{fig:short-psu112}.
We will take the sites of the left and right spin-chains to transform in identical modules. At the very end of our construction, we will see that self-consistency dictates the left and right algebras to be in different gradings, as it was the case in the coset construction of chapter~\ref{ch:sigmamodel}.

\subsection*{The ground state}

The states of the left- and right-moving spin-chains of length
$\ell$ transform in the $\ell$-fold tensor product of the above representation. The ground state of the full spin-chain is given by
\begin{equation}
\label{eq:spinchain-gs}
  \ket{0}_{\ell} = \Ket{(\phi^{(0)}_+)^{\ell}} \otimes \Ket{(\phi^{(0)}_+)^{\ell}}.
\end{equation}
This is the highest weight state of the short  $(\tfrac{\ell}{2};\tfrac{\ell}{2}) \otimes (\tfrac{\ell}{2};\tfrac{\ell}{2})$ representation of the superalgebra 
$\psu(1,1|2)_{\L} \oplus \psu(1,1|2)_{\R}$. Such a choice preserves as much supersymmetry as possible.
Specifically, the ground state is preserved by eight supercharges $\mathbf{Q}^{j\,\L,\R}$ and $\overline{\mathbf{Q}}{}^{j\,\L,\R}$, with $j=1,2$, as well as two central charges $\mathbf{H}^{\L,\R}$. In terms of the $\psu(1,1|2)$ generators they are given by
\begin{equation}
  \begin{gathered}
    \mathbf{Q}^{1\L} = \mathbf{Q}_{-++}^{\smallL} , \qquad
    \mathbf{Q}^{2\L} = -\mathbf{Q}_{-+-}^{\smallL} , \qquad
    \overline{\mathbf{Q}}{}^{1\L} = \mathbf{Q}_{+--}^{\smallL} , \qquad
    \overline{\mathbf{Q}}{}^{2\L} = \mathbf{Q}_{+-+}^{\smallL} ,\\
    \mathbf{Q}^{1\R} = \mathbf{Q}_{-++}^{\smallR} , \qquad
    \mathbf{Q}^{2\R} = -\mathbf{Q}_{-+-}^{\smallR} , \qquad
    \overline{\mathbf{Q}}{}^{1\R} =\mathbf{Q}_{+--}^{\smallR} , \qquad
    \overline{\mathbf{Q}}{}^{2\R} =\mathbf{Q}_{+-+}^{\smallR} ,\\
  \end{gathered}
\end{equation}
and
\begin{equation}
    \mathbf{H}^{\L} = \mathbf{L}_3^{\smallL}- \mathbf{J}_3^{\smallL},
\qquad
    \mathbf{H}^{\R} = \mathbf{L}_3^{\smallR}- \mathbf{J}_3^{\smallR}.
\end{equation}
This forms two (one left, one right) copies of the $(\su(1|1)^2)/ \u(1)$ algebra, where the quotient is due to the fact that we have the same Hamiltonian for $j=1,2$:
\begin{equation}\label{eq:su11-su11-algebra}
  \acomm{\mathbf{Q}^{j\,\L}}{\overline{\mathbf{Q}}{}^{k\,\L}} = \delta^{jk}  \mathbf{H}^{\L}\,,
\qquad
  \acomm{\mathbf{Q}^{j\,\R}}{\overline{\mathbf{Q}}{}^{k\,\R}} = \delta^{jk}  \mathbf{H}^{\R}\,.
\end{equation}
The charges $\mathbf{H}^{\L}$ and $\mathbf{H}^{\R}$ are the left- and right-moving spin-chain Hamiltonians. Let us define
\begin{equation}
  \mathbf{H} = \mathbf{H}^{\L} + \mathbf{H}^{\R} , \qquad
  \mathbf{M} = \mathbf{H}^{\L} - \mathbf{H}^{\R} ,
\end{equation}
where the positive-definite combination $\mathbf{H}$ has the interpretation of the spin-chain Hamiltonian, and may depend on the momenta of the spin-chain excitations. The central charge $\mathbf{M}$ measures an angular momentum in $\AdS_3\times\S^3$ and should be quantised.

This symmetry algebra appears \emph{at small coupling}, when the~L and R sectors are decoupled. It also coincides with the \emph{on-shell symmetry algebra} of section~\ref{sec:stringsymm}. This picture is similar to what happens in~$\AdS_5\times\S^5$, where the small-coupling spin chain does not display dependence on the additional momentum-dependent central charges.

For later convenience, let us introduce two additional generators $\mathbf{V}_1^{\smallL,\smallR}$ and $\mathbf{V}_2^{\smallL,\smallR}$ acting as outer automorphisms. These can be constructed from the $\psu(1,1|2)$ generators $\mathbf{J}_3^{\smallL,\smallR}$ and the automorphisms $\mathbf{U}^{\smallL,\smallR}$ defined in eq.~\eqref{eq:Uautomorphism},
\begin{equation}
\label{eq:su11-automorphisms-def}
\begin{aligned}
\mathbf{V}_1^{\smallL,\smallR}
&=
- \mathbf{U}^{\smallL,\smallR}
- \mathbf{J}_3^{\smallL,\smallR}\,,
\qquad\quad
&\mathbf{V}_2^{\smallL,\smallR}
&=
+ \mathbf{U}^{\smallL,\smallR}
- \mathbf{J}_3^{\smallL,\smallR}\,,\\
\mathbf{V}_1
&=
+\mathbf{V}_1^{\smallL}-\mathbf{V}_1^{\smallR}\,,
\qquad
&\mathbf{V}_2
&=+\mathbf{V}_2^{\smallL}-\mathbf{V}_2^{\smallR}\,.
\end{aligned}
\end{equation}
While all four left and right generators are automorphisms of~$\psu(1,1|2)^2$, only the combinations $\mathbf{V}_{1}$ and $\mathbf{V}_{2}$ annihilate the vacuum~\eqref{eq:spinchain-gs}.
Note also that $\mathbf{V}_1$ commutes with $\mathbf{Q}^{2\,\L,\R}$ and $\overline{\mathbf{Q}}{}^{2\,\L,\R}$, while $\mathbf{V}_2$ commutes with $\mathbf{Q}^{1\,\L,\R}$ and $\overline{\mathbf{Q}}{}^{1\,\L,\R}$. The commutation relations involving the supercharges then read
\begin{equation}
  \begin{aligned}
    \comm{\mathbf{V}_j}{\mathbf{Q}^{k\L}} &= - \delta_{j}^{k} \,\mathbf{Q}^{k\L} , \qquad &
    \comm{\mathbf{V}_j}{\overline{\mathbf{Q}}{}^{k\L}} &= + \delta_{j}^{k}\, \overline{\mathbf{Q}}{}^{k\L} , \\
    \comm{\mathbf{V}_j}{\mathbf{Q}^{k\R}} &= + \delta_{j}^{k} \,\mathbf{Q}_{k\R} , \qquad &
    \comm{\mathbf{V}_j}{\overline{\mathbf{Q}}{}^{k\R}} &= - \delta_{j}^{k}\, \overline{\mathbf{Q}}{}^{k\R} .
  \end{aligned}
\end{equation}
The generators $\mathbf{V}_j$ give useful  restrictions on the allowed deformations of the weak-coupling representations. Taking them into account, we regroup the symmetry algebra into two copies of $\u(1) \oplus \su(1|1)^2$, with the generators given by
\begin{equation}
\label{eq:su11sq-spin-chain}
\bigl\{\mathbf{Q}^{1\L}, \mathbf{Q}^{1\R}, \overline{\mathbf{Q}}{}^{1\L}, \overline{\mathbf{Q}}{}^{1\R}, \mathbf{H}^{\L},\mathbf{H}^{\R},\mathbf{V}_1\bigr\}
  \quad\text{and}\quad
  \bigl\{\mathbf{Q}^{2\L}, \mathbf{Q}^{2\R}, \overline{\mathbf{Q}}{}^{2\L}, \overline{\mathbf{Q}}{}^{2\R}, \mathbf{H}^{\L},\mathbf{H}^{\R},\mathbf{V}_2\bigr\}  ,
\end{equation}
respectively. This splitting is familiar from section~\ref{sec:stringsymm}, where it corresponded to a tensor-product structure in the representation of the excitations. We will see that the same holds~here.

\subsection*{Excitations at weak coupling}

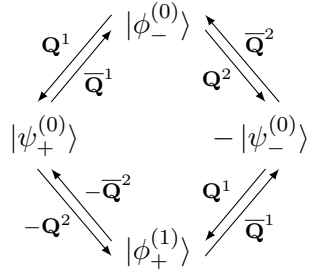
\begin{figure}
  \centering
  
  \begin{tikzpicture}[%
    box/.style={outer sep=1pt},
    Q node/.style={inner sep=1pt,outer sep=0pt},
    arrow/.style={-latex}
    ]%

    \node [box] (PhiM) at ( 0  , 1.5cm) {\small $\ket{\phi_-^{(0)}}$};
    \node [box] (PsiP) at (-1.5cm, 0cm) {\small $\ket{\psi_+^{(0)}}$};
    \node [box] (PsiM) at (+1.5cm, 0cm) {\small $\mathllap{-}\ket{\psi_-^{(0)}}$};
    \node [box] (PhiP) at ( 0  ,-1.5cm) {\small $\ket{\phi_+^{(1)}}$};

    \newcommand{\horshift}{0.09cm,0cm}
    \newcommand{\vershift}{0cm,0.10cm}
 
    \draw [arrow] ($(PhiM.west) +(\vershift)$) -- ($(PsiP.north)-(\horshift)$) node [pos=0.5,anchor=south east,Q node] {\scriptsize $\mathbf{Q}^1$};
    \draw [arrow] ($(PsiP.north)+(\horshift)$) -- ($(PhiM.west) -(\vershift)$) node [pos=0.5,anchor=north west,Q node] {\scriptsize $\overline{\mathbf{Q}}^1$};

    \draw [arrow] ($(PsiM.south)-(\horshift)$) -- ($(PhiP.east) +(\vershift)$) node [pos=0.5,anchor=south east,Q node] {\scriptsize $\mathbf{Q}^1$};
    \draw [arrow] ($(PhiP.east) -(\vershift)$) -- ($(PsiM.south)+(\horshift)$) node [pos=0.5,anchor=north west,Q node] {\scriptsize $\overline{\mathbf{Q}}^1$};

    \draw [arrow] ($(PhiM.east) -(\vershift)$) -- ($(PsiM.north)-(\horshift)$) node [pos=0.5,anchor=north east,Q node] {\scriptsize $\mathbf{Q}^2$};
    \draw [arrow] ($(PsiM.north)+(\horshift)$) -- ($(PhiM.east) +(\vershift)$) node [pos=0.5,anchor=south west,Q node] {\scriptsize $\overline{\mathbf{Q}}^2$};

    \draw [arrow] ($(PsiP.south)-(\horshift)$) -- ($(PhiP.west) -(\vershift)$) node [pos=0.5,anchor=north east,Q node] {\scriptsize $-\mathbf{Q}^2$};
    \draw [arrow] ($(PhiP.west) +(\vershift)$) -- ($(PsiP.south)+(\horshift)$) node [pos=0.5,anchor=south west,Q node] {\scriptsize $-\overline{\mathbf{Q}}^2$};
  \end{tikzpicture}

  \caption{The action of the supercharges $\mathbf{Q}^j=\mathbf{Q}^{j\,\L,\R}$ and $\overline{\mathbf{Q}}^j=\overline{\mathbf{Q}}{}^{j\,\L,\R}$ on the either of the left- or right-bifundamental representation~\eqref{eq:bi-fund-fields}. It takes the same form of figure~\ref{fig:representation}.}
  \label{fig:representation2}
\end{figure}
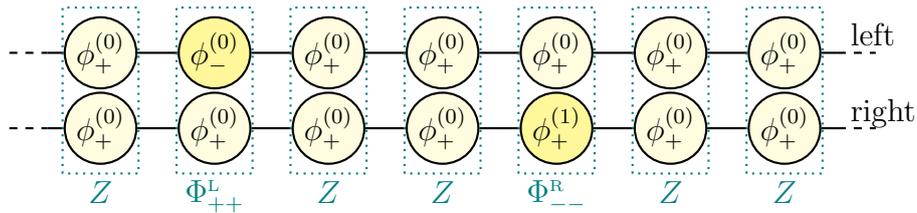
\begin{figure}
  \centering
\begin{tikzpicture}
  \begin{scope}[xshift=-0.95cm]
    \coordinate (c0) at (-5.7cm,0);
    \coordinate (c1) at (-5.2cm,0);
    \coordinate (c2) at (+5.2cm,0);
    \coordinate (c3) at (+5.7cm,0);

    \coordinate (c0b) at (-5.7cm,1cm);
    \coordinate (c1b) at (-5.2cm,1cm);
    \coordinate (c2b) at (+5.2cm,1cm);
    \coordinate (c3b) at (+5.7cm,1cm);

    \node (sp0) at (-4.5cm,-0.85cm) [teal]  {$Z $};
    \node (sp1) at (-3cm,-0.9cm)   [teal]  {$\Phi^{\smallL}_{++} $};
    \node (sp2) at (-1.5cm,-0.85cm) [teal]  {$Z $};
    \node (sp3) at (0cm,-0.85cm)    [teal]  {$Z $};
    \node (sp4) at (1.5cm,-0.9cm)  [teal]  {$\Phi^{\smallR}_{--} $};
    \node (sp5) at (3cm,-0.85cm)    [teal]  {$Z $};
    \node (sp6) at (4.5cm,-0.85cm)  [teal]  {$Z $};

    \node (ll) at (5.8cm,1.2cm)   {$\fixedspaceL{\text{right}}{\text{left}}$};
    \node (rr) at (5.8cm,0.2cm)   {$\fixedspaceL{\text{right}}{\text{right}}$};

    \node (sp10) at (-4.5cm,0cm) [spin] {$\phi^{(0)}_{+} $};
    \node (sp11) at (-3cm,0cm) [spin] {$\phi^{(0)}_{+} $};
    \node (sp12) at (-1.5cm,0cm) [spin] {$\phi^{(0)}_{+} $};
    \node (sp13) at (0cm,0cm) [spin] {$\phi^{(0)}_{+} $};
    \node (sp14) at (1.5cm,0cm) [spinex] {$\phi^{(\fixedspaceL{0}{1})}_{+} $};
    \node (sp15) at (3cm,0cm) [spin] {$\phi^{(0)}_{+} $};
    \node (sp16) at (4.5cm,0cm) [spin] {$\phi^{(0)}_{+} $};

    \node (sp20) at (-4.5cm,1cm) [spin] {$\phi^{(0)}_{+} $};
    \node (sp21) at (-3cm,1cm) [spinex] {$\phi^{(0)}_{-} $};
    \node (sp22) at (-1.5cm,1cm) [spin] {$\phi^{(0)}_{+} $};
    \node (sp23) at (0cm,1cm) [spin] {$\phi^{(0)}_{+} $};
    \node (sp24) at (1.5cm,1cm) [spin] {$\phi^{(0)}_{+} $};
    \node (sp25) at (3cm,1cm) [spin] {$\phi^{(0)}_{+} $};
    \node (sp26) at (4.5cm,1cm) [spin] {$\phi^{(0)}_{+} $};
    
    \draw [thick,dashed]  (c0) to (c1);
    \draw [thick]  (c1) to (sp10);
    \draw [thick]  (sp10) to (sp11);
    \draw [thick]  (sp11) to (sp12);
    \draw [thick]  (sp12) to (sp13);
    \draw [thick]  (sp13) to (sp14);
    \draw [thick]  (sp14) to (sp15);
    \draw [thick]  (sp15) to (sp16);
    \draw [thick]  (sp16) to (c2);
    \draw [thick,dashed]  (c2) to (c3);

    \draw [thick,dashed]  (c0b) to (c1b);
    \draw [thick]  (c1b) to (sp20);
    \draw [thick]  (sp20) to (sp21);
    \draw [thick]  (sp21) to (sp22);
    \draw [thick]  (sp22) to (sp23);
    \draw [thick]  (sp23) to (sp24);
    \draw [thick]  (sp24) to (sp25);
    \draw [thick]  (sp25) to (sp26);
    \draw [thick]  (sp26) to (c2b);
    \draw [thick,dashed]  (c2b) to (c3b);   

	\draw [thick,dotted,teal] (-5cm,-0.6cm) rectangle (-4cm,1.6cm);
	\draw [thick,dotted,teal] (-3.5cm,-0.6cm) rectangle (-2.5cm,1.6cm);
	\draw [thick,dotted,teal] (-2cm,-0.6cm) rectangle (-1cm,1.6cm);
	\draw [thick,dotted,teal] (-0.5cm,-0.6cm) rectangle (0.5cm,1.6cm);
	\draw [thick,dotted,teal] (1cm,-0.6cm) rectangle (2cm,1.6cm);
	\draw [thick,dotted,teal] (2.5cm,-0.6cm) rectangle (3.5cm,1.6cm);
	\draw [thick,dotted,teal] (4cm,-0.6cm) rectangle (5cm,1.6cm);

  \end{scope}
\end{tikzpicture}

  \caption{A pictorial representation of the $\psu(1,1|2)^2$ spin chain. We can think of it as two (left and right) spin chains, each with ground state~$\phi^{(0)}_{+}$. Excitations of the whole chain are considered as left- or right- moving depending on whether we excite a spin in the left or right~chain.}
  \label{fig:symmspinchain}
\end{figure}

To construct excited spin-chain states we replace one or more of the ground state sites by any other state in the same module. We can classify these excitations by their eigenvalues under the left and right spin-chain Hamiltonians $\mathbf{H}^{\L}$ and $\mathbf{H}^{\R}$ at zero coupling. Let us consider excitations in the left sector. Replacing one of the highest weight states $\phi^{(0)}_+$ by the scalar $\phi^{(n)}_-$ or $\phi^{(n)}_+$ increases the eigenvalue of $\mathbf{H}^{\smallL}$ by $n$ or $n+1$, respectively. Similarly, insertion of a fermion $\psi^{(n)}_{\pm}$ also adds $n$ to the energy. The lightest excitations are therefore
\begin{equation}
  \phi_-^{(0)}\, , \qquad \psi_+^{(0)}\, , \qquad \psi_-^{(0)}\, ,  \qquad \phi_+^{(1)}\,.
\end{equation}
It is easy to see that the charges of any heavier excitation can be reproduced by considering a combination of the four states above. These states form a four-dimensional bifundamental representation of either of the two $\psu(1|1)^2$ algebras~\eqref{eq:su11-su11-algebra}, as illustrated in figure~\ref{fig:representation2}. To emphasise this we introduce the notation
\begin{equation}\label{eq:bi-fund-fields}
\begin{aligned}
  \Phi_{+{+}}^{\smallL} \!=\! \phi_-^{(0)}\!\otimes\!\phi_+^{(0)} , \quad
  \Phi_{-{-}}^{\smallL} \!=\! \phi_+^{(1)}\!\otimes\!\phi_+^{(0)} , \quad
  \Phi_{-{+}}^{\smallL} \!=\! \psi_+^{(0)}\!\otimes\!\phi_+^{(0)} , \quad
  \Phi_{+{-}}^{\smallL} \!=\! -\psi_-^{(0)}\!\otimes\!\phi_+^{(0)} ,\\
  \Phi_{+{+}}^{\smallR} \!=\! \phi_+^{(0)}\!\otimes\!\phi_-^{(0)} , \quad
  \Phi_{-{-}}^{\smallR} \!=\! \phi_+^{(0)}\!\otimes\!\phi_+^{(1)} , \quad
  \Phi_{-{+}}^{\smallR} \!=\! \phi_+^{(0)}\!\otimes\!\psi_+^{(0)} , \quad
  \Phi_{+{-}}^{\smallR} \!=\! -\phi_+^{(0)}\!\otimes\!\psi_-^{(0)} ,
\end{aligned}
\end{equation}
where the tensor product is over left and right sites.
As before,the excitations $\Phi_{\pm\pm}$ are bosons while $\Phi_{\pm\mp}$ are fermions in either the left and right sector. We will also defined the short-hand for a vacuum site,
\begin{equation}
Z=\phi^{(0)}_+\otimes \phi^{(0)}_+\,.
\end{equation}
In figure~\ref{fig:symmspinchain} the structure of the spin chain is represented pictorially.

\subsubsection*{Fundamental representations}
As we did in the string picture, we can make the bifundamental nature of the representation above more explicit. Let us consider the left module. We introduce a fundamental $\su(1|1)$ representation with basis $(\phi^{\smallL}|\psi^{\smallL})$, and the generators $\mathbf{q}$, $\bar{\mathbf{q}}$ and $\mathbf{h}$ acting as
\begin{equation}\label{eq:su11-repr}
  \mathbf{q}^{\smallL} \ket{\phi^{\smallL}} = a^{\smallL} \ket{\psi^{\smallL}} , \quad
  \bar{\mathbf{q}}^{\smallL} \ket{\psi^{\smallL}} = \bar{a}^{\smallL} \ket{\phi^{\smallL}} , \quad
  \mathbf{h}^{\smallL} = |a^{\smallL}|^2\, \mathbf{I}^{\smallL} .
\end{equation}
where~$\mathbf{I}^{\smallL}$ is the identity on the~$(\phi^{\smallL}|\psi^{\smallL})$ representation.
The representation for right-movers is similar,
\begin{equation}\label{eq:su11-repr=right}
  \mathbf{q}^{\smallR} \ket{\phi^{\smallR}} = a^{\smallR} \ket{\psi^{\smallR}} , \quad
  \bar{\mathbf{q}}^{\smallR} \ket{\psi^{\smallR}} = \bar{a}^{\smallR} \ket{\phi^{\smallR}} , \quad
  \mathbf{h}^{\smallR} = |a^{\smallR}|^2\, \mathbf{I}^{\smallR} .
\end{equation}
We can also write down the action of the automorphism generators~$\mathbf{v}^{\smallL,\smallR}$ and~$\mathbf{v}$, which correspond to~$\mathbf{V}_{j}^{\smallL,\smallR}$ and~$\mathbf{V}_j$ respectively. This is found from the charges written in table~\ref{tab:charges-v1-v2}. The parameter~$v$ appearing there is a label of the representation, and as we will see it is natural to take it to be the same in the left and right ones. Looking at the eigenvalues, we see that indeed~$\mathbf{v}$ annihilates the vacuum, whereas each of the~$\mathbf{v}^{\smallL,\smallR}$ do not, and instead measure its length.

\begin{table}
  \centering
  \begin{tabular}{clll}
    \toprule
    & $\phantom{+}\mathllap{2}\,\mathbf{v}^{\smallL}$ 
    & $\phantom{+}\mathllap{2}\,\mathbf{v}^{\smallR}$
    & $\phantom{+}\mathllap{2}\,\mathbf{v}$ \\
    \midrule
    $\ket{\fixedspaceL{\psi}{0}}$ & $-\ell$ & $-\ell$ & $\phantom{+}0$ \\
    $\ket{\fixedspaceL{\psi}{\phi}^{\smallL}}$ & $-\ell+v$ & $-\ell$ & $+v$ \\
    $\ket{\fixedspaceL{\psi}{\psi}^{\smallL}}$ & $-\ell+v-1$ & $-\ell$& $+v-1$ \\
    $\ket{\fixedspaceL{\psi}{\phi}^{\smallR}}$ & $-\ell$ & $-\ell+v$ & $-v$ \\
    $\ket{\fixedspaceL{\psi}{\psi}^{\smallR}}$ & $-\ell$ & $-\ell+v-1$& $-v+1$ \\
    \midrule
    $\mathbf{q}^{\smallL}$ & $-1$ & $\phantom{+}0$& $-1$ \\
    $\bar{\mathbf{q}}^{\smallL}$ & $+1$ & $\phantom{+}0$& $+1$ \\
    $\mathbf{q}^{\smallR}$ & $\phantom{+}0$ & $-1$& $+1$ \\
    $\bar{\mathbf{q}}^{\smallR}$ & $\phantom{+}0$ & $+1$& $-1$ \\
    \bottomrule
  \end{tabular}
  \caption{%
    Charges under the automorphisms $\mathbf{v}^{\smallL,\smallR}$ and $\mathbf{v}$ of the ground state, of the single excitation states in the left- and right-moving multiplets as well as of the supercharges. All the spin-chain states have length
    $\ell$. To make the table less cluttered, the charges have been rescaled.%
  }
  \label{tab:charges-v1-v2}
\end{table}

\section{The dynamical spin chain}
\label{sec:dynamicalSC}

At non-vanishing coupling $h>0$, the spin-chain Hamiltonian $\mathbf{H}$ should depend on~$h$ and on the momentum of the excitations. This requires the bifundamental representations discussed above to be deformed---a procedure akin to going from the one-particle, on-shell $p=0$ representation to a general \emph{off-shell} one.

This deformation should be done in such a way that the angular momentum~$\mathbf{M}$ remains quantised, \ie~undeformed. This can be done if we allow the right generators to act nontrivially on the left-moving excitations, and \textit{vice versa}, which is what we expect from chapters~\ref{ch:sigmamodel} and~\ref{ch:smatrix}.

\subsection*{Constructing the central extension}
Let us investigate how to centrally extend the algebra. For simplicity, let us focus on constructing the extension of~$\u(1) \oplus \su(1|1)^2$ in~\eqref{eq:su11sq-spin-chain}. The generalisation to the full algebra and tensor-product representation will be straightforward.

We want the left representation to be charged under all of the supercharges. Focusing on the highest weight state~$\ket{\phi^{\smallL}}$, we have two possibilities
\begin{equation}
\text{(I)\,:}\quad
\mathbf{q}^{\smallR}\ket{\phi^{\smallL}}\neq 0\,,
\qquad\quad
\text{or}
\qquad\quad
\text{(II)\,:}\quad
\bar{\mathbf{q}}^{\smallR}\ket{\phi^{\smallL}}\neq 0\,.
\end{equation}
In case (I), in the right-hand side there should be a fermion whose charge under $\mathbf{v}$ is
\begin{equation}
\mathbf{v}\,\mathbf{q}^{\smallR}\ket{\phi^{\smallL}}=
\frac{1}{2}\big(v+1\big)\,\mathbf{q}^{\smallR}\ket{\phi^{\smallL}}\,.
\end{equation}
Looking back at table~\ref{tab:charges-v1-v2}, we see that there are no such fermionic states on the vacuum~$\ket{0}_{\ell}$ preserved by~$\mathbf{v}$. On the other hand, in case (II) we have
\begin{equation}
\mathbf{v}\,\bar{\mathbf{q}}^{\smallR}\ket{\phi^{\smallL}}=
\frac{1}{2}\big(v-1\big)\,\bar{\mathbf{q}}^{\smallR}\ket{\phi^{\smallL}}\,,
\end{equation}
so that the state $\ket{\psi^{\smallL}}$ is a good candidate to appear on the right hand side. Therefore, let us restrict to case (II), and further investigate the central extension by looking at the action of $\mathbf{v}^{\smallL}$ and $\mathbf{v}^{\smallR}$.  We have
\begin{equation}
\begin{aligned}
&\mathbf{v}^{\smallL}\,\bar{\mathbf{q}}^{\smallR}\ket{\phi^{\smallL}}=
\frac{1}{2}\big(v-1-(\ell-1)\big)\,\bar{\mathbf{q}}^{\smallR}\ket{\phi^{\smallL}}\,,
\\
&\mathbf{v}^{\smallR}\,\bar{\mathbf{q}}^{\smallR}\ket{\phi^{\smallL}}=
\frac{1}{2}\big(-(\ell-1)\big)\,\bar{\mathbf{q}}^{\smallR}\ket{\phi^{\smallL}}\,.
\end{aligned}
\end{equation}
We can interpret this by saying that~$\bar{\mathbf{q}}^{\smallR}$ acts on a state in the left-representation of length~$\ell$ by exchanging a boson with fermion and \emph{reducing the length} of the chain by one. If we write separately the left and right sites of the chain in a tensor product form, we have
\begin{equation}
\bar{\mathbf{q}}^{\smallR} \ket{\phi^{\smallL}} = \bar{\mathbf{q}}^{\smallR}\big(\Ket{\phi\,Z^{\ell-1}}\big)
\approx
\Ket{\psi\,Z^{\ell-2}} \,.
\end{equation}
More compactly, we can denote this action by $\bar{\mathbf{q}}^{\smallR}\Ket{\phi} \approx\Ket{\psi\,Z^{-}}$, to indicate that the state on the right hand side has been constructed on a vacuum of a length shorter by one with respect to the one on the left hand side. Still, the vacuum is of the form~\eqref{eq:spinchain-gs}, which is why this action preserves~$\mathbf{v}$.%
\footnote{%
The central extension (I) can also be understood in terms of length-changing effects, that however do not preserve~\eqref{eq:spinchain-gs}. In fact, they correspond to adding one site to the left spin chain and removing one to the right one, or \textit{vice versa}. This would force us to consider a much larger set of vacua $\ket{0}=\ket{Z^{L}}\otimes\ket{Z^{\tilde{L}}}$ which has no analogue in the string theory.
This central extension is further discussed in~\cite{Borsato:2012ud}.
} 
It is also convenient to introduce a symbol~$Z^+$ corresponding to the~\emph{insertion} of an additional vacuum site. We can finally write the action of centrally extended algebra on the left representation as
\begin{equation}
\begin{aligned}
&\mathbf{q}^{\smallL}\,\Ket{\phi^{\smallL}}
=a^{\smallL}\,\Ket{\psi^{\smallL}}\,,
\qquad
&&
\bar{\mathbf{q}}^{\smallL}\,\Ket{\psi^{\smallL}}
=\bar{a}^{\smallL}\,\Ket{\phi^{\smallL}}\,,\\
&\bar{\mathbf{q}}^{\smallR}\,\Ket{\phi^{\smallL}}
=\bar{b}^{\smallL}\,\Ket{\psi^{\smallL}Z^{-}}\,,
\qquad
&&
\mathbf{q}^{\smallR}\,\Ket{\psi^{\smallL}}
=b^{\smallL}\,\Ket{\phi^{\smallL}Z^{+}}\,,
\end{aligned}
\end{equation}
and similarly on the right one as
\begin{equation}
\begin{aligned}
&\mathbf{q}^{\smallR}\,\Ket{\phi^{\smallR}}
=a^{\smallR}\,\Ket{\psi^{\smallR}}\,,
\qquad
&&
\bar{\mathbf{q}}^{\smallR}\,\Ket{\psi^{\smallR}}
=\bar{a}^{\smallR}\,\Ket{\phi^{\smallR}}\,,\\
&\bar{\mathbf{q}}^{\smallL}\,\Ket{\phi^{\smallR}}
=\bar{b}^{\smallR}\,\Ket{\psi^{\smallR}Z^{-}}\,,
\qquad
&&
\mathbf{q}^{\smallL}\,\Ket{\psi^{\smallR}}
=b^{\smallR}\,\Ket{\phi^{\smallR}Z^{+}}\,,
\end{aligned}
\end{equation}
This action is very similar to~\eqref{eq:supercharges-1part-matrix}, but the algebra is now realised in terms of a \emph{dynamical spin chain}, where the symmetry generators may add or remove sites. Dynamical effects are a common feature of the spin chains appearing in $\AdS/\CFT$~\cite{Beisert:2003yb,Beisert:2003ys,Beisert:2005tm}, see also ref.~\cite{Rej:2010ju}. In the one-particle representation we can ignore length-changing effects as long as we restrict to \emph{asymptotic states} with~$\ell\to\infty$, which we will always do in this chapter.
Then, the algebra that we constructed closes to a central extension of $\su(1|1)^2$
\begin{equation}
\label{eq:smallalgebra-nonlin3}
\begin{aligned}
&\big\{\mathbf{q}^{\smallL},\bar{\mathbf{q}}^{\smallL}\big\}=\mathbf{h}^{\smallL}\,,
&\qquad &
\big\{\mathbf{q}^{\smallR},\bar{\mathbf{q}}^{\smallR}\big\}=\mathbf{h}^{\smallR}\,,\\
&\big\{\mathbf{q}^{\smallL},{\mathbf{q}}^{\smallR}\big\}=\mathbf{c}\,,
&\qquad &
\big\{\bar{\mathbf{q}}^{\smallL},\bar{\mathbf{q}}^{\smallR}\big\}=\bar{\mathbf{c}}\,,\\
\end{aligned}
\end{equation}
where on the one-particle representation
\begin{equation}
\mathbf{c}= a^{\smallL}b^{\smallL}\mathbf{I}^{\smallL}
+a^{\smallR}b^{\smallR}\mathbf{I}^{\smallR}\,,
\qquad
\bar{\mathbf{c}}= \bar{a}^{\smallL}\bar{b}^{\smallL}\mathbf{I}^{\smallL}
+\bar{a}^{\smallR}\bar{b}^{\smallR}\mathbf{I}^{\smallR}\,.
\end{equation}
Eq.~\eqref{eq:smallalgebra-nonlin3} defines the same algebra we found from analysis of the off-shell symmetries of asymptotic string states~\eqref{eq:smallalgebra-nonlin2}.

\subsection*{Magnons}
Since the supercharges and in particular~$\mathbf{H}$ are momentum-dependent we will consider spin-chain states in which the excitations carry specific momenta---these are the objects that ultimately we will want to scatter. A one-excitation state can then be written as a plane wave%
\footnote{%
We have the possibility of choosing the plane-wave coefficient to be~$e^{\pm ipn}$. Here we pick the negative sign, in contrast with the original choice of ref.~\cite{Borsato:2012ud}, to more easily compare with the string theory results. The two resulting S~matrices are related by a change of basis, as we will discuss in the summary to this chapter.
}
\begin{equation}
\label{eq:onemagnonansatz}
  \ket{\mathcal{X}_p} = \sum_{n=1}^{\ell} e^{-ipn} \ket{ Z^{n-1} \mathcal{X} Z^{\ell-n} } ,
\end{equation}
where $\mathcal{X}$ is any left or right excitation. Of course we can always think of localizing an excitation by constructing an appropriate wave packet.
It is now straightforward to generalise this form to the case of multiple excitations. For two excitations we have~\eg
\begin{equation}
\label{eq:spin2part}
  \ket{\mathcal{X}_p\mathcal{Y}_q} = \sum_{n_1< n_2}^{\ell} e^{-i(pn_1+qn_2)} \ket{ ZZ\cdots ZZ\mathcal{X}ZZ\cdots ZZ\mathcal{Y}ZZ\cdots ZZ },
\end{equation}
with $p>q$ and the excitations sitting at positions~$n_1$ and~$n_2$.
 We restrict to \emph{asymptotic} states, where the spin chain is considered to be very long, $\ell\to\infty$, and the excitations are well separated. The interactions are then described by the spin-chain S~matrix~$\mathcal{S}(p_1,p_2)$ permuting the order of excitations along the chain. 

The \emph{length-changing} action on the spin-chain excitations takes a simple form on the magnons. Adding or removing a vacuum site by~$Z^{\pm}$ in the plane-wave ansatz we get
\begin{equation}
\label{eq:magnon-commutation}
  \ket{Z^\pm \mathcal{X}_p} = e^{\pm ip} \ket{\mathcal{X}_p Z^\pm}\,,
\end{equation}
\ie~communtation with the length-changing effects result in a momentum-dependent phase.
By these relations we can always shift any insertions of $Z^\pm$ through all excitations and collect them at the right end of the state, and since we are dealing with asymptotic states, identify $\ket{\mathcal{X}_p Z^\pm}=\ket{\mathcal{X}_p}$. Using the identification of length-changing effects with phase shifts we will obtain a non-trivial coproduct similar to the one that in the NLSM appeared due to the non-local field~$x_{-}$.

\subsection*{Charge action on multiparticle states}
Suppose now that we want to act with a (super)charge on an asymptotic state containing two magnons, for instance $\ket{\phi_p^{\smallL}\phi_q^{\smallL}}$. How to do this follows from the natural action of the charges in the spin-chain representation, whereby a charge acts separately on every site of the chain as a derivation.
 To take statistics into account, such action should be graded, so that every time a supercharge is anticommuted (from the left, in our convention) past a fermionic  site in the chain, we pick up a minus sign. Therefore, we find \eg
\begin{equation}
\mathbf{q}^{\smallL} \ket{\phi_p^{\smallL}\phi_q^{\smallL}}
=a^{\smallL}_{p}\ket{\psi_p^{\smallL}\phi_q^{\smallL}}
+a^{\smallL}_{q}\ket{\phi_p^{\smallL}\psi_q^{\smallL}}\,,
\end{equation}
where we made explicit the dependence of the representation coefficients on the momentum, while for two fermions it would be
\begin{equation}
\bar{\mathbf{q}}^{\smallL} \ket{\psi_p^{\smallL}\psi_q^{\smallL}}
=\bar{a}^{\smallL}_{p}\ket{\phi_p^{\smallL}\psi_q^{\smallL}}
-\bar{a}^{\smallL}_{q}\ket{\psi_p^{\smallL}\phi_q^{\smallL}}\,.
\end{equation}
It is particularly interesting to look at the action of the supercharges that give rise to the central extension:
\begin{equation}
\label{eq:coproduct-deformed-sc}
\begin{aligned}
&\mathbf{q}^{\smallR}\ket{\psi_p^{\smallL}\psi_q^{\smallL}}
&&=b^{\smallL}_{p}\ket{\phi_p^{\smallL}Z^+\psi_q^{\smallL}}
-b^{\smallL}_{q}\ket{\psi_p^{\smallL}\phi_q^{\smallL}Z^+}\\
&&&=b^{\smallL}_{p}e^{iq} \ket{\phi_p^{\smallL}\psi_q^{\smallL}Z^+}
-b^{\smallL}_{q}\ket{\psi_p^{\smallL}\phi_q^{\smallL}Z^+}\\
&&&\approx b^{\smallL}_{p}e^{iq}\ket{\phi_p^{\smallL}\psi_q^{\smallL}}
-b^{\smallL}_{q}\ket{\psi_p^{\smallL}\phi_q^{\smallL}}\,,
\end{aligned}
\end{equation}
where in the last line we used that we are dealing with asymptotic states. We have that \emph{length-changing effects induce a non-trivial coproduct} on asymptotic multi-magnon states. This will dictate a specific form for the coefficient of the one-particle representation, as well as for the form of the two-particle one.

\subsection*{One-particle representation coefficients}
Let us consider the left-moving representation. We already know that the coefficients of the undeformed algebra, which coincide with ones we evaluated at zero momentum, satisfy
\begin{equation}
|a^{\smallL}_{p=0}|^2=1\,,
\qquad
|b^{\smallL}_{p=0}|^2=0\,.
\end{equation}
The action of $\mathbf{c}$ and $\bar{\mathbf{c}}$ on a one-particle state is given by~$a_p^{\smallL}b_p^{\smallL}$ and~$\bar{a}_p^{\smallL}\bar{b}_p^{\smallL}$ respectively, that  should vanish at~$p=0$, too. The action on the two particle states is more interesting: following the discussion of the previous subsection we get that it must be
\begin{equation}
\label{eq:consistency-2part}
a_p^{\smallL}{b}_p^{\smallL}\,e^{iq}
+ {a}_q^{\smallL}{b}_q^{\smallL}=0
\qquad
\text{if}
\qquad
p+q=0\,,
\end{equation}
Similar equations hold for the right-moving excitations by exchanging~$\bigL\leftrightarrow\bigR$. Finally, we must require that the angular momentum~$\mathbf{m}$ remains quantised for any value of~$p$, which gives
\begin{equation}
|a_p^{\smallL}|^2-|b_p^{\smallL}|^2=+1\,,
\qquad
|b_p^{\smallR}|^2-|a_p^{\smallR}|^2=-1\,,
\end{equation}
on the left and right representations respectively.
Condition~\eqref{eq:consistency-2part} can be solved by setting
\begin{equation}
\label{eq:ab-param}
a_p^{\smallL}b_p^{\smallL}=i\,\frac{h}{2} (e^{ip}-1)
\qquad\Rightarrow
\quad
\mathbf{c}=i\,\frac{h}{2} (e^{ip}-1)\,\mathbf{I}\,,
\end{equation}
and similarly for~$\bar{\mathbf{c}}$.
We therefore find the non-linear form of the central charges from the length-changing effects, rather than from the presence of~$x_{-}$. While in the string-theoretical description the presence of the complex exponential was due to the choice of the light-cone geodesics, here it arises from the plane-wave ansatz  for the magnons.

Together with their conjugates, the parameters~\eqref{eq:ab-param} can be used to fix the dispersion relation. In fact, as in section~\ref{sec:fullrepr} we are dealing with short representations of $\su(1|1)^2_{\text{c.e.}}$ or equivalently~$\psu(1|1)^4_{\text{c.e.}}$, so that we have the shortening condition
\begin{equation}
\mathbf{H}^2=\mathbf{M}^2+\overline{\mathbf{C}}\,\mathbf{C}= \mathbf{I}+\overline{\mathbf{C}}\,\mathbf{C}\,.
\end{equation}
In terms of $a_p$ and~$b_p$ we can find the dispersion relation by taking the positive branch of the square root of the shortening condition:
\begin{equation}
\omega(p)=|a_p^{\smallL}|^2+|b_p^{\smallL}|^2=\sqrt{1+|a_p^{\smallL}|^2|b_p^{\smallL}|^2}=\sqrt{1+4h^2\sin^2\big(\frac{p}{2}\big)}\,.
\end{equation}
Taking into account all of these conditions, we can parametrise the one-particle representation as we did in~\eqref{eq:repr-coeffs}
\begin{equation}
\label{eq:repr-coeffs-spinchain}
\begin{aligned}
&a_{\smallL}=a_{\smallR}=e^{\frac{1}{4}i\,p}\eta_{p}\,,
\qquad
&\bar{a}_{\smallL}=\bar{a}_{\smallR}=e^{-\frac{3}{4}i\,p}\,\eta_p\,,\\
&b_{\smallL}=b_{\smallR}=-\frac{e^{-\frac{3}{4}i\,p}}{x^{-}_p}\eta_{p}\,,
\qquad
&\bar{b}_{\smallL}=\bar{b}_{\smallR}=-\frac{e^{\frac{1}{4}i\,p}}{x^{+}_p}\,\eta_p\,,
\end{aligned}
\end{equation}
where the Zhukovski parameters~$x^{\pm}_p$ and $\eta_p$ are given by eqs.~\eqref{eq:zhukovski-def} and~\eqref{eq:eta-def}. Once again the symmetry between the left and right representations is reflected by having the same choice of parameters.

\subsection*{Two-particle representation}
As we have seen, using the spin-chain picture we can automatically build the two-particle representation on the space of magnons. It is useful to work with matrices, and to this end we introduce a basis~$\mathscr{B}$ which takes the same form of~\eqref{eq:B-basis}, where now the excitations are interpreted as magnons. Then it is easy to check that the charges~$\mathbf{q}^{\smallL,\smallR}$ and their conjugates have the same form as in~\eqref{eq:supercharges-1part-matrix}. What \emph{is different}, however, is the action on the two-particle states. Before we had the expression~\eqref{eq:16x16charges} whereby the coproduct would be deformed by a factor of~$e^{\pm \frac{i}{2}\,p}$ when the charge acts on the second-particle space.  Here, the momentum-dependence by~$e^{\pm i q}$ is on the first-magnon space, and \emph{only on one of the two representations}, as in~\eqref{eq:coproduct-deformed-sc}.

To express this, let us introduce two matrices~$\mathscr{L}_p$ and~$\mathscr{R}_p$ that act on the left and right representation exclusively,
\begin{equation}
\mathscr{L}_p=
e^{ip}\,\mathbf{I}^{\smallL}+\mathbf{I}^{\smallR}\,,
\qquad
\mathscr{R}_p=
\mathbf{I}^{\smallL}+e^{ip}\,\mathbf{I}^{\smallR}\,.
\end{equation} 
Then, on a two-magnon state we have
\begin{equation}
\label{eq:supercharges-2part-sc}
\begin{aligned}
&\left(\mathbf{q}^{\smallL}(p,q)\right)_{(12)}\!\!\!
&=&
\left(\mathscr{R}_{q}^{\phantom{\dagger}}\,\mathbf{q}^{\smallL}(p)\right)\otimes\mathbf{I}
+\Sigma\otimes \mathbf{q}^{\smallL}(q)\,,\\
&\left(\mathbf{q}^{\smallR}(p,q)\right)_{(12)}\!\!\!
&=&
\left(\mathscr{L}_{q}^{\phantom{\dagger}}\,\mathbf{q}^{\smallR}(p)\right)\otimes\mathbf{I}
+\Sigma\otimes \mathbf{q}^{\smallR}(q)\,,\\
&\left(\bar{\mathbf{q}}^{\smallL}(p,q)\right)_{(12)}\!\!\!
&=&
\left(\mathscr{R}_{q}^{\dagger}\,\bar{\mathbf{q}}^{\smallL}(p)\right)\otimes\mathbf{I}
+\Sigma\otimes \bar{\mathbf{q}}^{\smallL}(q)\,,\\
&\left(\bar{\mathbf{q}}^{\smallR}(p,q)\right)_{(12)}\!\!\!
&=&
\left(\mathscr{L}_{q}^{\dagger}\,\bar{\mathbf{q}}^{\smallR}(p)\right)\otimes\mathbf{I}
+\Sigma\otimes \bar{\mathbf{q}}^{\smallR}(q)\,.
\end{aligned}
\end{equation}
Here we made the fermion signs explicit by means of the matrix~$\Sigma$, see eq.~\eqref{eq:Sigma-def}.

\section{The spin-chain S~matrix}
\label{sec:spinchainSmat}
Following our discussion, we are now in a position to derive the two-body spin-chain S~matrix, which should be invariant under the dynamical symmetry algebra that we constructed. As in chapter~\ref{ch:smatrix}, it is easier to first  derive our results for~$\su(1|1)^2_{\text{c.e.}}$.

We have two interpretations for such an S~matrix: on the one hand, we can think of it as an operator~$\mathcal{S}$ that acts on pairs of spin-chain sites, possibly inserting or removing vacuum sites. On the other hand, we reduce its action to the one of a $16\times 16$  matrix~$\mathbf{S}$ (or~$\check{\mathbf{S}}$, in the convention of chapter~\ref{ch:smatrix}) on the vector space $\mathscr{V}_{\text{magn}}(p)\otimes\mathscr{V}_{\text{magn}}(q)$ where now the symmetries have a momentum-dependent coproduct~\eqref{eq:supercharges-2part-sc}. The former condition is useful to re-derive some properties of the spin chain that we first obtained from the ZF algebra, while the latter is more suitable for explicit calculations.

\subsection*{Properties of the S~matrix}
The two-body spin-chain S~matrix has very similar properties to the two-particle QFT S~matrix we investigated in chapter~\ref{ch:smatrix}, and as before they will be useful to explicitly find it, up to the scalar factors.

\subsubsection*{Symmetries}
The crucial ingredient to find the form of the S~matrix is requiring that it respects the~$\su(1|1)^2_{\text{c.e.}}$ (or $\su(1|1)^4_{\text{c.e.}}$) symmetry, \ie
\begin{equation}
\mathcal{S}_{(12)}\,\mathbf{Q}
=\mathbf{Q}\,\mathcal{S}_{(12)}\,.
\end{equation}
This can be written for the magnon S~matrix as
\begin{equation}
\check{\mathbf{S}}_{(12)}(p,q)\;\mathbf{Q}_{(12)}(p,q)
=\mathbf{Q}_{(12)}(q,p)\;\check{\mathbf{S}}_{(12)}(p,q)\,,
\end{equation}
or equivalently
\begin{equation}
{\mathbf{R}}_{(12)}(p,q)\;\mathbf{Q}_{(12)}(p,q)
=\mathbf{Q}_{(21)}(q,p)\;{\mathbf{R}}_{(12)}(p,q)\,,
\end{equation}
with the action of the supercharges on the two-magnon state is given by~\eqref{eq:supercharges-2part-sc}.
Additionally, we will once again require left-right symmetry, \ie\ that elements of the S~matrix that differ only by relabeling~$\bigL\leftrightarrow\bigR$ should be equal.

\subsubsection*{Braiding unitarity and physical unitarity}
The concept of braiding unitarity is quite natural in the spin-chain formalism: before, the S~matrix was exchanging two ZF creation operators, whence~\eqref{eq:braidingunitarity-Scheck} followed by iterating the exchange twice. Here we have that the twofold exchange of two excitations should be inconsequential,
\begin{equation}
\mathcal{S}_{(12)}\,
\mathcal{S}_{(12)}=\mathbf{I}\,,
\end{equation} 
where the subscript indices indicates the spaces where the matrix acts. This is depicted in figure~\ref{fig:unitarity}, and in terms of a matrix formulation reads
\begin{equation}
\check{\mathbf{S}}(p,q)\,\check{\mathbf{S}}(q,p)
=\mathbf{I}
=\mathbf{R}(p,q)\,\mathbf{R}(q,p)\,.
\end{equation}

Physical unitarity 	is once again just a natural reality condition on the scattering elements, which can be simply phrased in terms of a matrix representation of~$\mathcal{S}$,
\begin{equation}
\check{\mathbf{S}}^{\dagger}\;
\check{\mathbf{S}}=
\check{\mathbf{S}}\;
\check{\mathbf{S}}^{\dagger}=
\mathbf{I}=
{\mathbf{R}}^{\dagger}\;
{\mathbf{R}}=
{\mathbf{R}}\;
{\mathbf{R}}^{\dagger}\,.
\end{equation}

\subsubsection*{Yang-Baxter equation and multiparticle scattering}
\begin{figure}
  \centering
  \subfloat[Unitarity]{%
\label{fig:unitarity}
\begin{tikzpicture}
  \begin{scope}[xshift=-0.95cm]
    \coordinate (i1) at (-0.45cm,0);
    \coordinate (i2) at (+0.45cm,0);

    \node (v1) at (0,0.75cm) [S-mat] {$\scriptstyle \mathcal{S}_{(12)}$};

    \coordinate (m1) at (-0.5cm,1.5cm);
    \coordinate (m2) at (+0.5cm,1.5cm);

    \node (v2) at (0,2.25cm) [S-mat] {$\scriptstyle \mathcal{S}_{(12)}$};

    \coordinate (o1) at (-0.45cm,3cm);
    \coordinate (o2) at (+0.45cm,3cm);

    \draw [very thick,red]  [out=90,in=270-45] (i1) to (v1);
    \draw [very thick,blue] [out=90,in=270+45] (i2) to (v1);

    \draw [very thick,violet] [out=90+45,in=270] (v1) to (m1);
    \draw [very thick,olive]  [out=90-45,in=270] (v1) to (m2);

    \draw [very thick,violet] [out=90,in=270-45] (m1) to (v2);
    \draw [very thick,olive]  [out=90,in=270+45] (m2) to (v2);

    \draw [very thick,red]  [out=90+45,in=270] (v2) to (o1);
    \draw [very thick,blue] [out=90-45,in=270] (v2) to (o2);
  \end{scope}
  \node at (0,1.5cm) {$=$};
  \begin{scope}[xshift=+0.95cm]
    \coordinate (i1) at (-0.55cm,0);
    \coordinate (i2) at (+0.3cm,0);

    \coordinate (o1) at (-0.55cm,3cm);
    \coordinate (o2) at (+0.3cm,3cm);

    \draw [very thick,red]  (i1) to (o1);
    \draw [very thick,blue] (i2) to (o2);
  \end{scope}
\end{tikzpicture}
}%
\hspace{2.5cm}
  \subfloat[Yang-Baxter equation]{%
    \label{fig:Yang-Baxter}
\begin{tikzpicture}
  \begin{scope}[xshift=-2.6cm]
    \coordinate (i1) at (-0.9cm,0);
    \coordinate (i2) at (-0,    0);
    \coordinate (i3) at (+0.9cm,0);

    \coordinate (o1) at (-0.9cm,3cm);
    \coordinate (o2) at (-0,    3cm);
    \coordinate (o3) at (+0.9cm,3cm);

    \node (v1) at (-0.45cm,0.75cm) [S-mat] {$\scriptstyle \mathcal{S}_{(12)}$};
    \node (v2) at ( 0.3cm, 1.5cm)    [S-mat] {$\scriptstyle \mathcal{S}_{(23)}$};
    \node (v3) at (-0.45cm,2.25cm) [S-mat] {$\scriptstyle \mathcal{S}_{(12)}$};

    \draw [very thick,red]   [out=90,in=270-45] (i1) to (v1);
    \draw [very thick,blue]  [out=90,in=270+45] (i2) to (v1);
    \draw [very thick,cyan] [out=90,in=270+45] (i3) to (v2);

    \draw [very thick,olive]   [out=90-45,in=270-45] (v1) to (v2);
    \draw [very thick,violet]  [out=90+45,in=270-45] (v1) to (v3);
    \draw [very thick,cyan] [out=90+45,in=270+45] (v2) to (v3);

    \draw [very thick,cyan] [out=90+45,in=270] (v3) to (o1);
    \draw [very thick,blue]  [out=90-45,in=270] (v3) to (o2);
    \draw [very thick,red]   [out=90-45,in=270] (v2) to (o3);
  \end{scope}
  \node at (-1.3cm,1.5cm) {$=$};
  \begin{scope}[xshift=0cm]
    \coordinate (i1) at (-0.9cm,0);
    \coordinate (i2) at (-0,    0);
    \coordinate (i3) at (+0.9cm,0);

    \coordinate (o1) at (-0.9cm,3cm);
    \coordinate (o2) at (-0,    3cm);
    \coordinate (o3) at (+0.9cm,3cm);

    \node (v) at (+0cm,1.5cm) [S-mat] {$\ \;\scriptstyle \mathcal{S}_{(123)}\ $};

    \draw [very thick,red]    [out=90,in=270-45] (i1) to (v);
    \draw [very thick,blue]   [out=90,in=270] (i2) to (v);
    \draw [very thick,cyan] [out=90,in=270+45] (i3) to (v);

    \draw [very thick,cyan] [out=90+45,in=270] (v) to (o1);
    \draw [very thick,blue]  [out=90,in=270] (v) to (o2);
    \draw [very thick,red]   [out=90-45,in=270] (v) to (o3);
  \end{scope}
\node at (1.3cm,1.5cm) {$=$};
  \begin{scope}[xshift=+2.6cm]
    \coordinate (i1) at (-0.9cm,0);
    \coordinate (i2) at (-0,    0);
    \coordinate (i3) at (+0.9cm,0);

    \coordinate (o1) at (-0.9cm,3cm);
    \coordinate (o2) at (-0,    3cm);
    \coordinate (o3) at (+0.9cm,3cm);

    \node (v1) at (+0.45cm,0.75cm) [S-mat] {$\scriptstyle \mathcal{S}_{(23)}$};
    \node (v2) at (-0.3cm, 1.5cm)  [S-mat] {$\scriptstyle \mathcal{S}_{(12)}$};
    \node (v3) at (+0.45cm,2.25cm) [S-mat] {$\scriptstyle \mathcal{S}_{(23)}$};

    \draw [very thick,red]    [out=90,in=270-45] (i1) to (v2);
    \draw [very thick,blue]   [out=90,in=270-45] (i2) to (v1);
    \draw [very thick,cyan] [out=90,in=270+45] (i3) to (v1);

    \draw [very thick,cyan] [out=90+45,in=270+45] (v1) to (v2);
    \draw [very thick,violet]   [out=90-45,in=270+45] (v1) to (v3);
    \draw [very thick,olive]    [out=90-45,in=270-45] (v2) to (v3);

    \draw [very thick,cyan] [out=90+45,in=270] (v2) to (o1);
    \draw [very thick,blue]  [out=90+45,in=270] (v3) to (o2);
    \draw [very thick,red]   [out=90-45,in=270] (v3) to (o3);
  \end{scope}
\end{tikzpicture}%
}
  \caption{Consistency conditions on the spin-chain S~matrix. Unitarity (left~panel) requires that acting twice with $\mathcal{S}_{(12)}$ on a two-particle state gives back the original state. The Yang-Baxter equation (right~panel) resolves the ambiguity in decomposing the scattering of three excitations. Note how, with respect to figure~\ref{fig:YBE-Rmat}, this picture takes into account how~$\mathcal{S}$ permutes the spin-chain excitations.}
  \label{fig:unitarity-YBE}
\end{figure}
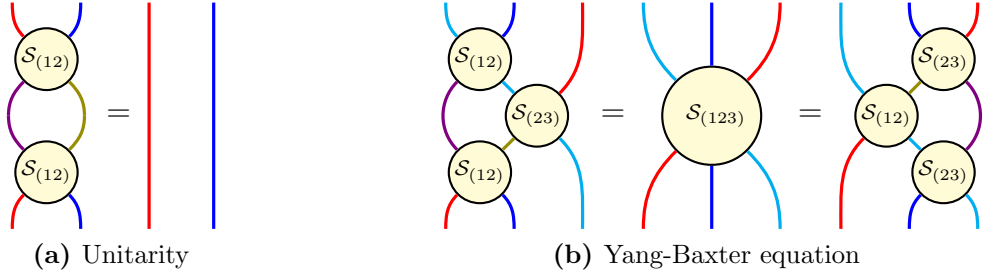

We now want to extend the action of the two-body S~matrix to $M$ sites. Once again this can be done in several inequivalent ways, as illustrated in figure~\ref{fig:Yang-Baxter} in the case of $M=3$. Requiring the equivalence of the two pictures we get again the Yang-Baxter equation
\begin{equation}
\label{eq:YB-spinchain}
\mathcal{S}_{(12)} \, \mathcal{S}_{(23)} \, \mathcal{S}_{(12)}
=
 \mathcal{S}_{(23)} \, \mathcal{S}_{(12)} \, \mathcal{S}_{(23)} .
\end{equation}
We should not forget that the spin-chain S~matrix will naturally feature length-changing processes, so that rewriting the above equation on the space of magnons requires some care. If we can restrict to processes where no length-changing effects arise, it is easy to rewrite the YB equation \eg\ for~$\check{\mathbf{S}}$
\begin{equation}
\label{eq:YB-scheck-sc}
\text{(no length-changing)}\quad
\check{\mathbf{S}}_{qr}\otimes\mathbf{I}
\,\cdot\,
\mathbf{I}\otimes\check{\mathbf{S}}_{pr}
\,\cdot\,
\check{\mathbf{S}}_{pq}\otimes\mathbf{I}
=
\mathbf{I}\otimes\check{\mathbf{S}}_{pq}
\,\cdot\,
\check{\mathbf{S}}_{pr}\otimes\mathbf{I}
\,\cdot\,
\mathbf{I}\otimes\check{\mathbf{S}}_{qr}\,,
\end{equation}
where the subscripts indicate the dependence on the magnon momenta $p,q$ and~$r$. This has the same form as~\eqref{eq:YB-scheck}.

More generally, however, this is not the case. Let us assume that there exists a process where two magnons $\mathcal{X}_p$ and $\mathcal{Y}_q$ scatter giving~$\widetilde{\mathcal{Y}}_q$ and $ \widetilde{\mathcal{X}}_p$, and producing length-changing effects, \ie
\begin{equation}
\label{eq:example-scattering-LC}
 \ket{\mathcal{X}_p \mathcal{Y}_q} \mapsto \#\ket{\widetilde{\mathcal{Y}}_q \widetilde{\mathcal{X}}_p\, Z^{\pm}}\,.
\end{equation}
When we take equation~\eqref{eq:YB-spinchain} and project it on the asymptotic states, we want to write all of the vacuum sites~\emph{to the right} of the excitations. When the process~\eqref{eq:example-scattering-LC} involves the two leftmost magnons, this means that the symbol~$Z^{\pm}$ should be commuted with the last magnon. If this rightmost magnon has momentum~$r$, we have to account for this by writing an explicit factor of~$e^{\pm i r}$ in the Yang-Baxter equation.
Therefore \emph{in presence of length-changing effects the Yang-Baxter equation is twisted} and reads
\begin{equation}
\label{eq:YB-mat-twist}
  \mathbf{I}\otimes\check{\mathbf{S}}_{pq}  \cdot 
  \left(\mathscr{F}_q^{\phantom{\dagger}} \check{\mathbf{S}}_{pr}\mathscr{F}_q^{\dagger}\right) \otimes \mathbf{I} \, \cdot \,
  \mathbf{I}\otimes\check{\mathbf{S}}_{qr}
   =
  \left(\mathscr{F}_p^{\phantom{\dagger}} \check{\mathbf{S}}_{qr}\mathscr{F}_p^{{\dagger}}\right) \otimes \mathbf{I} \, \cdot \,
  \mathbf{I}\otimes\check{\mathbf{S}}_{pr}  \cdot 
  \left(\mathscr{F}_r^{\phantom{\dagger}} \check{\mathbf{S}}_{pq}\mathscr{F}_r^{{\dagger}}\right) \otimes \mathbf{I} ,
\end{equation}
where the matrix $\mathscr{F}$ implements a twist depending on the momentum of the rightmost magnon. Again, once the Yang-Baxter equation is established to resolve the case of~$M=3$ sites, any~$M>3$ case follows. 

\subsection*{S-matrix elements}
Using all of the properties listed above it is easy to find the whole S~matrix up to two scalar factors~$\mathscr{S}_{pq}$ and~$\widetilde{{\mathscr{S}}}_{pq}$. Once again symmetry and unitarity requirements narrow down the solutions to two possibilities: a pure-transmission and a pure-reflection S~matrix, in the sense discussed in section~\ref{sec:stringsmat}. As we saw there, the physical choice is the pure-transmission one. Even if the form (\ie~the non-vanishing elements) of such a matrix is the same as in chapter~\ref{ch:smatrix}, the explicit expressions for the elements differ.
Let us write down those expressions, splitting them in the same-chirality and opposite-chirality sectors.

\subsubsection*{Same-chirality scattering}
Let us consider the scattering of two left magnons. The non-vanishing scattering processes are, as before,
\begin{equation}
\label{eq:SCscattLL}
  \begin{aligned}
    \mathcal{S} \ket{\fixedspaceL{\psi_p^{\smallL} \psi_q^{\smallL}}{\phi_p^{\smallL} \phi_q^{\smallL}}} 
    &= \fixedspaceR{\Del^{\smallLL}_{pq}}{\Ael^{\smallLL}_{pq}} \ket{\fixedspaceL{\psi_p^{\smallL} \psi_q^{\smallL}}{\phi_q^{\smallL} \phi_p^{\smallL}}} , \qquad &
    \mathcal{S} \ket{\fixedspaceL{\psi_p^{\smallL} \psi_q^{\smallL}}{\phi_p^{\smallL} \psi_q^{\smallL}}} 
    &= \fixedspaceR{\Del^{\smallLL}_{pq}}{\Bel^{\smallLL}_{pq}} \ket{\fixedspaceL{\psi_p^{\smallL} \psi_q^{\smallL}}{\psi_q^{\smallL} \phi_p^{\smallL}}} + \Cel^{\smallLL}_{pq} \ket{\fixedspaceL{\psi_p^{\smallL} \psi_q^{\smallL}}{\phi_q^{\smallL} \psi_p^{\smallL}}}, \\
    \mathcal{S} \ket{\fixedspaceL{\psi_p^{\smallL} \psi_q^{\smallL}}{\psi_p^{\smallL} \psi_q^{\smallL}}} &= \fixedspaceR{\Del^{\smallLL}_{pq}}{\Fel^{\smallLL}_{pq}} \ket{\fixedspaceL{\psi_p^{\smallL} \psi_q^{\smallL}}{\psi_q^{\smallL} \psi_p^{\smallL}}} , \qquad &
    \mathcal{S} \ket{\fixedspaceL{\psi_p^{\smallL} \psi_q^{\smallL}}{\psi_p^{\smallL} \phi_q^{\smallL}}} 
    &= \fixedspaceR{\Del^{\smallLL}_{pq}}{\Del^{\smallLL}_{pq}} \ket{\fixedspaceL{\psi_p^{\smallL} \psi_q^{\smallL}}{\phi_q^{\smallL} \psi_p^{\smallL}}} + \Eel^{\smallLL} \ket{\fixedspaceL{\psi_p^{\smallL} \psi_q^{\smallL}}{\psi_q^{\smallL} \phi_p^{\smallL}}}. \\
  \end{aligned}
\end{equation}
The matrix elements now read
\begin{equation}
\label{eq:spinchainSmat-el-LL}
\begin{aligned}
&\Ael^{\smallLL}_{pq}=
\frac{1}{\mathscr{S}^{\smallLL}_{pq}}
\frac{x^{+}_p-x^{-}_q}{x^{-}_p-x^{+}_q}\,,
\qquad 
&&\Bel^{\smallLL}_{pq}=
\frac{1}{\mathscr{S}^{\smallLL}_{pq}}
\frac{x^{-}_p-x^{-}_q}{x^{-}_p-x^{+}_q}\,,
\\
&\Cel^{\smallLL}_{pq}=
\frac{e^{\frac{i}{4}(p-3q)}}{\mathscr{S}^{\smallLL}_{pq}} 
\frac{\tfrac{2i}{h}\,\eta_p\,\eta_q}{x^{-}_p-x^{+}_q}
\,,
\qquad 
&&\Del^{\smallLL}_{pq}=
\frac{1}{\mathscr{S}^{\smallLL}_{pq}}
\frac{x^{+}_p-x^{+}_q}{x^{-}_p-x^{+}_q}\,,
\\
&\Eel^{\smallLL}_{pq}=
\frac{e^{\frac{i}{4}(q-3p)}}{\mathscr{S}^{\smallLL}_{pq}} 
\frac{\tfrac{2i}{h}\,\eta_p\,\eta_q}{x^{-}_p-x^{+}_q}
\,,
\qquad 
&&\Fel^{\smallLL}_{pq}=
\frac{-1}{\mathscr{S}^{\smallLL}_{pq}}\,.
\end{aligned}
\end{equation}
Whether to insert the scalar factor~$\mathscr{S}^{\smallLL}_{pq}$ or its inverse is arbitrary. The present choice has the advantage that it makes the diagonal matrix element~$\Ael^{\smallLL}_{pq}$ precisely equal to \emph{the inverse} of its string-theoretical counterpart, see eq.~\eqref{eq:stringSmatLL}.
Again, in the RR sector we find the same expressions with a scalar factor~$\mathscr{S}^{\smallRR}_{pq}$ and once again LR symmetry implies
\begin{equation}
\mathscr{S}^{\smallLL}_{pq}
=
\mathscr{S}^{\smallRR}_{pq}
=
\mathscr{S}_{pq}\,.
\end{equation}

\subsubsection*{Opposite-chirality scattering}
If we now consider processes of LR chirality we find
\begin{equation}
\label{eq:spincahinSmat-LR}
  \begin{aligned}
    \mathcal{S} \ket{\fixedspaceL{\psi_p^{\smallL}\psi_q^{\smallR}}{\phi_p^{\smallL} \phi_q^{\smallR}}} 
    &= \fixedspaceR{\Ael^{\smallLR}_{pq}}{\Ael^{\smallLR}_{pq}} \ket{\fixedspaceL{\psi_q^{\smallR}\psi_p^{\smallL}}{\phi_q^{\smallR}\phi_p^{\smallL}}} + \fixedspaceR{\Ael^{\smallLR}_{pq}}{\Bel^{\smallLR}_{pq}} \ket{\psi_q^{\smallR} \psi_p^{\smallL}Z^{-}}, \qquad &
    \mathcal{S} \ket{\fixedspaceL{\psi_p^{\smallL}\psi_q^{\smallR}}{\phi_p^{\smallL}\psi_q^{\smallR}}} 
    &= \fixedspaceR{\Ael^{\smallLR}_{pq}}{\Cel^{\smallLR}_{pq}} \ket{\fixedspaceL{\psi_q^{\smallR}\psi_p^{\smallL}}{\psi_q^{\smallR}\phi_p^{\smallL}}}, \\
    \mathcal{S} \ket{\fixedspaceL{\psi_p^{\smallL}\psi_q^{\smallR}}{\psi_p^{\smallL}\psi_q^{\smallR}}} 
    &= \fixedspaceR{\Ael^{\smallLR}_{pq}}{\Eel^{\smallLR}_{pq}} \ket{\fixedspaceL{\psi_q^{\smallR}\psi_p^{\smallL}}{\psi_q^{\smallR}\psi_p^{\smallL}}} + \fixedspaceR{\Ael^{\smallLR}_{pq}}{\Fel^{\smallLR}_{pq}} \ket{\phi_q^{\smallR}\phi_p^{\smallL}Z^{+}}, \qquad &
    \mathcal{S} \ket{\fixedspaceL{\psi_p^{\smallL}\psi_q^{\smallR}}{\psi_p^{\smallL}\phi_q^{\smallR}}} 
    &= \fixedspaceR{\Ael^{\smallLR}_{pq}}{\Del^{\smallLR}_{pq}} \ket{\fixedspaceL{\psi_q^{\smallR}\psi_p^{\smallL}}{\phi_q^{\smallR}\psi_p^{\smallL}}},
\end{aligned}
\end{equation}
while for RL
\begin{equation}
\label{eq:spinchainSmat-LR}
\begin{aligned}
        \mathcal{S} \ket{\fixedspaceL{\psi_p^{\smallR}\psi_q^{\smallL}}{\phi_p^{\smallR} \phi_q^{\smallL}}} 
    &= \fixedspaceR{\Ael^{\smallRL}_{pq}}{\Ael^{\smallRL}_{pq}} \ket{\fixedspaceL{\psi_q^{\smallL}\psi_p^{\smallR}}{\phi_q^{\smallL}\phi_p^{\smallR}}} + \fixedspaceR{\Ael^{\smallRL}_{pq}}{\Bel^{\smallRL}_{pq}} \ket{\psi_q^{\smallL} \psi_p^{\smallR}Z^{-}}, \qquad &
    \mathcal{S} \ket{\fixedspaceL{\psi_p^{\smallR}\psi_q^{\smallL}}{\phi_p^{\smallR}\psi_q^{\smallL}}} 
    &= \fixedspaceR{\Ael^{\smallRL}_{pq}}{\Cel^{\smallRL}_{pq}} \ket{\fixedspaceL{\psi_q^{\smallL}\psi_p^{\smallR}}{\psi_q^{\smallL}\phi_p^{\smallR}}}, \\
    \mathcal{S} \ket{\fixedspaceL{\psi_p^{\smallR}\psi_q^{\smallL}}{\psi_p^{\smallR}\psi_q^{\smallL}}} 
    &= \fixedspaceR{\Ael^{\smallRL}_{pq}}{\Eel^{\smallRL}_{pq}} \ket{\fixedspaceL{\psi_q^{\smallL}\psi_p^{\smallR}}{\psi_q^{\smallL}\psi_p^{\smallR}}} + \fixedspaceR{\Ael^{\smallRL}_{pq}}{\Fel^{\smallRL}_{pq}} \ket{\phi_q^{\smallL}\phi_p^{\smallR}Z^+}, \qquad &
    \mathcal{S} \ket{\fixedspaceL{\psi_p^{\smallR}\psi_q^{\smallL}}{\psi_p^{\smallR}\phi_q^{\smallL}}} 
    &= \fixedspaceR{\Ael^{\smallRL}_{pq}}{\Del^{\smallRL}_{pq}} \ket{\fixedspaceL{\psi_q^{\smallL}\psi_p^{\smallR}}{\phi_q^{\smallL}\psi_p^{\smallR}}}.\\
  \end{aligned}
\end{equation}
Notice that we have highlighted the presence of length-changing effects. They are responsible for the presence of a twist~$\mathscr{F}$ in multi-magnon scattering events, and notably in the Yang-Baxter equation~\eqref{eq:YB-spinchain}. In fact, from~\eqref{eq:spincahinSmat-LR} we can write the explicit form of the matrix~$\mathscr{F}_p$ as
\begin{equation}
\label{eq:Ftwist}
\mathscr{F}_p=\mathscr{U}_p\otimes\mathscr{U}_p\,,
\end{equation}
where~$\mathscr{U}_p$ is a diagonal matrix
\begin{equation}
\label{eq:Utwist}
\mathscr{U}_p=\text{diag}\left(
e^{\frac{i}{2}p},\,1,\,e^{\frac{i}{2}p},\,1
\right).
\end{equation}
We then find that the twisted YB equation is satisfied, upon using the form of the matrix elements~\eqref{eq:spinchainSmat-el-LL} and
\begin{equation}
\begin{aligned}
&\Ael^{\smallLR}_{pq}=
\frac{e^{-iq}}{\mathscr{S}^{\smallLR}_{pq}} 
\frac{1-x^{-}_px^{+}_q}{1-x^{-}_px^{-}_q}\,,
\qquad 
&&\Bel^{\smallLR}_{pq}=
\frac{e^{-\frac{3i}{4}(p+q)}}{\mathscr{S}^{\smallLR}_{pq}}
\frac{-\tfrac{2i}{h}\,\eta_p\,\eta_q}{1-x^{-}_px^{-}_q}\,,
\\
&\Cel^{\smallLR}_{pq}=
\frac{e^{-i(p+q)}}{\mathscr{S}^{\smallLR}_{pq}} 
\frac{1-x^{+}_px^{+}_q}{1-x^{-}_px^{-}_q}\,,
\qquad 
&&\Del^{\smallLR}_{pq}=
\frac{1}{\mathscr{S}^{\smallLR}_{pq}}\,,
\\
&\Eel^{\smallLR}_{pq}=
\frac{-e^{ip}}{\mathscr{S}^{\smallLR}_{pq}} 
\frac{1-x^{+}_px^{-}_q}{1-x^{-}_px^{-}_q}\,,
\qquad 
&&\Fel^{\smallLR}_{pq}=
\frac{e^{-\frac{3i}{4}(p+q)}}{\mathscr{S}^{\smallLR}_{pq}}
\frac{\tfrac{2i}{h}\,\eta_p\,\eta_q}{1-x^{-}_px^{-}_q}\,,
\end{aligned}
\end{equation}
with the remaining ones following by left-right symmetry.

As before, we can use a single scalar factor for the LR and RL sectors by the definition
\begin{equation}
\begin{aligned}
\mathscr{S}^{\smallLR}_{pq}=
&\,
\widetilde{\mathscr{S}}_{pq}
e^{-\frac{i}{2}(p+q)}\left(\frac{1-x^{+}_px^{+}_q}{1-x^{-}_px^{-}_q}\right)^{+1/2},\\
\qquad
\mathscr{S}^{\smallRL}_{pq}=
&\,
\widetilde{\mathscr{S}}_{pq}
e^{+\frac{i}{2}(p+q)}\left(\frac{1-x^{+}_px^{+}_q}{1-x^{-}_px^{-}_q}\right)^{-1/2},
\end{aligned}
\end{equation}
After which LR symmetry becomes manifest upon relabeling~$\bigL\leftrightarrow\bigR$.

\section{Comparing with the worlsheet S~matrix}
The S-matrix elements we just found differ from the ones of chapter~\ref{ch:smatrix}. Firstly, they seem to be related to the~\emph{inverse} scattering processes. This can be explained by our ansatz~\eqref{eq:onemagnonansatz} for the magnon wave~functions. The choice of a negative sign in the phase factor is non-standard~\cite{Gaudin} and leads to a two-magnon scattering wave~function of the form
\begin{equation}
\label{eq:spin2partScatt}
  \Psi(n_1,n_2) = e^{-i(p n_1 + q n_2)} + S(p,q)\, e^{-i(p n_2 + q n_1)} ,
\end{equation}
with $p>q$.
We can relate this to the standard form~\eqref{eq:2part-wavefunction-QFT} by flipping the sign of~$p$ and~$q$ which however reverse their order, yielding~$S(q,p)\approx S(p,q)^{-1}$.

Still, the~$\check{\mathbf{S}}$ that we found in the previous section cannot be just the inverse of the worldsheet one, as the latter satisfies Yang-Baxter equation rather than its twisted version~\eqref{eq:YB-mat-twist}. This discrepancy is due to the form of the spin-chain coproduct~\eqref{eq:supercharges-2part-sc}, as it was well understood already in the case of~$\AdS_5/\CFT_4$~\cite{Arutyunov:2006yd}.

\subsection*{From the spin-chain to the worldsheet coproduct}
Any coproduct can be modified by a non-local, momentum dependent change of basis in the two-excitation space~\cite{Plefka:2006ze, Arutyunov:2006yd,Torrielli:2010kq, Torrielli:2011gg}. As it was shown in ref.~\cite{Arutyunov:2006yd}, such transformations appear naturally from changes of basis of ZF algebra operators. In the case of the spin chain coproduct, let us consider the following change of basis on~$\mathscr{V}_{\text{magn}}(p)\otimes\mathscr{V}_{\text{magn}}(q)$, acting on the charges as
\begin{equation}
\label{eq:Utransform-charges}
\mathbf{Q}_{(12)}(p,q)\to
\mathscr{U}_q^{\dagger}\otimes \mathbf{I}
\cdot \mathbf{Q}_{(12)}(p,q) \cdot
\mathscr{U}_q^{\phantom{\dagger}}\otimes \mathbf{I}\,,
\end{equation}
and, in the case of the~$\su(1|1)^2_{\text{c.e.}}$ spin-chain S~matrix, let us take $\mathscr{U}_p$ as in eq.~\eqref{eq:Utwist}.
We then find a new form for the two-particle supercharges
\begin{equation}
\label{eq:charges-frame-redef}
\begin{aligned}
&\mathscr{U}_q^{\dagger}\otimes \mathbf{I}\cdot
\left(\mathbf{q}^{\smallL,\smallR}(p,q)\right)_{(12)}
\cdot\mathscr{U}_q^{\phantom{\dagger}}\otimes \mathbf{I}\!\!\!
&=&\;
\mathbf{q}^{\smallL,\smallR}(p) \otimes\mathbf{I}\,e^{+\frac{i}{2}q}
+\Sigma\otimes \mathbf{q}^{\smallL,\smallR}(q)\,,\\
&
\mathscr{U}_q^{\dagger}\otimes \mathbf{I}\cdot
\left(\bar{\mathbf{q}}^{\smallL,\smallR}(p,q)\right)_{(12)}
\cdot\mathscr{U}_q^{\phantom{\dagger}}\otimes \mathbf{I}\!\!\!
&=&\;
\bar{\mathbf{q}}^{\smallL,\smallR}(p)\otimes\mathbf{I}\,e^{-\frac{i}{2}q}
+\Sigma\otimes \bar{\mathbf{q}}^{\smallL,\smallR}(q)\,,
\end{aligned}
\end{equation}
which is now identical for L and R supercharges, but still different from the worldsheet one~\ref{eq:coproduct-deformed}. The transformation~\eqref{eq:Utransform-charges} induces a change on the~$\mathbf{R}$ and~$\check{\mathbf{S}}$ matrices,
\begin{equation}
\begin{aligned}
&\mathbf{R}(p,q)\to
\mathscr{U}_q^{\dagger}\otimes \mathbf{I}
\cdot \mathbf{R}(p,q) \cdot
\mathscr{U}_q^{\phantom{\dagger}}\otimes \mathbf{I}\,,\\
&\fixedspaceR{\mathbf{R}}{\check{\mathbf{S}}}
(p,q)\to
\mathscr{U}_p^{\dagger}\otimes \mathbf{I}
\cdot \fixedspaceR{\mathbf{R}}{\check{\mathbf{S}}}(p,q) \cdot
\mathscr{U}_q^{\phantom{\dagger}}\otimes \mathbf{I}\,,
\end{aligned}
\end{equation}
and the resulting matrices are precisely the ones that we would have found from the invariances
\begin{equation}
\label{eq:smatrix-frame-redef}
\begin{aligned}
&{\mathbf{R}}(p,q)\,\mathbf{Q}_{(12)}(p,q)
=\mathbf{Q}_{(21)}(q,p)\,{\mathbf{R}}(p,q)\,,\\
&\fixedspaceR{\mathbf{R}}{\check{\mathbf{S}}}(p,q)
\,\mathbf{Q}_{(12)}(p,q)
=\mathbf{Q}_{(12)}(q,p)\,
\fixedspaceR{\mathbf{R}}{\check{\mathbf{S}}}(p,q)\,,
\end{aligned}
\end{equation}
had we considered the charges on the right hand side of~\eqref{eq:charges-frame-redef}.
In fact, those transformed scattering matrices would obey the usual untwisted Yang-Baxter equation. To see this one can plug the right hand side of~\eqref{eq:smatrix-frame-redef} in the twisted Yang-Baxter equation~\eqref{eq:YB-mat-twist} and use the fact that the twist matrix~$\mathscr{F}_p$ is precisely given by~\eqref{eq:Ftwist}.

It is worth exploring further the symmetry invariance condition in the new frame, which explicitly takes the form
\begin{equation}
\begin{aligned}
&\mathscr{U}_q^{\dagger}\otimes \mathbf{I}
\cdot \mathbf{R}_{pq} \cdot
\mathscr{U}_q^{\phantom{\dagger}}\otimes \mathbf{I}
\cdot\left(\mathbf{q}^{\smallL,\smallR}(p) \otimes\mathbf{I}\,e^{\frac{i}{2}q}
+\Sigma\otimes \mathbf{q}^{\smallL,\smallR}(q)\right)\\
&\qquad\qquad
=\left(\mathbf{q}^{\smallL,\smallR}(q) \otimes\Sigma+
e^{\frac{i}{2}q}\,\mathbf{I}\otimes \mathbf{q}^{\smallL,\smallR}(p)\right)
\cdot\mathscr{U}_q^{\dagger}\otimes \mathbf{I}
\cdot \mathbf{R}_{pq} \cdot
\mathscr{U}_q^{\phantom{\dagger}}\otimes \mathbf{I}\,,
\end{aligned}
\end{equation}
where we used the coproduct appearing in the right hand side of~\eqref{eq:charges-frame-redef}, and a similar equation holds for the conjugates~$\bar{\mathbf{q}}^{\smallL,\smallR}$. Using the graded permutation matrix~$\Pi^g$ and introducing the short-hand notation
\begin{equation}
\widetilde{\mathbf{R}}_{pq}=
\Pi^g\cdot
\mathscr{U}_q^{\dagger}\otimes \mathbf{I}
\cdot \mathbf{R}_{pq} \cdot
\mathscr{U}_q^{\phantom{\dagger}}\otimes \mathbf{I}\cdot\Pi^g\,,
\end{equation}
we can finally write%
\footnote{%
We use that $(\Pi^g)^2=\mathbf{I}$ and that for any supercharge~$\mathbf{Q}$ we have $\Pi^g\cdot \mathbf{I}\otimes \mathbf{Q}\cdot \Pi^g=\mathbf{Q}\otimes\Sigma$ and $\Pi^g\cdot \Sigma\otimes \mathbf{Q}\cdot \Pi^g=\mathbf{Q}\otimes\mathbf{I}$.
}
\begin{equation}
\widetilde{\mathbf{R}}_{pq} 
\cdot\left(\mathbf{q}^{\smallL,\smallR}(q) \otimes\mathbf{I}\
+e^{\frac{i}{2}q}\,\Sigma\otimes \mathbf{q}^{\smallL,\smallR}(p)\right)
=\left(\mathbf{q}^{\smallL,\smallR}(p) \otimes\Sigma\,e^{\frac{i}{2}q}+
\,\mathbf{I}\otimes \mathbf{q}^{\smallL,\smallR}(q)\right)
\cdot \widetilde{\mathbf{R}}_{pq} \,.
\end{equation}
Comparing this with the QFT invariance property of~$\mathbf{R}$~\eqref{eq:R-matrix-invariance}, we see that they coincide upon swapping the momenta~$p\leftrightarrow q$. We have therefore established the relation between the worldsheet and spin-chain (magnon) S~matrices,
\begin{equation}
\label{eq:Smatrix-comparison}
\mathbf{R}_{pq}\Big|_{\text{worldsheet}}=\Pi^g\cdot
\mathscr{U}_p^{\dagger}\otimes \mathbf{I}
\cdot \mathbf{R}_{qp} \cdot
\mathscr{U}_p^{\phantom{\dagger}}\otimes \mathbf{I}\cdot\Pi^g\Big|_{\text{spin chain}},
\end{equation}
which confirms that the S~matrix computed in this chapter and the one of chapter~\ref{ch:smatrix} describe scattering processes ``inverse'' to each other.
It is easy to explicitly check that this identity holds using the explicit form of the S~matrix elements given here and in the previous chapters.

\subsection*{Scalar factors and crossing}
We have established that the matrix part of~$\mathbf{R}_{pq}$ can be computed equivalently in a spin-chain or worldsheet picture. However, in our previous discussion we have also exploited crossing invariance of the scalar factors to constrain their form. Crossing symmetry is seemingly a purely field theoretical property, and is not obvious that anything similar can be imposed purely in a spin chain picture.%
\footnote{%
However, once the spin-chain coproduct~\eqref{eq:supercharges-2part-sc} has been established, this could be used to define a Hopf algebra structure where the notion of crossing symmetry is naturally related to the antipode operation, see appendix~B in ref.~\cite{Borsato:2013qpa}.
}

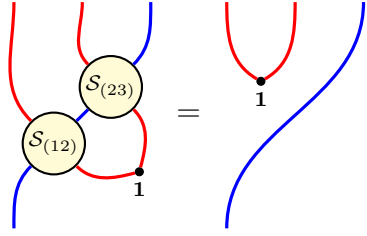
\begin{figure}
  \centering
  \begin{tikzpicture}
    \begin{scope}[xshift=-1.4cm]
      \coordinate (i1) at (-0.9cm,0);

      \coordinate (o1) at (-0.9cm,3cm);
      \coordinate (o2) at (-0,    3cm);
      \coordinate (o3) at (+0.9cm,3cm);

      \node (v1) at (-0.375cm,1.125cm) [S-mat] {$\scriptstyle \mathcal{S}_{(12)}$};
      \node (v2) at (+0.375cm,1.875cm) [S-mat] {$\scriptstyle \mathcal{S}_{(23)}$};

      \node (v3) at ($(0,1.5cm)+(0.75cm,-0.75cm)$)  [circle,draw=black,fill=black,minimum size=3pt,inner sep=0] {};

      \draw [very thick,blue] [out=90,in=270-45] (i1) to (v1);
      \draw [very thick,blue] (v1) to (v2);
      \draw [very thick,blue] [out=90-45,in=270] (v2) to (o3);

      \draw [very thick,red] [out=270,in=135] (o1) to (v1);
      \draw [very thick,red] [out=135,in=270] (v2) to (o2);

      \draw [very thick,red] [out=270+45,in=180+45-30] (v1) to (v3);
      \draw [very thick,red] [out=0+45+30,in=315] (v3) to (v2);

      \node at (v3) [anchor=north] {$\scriptstyle \mathbf{1}$};
    \end{scope}
    \node at (0,1.5cm) {$=$};
    \begin{scope}[xshift=+1.4cm]
      \coordinate (i1) at (-0.9cm,0);

      \coordinate (o1) at (-0.9cm,3cm);
      \coordinate (o2) at (-0,    3cm);
      \coordinate (o3) at (+0.9cm,3cm);

      \node (v) at (-0.45cm,1.95cm) [circle,draw=black,fill=black,minimum size=3pt,inner sep=0] {};

      \draw [very thick,blue] [out=90,in=270] (i1) to (o3);
      \draw [very thick,red] [out=270,in=90+60] (o1) to (v);
      \draw [very thick,red] [out=90-60,in=270] (v) to (o2);

      \node at (v) [anchor=north] {$\scriptstyle \mathbf{1}$};
    \end{scope}
  \end{tikzpicture}

  \caption{The scattering of a fundamental excitation with a singlet is trivial. We depict the singlet as composed of two excitation, one of which moves ``backward'' due to analytical continuation to the crossed region.}
  \label{fig:singlet}
\end{figure}

An ingenious observation due to Beisert~\cite{Beisert:2005tm}, see also ref.~\cite{Vieira:2010kb}, shows that this is actually the case. Suppose that we can construct a two-magnon singlet state~$\Ket{\mathbf{1}_{q_1q_2}}$ that is annihilated by all of the supercharges, \ie\ in our case a singlet satisfying
\begin{equation}
\mathbf{q}^{\smallL,\smallR}\Ket{\mathbf{1}_{q_1q_2}}=0\,,
\qquad
\bar{\mathbf{q}}^{\smallL,\smallR}\Ket{\mathbf{1}_{q_1q_2}}=0\,.
\end{equation}
By $\su(1|1)^2_{\text{c.e.}}$ symmetry, scattering any magnon $\ket{\mathcal{X}_{p}}$ with it should have no consequences. On the other hand, we can think of scattering \emph{its constituent magnons} separately with $\ket{\mathcal{X}_{p}}$, as illustrated in figure~\ref{fig:singlet}, which is reminiscent of the bootstrap condition, see figure~\ref{fig:boundstates}.
This implies that the product of certain pairs of S-matrix elements should give one, when they are evaluated at momenta $p,q_1$ and $p,q_2$. This is a constraint on the scalar factors. Moreover, since the state $\Ket{\mathbf{1}_{q_1q_2}}$ has zero momentum and energy, it must be
\begin{equation}
q_1=-q_2\,,
\qquad
\omega(q_1)=-\omega(q_2)\,.
\end{equation}
The only non-trivial solutions to these equations imply that one of the two momenta has been continued to the crossed region, \ie
\begin{equation}
q_1=q_2^{\mathsf{c}}
\qquad\text{or}\qquad
q_2=q_1^{\mathsf{c}}\,.
\end{equation}

Focusing for the moment on the latter choice, in the case of~$\psu(1|1)^2_\text{c.e.}$ we have that the singlet takes the form
\begin{equation}
\begin{aligned}
\Ket{\mathbf{1}_{qq^{\mathsf{c}}}}
&=
\xi^{\smallLR}\Ket{\mathbf{1}^{\smallLR}_{qq^{\mathsf{c}}}}+
\xi^{\smallRL}\Ket{\mathbf{1}^{\smallRL}_{qq^{\mathsf{c}}}}\\
&=
\xi^{\smallLR}\left(e^{-\frac{i}{2}q}\Ket{\phi_{q}^{\smallL}\phi_{q^{\mathsf{c}}}^{\smallR}}
+i \Ket{\psi_{q}^{\smallL}\psi_{q^{\mathsf{c}}}^{\smallR}}\right)
+
\xi^{\smallRL}\left(e^{-\frac{i}{2}q}\Ket{\phi_{q}^{\smallR}\phi_{q^{\mathsf{c}}}^{\smallL}}
+i \Ket{\psi_{q}^{\smallR}\psi_{q^{\mathsf{c}}}^{\smallL}}\right)
.
\end{aligned}
\end{equation}
The fact that the singlet couples the L and R representations is not surprising, since in particular it should be annihilated by~$\mathbf{m}$. This whole discussion is more and more reminiscent of the one of crossing of section~\ref{sec:crossing}. In fact, by an argument of Arutyunov and Frolov~\cite{Arutyunov:2009ga} it is not hard to see that the two are equivalent. If we remove the length-changing effects we have that
\begin{equation}
\Ket{\mathbf{1}_{qq^{\mathsf{c}}}}\to
\mathscr{U}_{q^{\mathsf{c}}}^{\dagger}\,
\Ket{\mathbf{1}_{qq^{\mathsf{c}}}}
= \mathbf{A}^{\dagger}(q)\;\mathscr{C}\;\mathbf{A}^{\dagger\,\text{t}}(q^{\mathsf{c}})\,,
\end{equation}
where the last expression in terms of the ZF creation operation are contracted to form a scalar%
\footnote{
Recall that $\mathbf{A}^{\dagger}$ is a row vector so that~$\mathbf{A}^{\dagger\,\text{t}}$ is a column vector.
}.
Then triviality of scattering with $\Ket{\mathcal{X}_p}$ amounts to the statement
\begin{equation}
\label{eq:singlet-scatt}
\mathbf{A}_{(1)}^{\dagger}(p)
\Big(\mathbf{A}_{(2)}^{\dagger}(q)\;\mathscr{C}_{(2)}\;\mathbf{A}_{(2)}^{\dagger\,\text{t}}(q^{\mathsf{c}})\Big)
=\left(\mathbf{A}_{(2)}^{\dagger}(q)\;\mathscr{C}_{(2)}\;\mathbf{A}_{(2)}^{\dagger\,\text{t}}(q^{\mathsf{c}})\right)\mathbf{A}_{(1)}^{\dagger}(p)\,.
\end{equation}
It is straightforward to use the ZF algebra relations to find
\begin{equation}
\begin{aligned}
&\mathbf{A}_{(1)}^{\dagger}(p)
\;\mathbf{A}_{(2)}^{\dagger}(q)\;\mathscr{C}_{(2)}\;\mathbf{A}_{(2)}^{\dagger\,\text{t}}(q^{\mathsf{c}})\\
&\qquad
=\mathbf{A}_{(2)}^{\dagger}(q)\;\mathbf{A}_{(1)}^{\dagger}(p)
\;\mathbf{R}_{(12)}(p,q)\;\mathscr{C}_{(2)}\;\mathbf{A}_{(2)}^{\dagger\,\text{t}}(q^{\mathsf{c}})\\
&\qquad
=\mathbf{A}_{(2)}^{\dagger}(q)
\Big(\mathbf{A}_{(1)}^{\dagger}(p)
\;\mathbf{A}_{(2)}^{\dagger}(q^{\mathsf{c}})\;
\mathscr{C}_{(2)}\; \mathbf{R}_{(12)}^{\text{t}_2}(p,q)\Big)^{\text{t}_2}\\
&\qquad
=\mathbf{A}_{(2)}^{\dagger}(q)
\Big(\mathbf{A}_{(2)}^{\dagger}(q^{\mathsf{c}})\;\mathbf{A}_{(1)}^{\dagger}(p)
\;\mathbf{R}_{(12)}(p,q^{\mathsf{c}})\;
\mathscr{C}_{(2)}^{\text{t}_2}\; \mathbf{R}_{(12)}^{\text{t}_2}(p,q)\Big)^{\text{t}_2}\\
&\qquad
=\mathbf{A}_{(2)}^{\dagger}(q)\;
\Big(\mathbf{R}_{(12)}(p,q)\;\mathscr{C}_{(2)}\;
\mathbf{R}_{(12)}^{\text{t}_2}(p,q^{\mathsf{c}})\Big)
\mathbf{A}_{(2)}^{\dagger\,\text{t}}(q^{\mathsf{c}})\;\mathbf{A}_{(1)}^{\dagger}(p)\,.
\end{aligned}
\end{equation}
The expression in the last line equals the right hand side of~\eqref{eq:singlet-scatt} if 
\begin{equation}
\mathbf{R}_{(12)}(p,q)\;\mathscr{C}_{(2)}\;
\mathbf{R}_{(12)}^{\text{t}_2}(p,q^{\mathsf{c}})=\mathscr{C}_{(2)}\,.
\end{equation}
This is precisely equivalent%
\footnote{
It may appear strange that, while matching the S~matrices also required exchanging $p\leftrightarrow q$ and involved graded permutations, none of this is necessary to reproduce the crossing equation. This is simply due to the fact that crossing invariance must hold (or, can be imposed) for~$\mathbf{R}$, its inverse,  and its graded permutation.
}
 to the crossing equations~\eqref{eq:crossing2var}. If we instead chose to set~$q_1=q_2^{\mathsf{c}}$ we can derive the crossing equation in the first variable instead. In that case we have
\begin{equation}
\begin{aligned}
\Ket{\mathbf{1}_{q^{\mathsf{c}}q}}
&=
\xi^{\smallLR}\Ket{\mathbf{1}^{\smallLR}_{q^{\mathsf{c}}q}}+
\xi^{\smallRL}\Ket{\mathbf{1}^{\smallRL}_{q^{\mathsf{c}}q}}\\
&=
\xi^{\smallLR}\left(e^{\frac{i}{2}q}\Ket{\phi_{q^{\mathsf{c}}}^{\smallL}\phi_{q}^{\smallR}}
-i \Ket{\psi_{q^{\mathsf{c}}}^{\smallL}\psi_{q}^{\smallR}}\right)
+
\xi^{\smallRL}\left(e^{\frac{i}{2}q}\Ket{\phi_{q^{\mathsf{c}}}^{\smallR}\phi_{q}^{\smallL}}
-i \Ket{\psi_{q^{\mathsf{c}}}^{\smallR}\psi_{q}^{\smallL}}\right)
,
\end{aligned}
\end{equation}
which we can rewrite as
\begin{equation}
\Ket{\mathbf{1}_{q^{\mathsf{c}}q}}\to
\mathscr{U}_{q}^{\dagger}\,
\Ket{\mathbf{1}_{q^{\mathsf{c}}q}}
= \mathbf{A}^{\dagger}(q^{\mathsf{c}})\;\mathscr{C}^{\dagger}\;\mathbf{A}^{\dagger\,\text{t}}(q)\,,
\end{equation}
which leads to a similar discussion as above and reproduces~\eqref{eq:crossing1var}.
 
Finally, to further confirm our derivation, it is easy to check explicitly that imposing
\begin{equation}
\mathcal{S}_{(23)}\;\mathcal{S}_{(12)}
\Ket{\mathcal{X}_{p}\,\mathbf{1}_{qq^{\textsf{c}}}}=
\Ket{\mathbf{1}_{qq^{\textsf{c}}}\,\mathcal{X}_{p}}\,,
\qquad
\mathcal{S}_{(23)}\;\mathcal{S}_{(12)}
\Ket{\mathcal{X}_{p}\,\mathbf{1}_{q^{\textsf{c}}q}}=
\Ket{\mathbf{1}_{q^{\textsf{c}}q}\,\mathcal{X}_{p}}\,,
\end{equation}
results precisely in the equations~\eqref{eq:crossing12} for the dressing factors. As always, all of this can be straightforwardly extended to the~$\psu(1|1)^4_{\text{c.e.}}$ S~matrix, see also ref.~\cite{Borsato:2013qpa}.

\section{Chapter summary}
In this chapter we have seen how the all-loop S~matrix and dispersion relation of chapter~\ref{ch:smatrix} can be equivalently found from a spin-chain picture.
 
Our derivation built on the weakly-coupled spin chain description of ref.~\cite{OhlssonSax:2011ms}. For $h\ll1$, we can think of \emph{two} spin~chains, one containing left-moving excitations transforming under~$\psu(1,1|2)_{\L}$, and the other containing right-moving ones transforming under~$\psu(1,1|2)_{\R}$, as in figure~\ref{fig:symmspinchain}. We have then shown how from the~$\psu(1,1|2)^2$ representations it is possible to construct a $\tfrac{1}{2} $-supersymmetric vacuum, and excitations above it that transform under a~$\psu(1|1)^4$ residual symmetry. In fact, this weak-coupling symmetry is precisely equal to the on-shell symmetry of worldsheet excitations.

Moreover, we have extended our construction to arbitrary values of the coupling. This results in the excitations of a given chirality being charged under both left and right supercharges, together with a central extension of the residual symmetries and a \emph{dynamical} spin chain, where the number of sites can fluctuate. Using these symmetries, supplemented by a discrete left-right one, we have been able to determine the two-magnon S~matrix, which in fact satisfies an analogue of the Yang-Baxter equation and can be used to construct many-magnon S~matrices in a consistent way.

The construction, including the form of the~$\psu(1,1)^4_{\text{c.e.}}$ symmetry, was strongly reminiscent of what we did in the worldsheet theory in chapter~\ref{ch:smatrix}. Still, an apparent difference was in the form of the coproduct~\eqref{eq:supercharges-2part-sc} that defines the action of the symmetries on several spin-chain sites. We have shown explicitly that, up to a change of the two-particle basis, such a difference can be reabsorbed and in fact the worldsheet and spin-chain S~matrices are equivalent~\eqref{eq:Smatrix-comparison}. This was expected  based on what happens in~$\AdS_5/\CFT_4$, see ref.~\cite{Arutyunov:2006yd}.
The final ingredient to put the spin-chain and worldsheet pictures on the same footing was crossing symmetry, which would appear to be a genuinely field-theoretical feature. We show that an equivalent notion holds for our spin-chain, similarly to was argued by Beisert in ref.~\cite{Beisert:2005tm}.

Let us conclude by mentioning that the notations used here for the spin-chain symmetry algebra and S~matrix differ from the ones of our original works~\cite{Borsato:2012ud,Borsato:2013qpa}. The present choice was meant to further emphasise the similarities with the worldsheet analysis, in particular by taking the very same form of the supercharges on the one-magnon representation as it was for the one-particle representation in chapter~\ref{ch:smatrix}.
Relating this choice with~\cite{Borsato:2012ud,Borsato:2013qpa} amounts to another change of the two-particle basis, see also ref.~\cite{Arutyunov:2006yd}.

\chapter{The all-loop Bethe ansatz equations}
\label{ch:betheansatz}
Once the two-body S~matrix and the dispersion relation are known, it only remains to impose that the spatial dimension of the worldsheet, or equivalently the spin~chain, are given by a circle of finite length~$\ell$. This results in the Bethe-Yang equations for the QFT, or in the asymptotic%
\footnote{%
As we mentioned in the introduction, these equations are asymptotic because they ignore wrapping effects. With this \textit{caveat} in mind, we will often interchangeably refer to (asymptotic) Bethe ansatz and Bethe-Yang equations when comparing the spin-chain and worldsheet approaches.
}
 Bethe ansatz equations (BAE) for the spin chain. As we will argue, the equations are equivalent in the two frameworks, so that the two theories will have the same spectrum of momenta and, given that the dispersion relation is the same, of energy.
There are several ways of deriving the BAE. Here we will focus on what is perhaps the most intuitive way from the physical point of view, that is the coordinate Bethe ansatz. We will work it out for both the spin-chain and the worldsheet picture.

\section{Bethe ansatz essentials}
Before working out the BAE for S~matrix that we found in the previous chapters, let us illustrate the idea behind the Bethe ansatz on the simplest possible example, and postpone the more complicated cases to the next sections.
We consider here the $\su(1|1)^2_{\text{c.e.}}$ S~matrix and \emph{restrict to a single type of excitation}, \eg~$\mathcal{X}=\phi^{\smallL}$. Even if this truncation violates crossing symmetry, it is consistent from the scattering point of view, since~$\phi^{\smallL}_p$ scatters diagonally with itself. Then, the S~matrix reduces to a number. 

\subsection*{Imposing periodicity}
The spin-chain picture is perhaps the easiest to visualise. 
Let us therefore start by considering the $M$-magnon asymptotic wave~function for an integrable theory. Consider an asymptotic state
\begin{equation}
\label{eq:BA-spinchain-state}
\Ket{\mathcal{X}_{p_1}\cdots\mathcal{X}_{p_M}}
=
\sum_{n_1\ll \cdots \ll n_M}
e^{-i (p_1 n_1+\dots + p_M n_M)}
\ket{Z\cdots Z\mathcal{X}_{(n_1)}Z\cdots Z\mathcal{X}_{(n_M)}Z\cdots}\,,
\end{equation}
with $p_1>\cdots>p_M$, which is a natural generalisation of the two-magnon state~\eqref{eq:spin2part}. By the same reasoning that gave us factorised scattering in chapter~\ref{ch:smatrix}, it is clear now that the magnons will undergo pairwise scattering along the one-dimensional chain. For instance, if the first and second magnon scatter, we will get to an ansatz where the magnon with momentum $p_1$ is in the region indexed by $n_2$, and \textit{vice versa}. In general, the multimagnon wave function will be a combination
\begin{equation}
\label{eq:asympt-wavefunct}
\Ket{\Psi(p_1,\dots, p_M)}=\sum_{\pi\in \text{S}_M} \chi(p_{\pi(1)},\dots p_{\pi(M)}) 
\Ket{\mathcal{X}_{p_{\pi(1)}}\cdots\mathcal{X}_{p_{\pi(M)}}}\,,
\end{equation}
where $\pi\in \text{S}_M$ is a permutation. This wave~function is still asymptotic, in the sense that it does not explicitly depend on regions where two magnons come close, \ie~where scattering happens. Interactions are encoded in the coefficients~$\chi$, which will involve S-matrix elements. Comparing with the two-magnon case~\eqref{eq:spin2part} and~\eqref{eq:spin2partScatt} it is immediate to see that if the permutation involves only two indices, it will be
\begin{equation}
\label{eq:wavefunct-coeffs}
\chi(p_1,\dots p_{j+1},p_{j},\dots p_{M})=
S(p_j,p_{j+1})\,
\chi(p_1,\dots p_{j},p_{j+1},\dots p_{M})\,.
\end{equation}
If we now take into account factorised scattering, we can immediately extend this to an~\emph{arbitrary} permutation, so that we can rewrite
\begin{equation}
\begin{aligned}
&\Ket{\Psi(p_1,\dots, p_M)}
=
\sum_{\pi\in \text{S}_M} \mathcal{S}_{\pi}
\Ket{\mathcal{X}_{p_{1}}\cdots\mathcal{X}_{p_{M}}}\,,\\
&\mathcal{S}_{\pi}\Ket{\mathcal{X}_{p_{1}}\cdots\mathcal{X}_{p_{M}}}
=
\prod_{(j,k)\in\pi}\!\! S(p_j,p_k)\;
\Ket{\mathcal{X}_{p_{\pi(1)}}\cdots\mathcal{X}_{p_{\pi(M)}}}\,,
\end{aligned}
\end{equation}
where we set~$\chi(p_{1},\dots  p_{M})=1$.
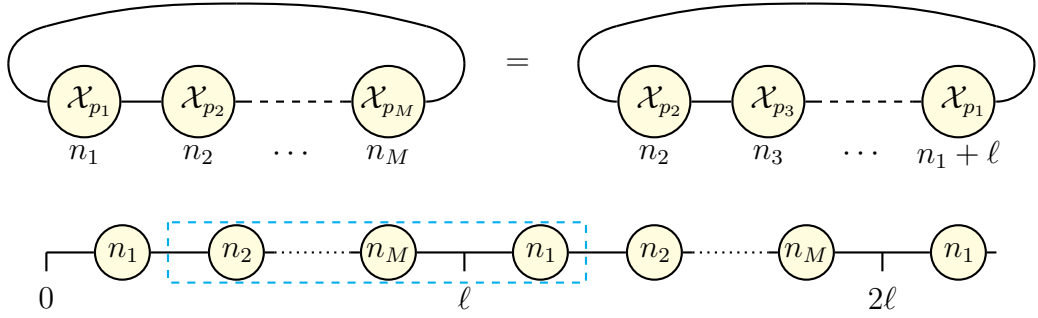
\begin{figure}
  \centering
\begin{tikzpicture}
\begin{scope}
  \begin{scope}[xshift=-5.5cm]
    \coordinate (c0) at (2.1cm,0);
    \coordinate (c1) at (3.4cm,0);
	\coordinate (v0) at (-0.5cm,0);
	\coordinate (v1) at (-1cm,0.5cm);
	\coordinate (v2) at (-0.5cm,1cm);
	\coordinate (v3) at (2cm,1.3cm);
	\coordinate (v4) at (4.5cm,1cm);
	\coordinate (v5) at (5cm,0.5cm);
	\coordinate (v6) at (4.5cm,0cm);

    \node (sp11) at (0cm,0cm) [spin] {$\fixedspaceR{\mathcal{X}_{p_{M}}}{\mathcal{X}_{p_{1}}}$};
    \node (sp12) at (1.5cm,0cm) [spin] {$\fixedspaceR{\mathcal{X}_{p_{M}}}{\mathcal{X}_{p_{2}}}$};
    \node (sp14) at (4cm,0cm) [spin] {$\fixedspaceR{\mathcal{X}_{p_{M}}}{\mathcal{X}_{p_{M}}} $};
    
    \node at (0cm,-0.7cm)   {$n_1$};
    \node at (1.5cm,-0.7cm)  {$n_2$};
    \node at (2.75cm,-0.7cm)  {$\cdots $};
    \node at (4cm,-0.7cm) {$n_M $};

    \draw [thick]  (v0) to (sp11);
    \draw [thick]  (sp11) to (sp12);
    \draw [thick]  (sp12) to (c0);
    \draw [thick,dashed]  (c0) to (c1);
    \draw [thick]  (c1) to (sp14);
    \draw [thick]  (sp14) to (v6);

    \draw [thick] [out=180,in=270] (v0) to (v1);
    \draw [thick] [out=90,in=195]  (v1) to (v2);
    \draw [thick] [out=15,in=180]  (v2) to (v3);
    \draw [thick] [out=0,in=165]   (v3) to (v4);
    \draw [thick] [out=345,in=90]  (v4) to (v5);
    \draw [thick] [out=270,in=0]    (v5) to (v6);

  \end{scope}
  \node at (0.2cm,0.5cm)  {$=$};
  \begin{scope}[xshift=2cm]
    \coordinate (c0) at (2.1cm,0);
    \coordinate (c1) at (3.4cm,0);
	\coordinate (v0) at (-0.5cm,0);
	\coordinate (v1) at (-1cm,0.5cm);
	\coordinate (v2) at (-0.5cm,1cm);
	\coordinate (v3) at (2cm,1.3cm);
	\coordinate (v4) at (4.5cm,1cm);
	\coordinate (v5) at (5cm,0.5cm);
	\coordinate (v6) at (4.5cm,0cm);

    \node (sp11) at (0cm,0cm) [spin] {$\fixedspaceR{\mathcal{X}_{p_{M}}}{\mathcal{X}_{p_{2}}}$};
    \node (sp12) at (1.5cm,0cm) [spin] {$\fixedspaceR{\mathcal{X}_{p_{M}}}{\mathcal{X}_{p_{3}}}$};
    \node (sp14) at (4cm,0cm) [spin] {$\fixedspaceR{\mathcal{X}_{p_{M}}}{\mathcal{X}_{p_{1}}} $};

    \node at (0cm,-0.7cm)  {$n_2$};
    \node at (1.5cm,-0.7cm)  {$n_3$};
    \node at (2.75cm,-0.7cm)  {$\cdots $};
    \node at (4cm,-0.7cm)   {$n_1+\ell $};

    \draw [thick]  (v0) to (sp11);
    \draw [thick]  (sp11) to (sp12);
    \draw [thick]  (sp12) to (c0);
    \draw [thick,dashed]  (c0) to (c1);
    \draw [thick]  (c1) to (sp14);
    \draw [thick]  (sp14) to (v6);

    \draw [thick] [out=180,in=270] (v0) to (v1);
    \draw [thick] [out=90,in=195]  (v1) to (v2);
    \draw [thick] [out=15,in=180]  (v2) to (v3);
    \draw [thick] [out=0,in=165]   (v3) to (v4);
    \draw [thick] [out=345,in=90]  (v4) to (v5);
    \draw [thick] [out=270,in=0]    (v5) to (v6);

  \end{scope}
\end{scope}

\coordinate (zero) at (-6cm,-2cm);
\coordinate (ell) at (-0.5cm,-2cm);
\coordinate (ell2) at (5cm,-2cm);
\coordinate (zeroS) at (-6cm,-2.25cm);
\coordinate (ellS) at (-0.5cm,-2.25cm);
\coordinate (ell2S) at (5cm,-2.25cm);

\node at (-6cm,-2.6cm)  {$0$};
\node at (-0.5cm,-2.6cm)  {$\ell$};
\node at (5cm,-2.6cm)  {$2\ell$};

\coordinate (end) at (6.5cm,-2cm);
\coordinate (m2l) at (-3cm,-2cm);
\coordinate (m2r) at (-2cm,-2cm);
\coordinate (m5l) at (2.5cm,-2cm);
\coordinate (m5r) at (3.5cm,-2cm);

	\draw [thick,dashed,cyan] (-4.4cm,-1.6cm) rectangle (1.1cm,-2.4cm);

\node (m1) at (-5cm,-2cm) [spin] {$\fixedspaceR{n_{M}}{n_1\,}$};
\node (m2) at (-3.5cm,-2cm) [spin] {$\fixedspaceR{n_{M}}{n_2\,}$};
\node (m3) at (-1.5cm,-2cm) [spin] {$\fixedspaceR{n_{M}}{n_M}$};

\node (m4) at (0.5cm,-2cm) [spin] {$\fixedspaceR{n_{M}}{n_1\,}$};
\node (m5) at (2cm,-2cm) [spin] {$\fixedspaceR{n_{M}}{n_2\,}$};
\node (m6) at (4cm,-2cm) [spin] {$\fixedspaceR{n_{M}}{n_M}$};

\node (m7) at (6cm,-2cm) [spin] {$\fixedspaceR{n_{M}}{n_1\,}$};

    \draw [thick] (zero) to (m1);
    \draw [thick] (m1) to (m2);
    \draw [thick] (m2) to (m2l);
    \draw [thick,dotted] (m2l) to (m2r);
    \draw [thick] (m2r) to (m3);
    \draw [thick] (m3) to (m4);
    \draw [thick] (m4) to (m5);
    \draw [thick] (m5) to (m5l);
    \draw [thick,dotted] (m5l) to (m5r);
    \draw [thick] (m5r) to (m6);
    \draw [thick] (m6) to (m7);
    \draw [thick] (m7) to (end);
    
    \draw [thick] (zero) to (zeroS);
    \draw [thick] (ell) to (ellS);
    \draw [thick] (ell2) to (ell2S);
\end{tikzpicture}

  \caption{Illustrations of the periodicity condition leading to the Bethe ansatz equations. Above: the two ends of the chain have been identified, so that the first excitation $\mathcal{X}_{p_1}$ in position~$n_1$ can be thought as sitting at the end of the chain, in position~$n_1+\ell$. This can be realised by scattering~$\mathcal{X}_{p_1}$ through all of the~$\mathcal{X}_{p_j}$.
Below: equivalently, the wave-function should be $\ell$-periodic as a function on~$\mathbbm{R}$. The box identifies the periodicity condition evaluated as in~\eqref{eq:coord-shift}.
 }
  \label{fig:byperiodicity}
\end{figure}

We now will impose that this wave~function is $\ell$-periodic. Let us shift e.g. the first coordinate~$n_{1}$ by~$\ell$ and bring it all the way around the chain:
\begin{equation}
\label{eq:coord-shift}
(n_1,n_2,\dots n_{M-1},n_{M})\to
(n_2,n_3,\dots n_{M},n_1+\ell)\,,
\end{equation}
as depicted in figure~\ref{fig:byperiodicity}.
Periodicity is the statement that~$\Ket{\Psi}$ is invariant under such a transformation. This is not granted by the ansatz~\eqref{eq:asympt-wavefunct}. Using~\eqref{eq:wavefunct-coeffs} we in fact have that shifting~$n_1$  by~$\ell $ gives the periodicity condition%
\footnote{%
One may find the notation of eq.~\eqref{eq:periodicitycond} confusing, since quantum-mechanical states are defined up to an overall phase. However here we normalised  $\chi(p_{1},\dots  p_{M})=1$. Eq.~\eqref{eq:periodicitycond} should be then understood as a condition of spatial periodicity on the wave-function.
}
\begin{equation}
\label{eq:periodicitycond}
\Ket{\Psi(p_1,\dots p_M)}
=
\left(e^{-ip_1\ell}\prod_{j=2}^M
S(p_1,p_j)\right)\,
\Ket{\Psi(p_1,\dots p_M)}\,.
\end{equation}
We then have to require the expression in the big brackets to be equal to~$1$. Physically, this is a quantisation condition for the momentum~$p_1$. There are several such conditions: suppose that we repeat the shift~\eqref{eq:coord-shift} for $n_2$, then $n_3$, until $n_k$. We then get to the configuration
\begin{equation}
(n_{k},n_{k+1},\dots n_{M},n_1+\ell,\dots n_{k-1}+\ell,n_{k}+\ell)\,.
\end{equation}
Imposing periodicity for any~$k=1,\dots M$ results in~$k$ coupled equations for the magnon momenta
\begin{equation}
\label{eq:spinchain-BA}
\text{spin chain:}
\quad e^{-ip_k\ell}\prod_{j\neq k}^M
S(p_k,p_j)=1\,,
\qquad
k=1,\dots M\,.
\end{equation}
These are the celebrated~\emph{Bethe ansatz equations}. More precisely, the above derivation is referred to as the \emph{coordinate} Bethe ansatz.

Let us remark that in our derivation we assumed that~$\mathcal{X}_p$ has a single flavour, so that the scattering is diagonal and the S~matrix $S(p,q)$ is just a number. Extending the discussion above to the case of several flavours that scatter diagonally, \ie~by pure transmission of \emph{all} quantum numbers is completely straightforward---it only requires adding flavour indices to $\mathcal{X}_p$ and~$S(p,q)$. Our S~matrix is not so simple, because the fermion number is not always transmitted, for instance in the process
\begin{equation}
    \mathcal{S} \ket{\fixedspaceL{\psi_p^{\smallL} \psi_q^{\smallL}}{\phi_p^{\smallL} \psi_q^{\smallL}}} 
    = \fixedspaceR{\Del^{\smallLL}_{pq}}{\Bel^{\smallLL}_{pq}} \ket{\fixedspaceL{\psi_p^{\smallL} \psi_q^{\smallL}}{\psi_q^{\smallL} \phi_p^{\smallL}}} + \Cel^{\smallLL}_{pq} \ket{\fixedspaceL{\psi_p^{\smallL} \psi_q^{\smallL}}{\phi_q^{\smallL} \psi_p^{\smallL}}}.
\end{equation}
In order to impose periodicity we will first have to diagonalise the action of the S~matrix by the so-called nested Bethe ansatz. This is a technical complication that we will address in the next section in the simple case  of the~$\su(1|1)^2_{\text{c.e.}}$ S~matrix.

\subsubsection*{Bethe ansatz from the worldsheet}
It is easy to see that the reasoning above can be repeated almost \textit{verbatim} from the point of view of the ZF algebra and S~matrix. In that case, the asymptotic wave~function would be 
\begin{equation}
\Ket{\Psi(p_1,\dots p_M)}=
\sum_{\pi\in \text{S}_M} \chi(p_{\pi(1)},\dots p_{\pi(M)}) 
\Ket{p_{\pi(1)},\dots p_{\pi(M)}}\,,
\end{equation}
where each asymptotic state would read, in position space,
\begin{equation}
\label{eq:worldsheet-state}
\Ket{p_{1},\dots p_{M}}
=
\int\limits_{\sigma_1\ll\dots \ll \sigma_M}
 \de \sigma_1\cdots \de \sigma_M
e^{i(p_1\sigma_1+\cdots+p_M\sigma_M)}\mathbf{A}^{\dagger}(\sigma_1)\cdots \mathbf{A}^{\dagger}(\sigma_M)\Ket{0}\,.
\end{equation}
Note that we kept an ordering such that the particle of momentum~$p_1$ is the leftmost one. We can now require periodicity under shifts of $\sigma_j\to \sigma_j +\ell$, obtaining the Bethe~ansatz
\begin{equation}
\label{eq:worldsheet-BA}
\text{worldsheet:}
\quad e^{ip_k\ell}\prod_{j\neq k}^M
S(p_k,p_j)=1\,,
\qquad
k=1,\dots M\,.
\end{equation}
The different sign in the plane-wave coefficient resulted in different quantisation conditions. As discussed in the previous chapter, that same sign choice makes the spin chain S~matrix be the inverse of the worldsheet one. Indeed, equations~\eqref{eq:worldsheet-BA} and~\eqref{eq:spinchain-BA} are related by inverting~$S(p,q)$.

\subsection*{The level-matching condition}
In chapter~\ref{ch:sigmamodel} we have found that in string theory the only physical states are the ones satisfying the constraint
\begin{equation}
\mathbf{P}\,\Ket{\text{physical}}=0\,,
\end{equation}
in the case of zero winding. This results in a condition on the momenta of the $M$-particle state
\begin{equation}
p_1+\dots +p_M=0\,.
\end{equation}
If we allow for winding it is not hard to see that the right hand side of the constraint should not vanish, but rather equal~$2\pi W$, where the integer~$W$ is the  winding number.

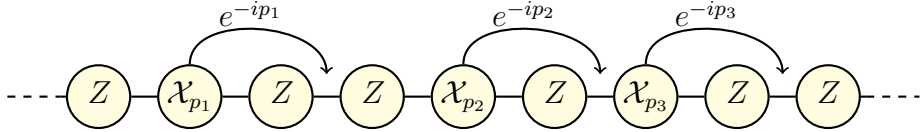
\begin{figure}
  \centering
\begin{tikzpicture}
  \begin{scope}[xshift=-3.6cm]
    \coordinate (c0) at (-0.5cm,0);
    \coordinate (c0d) at (-1.2cm,0);
    \coordinate (c1) at (10.1cm,0);
    \coordinate (c1d) at (10.8cm,0);

    \coordinate (j1) at (3.0cm,0.3cm);
    \coordinate (j2) at (6.6cm,0.3cm);
    \coordinate (j3) at (9.0cm,0.3cm);

    \node (sp11) at (0.0cm,0cm) [spin] {$\fixedspaceR{\mathcal{X}_{p_{1}}}{Z\ }$};
    \node (sp12) at (1.2cm,0cm) [spin] {$\fixedspaceR{\mathcal{X}_{p_{1}}}{\mathcal{X}_{p_{1}}}$};
    \node (sp13) at (2.4cm,0cm) [spin] {$\fixedspaceR{\mathcal{X}_{p_{1}}}{Z\ }$};
    \node (sp14) at (3.6cm,0cm) [spin] {$\fixedspaceR{\mathcal{X}_{p_{1}}}{Z\ }$};
    \node (sp15) at (4.8cm,0cm) [spin] {$\fixedspaceR{\mathcal{X}_{p_{1}}}{\mathcal{X}_{p_{2}}}$};
    \node (sp16) at (6.0cm,0cm) [spin] {$\fixedspaceR{\mathcal{X}_{p_{1}}}{Z\ }$};
    \node (sp17) at (7.2cm,0cm) [spin] {$\fixedspaceR{\mathcal{X}_{p_{1}}}{\mathcal{X}_{p_{3}}}$};
    \node (sp18) at (8.4cm,0cm) [spin] {$\fixedspaceR{\mathcal{X}_{p_{1}}}{Z\ }$};
    \node (sp19) at (9.6cm,0cm) [spin] {$\fixedspaceR{\mathcal{X}_{p_{1}}}{Z\ }$};
    
    \draw [thick]  (sp11) to (sp12);
    \draw [thick]  (sp12) to (sp13);
    \draw [thick]  (sp13) to (sp14);
    \draw [thick]  (sp14) to (sp15);
    \draw [thick]  (sp15) to (sp16);
    \draw [thick]  (sp16) to (sp17);
    \draw [thick]  (sp17) to (sp18);
    \draw [thick]  (sp18) to (sp19);
    
    \draw [thick,dashed]  (c0d) to (c0);
    \draw [thick]  (c0) to (sp11);
    \draw [thick]  (sp19) to (c1);
    \draw [thick,dashed]  (c1) to (c1d);

    \draw [thick]  [out=90,in=90,->]  (sp12) to (j1);
    \draw [thick]  [out=90,in=90,->]  (sp15) to (j2);
    \draw [thick]  [out=90,in=90,->]  (sp17) to (j3);
    
    \node at (2cm,1.1cm)  {$e^{-ip_1} $};
    \node at (5.6cm,1.1cm)  {$e^{-ip_2} $};
    \node at (8cm,1.1cm)  {$e^{-ip_3} $};
    
  \end{scope}
\end{tikzpicture}

  \caption{By shifting each magnon to the right by one site, we produce a shift of  $e^{-i(p_1+\dots+p_M)}$. On the other hand, for a cyclic chain this produces no effect, so that consistency dictates the level-matching condition~\eqref{eq:spinchainLM}.}
  \label{fig:levelmatching}
\end{figure}

This string-theoretical requirement has an interesting equivalent in the spin-chain picture. Consider an $M$-magnon asymptotic state in the spin chain, and imagine of shifting each magnon \eg~to the left by one site, as in figure~\ref{fig:levelmatching}. On the one hand, cyclicity of the chain requires this transformation to leave it invariant. On the other hand, by eq.~\eqref{eq:magnon-commutation} this results in~$M$ phase shifts of the form~$e^{-i p_j}$. We then have an additional requirement
\begin{equation}
\label{eq:spinchainLM}
e^{i(p_1+\dots + p_M)}=1\,.
\end{equation}
We then find that the spin chain and worldsheet pictures are completely equivalent, \textit{mutatis mutandis}, from the Bethe ansatz point of view. It is then just a matter of convenience to derive our equations with reference to one or the other framework.

\section{Nested Bethe ansatz}
\label{sec:nesting}
In this subsection we will illustrate how the nesting procedure works for the simple case of the~$\su(1|1)^2_{\text{c.e.}}$ S~matrix. In what follows, for  definiteness, we will work \emph{from the spin-chain point of view}, and only at the end of the day comment on the worldsheet picture.

The strategy will be to split the action of the S~matrix in several steps: first, consider a set of ``level-I excitations'' on the usual vacuum, that have the property of scattering diagonally among themselves.
For these excitations, the coordinate Bethe ansatz can be straightforwardly implemented as we illustrated in the previous section. In order to incorporate the remaining excitations, we construct a new ``level-II'' vacuum. This is a state consisting of level-I excitations only. We can now consider level-II excitations on this vacuum that scatter trivially among themselves. If needed, we can use those to construct a level-III vacuum, and so on. In this way, level by level, the scattering is very simple. Of course, we will have to require that \eg~the level-II excitations propagate on the level-II vacuum in a way that is compatible with the dynamics of the fundamental S~matrix. This will ensure that different levels are ``glued'' in a consistent way.

\subsection*{Level-I excitations}
For definiteness, here we work with the spin-chain S~matrix. Looking at the scattering elements in section~\ref{sec:spinchainSmat}, we see that there are several non-diagonal scattering processes. The level-I vacuum is the usual one,
\begin{equation}
\Ket{0}^{\I}=\Ket{Z^\ell}\,.
\end{equation}
We have two possible choices of processes that scatter diagonally among themselves:
\begin{equation}
\label{eq:BA-excit-set}
V^{\I}_{A}=\big\{\phi^{\smallL},\psi^{\smallR}\big\}\,,
\qquad
V^{\I}_{B}=\big\{\phi^{\smallR},\psi^{\smallL}\big\}\,.
\end{equation}
We can think of this as a choice of the highest weight states in the left and right representations, or equivalently of the grading of the superalgebra. Clearly the choice $A$ corresponds to having as lowering operators~$\mathbf{q}^{\smallL}$ and~$\bar{\mathbf{q}}^{\smallR}$, while $B$ corresponds to picking $\bar{\mathbf{q}}^{\smallL}$ and~$\mathbf{q}^{\smallR}$. Let us for the moment pick the choice~$A$.

A level-I state is a collection of excitations of~$V^{\I}_{A}$, \eg
\begin{equation}
\Ket{\mathcal{X}^{\a_1}_{p_1},\dots \mathcal{X}^{\a_M}_{p_M}}^{\I}
=\Ket{ZZ\mathcal{X}^{\a_1}_{p_1}ZZ\cdots ZZ\mathcal{X}^{\a_M}_{p_M}ZZ\cdots}\,,
\end{equation}
like in~\eqref{eq:BA-spinchain-state}. There will be~$M^{\I}_{\smallL}$ left excitations and~$M^{\I}_{\smallR}$ right ones, distinguished by the flavour label~$\alpha$, so that
\begin{equation}
M=M^{\I}_{\smallL}+M^{\I}_{\smallR}\,.
\end{equation}
Out of such states we can as before construct the asymptotic level-one wave~function by acting with the multiparticle S~matrix~$\mathcal{S}_{\pi}$
\begin{equation}
\Ket{\Psi(p_1,\dots, p_M)}^{\I}_{\a_1,\dots \a_M}
=
\sum_{\pi\in \text{S}_M} \mathcal{S}_{\pi}
\Ket{\mathcal{X}^{\a_1}_{p_1},\dots \mathcal{X}^{\a_M}_{p_M}}^{\I}\,.
\end{equation}
Now also the S~matrix will carry flavour indices~$\a$ that can take values~L and~R. Still, by construction~$\check{\mathbf{S}}_{\pi}$ acts diagonally on the level-one states,
\begin{equation}
\mathcal{S}_{\pi}
\Ket{\mathcal{X}^{\a_1}_{p_1},\dots \mathcal{X}^{\a_M}_{p_M}}^{\I}=
S_{\pi}^{\I}(p_1,\dots p_M)
\Ket{\mathcal{X}^{\a_{\pi(1)}}_{p_{\pi(1)}},\dots \mathcal{X}^{\a_{\pi(M)}}_{p_{\pi(M)}}}^{\I}\,,
\end{equation}
where the phase~$S_{\pi}$ factorises
\begin{equation}
S_{\pi}(p_1,\dots p_M)=\prod_{(j,k)\in\pi}
S^{\I,\I}_{\a_j\a_k}(p_j,p_k)\,.
\end{equation}
The level-one S-matrix elements~$S^{\I,\I}_{\a_j\a_k}$ can be immediately read off the fundamental S~matrix elements of section~\ref{sec:spinchainSmat},
\begin{equation}
S^{\I,\I}_{\smallLL}=\Ael^{\smallLL}\,,
\qquad
S^{\I,\I}_{\smallRR}=\Fel^{\smallRR}\,,
\qquad
S^{\I,\I}_{\smallLR}=\Cel^{\smallLR}\,,
\qquad
S^{\I,\I}_{\smallRL}=\Del^{\smallRL}\,.
\end{equation}

\subsection*{Level-II excitations}
The level-II vacuum is a collection of level-I excitations
\begin{equation}
\Ket{0}^{\II}=\Ket{\mathcal{X}^{\a_1}_{p_1}\cdots \mathcal{X}^{\a_M}_{p_M}}\,.
\end{equation}
Level-II excitations can be constructed by acting on~$\Ket{0}^{\II}$ with the lowering operators. Acting by $\mathbf{q}^{\smallL}$ turns $\phi^{\smallL}$ and~$\psi^{\smallR}$ into~$\psi^{\smallL}$ and~$\phi^{\smallR}$ respectively%
\footnote{%
Since we are working in the spin chain picture, we should not forget to insert the appropriate length-changing effects.
}. However we want to treat~$\ket{0}^{\II} $ as a vacuum,  so that  the scattering properties its the level-II excitations should be blind to the underlying level-I excitations.
The effect of ${\mathbf{q}}^{\smallR}$ would be similar, up to exchanging the left and right modules.

The generic form of a level-II state containing a single excitation is
\begin{equation}
\label{eq:BA-levelII-planewave}
\begin{aligned}
&\Ket{\mathcal{Y}^{\smallL}_y}^{\II}=\sum_{k=1}^M\chi_k(y^{\smallL})\,\Ket{\mathcal{X}^{\a_1}_{p_1}\cdots
\mathcal{Y}^{\smallL}_{y}
\cdots\mathcal{X}^{\a_M}_{p_M}}\,,
\\
&\chi_k(y^{\smallL})=f(y^{\smallL},p_k)\,\prod_{j=1}^{k-1} S^{\II,\I}_{\a_k\a_j}(y^{\smallL},p_j)\,,
\end{aligned}
\end{equation}
where for brevity we wrote $\mathcal{Y}^{\smallL} = \mathbf{q}^{\smallL}\mathcal{X}^{\a_{k}}_{p_k}$. The superscript L in $\mathcal{Y}^{\smallL}$ is meant to remind us that we are acting on the vacuum by a left lowering operator.
The coefficient~$f(y^{\smallL},p_k)$ stands for the creation of a level-II excitation on top of a level-I one which had momentum~$p_k$, while~$S^{\II,\I}_{\a_k\a_j}(y^{\smallL},p_j)$ accounts for the scattering of the level-II excitation with level-I ones, that is needed to position the former at site~$k$. These functions cannot be arbitrary if we want this description to be compatible with the one at level~I. To this end, we require that creating a level-II excitation and scattering the underlying level-I excitation are commuting operations:
\begin{equation}
\label{eq:BA-compatibility}
\mathcal{S}_{\pi}
\Ket{y^{\smallL}}^{\II}=
S_{\pi}^{\I}(p_1,\dots p_M)\Ket{y^{\smallL}}^{\II}_{\pi}\,,
\end{equation}
where $\Ket{y^{\smallL}}^{\II}_{\pi}$ is the level-II state constructed on the permuted vacuum,
\begin{equation}
\begin{aligned}
&\Ket{y^{\smallL}}^{\II}_{\pi}=\sum_{k=1}^M\chi_{k,\pi}(y^{\smallL})\,
\Ket{\mathcal{X}^{\a_{\pi(1)}}_{p_{\pi(1)}}\cdots
\mathcal{Y}^{\smallL}_{y}
\cdots\mathcal{X}^{\a_{\pi(M)}}_{p_{\pi(M)}}}\,
,
\\
&\chi_{k,\pi}(y^{\smallL})=f(y^{\smallL},p_{\pi(k)})\,\prod_{j=1}^{k-1} S^{\II,\I}_{\a_{\pi(k)}\a_{\pi(j)}}(y^{\smallL},p_{\pi(j)})\,.
\end{aligned}
\end{equation}
On the other hand,~$\mathcal{S}_{\pi}
\Ket{y}^{\II}$ can be computed just by acting on the excitations with the S~matrix of section~\ref{sec:spinchainSmat}, regardless of whether they are in the first or second level.

\subsection*{Level-II excitations: propagation}
To be more specific, let us consider a level-II vacuum consisting of two $\phi^{\smallL}$ bosons, and one level-II excitation on it,
\begin{equation}
\Ket{y^{\smallL}}^{\II}_{\smallLL}=
f_{\smallL}(y^{\smallL},p)\Ket{\psi^{\smallL}_p\phi^{\smallL}_q}+
f_{\smallL}(y^{\smallL},q)S^{\II,\I}_{\smallLL}(y^{\smallL},p)\Ket{\phi^{\smallL}_p\psi^{\smallL}_q}\,.
\end{equation}
The generalisation to a longer vacuum will be straightforward owing to the factorisation property.
Note how both of the expressions on the right hand side of the non-diagonal scattering processes appear in the ansatz. After scattering, the compatibility condition~\eqref{eq:BA-compatibility} mandates that this should be proportional to
\begin{equation}
\Ket{y}^{\II}_{\smallLL,\pi}=
f_{\smallL}(y^{\smallL},q)\Ket{\psi^{\smallL}_q\phi^{\smallL}_p}+
f_{\smallL}(y^{\smallL},p)S^{\II,\I}_{\smallLL}(y^{\smallL},q)\Ket{\phi^{\smallL}_q\psi^{\smallL}_p}\,,
\end{equation}
with a proportionality constant equal to~$S^{\I,\I}_{\smallLL}(p,q)=\Ael^{\smallLL}_{pq}$. On the other hand, we can act with~$\mathcal{S}_{pq}$ on~$\Ket{\psi^{\smallL}_p\phi^{\smallL}_q}$ and~$\Ket{\phi^{\smallL}_p\psi^{\smallL}_q}$, as in~\eqref{eq:SCscattLL}. The resulting expression can be proportional to~$\Ket{y^{\smallL}}^{\II}_{\smallLL,\pi}$ and  in fact equate~$\Ael^{\smallLL}_{pq}\Ket{y}^{\II}_{\smallLL,\pi}$, provided that the level-II coefficients satisfy
\begin{equation}
\label{eq:BAcompat-explicit}
\begin{aligned}
&S^{\II,\I}_{\smallLL}(y^{\smallL},q)
f_{\smallL}(y^{\smallL},p)\;\Ael^{\smallLL}_{pq}=
f_{\smallL}(y^{\smallL},p)\;\Del^{\smallLL}_{pq}+
f_{\smallL}(y^{\smallL},q)S^{\II,\I}_{\smallLL}(y^{\smallL},p) \;\Cel^{\smallLL}_{pq}\,,\\
&\phantom{S^{\II,\I}_{\smallLL}(y^{\smallL},q)}
 f_{\smallL}(y^{\smallL},q)\;\Ael^{\smallLL}_{pq}=
f_{\smallL}(y^{\smallL},p)\;\Eel^{\smallLL}_{pq}+
f_{\smallL}(y^{\smallL},q)S^{\II,\I}_{\smallLL}(y^{\smallL},p) \;\Bel^{\smallLL}_{pq}\,.
\end{aligned}
\end{equation}
Using the explicit form of the S-matrix elements~\eqref{eq:spinchainSmat-el-LL}, we find that it must be
\begin{equation}
f_{\smallL}(y^{\smallL},p)=g_{\smallL}(y^{\smallL})\frac{\eta_p\,e^{\frac{i}{4}p}}{k_{\smallL}(y^{\smallL})-x^-_p}\,,
\qquad
S^{\II,\I}_{\smallLL}(y^{\smallL},p)=\frac{k_{\smallL}(y^{\smallL})-x^+_p}{k_{\smallL}(y^{\smallL})-x^-_p}\,,
\end{equation}
where~${g}_{\smallL}(y)$ and~$k_{\smallL}(y)$ are arbitrary functions of~$y$. Note that at this nesting level none of these functions can depend on the dressing factors.

A similar calculation in the right sector gives the ansatz
\begin{equation}
\begin{aligned}
\Ket{y^{\smallL}}^{\II}_{\smallRR}=
f_{\smallR}(y^{\smallL},p)\Ket{\phi^{\smallR}_pZ^{+}\psi^{\smallR}_q}-
f_{\smallR}(y^{\smallL},q)S^{\II,\I}_{\smallRR}(y^{\smallL},p)\Ket{\psi^{\smallR}_p\phi^{\smallR}_qZ^{+}}\,,\\
\Ket{y^{\smallL}}^{\II}_{\smallRR,\pi}=
f_{\smallR}(y^{\smallL},q)\Ket{\phi^{\smallR}_qZ^{+}\psi^{\smallR}_p}-
f_{\smallR}(y^{\smallL},p)S^{\II,\I}_{\smallRR}(y^{\smallL},q)\Ket{\psi^{\smallR}_q\phi^{\smallR}_pZ^{+}}\,,
\end{aligned}
\end{equation}
where now $\psi^{\smallR}$ is the level-I excitation, and length-changing effects appear. The minus signs take into account the fact that the underlying vacuum is fermionic. With this choice it will be easier to impose the periodicity condition at the end of the day, since all fermion signs will be accounted for.
The consistency condition now results in the equations
\begin{equation}
\begin{aligned}
S^{\II,\I}_{\smallRR}(y^{\smallL},q)
f_{\smallR}(y^{\smallL},p)\;\Fel^{\smallRR}_{pq}=-
f_{\smallR}(y^{\smallL},p)e^{iq}\Cel^{\smallRR}_{pq}+
f_{\smallR}(y^{\smallL},q)S^{\II,\I}_{\smallRR}(y,p) \;\Del^{\smallRR}_{pq}\,,\\
 f_{\smallR}(y^{\smallL},q)e^{ip}\Fel^{\smallRR}_{pq}=+
f_{\smallR}(y^{\smallL},p)e^{iq}\Bel^{\smallRR}_{pq}-
f_{\smallR}(y^{\smallL},q)S^{\II,\I}_{\smallRR}(y,p) \;\Eel^{\smallRR}_{pq}\,.
\end{aligned}
\end{equation}
which are solved by
\begin{equation}
f_{\smallR}(y^{\smallL},p)=-i\frac{g_{\smallR}}{x^{-}_p}(y)\frac{\eta_p\,e^{-\frac{3}{4}ip}}{1-\frac{1}{k_{\smallR}(y^{\smallL})\,x^+_p}}\,,
\qquad
S^{\II,\I}_{\smallRR}(y^{\smallL},p)=\frac{1-\frac{1}{k_{\smallR}(y^{\smallL})\,x^-_p}}{1-\frac{1}{k_{\smallR}(y^{\smallL})\,x^+_p}}\,.
\end{equation}

There is one last consistency condition the we should impose, which is the one arising when the vacuum contains states of the left \emph{and} right modules. For two states, this gives a level-II excitation
\begin{equation}
\begin{aligned}
\Ket{y^{\smallL}}^{\II}_{\smallLR}=
+f_{\smallL}(y^{\smallL},p)\Ket{\psi^{\smallL}_p\psi^{\smallR}_q}+
f_{\smallR}(y^{\smallL},q)S^{\II,\I}_{\smallRL}(y^{\smallL},p)\Ket{\phi^{\smallL}_p\phi^{\smallR}_qZ^+}\,,\\
\Ket{y^{\smallL}}^{\II}_{\smallLR,\pi}=
-f_{\smallR}(y^{\smallL},q)\Ket{\psi^{\smallR}_q\psi^{\smallL}_p}+
f_{\smallL}(y^{\smallL},p)S^{\II,\I}_{\smallLR}(y^{\smallL},q)\Ket{\phi^{\smallR}_qZ^+\phi^{\smallL}_p}\,,
\end{aligned}
\end{equation}
when we take a left and right state in this order. This results in the equations
\begin{equation}
\begin{aligned}
S^{\II,\I}_{\smallLR}(y^{\smallL},q)
f_{\smallL}(y^{\smallL},p)e^{ip}\;\Cel^{\smallLR}_{pq}&= 
f_{\smallL}(y^{\smallL},p)\;\Eel^{\smallLR}_{pq}+
f_{\smallR}(y^{\smallL},q)S^{\II,\I}_{\smallRL}(y^{\smallL},p) \;\Bel^{\smallLR}_{pq}\,,\\
-f_{\smallR}(y^{\smallL},q)\;\Cel^{\smallLR}_{pq}&=
f_{\smallL}(y^{\smallL},p)\;\Fel^{\smallLR}_{pq}+
f_{\smallR}(y^{\smallL},q)S^{\II,\I}_{\smallRL}(y^{\smallL},p) \;\Ael^{\smallLR}_{pq}\,.\;
\end{aligned}
\end{equation}
These conditions can be solved by imposing
\begin{equation}
\begin{aligned}
&k_{\smallL}(y^{\smallL})=k_{\smallR}(y^{\smallL})=y^{\smallL}\,,
\qquad	
&&g_{\smallL}(y^{\smallL})=-i\,y\;g_{\smallR}(y^{\smallL})\,,\\
&S^{\II,\I}_{\smallRL}(y^{\smallL},p)=S^{\II,\I}_{\smallLL}(y^{\smallL},p)\,,
\qquad
&&S^{\II,\I}_{\smallLR}(y^{\smallL},p)=S^{\II,\I}_{\smallRR}(y^{\smallL},p)\,.
\end{aligned}
\end{equation}
The two last equations can be interpreted by saying that when we act with the lowering operator~$\mathbf{q}^{\smallL}$ in the plane-wave ansatz~\eqref{eq:BA-levelII-planewave} we create a well-defined   excitation. Its scattering does not depend on the underlying vacuum that it is created out of, but only on what it scatters with.
This confirms that the lowering operators are blind to the constituents of~$\ket{0}^{\II}  $.

So far we have exclusively considered excitations created by~$\mathbf{q}^{\smallL}$. We can create different level-II excitations by acting with~$\bar{\mathbf{q}}^{\smallR}$ instead. We will denote the corresponding one-particle excitation by~$\ket{y_{\smallR}}^{\II}$. The computations follow the same pattern, up to appropriately taking into account the form of the S-matrix elements and the length-changing effects. We have, in particular
\begin{equation}
\begin{aligned}
&S^{\II,\I}_{\smallLL}(y^{\smallR},p)=
\frac{1-\frac{1}{y^{\smallR}x^+_p}}{1-\frac{1}{y^{\smallR}x^-_p}}\,,
\qquad	
&&S^{\II,\I}_{\smallRR}(y^{\smallR},p)=
\frac{y^{\smallR}-x^-_p}{y^{\smallR}-x^+_p}\,,\\
&S^{\II,\I}_{\smallRL}(y^{\smallR},p)=S^{\II,\I}_{\smallLL}(y^{\smallR},p)\,,
\qquad
&&S^{\II,\I}_{\smallLR}(y^{\smallR},p)=S^{\II,\I}_{\smallRR}(y^{\smallR},p)\,,
\end{aligned}
\end{equation}
where as usual fermion signs have been accounted for.

\subsection*{Level-II excitations: scattering }
The preceding discussion covers all of the dynamics involving a single level-II excitation on an arbitrary vacuum, owing to factorisation of scattering. However, as soon as we include another excitation, a new dynamics arises: scattering of two level-II excitations. 

A state containing two level-II excitations obtained by acting with~$\mathbf{q}^{\smallL}$ takes the form
\begin{equation}
\begin{aligned}
&\Ket{y^{\smallL}_1,y^{\smallL}_2}^{\II}=
\sum_{k<l}^M\chi_k(y^{\smallL}_1)\,\chi_l(y^{\smallL}_2)\,\,\Ket{\mathcal{X}^{\a_{1}}_{p_{1}}\cdots
\mathcal{Y}^{\smallL}_{y_1}\cdots
\mathcal{Y}^{\smallL}_{y_2}
\cdots\mathcal{X}^{\a_{M}}_{p_{M}}},
\end{aligned}
\end{equation}
where~$\mathcal{Y}^{\smallL}_{y_1}$ sits at position~$k$ and~$\mathcal{Y}^{\smallL}_{y_2}$ at position~$l$.
Once again, this ansatz is subject to a condition similar to~\eqref{eq:BA-compatibility}, reading
\begin{equation}
\label{eq:BA-twopart-levelII}
\check{\mathbf{S}}_{\pi}
\Ket{y^{\smallL}_1,y^{\smallL}_2}^{\II}=
S_{\pi}^{\I}(p_1,\dots p_M)\Ket{y^{\smallL}_1,y^{\smallL}_2}^{\II}_{\pi}\,,
\end{equation}
Clearly the only problems may arise from the scattering of $y^{\smallL}_1$ and~$y^{\smallL}_2$. Let us consider the case where we have two left-excitations on a length-two vacuum. We make an ansatz
\begin{equation}
  \begin{aligned}
    \ket{y^{\smallL}_1 y^{\smallL}_2}^{\II}_{\smallLL} &= f_{\smallL}(y^{\smallL}_1,p) f_{\smallL}(y^{\smallL}_2,q)\; S^{\II,\I}_{\smallLL}(y^{\smallL}_2,p)\, \ket{\psi^{\smallL}_p \psi^{\smallL}_q} \\
    &\qquad - f_{\smallL}(y^{\smallL}_2,p) f_{\smallL}(y^{\smallL}_1,q)\; S^{\II,\I}_{\smallLL}(y^{\smallL}_1,p) S^{\II,\II}_{\smallLL}(y^{\smallL}_1,y^{\smallL}_2)\, \ket{\psi^{\smallL}_p \psi^{\smallL}_q} , \\
    \ket{y^{\smallL}_1 y^{\smallL}_2}^{\II}_{\smallLL,\pi} &= f_{\smallL}(y^{\smallL}_1,q) f_{\smallL}(y^{\smallL}_2,p)\; S^{\II,\I}_{\smallLL}(y^{\smallL}_2,q)\, \ket{\psi^{\smallL}_q \psi^{\smallL}_p} \\
    &\qquad - f_{\smallL}(y^{\smallL}_2,q) f_{\smallL}(y^{\smallL}_1,p)\; S^{\II,\I}_{\smallLL}(y^{\smallL}_1,q) S^{\II,\II}_{\smallLL}(y^{\smallL}_1,y^{\smallL}_2)\, \ket{\psi^{\smallL}_q \psi^{\smallL}_p} ,
  \end{aligned}
\end{equation}
where again we accounted for the minus sign.
The only new ingredient is the undetermined factor~$S^{\II,\II}_{\smallLL}(y^{\smallL}_1,y^{\smallL}_2)$ that represents the scattering of the two level-II excitations. It is easy to see that the consistency condition is solved by
\begin{equation}
S^{\II,\II}_{\smallLL}(y^{\smallL}_1,y^{\smallL}_2)=1\,,
\end{equation}
\ie~the scattering is trivial. In fact, it is immediate to check that all of the level-II scattering matrices are trivial, including the ones scattering~$\ket{y^{\smallR}}$ with~$\ket{y^{\smallL}}$ or with itself. Using factorisation of scattering, all our considerations extend to any level-II state with $M_{\smallL}^{\II}$ left and $M_{\smallR}^{\II}$ right excitations.

Since we have taken into account all of the fundamental excitations and diagonalised all scattering processes, the nesting procedure terminates here. Had we to include additional non-diagonal processes, it would be necessary to construct a level-III vacuum, and so on.

\subsection*{Bethe ansatz equations}
The Bethe ansatz equations arise now out of imposing periodicity for each set of excitations.
Looking at~\eqref{eq:BA-twopart-levelII} we see that we chose an ordering of the level-II magnons. As discussed in the previous section, the description should be unchanged if we rearrange them using the fact that the chain is periodic. It is easy to see that this gives the conditions
\begin{equation}
\begin{aligned}
1=
\prod_{j=1}^{M^{\I}_{\smallL}}S^{\II,\I}_{\smallLL}(y_{k}^{\smallL},p_j)\;
\prod_{j=1}^{M^{\I}_{\smallR}}S^{\II,\I}_{\smallRR}(y_{k}^{\smallL},p_j)\,,
\quad
k=1,\dots M^{\II}_{\smallL}\,,\\
1=
\prod_{j=1}^{M^{\I}_{\smallL}}S^{\II,\I}_{\smallLL}(y_{k}^{\smallR},p_j)\;
\prod_{j=1}^{M^{\I}_{\smallR}}S^{\II,\I}_{\smallRR}(y_{k}^{\smallR},p_j)\,,
\quad
k=1,\dots M^{\II}_{\smallR}\,,
\end{aligned}
\end{equation}
where we used that $S^{\II,\II}=1$. Note that we do not have to take into account any additional fermion sign, due to our convention. One pictorial way to interpret this equation is to take one level-II excitation and carry it along the chain: by periodicity, the phase arising from the whole scattering sequence should be one.

Fundamental excitations obey a slightly modified version of~\eqref{eq:spinchain-BA}, which now includes also higher level excitations:
\begin{equation}
\begin{aligned}
e^{i p L}=
\prod_{j\neq k}^{M^{\I}_{\smallL}}S^{\I,\I}_{\smallLL}(p^{\smallL}_k,p^{\smallL}_j)
\prod_{j=1}^{M^{\I}_{\smallR}}S^{\I,\I}_{\smallLR}(p^{\smallL}_k,p^{\smallR}_j)
\prod_{j=1}^{M^{\II}_{\smallL}}S^{\I,\II}_{\smallLL}(p^{\smallL}_k,y^{\smallL}_j)
\prod_{j=1}^{M^{\II}_{\smallR}}S^{\I,\II}_{\smallLR}(p^{\smallL}_k,y^{\smallR}_j)\,,\\
e^{i p L}=
\prod_{j\neq k}^{M^{\I}_{\smallR}}S^{\I,\I}_{\smallRR}(p^{\smallR}_k,p^{\smallR}_j)
\prod_{j=1}^{M^{\I}_{\smallL}}S^{\I,\I}_{\smallRL}(p^{\smallR}_k,p^{\smallL}_j)
\prod_{j=1}^{M^{\II}_{\smallR}}S^{\I,\II}_{\smallRR}(p^{\smallR}_k,y^{\smallR}_j)
\prod_{j=1}^{M^{\II}_{\smallL}}S^{\I,\II}_{\smallRL}(p^{\smallR}_k,y^{\smallL}_j)\,,
\end{aligned}
\end{equation}
for $k=1,\dots M^{\I}_{\smallL,\smallR} $.
Note that we have introduced explicit labels for the momenta of left and right excitations. For brevity, in terms of the Zukhovski variables, we set
\begin{equation}
\label{eq:BA-shorthand}
x_j^{\pm}=x_{j}^{\pm}(p^{\smallL}),
\qquad
\tilde{x}_j^{\pm}=x_{j}^{\pm}(p^{\smallR}),
\qquad
y=y^{\smallL},
\qquad
\tilde{y}=y^{\smallR}.
\end{equation}
Then we can write the full set of Bethe equations as
\begin{align}
    \begin{split}
    \label{eq:bethe-equations}
      \left(\frac{x_k^+}{x_k^-}\right)^{\ell} &=
      \prod_{\substack{j = 1\\j \neq k}}^{M^{\I}_{\smallL}} \frac{x_k^+ - x_j^-}{x_k^- - x_j^+} \mathscr{S}^{\smallLL}_{kj}
      \prod_{j=1}^{M^{\II}_{\smallL}} \frac{x_k^- - y_j}{x_k^+ - y_j} 
      \prod_{j=1}^{M^{\I}_{\smallR}} \sqrt{\frac{1- \frac{1}{x_k^+ \tilde{x}_j^+}}{1- \frac{1}{x_k^- \tilde{x}_j^-}}} \mathscr{S}^{\smallLR}_{k\tilde{\jmath}}
      \prod_{j=1}^{M^{\II}_{\smallR}} \frac{1 - \frac{1}{x_k^- \tilde{y}_j}}{1- \frac{1}{x_k^+ \tilde{y}_j}} ,
    \end{split} \\
    1 &= 
    \prod_{j=1}^{K^{\I}_{\smallL}} \frac{y_k - x_j^+}{y_k - x_j^-}
    \prod_{j=1}^{K^{\I}_{\smallR}} \frac{1 - \frac{1}{y_k \tilde{x}_j^-}}{1- \frac{1}{y_k \tilde{x}_j^+}} \,, \\
%
%
    \begin{split}
    \label{eq:BAE11:Right}
      \left(\frac{\tilde{x}_k^+}{\tilde{x}_k^-}\right)^{\ell} &=
       \prod_{\substack{j = 1\\j \neq k}}^{M^{\I}_{\smallR}} \mathscr{S}^{\smallRR}_{\tilde{k}\tilde{\jmath}}
      \prod_{j=1}^{M^{\II}_{\smallR}} \frac{\tilde{x}_k^+ - \tilde{y}_j}{\tilde{x}_k^- - \tilde{y}_j}
      \prod_{j=1}^{M^{\I}_{\smallL}} \sqrt{\frac{1- \frac{1}{\tilde{x}_k^- x_j^-}}{1- \frac{1}{\tilde{x}_k^+ x_j^+}}} \mathscr{S}^{\smallRL}_{\tilde{k}j}
      \prod_{j=1}^{M^{\II}_{\smallL}} \frac{1 - \frac{1}{\tilde{x}_k^+ y_j}}{1- \frac{1}{\tilde{x}_k^- y_j}}\,,
    \end{split} \\
    1 &= 
     \prod_{j=1}^{M^{\II}_{\smallR}} \frac{\tilde{y}_k - \tilde{x}_j^-}{\tilde{y}_k - \tilde{x}_j^+}
    \prod_{j=1}^{M^{\II}_{\smallL}} \frac{1 - \frac{1}{\tilde{y}_k x_j^+}}{1- \frac{1}{\tilde{y}_k x_j^-}}\,.
\end{align}
These equations should be supplemented by the level matching condition
\begin{equation}
1=\prod_{j=1}^{M^{\I}_{\smallL}}\frac{x^{+}_j}{x^{-}_j}\;
\prod_{j=1}^{M^{\I}_{\smallR}}\frac{\tilde{x}^{+}_j}{\tilde{x}^{-}_j}\,.
\end{equation}
By these equations, we can find the asymptotic spectrum of the theory. In fact, once we specify a state by its classical dimension and angular momenta, the labels $M_{\smallL,\smallR}^{\I}$, $M_{\smallL,\smallR}^{\II}$ and~$\ell$ will be  fixed. Then it only remains to solve the above equations: first for the auxiliary roots $y,\tilde{y}$ and then for the momenta. Using those, we can compute the energy by the dispersion relation
\begin{equation}
E(\{x^{\pm},\tilde{x}^{\pm}\})=\sum_{j=1}^{M_{\smallL}^{\I}} \sqrt{1+4h^2\sin^2\frac{{p}_j}{2}} 
+
\sum_{j=1}^{M_{\smallR}^{\I}} \sqrt{1+4h^2\sin^2\frac{\tilde{p}_j}{2}}\,.
\end{equation}
Since the S~matrix and Bethe ansatz that we just considered now does not correspond to the physical theory we are interested in, we will postpone the identification of the excitation numbers with physical charges to the next section, where we will consider the full~$\psu(1,1|2)^2$ Bethe ansatz. Before moving to that, remark that at the beginning of the nesting procedure we choose the set of excitations $V_A^{\I}$ in~\eqref{eq:BA-excit-set}. The other choice quite clearly would lead to a set of equation differing by $\bigL\leftrightarrow\bigR$, and hopefully this will leave the spectrum unchanged. This is actually the case, as we show in appendix~\ref{app:duality} by a \emph{duality transformation} of the equations~\cite{Essler:1992nk,Beisert:2005fw}.

\subsection*{Worldsheet picture}
In the worldsheet picture, the derivation of the Bethe ansatz equations (more properly, of the Bethe-Yang equations) follows the same logic as above. The nesting procedure is exactly the same, since the form of the S~matrix (\ie~its non-vanishing entries) is the same. Again, we have two choices of level-I excitations, given by~$V^{\I}_A, V^{\I}_B$ as in~\eqref{eq:BA-excit-set}. For the choice~$V^{\I}_A$, the level-I S~matrices are as before
\begin{equation}
S^{\I,\I}_{\smallLL}=\Ael^{\smallLL}\,,
\qquad
S^{\I,\I}_{\smallRR}=\Fel^{\smallRR}\,,
\qquad
S^{\I,\I}_{\smallLR}=\Cel^{\smallLR}\,,
\qquad
S^{\I,\I}_{\smallRL}=\Del^{\smallRL}\,.
\end{equation}
Now, however, these elements should be read off the \emph{worldsheet} fundamental S~matrix of section~\ref{sec:stringsmat}. As we discussed at length in the previous chapter, those elements differ from their spin-chain counterparts for being~\emph{their inverse}%
\footnote{%
Looking at formula~\eqref{eq:Smatrix-comparison} we see that a graded permutation appears in the map between the spin-chain and worldsheet frame. Since we are dealing with diagonal elements and due to left-right symmetry, this plays no role.
}
 and featuring additional factors of the form~$e^{ip}$.
 
The diagonalisation works in the same way. In fact, even the compatibility equations that fix the higher-level S-matrix elements~$S^{\II,\I}$, take essentially the same form as above, \eg~as in~\eqref{eq:BAcompat-explicit}. In the worldsheet picture however we have no length-changing effects, and the non-trivial coproduct is entirely encoded in the form of the S~matrix elements of section~\ref{sec:stringsmat}. Unsurprisingly, we find the inverse of the spin-chain results up to some factors of~$e^{ip}$. More specifically, we have for the level-II excitations coming from~$\mathbf{q}^{\smallL}$
\begin{equation}
\begin{aligned}
&S^{\II,\I}_{\smallLL}(y^{\smallL},p)=
\frac{y^{\smallL}-x^-_p}{y^{\smallL}-x^+_p}\,e^{+\frac{i}{2}p}\,,
\qquad	
&&S^{\II,\I}_{\smallRR}(y^{\smallL},p)=
\frac{1-\frac{1}{y^{\smallL}x^+_p}}{1-\frac{1}{y^{\smallL}x^-_p}}\,e^{+\frac{i}{2}p}\,,\\
&S^{\II,\I}_{\smallRL}(y^{\smallL},p)=S^{\II,\I}_{\smallLL}(y^{\smallL},p)\,,
\qquad
&&S^{\II,\I}_{\smallLR}(y^{\smallL},p)=S^{\II,\I}_{\smallRR}(y^{\smallL},p)\,,
\end{aligned}
\end{equation}
and for the ones coming from~$\bar{\mathbf{q}}^{\smallR}$
\begin{equation}
\begin{aligned}
&S^{\II,\I}_{\smallLL}(y^{\smallR},p)=
\frac{1-\frac{1}{y^{\smallR}x^-_p}}{1-\frac{1}{y^{\smallR}x^+_p}}\,e^{-\frac{i}{2}p}\,,
\qquad	
&&S^{\II,\I}_{\smallRR}(y^{\smallR},p)=
\frac{y^{\smallR}-x^+_p}{y^{\smallR}-x^-_p}\,e^{-\frac{i}{2}p}\,,\\
&S^{\II,\I}_{\smallRL}(y^{\smallR},p)=S^{\II,\I}_{\smallLL}(y^{\smallR},p)\,,
\qquad
&&S^{\II,\I}_{\smallLR}(y^{\smallR},p)=S^{\II,\I}_{\smallRR}(y^{\smallR},p)\,,
\end{aligned}
\end{equation}
while again for all the excitations it is
\begin{equation}
S^{\II,\II}=1\,.
\end{equation}
Using as before the short-hand notation~\eqref{eq:BA-shorthand}, we can write the Bethe-Yang equations as
\begin{align}\label{eq:bethe-equations-full}
    \begin{split}
      \left(\frac{x_k^+}{x_k^-}\right)^{\ell+\delta}\!\!\!\!\!\!\!\! &=e^{\frac{i}{2}P_{\text{tot.}}}\!\!
      \prod_{\substack{j = 1\\j \neq k}}^{M^{\I}_{\smallL}} \frac{x_k^+ - x_j^-}{x_k^- - x_j^+} \mathscr{S}^{\smallLL}_{kj}
      \prod_{j=1}^{M^{\II}_{\smallL}} \frac{x_k^- - y_j}{x_k^+ - y_j} 
      \prod_{j=1}^{M^{\I}_{\smallR}} \sqrt{\frac{1- \frac{1}{x_k^+ \tilde{x}_j^+}}{1- \frac{1}{x_k^- \tilde{x}_j^-}}} \mathscr{S}^{\smallLR}_{k\tilde{\jmath}}
      \prod_{j=1}^{M^{\II}_{\smallR}} \frac{1 - \frac{1}{x_k^- \tilde{y}_j}}{1- \frac{1}{x_k^+ \tilde{y}_j}} ,
    \end{split}
 \end{align}
 \begin{align}
    1 &= 
    e^{-\frac{i}{2}P_{\text{tot.}}}\prod_{j=1}^{M^{\I}_{\smallL}} \frac{y_k - x_j^+}{y_k - x_j^-}
    \prod_{j=1}^{M^{\I}_{\smallR}} \frac{1 - \frac{1}{y_k \tilde{x}_j^-}}{1- \frac{1}{y_k \tilde{x}_j^+}} \,,\\
%
%
    \begin{split}
      \left(\frac{\tilde{x}_k^+}{\tilde{x}_k^-}\right)^{\ell+\delta}\!\!\!\!\!\!\!\! &=\;
       \prod_{\substack{j = 1\\j \neq k}}^{M^{\I}_{\smallR}} \mathscr{S}^{\smallRR}_{\tilde{k}\tilde{\jmath}}
      \prod_{j=1}^{M^{\II}_{\smallR}} \frac{\tilde{x}_k^+ - \tilde{y}_j}{\tilde{x}_k^- - \tilde{y}_j}
      \prod_{j=1}^{M^{\I}_{\smallL}} \sqrt{\frac{1- \frac{1}{\tilde{x}_k^- x_j^-}}{1- \frac{1}{\tilde{x}_k^+ x_j^+}}} \mathscr{S}^{\smallRL}_{\tilde{k}j}
      \prod_{j=1}^{M^{\II}_{\smallL}} \frac{1 - \frac{1}{\tilde{x}_k^+ y_j}}{1- \frac{1}{\tilde{x}_k^- y_j}}\,,
    \end{split} \\
    1 &= 
     e^{+\frac{i}{2}P_{\text{tot.}}}\prod_{j=1}^{M^{\II}_{\smallR}} \frac{\tilde{y}_k - \tilde{x}_j^-}{\tilde{y}_k - \tilde{x}_j^+}
    \prod_{j=1}^{M^{\II}_{\smallL}} \frac{1 - \frac{1}{\tilde{y}_k x_j^+}}{1- \frac{1}{\tilde{y}_k x_j^-}}\,,
\end{align}
where now~$\ell$ has the interpretation of the worldsheet size%
\footnote{%
Strictly speaking, in this simple exercise the Bethe equations are not supposed to be matched with a genuine worldsheet theory---we will use the full~$\psu(1|1)^4_{\text{c.e.}}$ S~matrix for that purpose.
}
 and we introduced the shift
\begin{equation}
\delta=-\tfrac{1}{2}M^{\I}_{\smallL}+\tfrac{1}{2}M^{\II}_{\smallL}-\tfrac{1}{2}M^{\II}_{\smallR}\,.
\end{equation}
In order to rewrite the equations in terms of this quantity and the total momentum~$P_{\text{tot.}}$, we observed that whenever we multiply the momentum-carrying terms by an expression such as~$e^{i(p\, f(q)-q\,f(p))}$, we can collect 
\begin{equation}
\label{eq:BAframefactors}
\prod_{j\neq k}^{M^{\I}_{\smallL}}e^{i(p_k f(p_j)-p_j\,f(p_k))}
\prod_{j=1}^{M^{\I}_{\smallR}}e^{i(p_k f(p_j)-p_j\,f(p_k))}
=e^{i p_k\, F_{\text{tot.}}}\,e^{-i P_{\text{tot.}}\,f(p_k)  }\,,
\end{equation}
where~$F_{\text{tot.}}$ is the sum of~$f(p_j)$ on all momentum-carrying excitations.
In this simple case, we can use this identity with~$ f(p)=1$. We then have that the only new features of the worldsheet Bethe ansatz are a shift in the notion of length with respect to the spin-chain one, and some (fractionary) powers of the total momentum.

\section{The~\texorpdfstring{$\psu(1,1|2)^2$}{psu(1,1|2)**2} Bethe ansatz}
Let us finally work out the Bethe equations for the full~$\psu(1|1)^4_{\text{c.e.}}$ S~matrix, again in a spin-chain picture. The derivation will be similar to the one we worked out in detail in the previous section, and we will only sketch the most conceptual points, referring the reader to ref.~\cite{Borsato:2013qpa} for more details.

\subsection*{Sketch of the nesting procedure}
Again, we want to construct asymptotic eigenstates of the multi-magnon S~matrix. 
We start from the level-I vacuum, which again  is just $\ket{0}^{\I} \equiv \ket{Z^\ell}$. 
To proceed with nesting, we need to choose a maximal set of excitations that scatter by pure transmission among each other. The structure of $\mathcal{S}$ leads to four possible choices
\begin{equation}
\label{eq:level-II-vacuum}
  \begin{aligned}
    V^{\II}_A &= \{ \Phi^{\smallL}_{+{+}},{\Phi}^{\smallR}_{-{-}}\}, \qquad \qquad 
     &V^{\II}_B= \{ \Phi^{\smallL}_{-{-}},{\Phi}^{\smallR}_{+{+}}\}, \\
    V^{\II}_C &= \{ \Phi^{\smallL}_{+{-}},{\Phi}^{\smallR}_{-{+}}\}, \qquad \qquad 
    &V^{\II}_D = \{ \Phi^{\smallL}_{-{+}},{\Phi}^{\smallR}_{+{-}}\}. 
\end{aligned}
\end{equation}
Each candidate level-II vacuum is composed of one left and one right excitation, that are either both bosonic or both fermionic. Correspondingly, we will have different choices of the lowering operators.
In the following we will choose the set $V^{\II}_A$ to construct the level-II vacuum. As we expect, the other possible choices are related by dualities, allowing us to write all-loop Bethe equations in four different ways, as we discuss in appendix~\ref{app:duality}.

Level-II excitations can be found as before by acting with the lowering operators on the level-II vacuum.
Now we have four supercharges at our disposal, \ie~$\mathbf{Q}^{1\smallL},\mathbf{Q}^{2\smallL}$ and~$\overline{\mathbf{Q}}{}^{1\smallR},\overline{\mathbf{Q}}{}^{2\smallR}$. In table~\ref{tab:lev-II-exc} we collect their action on the fields in $V^{\II}_A$, writing also the length-changing effect.
\begin{table}
  \centering
  \begin{tabular}{r@{\hspace{1.5em}}cccc}
    \toprule
    & $\mathbf{Q}^{1\smallL}$ & $\mathbf{Q}^{2\smallL}$ & $\overline{\mathbf{Q}}{}^{1\smallR}$ & $\overline{\mathbf{Q}}{}^{2\smallR}$  \\
    \midrule
    $\Phi_{+{+}}^{\smallL}$    &$\Phi_{-{+}}^{\smallL}$    &$\Phi_{+{-}}^{\smallL}$    &$\Phi_{-{+}}^{\smallL} Z^-$    &$\Phi_{+{-}}^{\smallL}Z^-$     \\
    ${\Phi}_{-{-}}^{\smallR}$ & ${\Phi}_{+{-}}^{\smallR}Z^+$    &${\Phi}_{-{+}}^{\smallR}Z^+$    &${\Phi}_{+{-}}^{\smallR}$    &${\Phi}_{-{+}}^{\smallR}$    \\
    \bottomrule
  \end{tabular}
  \caption{%
    Action of the lowering operators on the states of the level-II vacuum~$V^{\II}_A$, including length-changing effects in the spin-chain picture.
  }
  \label{tab:lev-II-exc}
\end{table}

The presence of additional fields and charges produces a new feature: the fields $\Phi_{-{-}}^{\smallL}, {\Phi}_{+{+}}^{\smallR}$ can be created by acting twice with the supercharges (\ie, respectively by consecutively applying $\mathbf{Q}^{1\smallL}$ and $\mathbf{Q}^{2\smallL}$ or $\Phi_{+{+}}^{\smallL}$ and $\overline{\mathbf{Q}}{}^{1\smallR}$ and $\overline{\mathbf{Q}}{}^{2\smallR}$ on $\bar{\Phi}_{-{-}}^{\smallR}$)
and therefore can be considered as composite excitations. As such, they will not explicitly appear in the Bethe ansatz, but rather be represented by pair of suitable excitations.

Apart from this, the calculations to diagonalise the S~matrix are exactly the same as the ones of the previous section. This is not surprising, since the lowering operators we should consider now are tensor products of the previous ones, and we deal again with  doublets of $\su(1|1)$---now we have \eg~$(\Phi_{+{+}}^{\smallL} | \Phi_{-{+}}^{\smallL})$ instead of $(\phi^{\smallL}|\psi^{\smallL})$.
In particular, we find once again non-trivial matrices  $S^{\II,\I}$, with the same functional form as before, and  trivial~$S^{\II,\II}$ matrix for all level-II excitations.

\subsection*{Bethe equations}
\label{sec:BAE}

As before, to obtain the Bethe ansatz equations we impose periodic boundary conditions on a spin-chain of finite length $\ell$ and use the S~matrix in its diagonal form. We have again two momentum-carrying excitations, \ie~the ones in~$V^{\I}_A$. We denote as before the corresponding Zukhovski variables by $x^\pm$ and $\tilde{x}^\pm$ for left and right excitations respectively.
Their number is denoted by $M^{\I}_{\smallL}$ and $M^{\I}_{\smallR}$. We have two auxiliary ``left'' roots denoted by $y_1,y_2$, corresponding respectively to the action of the supercharges $\mathbf{Q}^{1\smallL}$, $\mathbf{Q}^{2\smallL}$. The two auxiliary ``right'' roots are denoted by $y_{\tilde{1}},y_{\tilde{2}}$ and they correspond respectively to the action of the supercharges $\overline{\mathbf{Q}}{}^{1\smallR}$, $\overline{\mathbf{Q}}{}^{2\smallR}$. The number of the corresponding excitations is denoted by $M^{\II}_{1\smallL}$, $M^{\II}_{2\smallL}$, $M^{\II}_{1\smallR}$, and $M^{\II}_{2\smallR}$.
The Bethe equations then read
\begin{align}
\label{eq:BA-1}
    1 &= 
    \prod_{j=1}^{M^{\I}_{\smallL}} \frac{y_{1,k} - x_j^+}{y_{1,k} - x_j^-}
    \prod_{j=1}^{M^{\I}_{\smallR}} \frac{1 - \frac{1}{y_{1,k} \tilde{x}_j^-}}{1- \frac{1}{y_{1,k} \tilde{x}_j^+}} , \\
    \begin{split}
    \label{eq:BA-2}
      \left(\frac{x_k^+}{x_k^-}\right)^{\ell} &=
      \prod_{\substack{j = 1\\j \neq k}}^{K^{\I}_{\smallL}} \frac{x_k^+ - x_j^-}{x_k^- - x_j^+} \frac{1- \frac{1}{x_k^+ x_j^-}}{1- \frac{1}{x_k^- x_j^+}} \sigma^2(x_k,x_j)
      \prod_{j=1}^{M^{\II}_{1\smallL}} \frac{x_k^- - y_{1,j}}{x_k^+ - y_{1,j}}
      \prod_{j=1}^{M^{\II}_{2\smallL}} \frac{x_k^- - y_{2,j}}{x_k^+ - y_{2,j}}
      \\ &\phantom{\ = \ }\times
      \prod_{j=1}^{M^{\I}_{\smallR}} \frac{1- \frac{1}{x_k^+ \tilde{x}_j^+}}{1- \frac{1}{x_k^- \tilde{x}_j^-}} \frac{1- \frac{1}{x_k^+ \tilde{x}_j^-}}{1- \frac{1}{x_k^- \tilde{x}_j^+}} \tilde{\sigma}^2(x_k,\tilde{x}_j)
      \prod_{j=1}^{M^{\II}_{1\smallR}} \frac{1 - \frac{1}{x_k^- y_{\tilde{1},j}}}{1- \frac{1}{x_k^+ y_{\tilde{1},j}}}
      \prod_{j=1}^{M^{\II}_{2\smallR}} \frac{1 - \frac{1}{x_k^- y_{\tilde{2},j}}}{1- \frac{1}{x_k^+ y_{\tilde{2},j}}} ,
    \end{split}
    \end{align}
\begin{align}
    \label{eq:BA-3}
    1 &= 
    \prod_{j=1}^{M^{\I}_{\smallL}} \frac{y_{2,k} - x_j^+}{y_{2,k} - x_j^-}
    \prod_{j=1}^{M^{\I}_{\smallR}} \frac{1 - \frac{1}{y_{2,k} \tilde{x}_j^-}}{1- \frac{1}{y_{2,k} \tilde{x}_j^+}} ,\\
    \label{eq:BA-1b}
    1 &= 
    \prod_{j=1}^{M^{\I}_{\smallR}} \frac{y_{\tilde{1},k} - \tilde{x}_j^-}{y_{\tilde{1},k} - \tilde{x}_j^+}
    \prod_{j=1}^{M^{\I}_{\smallL}} \frac{1 - \frac{1}{y_{\tilde{1},k} x_j^+}}{1- \frac{1}{y_{\tilde{1},k} x_j^-}} , \\
    \begin{split}
    \label{eq:BA-2b}
      \left(\frac{\tilde{x}_k^+}{\tilde{x}_k^-}\right)^{\ell} &=
      \prod_{\substack{j = 1\\j \neq k}}^{M^{\I}_{\smallR}} \frac{\tilde{x}_k^- - \tilde{x}_j^+}{\tilde{x}_k^+ - \tilde{x}_j^-} \frac{1- \frac{1}{\tilde{x}_k^+ \tilde{x}_j^-}}{1- \frac{1}{\tilde{x}_k^- \tilde{x}_j^+}} \sigma^2(\tilde{x}_k,\tilde{x}_j)
      \prod_{j=1}^{M^{\II}_{1\smallR}} \frac{\tilde{x}_k^+ - y_{\tilde{1},j}}{\tilde{x}_k^- - y_{\tilde{1},j}}
      \prod_{j=1}^{M^{\II}_{2\smallR}} \frac{\tilde{x}_k^+ - y_{\tilde{2},j}}{\tilde{x}_k^- - y_{\tilde{2},j}}
      \\ &\phantom{\ = \ }\times
      \prod_{j=1}^{M^{\I}_{\smallL}} \frac{1- \frac{1}{\tilde{x}_k^- x_j^-}}{1- \frac{1}{\tilde{x}_k^+ x_j^+}} \frac{1- \frac{1}{\tilde{x}_k^+ x_j^-}}{1- \frac{1}{\tilde{x}_k^- x_j^+}} \tilde{\sigma}^2(\tilde{x}_k,x_j)
      \prod_{j=1}^{M^{\II}_{1\smallL}} \frac{1 - \frac{1}{\tilde{x}_k^+ y_{1,j}}}{1- \frac{1}{\tilde{x}_k^- y_{1,j}}}
      \prod_{j=1}^{M^{\II}_{2\smallL}} \frac{1 - \frac{1}{\tilde{x}_k^+ y_{3,j}}}{1- \frac{1}{\tilde{x}_k^- y_{2,j}}} ,
    \end{split} \\
    \label{eq:BA-3b}
    1 &= 
    \prod_{j=1}^{M^{\I}_{\smallR}} \frac{y_{\tilde{2},k} - \tilde{x}_j^-}{y_{\tilde{2},k} - \tilde{x}_j^+}
    \prod_{j=1}^{M^{\I}_{\smallL}} \frac{1 - \frac{1}{y_{\tilde{2},k} x_j^+}}{1- \frac{1}{y_{\tilde{2},k} x_j^-}} .
\end{align}
Note that, since we are dealing with the physical S~matrix for the massive modes, we normalised the level-I scattering factors in terms of the canonical dressing factors~$\sigma_{pq}$ and~$\widetilde{\sigma}_{pq}$.
The level matching condition is once again
\begin{equation}
\prod_j^{M^{\I}_{\smallL}} \frac{x^+_j}{x^-_j} \, \prod_j^{M^{\I}_{\smallR}} \frac{\tilde{x}^+_j}{\tilde{x}^-_j}=1.
\end{equation}
The total energy of a multi-excitation state that satisfies the Bethe equations and the level matching condition is given by
\begin{equation}
E(\{x^{\pm},\tilde{x}^{\pm}\})=\sum_{j=1}^{M_{\smallL}^{\I}} \sqrt{1+4h^2\sin^2\frac{{p}_j}{2}} 
+
\sum_{j=1}^{M_{\smallR}^{\I}} \sqrt{1+4h^2\sin^2\frac{\tilde{p}_j}{2}}\,.
\end{equation}

Let us now analyse the small-$h$ limit, which in the spin-chain picture is the most natural, since there the left and right chains decouple and the dynamical length-changing  effects are suppressed. We will come back to the large-$h$ limit in section~\ref{sec:comparisons}, when we will compare it with an independent result, the ``finite-gap'' equations.

\subsection*{Small-\texorpdfstring{$h$}{h} limit and Cartan matrix}
\label{sec:small-h-limit}
When we are dealing with non-dynamical spin chains having a Lie (super)algebra structure, the Bethe equations can immediately be written down using data from the algebra and its representations~\cite{Ogievetsky:1986hu,Minahan:2002ve}. In particular, in the case of~$\psu(1,1|2)^2$, we should write
\begin{equation}
  \label{eq:BE-one-loop}
  \left( \frac{u_{l,k} + \frac{i}{2} w_l}{u_{l,k} - \frac{i}{2} w_l} \right)^{\ell}
  = \prod_{\substack{j = 1\\j \neq k}}^{K_l} \frac{u_{l,k} - u_{l,j} + \frac{i}{2} A_{ll}}{u_{l,k} - u_{l,j} - \frac{i}{2} A_{ll}}\;
  \prod_{l' \neq l} \left(\prod_{j=1}^{K_{l'}} 
  \frac{u_{l,k} - u_{l',j} + \frac{i}{2} A_{ll'}}{u_{l,k} - u_{l,j} - \frac{i}{2} A_{ll'}}\right),
\end{equation}
where $w_l$ are weights, $A_{ll'}$ are elements of the Cartan matrix of $\psu(1,1|2)^2$, and~$K_l$ are excitation numbers pertaining to each Cartan element. Since, as discussed in appendix~\ref{app:psu112}, each copy of~$\psu(1,1|2)$ has rank~$3$, we expect to find 6 equations from this construction.

On the other hand, when $h\ll1$, we can expand the Zukhovski variables in terms of the spin-chain rapidity~$u$ as 
\begin{equation}
x^\pm\approx \frac{u_x\pm i/2}{h/2},\qquad
y\approx\frac{u_y}{h/2},
\end{equation}
in the left sector, where $u_i$ are finite as $h\to0$, and similarly  in the right sector. In these terms, we have the familiar formulae~(see \eg~\cite{Faddeev:1996iy})
\begin{equation}
p\approx \frac{u+\frac{i}{2}}{u-\frac{i}{2}}\,,
\qquad
E=1+\delta D\approx 1+\frac{2\,h^2}{1+4\,u^2}\,,
\end{equation}
where in the last formula we split the energy into a classical contribution and the correction due to the anomalous dimension, that for~$M$ excitations is
\begin{equation}\label{eq:anom-dim-def}
  \delta D(p) = E(p) - M = 
  -ih \sum_{k=1}^{M} \left(\frac{1}{x_k^-} - \frac{1}{x_k^+}\right) \,,
\end{equation}
where the sum is over all (left and right) momentum-carrying excitations.

\begin{figure}
  \centering
  
\begin{tikzpicture}
  \begin{scope}
    \coordinate (m) at (0cm,0cm);

    \node (v1L) at (-1.25cm, 1cm) [dynkin node] {};
    \node (v2L) at (-1.25cm, 0cm) [dynkin node] {$\scriptscriptstyle +1$};
    \node (v3L) at (-1.25cm,-1cm) [dynkin node] {};

    \draw [dynkin line] (v1L) -- (v2L);
    \draw [dynkin line] (v2L) -- (v3L);
    
    \node (v1R) at (+1.25cm, 1cm) [dynkin node] {};
    \node (v2R) at (+1.25cm, 0cm) [dynkin node] {$\scriptscriptstyle -1$};
    \node (v3R) at (+1.25cm,-1cm) [dynkin node] {};

    \draw [dynkin line] (v1L.south west) -- (v1L.north east);
    \draw [dynkin line] (v1L.north west) -- (v1L.south east);

    \draw [dynkin line] (v3L.south west) -- (v3L.north east);
    \draw [dynkin line] (v3L.north west) -- (v3L.south east);
    
    \draw [dynkin line] (v1R.south west) -- (v1R.north east);
    \draw [dynkin line] (v1R.north west) -- (v1R.south east);

    \draw [dynkin line] (v3R.south west) -- (v3R.north east);
    \draw [dynkin line] (v3R.north west) -- (v3R.south east);

    \draw [dynkin line] (v1R) -- (v2R);
    \draw [dynkin line] (v2R) -- (v3R);

    \begin{pgfonlayer}{background}
      \draw [inverse line] [out=  0+30,in= 120] (v1L) to (v2R);
      \draw [inverse line] [out=  0-30,in=240] (v3L) to (v2R);
      \draw [inverse line] [out=180-30,in= 60] (v1R) to (v2L);
      \draw [inverse line] [out=180+30,in=300] (v3R) to (v2L);
    \end{pgfonlayer}

    \draw [red phase,thick] [out=0,in=180] (v2L) to (v2R);

    \draw [blue phase,thick] [out=180-40,in= 180+40,loop] (v2L) to (v2L);
    \draw [blue phase,thick] [out=-40,in=40,loop] (v2R) to (v2R);
  \end{scope}

  \begin{scope}[xshift=+3cm,yshift=-0.75cm]
    \draw [dynkin line]  (0cm,1.5cm) -- (1cm,1.5cm) node [anchor=west,black] {\small Dynkin links};
    \draw [inverse line] (0cm+0.75pt,1.0cm) -- (1cm,1.0cm) node [anchor=west,black] {\small Fermionic inversion symmetry links};
    \draw [blue phase,thick]   (0cm,0.5cm) -- (1cm,0.5cm) node [anchor=west,black] {\small Dressing phase $\sigma_{pq}$};
    \draw [red phase,thick]    (0cm,0.0cm) -- (1cm,0.0cm) node [anchor=west,black] {\small Dressing phase $\widetilde{\sigma}_{pq}$};
  \end{scope}
\end{tikzpicture}

  \caption{The Dynkin diagram for $\psu(1,1|2)^2$ with the various interaction terms appearing in the Bethe ansatz indicated. The label $\pm 1$ inside the middle Dynkin nodes indicate the $\su(2)$ and $\sl(2)$ gradings of the left- and right-moving sectors.}
  \label{fig:bethe-equations}
\end{figure}
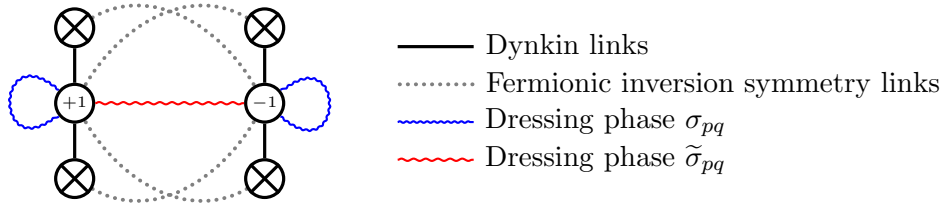

If we assume that the dressing phases $\sigma$ and $\tilde{\sigma}$ expand trivially in this limit%
\footnote{
In the next chapter we will return on this issue and see that this is not exactly the case.
},
 we indeed find that the Bethe ansatz takes the form~\eqref{eq:BE-one-loop} and we can read off the resulting Cartan matrix
\begin{equation}\label{eq:cartan}
  A = \begin{pmatrix}
    0  & -1 &0 &0 & 0 &0 \\
    -1  & +2 & -1 & 0 & 0 & 0 \\
    0 & -1 & 0 &0 & 0 &0 \\
    0 & 0 &0 & 0 & +1  & 0 \\
    0 & 0 & 0 & +1  & -2 & +1 \\
    0 & 0 &0 & 0 & +1 & 0
  \end{pmatrix}.
\end{equation}
\begin{figure}
  \centering

  \subfloat[\label{fig:dynkin-su22-su-main}]{
    \begin{tikzpicture}
      [
      thick,
      node/.style={shape=circle,draw,thick,inner sep=0pt,minimum size=5mm}
      ]

      \useasboundingbox (-1.5cm,-1cm) rectangle (1.5cm,1cm);

      \node (v1) at (-1.1cm, 0cm) [node] {};
      \node (v2) at (  0.0cm, 0cm) [node] {};
      \node (v3) at (  1.1cm, 0cm) [node] {};

      \draw (v1.south west) -- (v1.north east);
      \draw (v1.north west) -- (v1.south east);

      \draw (v3.south west) -- (v3.north east);
      \draw (v3.north west) -- (v3.south east);

      \draw (v1) -- (v2);
      \draw (v2) -- (v3);

      \node at (v2.south) [anchor=north] {$+1$};
    \end{tikzpicture}
  }
  \hspace{1cm}
  \subfloat[\label{fig:dynkin-su22-fff-main}]{
    \begin{tikzpicture}
      [
      thick,
      node/.style={shape=circle,draw,thick,inner sep=0pt,minimum size=5mm}
      ]

      \useasboundingbox (-1.5cm,-1cm) rectangle (1.5cm,1cm);

      \node (v1) at (-1.1cm, 0cm) [node] {};
      \node (v2) at (  0.0cm, 0cm) [node] {};
      \node (v3) at (  1.1cm, 0cm) [node] {};

      \draw (v1.south west) -- (v1.north east);
      \draw (v1.north west) -- (v1.south east);

      \draw (v2.south west) -- (v2.north east);
      \draw (v2.north west) -- (v2.south east);

      \draw (v3.south west) -- (v3.north east);
      \draw (v3.north west) -- (v3.south east);

      \draw (v1) -- (v2);
      \draw (v2) -- (v3);

      \node at (v2.south) [anchor=north] {$\pm 1$};
    \end{tikzpicture}
  }
  \hspace{1cm}
  \subfloat[\label{fig:dynkin-su22-sl-main}]{
    \begin{tikzpicture}
      [
      thick,
      node/.style={shape=circle,draw,thick,inner sep=0pt,minimum size=5mm}
      ]

      \useasboundingbox (-1.5cm,-1cm) rectangle (1.5cm,1cm);

      \node (v1) at (-1.1cm, 0cm) [node] {};
      \node (v2) at (  0.0cm, 0cm) [node] {};
      \node (v3) at (  1.1cm, 0cm) [node] {};

      \draw (v1.south west) -- (v1.north east);
      \draw (v1.north west) -- (v1.south east);

      \draw (v3.south west) -- (v3.north east);
      \draw (v3.north west) -- (v3.south east);

      \draw (v1) -- (v2);
      \draw (v2) -- (v3);

      \node at (v2.south) [anchor=north] {$-1$};
    \end{tikzpicture}
  }
  
  \caption{Three Dynkin diagrams for $\psu(1,1|2)$. From left to right, we depict the~$\su(2)$ grading, two fully fermionic gradings and the~$\sl(2)$ grading.}
  \label{fig:dynkin-su22-main}
\end{figure}
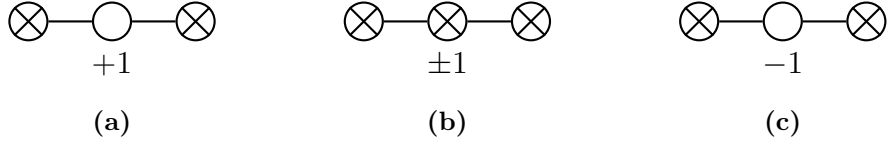
Comparing this with the Cartan matrices in appendix~\ref{app:psu112}, in particular equations~\eqref{eq:Cartan-su2} and~\eqref{eq:Cartan-sl2}, we see that this does correspond to $\psu(1,1|2)^2$, with different gradings for the two factors of the algebra. The left copy is in the~$\su(2)$ grading, while the right one is in the~$\sl(2)$ grading, represented by the Dynkin diagrams of figure~\ref{fig:dynkin-su22-su-main} and~\ref{fig:dynkin-su22-sl-main}.
This shows that there is a strict connection between the type of particles appearing in the Bethe ansatz and the nodes of the Dynkin diagrams. In the Cartan basis is natural to identify the nodes by $1,2,3$ for the left algebra and $\tilde{1},\tilde{2},\tilde{3}$ for the right one. We can identify the momentum-carrying excitations with the middle nodes,
\begin{equation}
\label{eq:labelcartan1}
M^{\I}_{\smallL}=K_2\,,
\qquad
M^{\I}_{\smallR}=K_{\tilde{2}}\,,
\end{equation}
and the remaining ones with the peripheral nodes
\begin{equation}
\label{eq:labelcartan2}
M^{\II}_{1\smallL}=K_1\,,\quad M^{\II}_{2\smallL}=K_3\,,
\qquad
M^{\II}_{1\smallR}=K_{\tilde{1}}\,,\quad M^{\II}_{2\smallR}=K_{\tilde{3}}\,.
\end{equation}
In this way it is easy to represent the Bethe equations pictorially, as in figure~\ref{fig:bethe-equations}.
There, the solid links are the ones that survive the~$h\to0$ limit (together with the self~interaction of the momentum-carrying nodes), and therefore can be read off directly from the Dynkin diagram. The dashed ones can be interpreted as arising from the $\mathbbm{Z}_4$-graded structure of the model, and finally the curly lines are couplings involving the dressing phases, which as discussed can appear only between momentum-carrying nodes.

Not surprisingly, different Dynkin diagrams would appear, should we choose different level-I excitations in nesting procedure. As we discuss in appendix~\ref{app:duality}, this  amounts to dualisation of the Bethe equations, \ie~to replacing a set of roots with an equivalent one. In particular, after dualisation of the nodes 1 and $\tilde{1}$ the Bethe equations are written in a different grading, where all the nodes of the Dynkin diagrams are fermionic. From the weak coupling expansion we get the Cartan matrix
\begin{equation}\label{eq:cartan-dual}
  \tilde{A} =
  \begin{pmatrix}
    0  & 1   &0 &0 & 0 &0 \\
    1  & 0 & -1 & 0 & 0 & 0 \\
    0 & -1 & 0 &0 & 0 &0 \\
    0 & 0 &0 & 0 & -1  & 0 \\
    0 & 0 & 0 & -1  & 0 & 1 \\
    0 & 0 &0 & 0 & 1 & 0
  \end{pmatrix},
\end{equation}
corresponding to the fermionic gradings in~\eqref{eq:Cartan-ferm}. If we had dualised the nodes 3 and $\tilde{3}$ instead we would have found the Cartan matrix $-\tilde{A}$. The consecutive dualisation of 1, $\tilde{1}$ and 3, $\tilde{3}$ gives the Cartan matrix $-A$. These last two choices are once again a manifestation of left-right symmetry.

\subsection*{Global charges}
By expanding the Bethe equations around large values of the spectral parameter~$u$ we should obtain the global charges of the symmetry algebra~\cite{Beisert:2005fw,OhlssonSax:2011ms}. In doing so we will assume again that the dressing factors $\sigma_{pq}$ and $\widetilde{\sigma}_{pq}$ do not contribute to the charges.%
\footnote{
In the next chapter we will see that the dressing factors do contribute, and in a troubling way.
}

As we have seen above, the left- and right-moving sectors of the Bethe equations are naturally written using different gradings of the $\psu(1,1|2)$ algebra. The Dynkin labels $r_1$, $r_2$ and $r_3$ for the left-movers therefore give the eigenvalues of the Cartan generators $\gen{h}_j$ given in~\eqref{eq:SC-basis-su2}, while the labels $r_{\tilde{1}}$, $r_{\tilde{2}}$ and $r_{\tilde{3}}$ for the right-movers correspond to the generators $\widetilde{\gen{h}}_j$ in~\eqref{eq:SC-basis-sl2}.
Expanding the Bethe equations around small values of the momentum (or equivalently large values of the rapidity~$u$) we find
\begin{equation}
  \begin{aligned}
    \label{eq:dynkinlab}
    r_1 &= r_3 = + K_2 + \tfrac{1}{2} \delta D , \qquad & 
    r_2 &= \ell+K_1-2K_2+K_3 , \\
    r_{\tilde{1}} &= r_{\tilde{3}} =- K_{\tilde{2}}  -\tfrac{1}{2} \delta D , \qquad &
    r_{\tilde{2}} &= \ell - K_{\tilde{1}} + 2K_{\tilde{2}} - K_{\tilde{3}} + \delta D,
  \end{aligned} 
\end{equation}
where the anomalous dimension $\delta D$ is given  by~\eqref{eq:anom-dim-def}.

A representation of $\psu(1,1|2)^2$ can be labeled by the eigenvalues of the highest weight state under the four generators $\mathbf{L}_3^{\L}$, $\mathbf{L}_3^{\R}$, $\mathbf{J}_3^{\L}$ and~$\mathbf{J}_3^{\R}$. It is useful to combine them into the charges
\begin{equation}\label{eq:cartan-generators}
  \begin{aligned}
    \mathbf{D} &= \mathbf{L}_3^{\L} + \mathbf{L}_3^{\R} , \qquad &
    \mathbf{J} &= \mathbf{J}_3^{\L} + \mathbf{J}_{3}^{\R} , \\
    \mathbf{L} &= \mathbf{L}_3^{\L} - \mathbf{L}_3^{\R} , \qquad &
    \mathbf{K} &= \mathbf{J}_3^{\L} - \mathbf{J}_3^{\R} .
  \end{aligned}
\end{equation}
The most important of these is the generator of dilatations~$\mathbf{D} $, that is the dual of the target-space energy. It is related to the spin-chain Hamiltonian~$\mathbf{H}$ by
\begin{equation}
  \mathbf{H} = \mathbf{D} - \mathbf{J} ,
\end{equation}
a relation that we have already seen in~\eqref{eq:HEJrel} in terms of the worldsheet theory.

We can now express the eigenvalues of the generators~\eqref{eq:cartan-generators} in terms of the excitation numbers $K_j$ as
\begin{equation}
\label{eq:BAcharges-phys}
  \begin{aligned}
    D &= + K_{\tilde{2}} + \tfrac{1}{2}( K_1 + K_3 - K_{\tilde{1}} - K_{\tilde{3}}) +\ell + \delta D , \\
    L &= - K_{\bar{2}} + \tfrac{1}{2}( K_1 + K_3 + K_{\bar{1}} + K_{\tilde{3}}) , \\
    J &= - K_{2} + \tfrac{1}{2}( K_1 + K_3 - K_{\tilde{1}} - K_{\tilde{3}}) + \ell , \\
    K &= - K_{2} + \tfrac{1}{2}( K_1 + K_3 + K_{\tilde{1}} + K_{\tilde{3}}) .
  \end{aligned}
\end{equation}
Note that the anomalous dimension $\delta D$ only contributes to the eigenvalue $\Delta$ of the dilatation operator. The eigenvalue of the Hamiltonian now takes the form
\begin{equation}
  E =\Delta - J= K_2 + K_{\tilde{2}} + \delta D,
\end{equation}
as expected from equation~\eqref{eq:anom-dim-def}.

\subsection*{Worldsheet picture}
As we have seen in the previous section, the Bethe ansatz in the worldsheet picture is not very different from the spin-chain one, and in fact can be found in an analogous way.

The resulting equations read
\begin{align}
\label{eq:BA-1-ws}
    1 &= e^{-\frac{i}{2}P_{\text{tot.}}}
    \prod_{j=1}^{M^{\I}_{\smallL}} \frac{y_{1,k} - x_j^+}{y_{1,k} - x_j^-}
    \prod_{j=1}^{M^{\I}_{\smallR}} \frac{1 - \frac{1}{y_{1,k} \tilde{x}_j^-}}{1- \frac{1}{y_{1,k} \tilde{x}_j^+}} , \\
    \begin{split}
    \label{eq:BA-2-ws}
      \left(\frac{x_k^+}{x_k^-}\right)^{\ell+\delta} &=e^{iP_{\text{tot.}}}
      \prod_{\substack{j = 1\\j \neq k}}^{K^{\I}_{\smallL}} \frac{x_k^+ - x_j^-}{x_k^- - x_j^+} \frac{1- \frac{1}{x_k^+ x_j^-}}{1- \frac{1}{x_k^- x_j^+}} \sigma^2(x_k,x_j)
      \prod_{j=1}^{M^{\II}_{1\smallL}} \frac{x_k^- - y_{1,j}}{x_k^+ - y_{1,j}}
      \prod_{j=1}^{M^{\II}_{2\smallL}} \frac{x_k^- - y_{2,j}}{x_k^+ - y_{2,j}}
      \\ &\phantom{\ =\ }\times
      \prod_{j=1}^{M^{\I}_{\smallR}} \frac{1- \frac{1}{x_k^+ \tilde{x}_j^+}}{1- \frac{1}{x_k^- \tilde{x}_j^-}} \frac{1- \frac{1}{x_k^+ \tilde{x}_j^-}}{1- \frac{1}{x_k^- \tilde{x}_j^+}} \tilde{\sigma}^2(x_k,\tilde{x}_j)
      \prod_{j=1}^{M^{\II}_{1\smallR}} \frac{1 - \frac{1}{x_k^- y_{\tilde{1},j}}}{1- \frac{1}{x_k^+ y_{\tilde{1},j}}}
      \prod_{j=1}^{M^{\II}_{2\smallR}} \frac{1 - \frac{1}{x_k^- y_{\tilde{2},j}}}{1- \frac{1}{x_k^+ y_{\tilde{2},j}}} ,
    \end{split} \\
    \label{eq:BA-3-ws}
    1 &= e^{-\frac{i}{2}P_{\text{tot.}}}
    \prod_{j=1}^{M^{\I}_{\smallL}} \frac{y_{2,k} - x_j^+}{y_{2,k} - x_j^-}
    \prod_{j=1}^{M^{\I}_{\smallR}} \frac{1 - \frac{1}{y_{2,k} \tilde{x}_j^-}}{1- \frac{1}{y_{2,k} \tilde{x}_j^+}} , 
    \end{align}
    \begin{align}
    \label{eq:BA-1b-ws}
    1 &= e^{+\frac{i}{2}P_{\text{tot.}}}
    \prod_{j=1}^{M^{\I}_{\smallR}} \frac{y_{\tilde{1},k} - \tilde{x}_j^-}{y_{\tilde{1},k} - \tilde{x}_j^+}
    \prod_{j=1}^{M^{\I}_{\smallL}} \frac{1 - \frac{1}{y_{\tilde{1},k} x_j^+}}{1- \frac{1}{y_{\tilde{1},k} x_j^-}} , \\
    \begin{split}
    \label{eq:BA-2b-ws}
      \left(\frac{\tilde{x}_k^+}{\tilde{x}_k^-}\right)^{\ell+\delta} &=
      \prod_{\substack{j = 1\\j \neq k}}^{M^{\I}_{\smallR}} \frac{\tilde{x}_k^- - \tilde{x}_j^+}{\tilde{x}_k^+ - \tilde{x}_j^-} \frac{1- \frac{1}{\tilde{x}_k^+ \tilde{x}_j^-}}{1- \frac{1}{\tilde{x}_k^- \tilde{x}_j^+}} \sigma^2(\tilde{x}_k,\tilde{x}_j)
      \prod_{j=1}^{M^{\II}_{1\smallR}} \frac{\tilde{x}_k^+ - y_{\tilde{1},j}}{\tilde{x}_k^- - y_{\tilde{1},j}}
      \prod_{j=1}^{M^{\II}_{2\smallR}} \frac{\tilde{x}_k^+ - y_{\tilde{2},j}}{\tilde{x}_k^- - y_{\tilde{2},j}}
      \\ &\phantom{\ =\ }\times
      \prod_{j=1}^{M^{\I}_{\smallL}} \frac{1- \frac{1}{\tilde{x}_k^- x_j^-}}{1- \frac{1}{\tilde{x}_k^+ x_j^+}} \frac{1- \frac{1}{\tilde{x}_k^+ x_j^-}}{1- \frac{1}{\tilde{x}_k^- x_j^+}} \tilde{\sigma}^2(\tilde{x}_k,x_j)
      \prod_{j=1}^{M^{\II}_{1\smallL}} \frac{1 - \frac{1}{\tilde{x}_k^+ y_{1,j}}}{1- \frac{1}{\tilde{x}_k^- y_{1,j}}}
      \prod_{j=1}^{M^{\II}_{2\smallL}} \frac{1 - \frac{1}{\tilde{x}_k^+ y_{3,j}}}{1- \frac{1}{\tilde{x}_k^- y_{2,j}}} ,
    \end{split} \\
    \label{eq:BA-3b-ws}
    1 &= e^{+\frac{i}{2}P_{\text{tot.}}}
    \prod_{j=1}^{M^{\I}_{\smallR}} \frac{y_{\tilde{2},k} - \tilde{x}_j^-}{y_{\tilde{2},k} - \tilde{x}_j^+}
    \prod_{j=1}^{M^{\I}_{\smallL}} \frac{1 - \frac{1}{y_{\tilde{2},k} x_j^+}}{1- \frac{1}{y_{\tilde{2},k} x_j^-}} ,
\end{align}
where now
\begin{equation}
\delta = -M^{\I}_{\smallL}+\tfrac{1}{2}M^{\II}_{1\smallL} +\tfrac{1}{2}M^{\II}_{1\smallL}  -\tfrac{1}{2}M^{\II}_{1\smallR} -\tfrac{1}{2}M^{\II}_{2\smallR}\,.
\end{equation}
By comparing with~\eqref{eq:BAcharges-phys}, we see that the difference between the spin-chain and worldsheet notion of length is given by the charge~$J$, which is the~$\S^3$ angular momentum. Moreover, the total-momentum contributions drop out of the equations upon imposition of the level-matching constraint---the spin-chain and worldsheet construction are equivalent.

Let us remark that~$\ell$ is not a gauge-invariant quantity. Going back to the (\eg~bosonic) string theory discussion of chapter~\ref{ch:sigmamodel}, we  have seen that~$\ell=P_+$ depends on the choice of ligthcone gauge-fixing. In a more general $a$-dependent gauge fixing~\cite{Arutyunov:2004yx,Arutyunov:2005hd,Arutyunov:2006gs}, we would have
\begin{equation}
\ell=P_+=a\,E^{\text{t.s.}}+(1-a)\,J=J+a\,E\,.
\end{equation}
Since the spectrum is supposed to be gauge invariant, the dependence on~$a$ should drop from the Bethe-Yang equation. 
 This can be understood in terms of an overall prefactor for the S~matrix, which in particular has a non-trivial momentum dependence in the form~$e^{i\,a\,(p\,\delta D_q-q\,\delta D_p)}$. It is easy to check that this factor drops out of the crossing equations, and therefore cannot be fixed based on our symmetry arguments. Moreover, from~\eqref{eq:BAframefactors} we see that this produces exactly the energy-dependent shift that can cancel the $a$-dependent part of~$\ell$.
A tree-level perturbative calculation~\cite{Rughoonauth:2012qd, Sundin:2013ypa, Hoare:2013pma} immediately shows that the factor appears in such a way as to cancel the gauge dependence, as it is expected from what happens in~$\AdS_5\times\S^5$, see \eg~ref.~\cite{Klose:2006zd}.

\section{Chapter summary}
In this chapter we have worked out the asymptotic Bethe ansatz equations for the all-loop S~matrix we found earlier. We outlined the procedure for doing so in a spin-chain and worldsheet picture, that turned out to be equivalent.

The Bethe equations can be nicely summarised in figure~\ref{fig:bethe-equations}, where the excitations are represented by the corresponding nodes of the~$\psu(1,1|2)^2$ Dynkin diagram. This is a manifestation of the~$\psu(1,1|2)^2$ symmetry of the spectrum, which in the Bethe ansatz construction is expected~\cite{Arutyunov:2011uz}, see also the discussion in chapter~\ref{ch:outlook}. As we have already seen a number of times, such a symmetry requires the two copies of the superalgebra to be in opposite gradings.

As we stressed in the introduction, these equations describe only the~\emph{asymptotic} spectrum, \ie~the spectrum when~$\ell$ is finite but large. This is due to wrapping effects: either particles wrapping around the worldsheet or the spin-chain non-local interactions wrapping around the chain. While these effects are exponentially suppressed in~$\ell$, a complete treatment of the spectral problem requires accounting for them. This can be done by the mirror thermodynamical Bethe ansatz (Mirror TBA), as we briefly sketched in section~\ref{sec:intro:string}. As we mentioned there, this requires having a handle over the whole asymptotic spectrum of the theory, including massless excitations and bound states. For this reason, such a description is for the moment beyond our reach.

\chapter{Comparison with perturbative calculations}
\label{ch:comparison}

It is finally time to compare the results of the integrability approach to perturbative computations. This will give us some insight on whether our assumptions were correct.

Since the dual~$\CFT_2$ is quite hard to evaluate perturbatively~\cite{Pakman:2009zz}, all the calculations so far available are from the string side. Therefore, they all require the limit~$h\to\infty$. This can be taken in several ways, depending on how the magnon momentum scales. We have already encountered in chapter~\ref{ch:sigmamodel} the BMN or near-plane-wave limit~\cite{Berenstein:2002jq}, whereby the momentum scales as~$p \sim \mathsfit{p}/h$ and the Zhukovsky variables expand~as
\begin{equation}
\label{eq:BMN-expansion}
  x^{\pm}_{\mathsfit{p}} = \left(1 \pm \frac{i \mathsfit{p}}{2 h}\right) \frac{(1+\omega_{\mathsfit{p}})}{\mathsfit{p}} + \order(1/h^2),\qquad \omega_{\mathsfit{p}}=\sqrt{1+{\mathsfit{p}}^2}\; .
\end{equation}
In this approximation, the theory becomes relativistic as the dispersion~$\omega_{\mathsfit{p}}$ shows. Another useful limit is the Maldacena-Swanson or near-flat-space (NFS) limit~\cite{Maldacena:2006rv}, whereby one uses light-cone kinematics and scales~$p \sim p_{-}/\sqrt{h}$. In this regime, loop calculations on the worldsheet are more feasible. Finally, it will be useful to consider the semiclassical limit, whereby we also take the number of excitations to infinity and scale the spectral parameters as
\begin{equation}
\label{eq:FG-expansion}
x^\pm =x\pm\frac{i}{h}\frac{x^2}{x^2-1}+O\left(\frac{1}{h^3}\right)\,.
\end{equation}

The expansion of the S-matrix elements would be straightforward were not for the dressing factors, that require particular care.

\section{Expansions of the dressing factors}\label{sec:expansions}
In chapter~\ref{ch:crossing} we solved the $\AdS_3$ crossing equations~(\ref{eq:crossing1}) in terms of~(\ref{eq:solution}).
In what follows we give the strong- and weak-coupling expansions of these all-loop phases.

\subsection*{Strong-coupling expansion}
The dressing phases admit an expansion in terms of local conserved charges $q_r(p)$~\cite{Arutyunov:2004vx}
\begin{equation}
\label{eq:theta-charges}
  \theta(p_1,p_2) = \sum_{r=1}^{\infty} \sum^\infty_{\substack{s>r \\ r+s=\text{odd}}} c_{r,s}(h) \left[ q_r(p_1)q_s(p_2)-q_r(p_2)q_s(p_1) \right]\,,
\end{equation}
where $c_{r,s}(h)$ are functions of the coupling constant $h$ with expansion
\begin{equation}
c_{r,s}(h)=h c_{r,s}^{(0)}+ c_{r,s}^{(1)}+ c_{r,s}^{(2)}h^{-1}+\dots
\label{eq:cexp}
\end{equation} 
and are antisymmetric in $r,s$. The phase $\widetilde{\theta}(p_1,p_2)$ has a similar expansion where the coefficients will be denoted $\tilde{c}_{r,s}(h)$. The expression above is similar to the corresponding one in $\AdS_5$, but unlike what happens in that case, we will need to include the $r=1$ terms. This new feature was first noted in ref.~\cite{Beccaria:2012kb}. For $r\geq 2$ the conserved charges are given by
\begin{equation}
\begin{aligned}
&q_r(p_k) = 
\newq_r(x_k^+)-\newq_r(x_k^-)= 
\frac{i}{r-1} \left[ \frac{1}{(x_k^+)^{r-1}} - \frac{1}{(x_k^-)^{r-1}}\right]\,,\\
& \newq_r(x)= \frac{i}{r-1} \frac{1}{x^{r-1}}\,,
\end{aligned}
\end{equation}
where we introduced the function $\newq_r(x_k)$ for later convenience.
For $r=1$ the charge is just the momentum
\begin{equation}
q_1(p_k) = \newq_1(x_k^+)-\newq_1(x_k^-)=-i \log{\left(\frac{x_k^+}{x_k^-}\right)}, \qquad \newq_1(x)\equiv i \log\left(\frac{1}{x}\right)\,.
\end{equation}
Expressing $\theta(p_1,p_2)$ in terms of $\chi$ (see equation~\eqref{eq:thchi}), we obtain the expansion
\begin{equation}
  \chi(x,y) = \sum_{r=1}^{\infty} \sum^\infty_{\substack{s>r \\ r+s=\text{odd}}} c_{r,s}(h) \left[ \newq_r(x)\newq_s(y) - \newq_r(y)\newq_s(x) \right].
  \label{eq:chiexp}
\end{equation}
with a corresponding expression for $\widetilde{\chi}$.
The coefficients $c_{r,s}$ and $\tilde{c}_{r,s}$ can be obtained by expanding the integrands through which $\chi$ and $\tilde{\chi}$ are defined. This expansion is in $h\gg 1 $ and then in $x,y\gg 1 $. Recall that our phases involve the Hern\'andez-L\'opez~\cite{Hernandez:2006tk} and Beisert-Eden-Staudacher phases~\cite{Beisert:2006ez}, as well as the phase~ $\chi^{-}$ of eq.~\eqref{eq:chi-}. The expansions for $\chi^{\BES}$ and $\chi^{\HL}$ are well known in the literature, and in particular we have
\begin{equation}
\chi^{\HL}(x,y)=\frac{2}{\pi}\sum_{r=2}^\infty\sum^\infty_{\substack{s>r \\ r+s=\text{odd}}}\frac{(r-1)(s-1)}{(r-s)(r+s-2)}\left[ \newq_r(x)\newq_s(y) - \newq_r(y)\newq_s(x) \right]
\,.
\label{eqchihlcrs}
\end{equation}
The expansion for $\chi^-$ (see equation~\eqref{eq:chi-}) is
\begin{equation}
  \begin{aligned}
    \chi^-(x,y) &= -\frac{1}{\pi}\sum_{r=2}^\infty\sum^\infty_{\substack{s>r \\ r+s=\text{odd}}}\frac{(r-1)^2+(s-1)^2}{(r-s)(r+s-2)}\left[ \newq_r(x)\newq_s(y) - \newq_r(y)\newq_s(x) \right]
    \\ &\phantom{{}={}}
    +\frac{1}{2\pi}\sum^\infty_{\substack{s>1 \\ s=\text{even}}}\left[ \newq_1(x)\newq_s(y) - \newq_1(y)\newq_s(x) \right]\,.
    \label{eqchimcrs}
  \end{aligned}
\end{equation}
Expanding~(\ref{eq:solution}) at large $h$ we find
\begin{equation}
\begin{aligned}
\chi(x,y) &= h\, \chi^{\AFS}(x,y) + \frac{1}{2} ( \chi^{\HL}(x,y) + \chi^{-}(x,y) ) + \mathcal{O}\big(\frac{1}{h}\big)\,, \\ 
\widetilde{\chi}(x,y) &= h\, \chi^{\AFS}(x,y) + \frac{1}{2} ( \chi^{\HL}(x,y) - \chi^{-}(x,y) ) + \mathcal{O}\big(\frac{1}{h}\big)\,,
\end{aligned}
\end{equation}
where we have extracted the $h$-scaling of each phase. At leading order both phases reduce to the Arutyunov-Frolov-Staudacher one~\cite{Arutyunov:2004vx}.
However,  at HL-order all three terms on the right hand side of equation~(\ref{eq:solution}) contribute and we find
\begin{equation}
\begin{aligned}
c_{r,s}^{(1)}&= +\frac{1}{2\pi} \frac{1-(-1)^{s+r}}{2}\left[\frac{s-r}{s+r-2}-\frac{1}{2}\big(\delta_{r,1}-\delta_{1,s}\big)\right]\,,\\
\tilde{c}_{r,s}^{(1)}&= -\frac{1}{2\pi} \frac{1-(-1)^{s+r}}{2}\left[\frac{s+r-2}{s-r}-\frac{1}{2}\big(\delta_{r,1}-\delta_{1,s}\big)\right]\,,
\end{aligned}
\end{equation}
for $s>r>0$. 
Finally, the higher order coefficients~$c^{(n)}_{r,s}=\tilde{c}^{(n)}_{r,s}$ with $n>1$ are exactly the same as in the expansion of the BES phase.

\subsubsection*{Semiclassical and near-flat-space limits}
In order to compare with perturbative results, it is convenient to write explicit expressions for our phases in the semiclassical limit~\eqref{eq:FG-expansion}.
Such an expansion for the BES phase is well known: the leading order~$O(1/h)$ is given by the AFS phase~\eqref{eq:AFS-xpxm}, which in our normalisation reads
\begin{equation}
\theta^{\AFS}(x,y)=\frac{2}{h}\frac{x-y}{(x^2-1)(x y-1)(y^2-1)}+O\left(\frac{1}{h^3}\right)\,,
\end{equation}
whereas the next-to-leading-order is given by the HL phase which can be found by expanding~\eqref{eq:DHM-HL} under the integral. Doing so also for~\eqref{eq:chi-}, we get to the expressions
\begin{equation}
\begin{aligned}
\theta(x,y)=\theta^{\AFS}(x,y)+\frac{1}{\pi h^2}\frac{x^2}{x^2-1}&\frac{y^2}{y^2-1}\big[\frac{(x+y)^2(1-\frac{1}{xy})}{(x^2-1)(x-y)(y^2-1)}\\
&\ +\frac{2}{(x-y)^2}\log\big(\frac{x+1}{x-1}\frac{y-1}{y+1}\big)\big]+O\left(\frac{1}{h^3}\right),\\
\widetilde{\theta}(x,y)=\theta^{\AFS}(x,y)+\frac{1}{\pi h^2}\frac{x^2}{x^2-1}&\frac{y^2}{y^2-1}\big[\frac{(xy+1)^2(\frac{1}{x}-\frac{1}{y})}{(x^2-1)(xy-1)(y^2-1)}\\
&\ + \frac{2}{(xy-1)^2}\log\big(\frac{x+1}{x-1}\frac{y-1}{y+1}\big)\big]+O\left(\frac{1}{h^3}\right).
\end{aligned}
\label{eq:FGsol}
\end{equation}
Let us also evaluate the dressing factors in the near-flat-space limit~\cite{Maldacena:2006rv}
\begin{equation}
\begin{aligned}
\theta(p_-,q_-)=\frac{p_- q_- (p_--q_-)}{8 h (p_-+q_-)}+\frac{p_-^2 q_-^2 \left(p_-^2+2 p_- q_- \log\frac{q_-}{p_-}-q_-^2\right)}{64 \pi  h^2
   (p_--q_-)^2}+O\left(\frac{1}{h^3}\right)\,,\\
\widetilde{\theta}(p_-,q_-)=\frac{p_- q_- (p_--q_-)}{8 h (p_-+q_-)}-\frac{p_-^2 q_-^2 \left(p_-^2-2 p_- q_- \log\frac{q_-}{p_-}-q_-^2\right)}{64 \pi  h^2
   (p_-+q_-)^2}+O\left(\frac{1}{h^3}\right)\,.
\end{aligned}
\label{eq:NFSsol}
\end{equation}

\subsection*{Weak-coupling expansion}
In this subsection we compute the weak-coupling expansion of the dressing phases. While this will not be comparable with any direct calculation, it naturally enters our spin-chain Bethe ansatz, in particular in the identification of the global charges.

The results for $\sigma^{\BES}$ are well known from $\AdS_5/\CFT_4$. The leading-order contribution to the dressing phase starts at $\order(h^6)$~\cite{Beisert:2006ez}, and comes from the $r=2$, $s=3$ terms in the expansion of $\chi^{\BES}$.\footnote{See equation~\eqref{eq:chiexp}, and recall that for the BES phase there is no $r=1$ term.} 
The $\AdS_3$ dressing phases~\eqref{eq:solution} contain extra terms besides the BES phase. The coefficients $c_{r,s}$ and $\tilde{c}_{r,s}$ that come from these extra contributions are all order $h^0$ (see equation~\eqref{eqchimcrs} and~\eqref{eqchihlcrs}). The coupling constant dependence comes only from the charges $q_r$ and $q_s$ in equation~\eqref{eq:chiexp} through~$x^{\pm}$. In fact, the leading contribution comes from the $r=1$ and $s=2$ term, and can be written
\begin{equation}
\begin{aligned}
\label{eq:weakcoupl-phases}
&\theta(p,q)= +i\,h\,c^{(1)}_{1,2}\;
\Big(p\,\delta D(q)-q\,\delta D(p)\Big)
+ \order(h^3) \,,\\
&\widetilde{\theta}(p,q)= -i\,h\,c^{(1)}_{1,2}\;
\Big(p\,\delta D(q)-q\,\delta D(p)\Big)
+ \order(h^3) \,,\\
\end{aligned}
\end{equation}
where $\delta D(p)=E(p)-1$ is the anomalous part of the dispersion relation. Note that the~$O(h^2)$ terms vanish.

The above result shows that the $r=1$ terms, which are novel to $\AdS_3$, contribute at order $h$ to the BA, and so should modify the energy of states in the weakly-coupled spin-chain at order $h^3$. 
Let us remark that \emph{a priori} we do not know how $h(\lambda)$ behaves at weak coupling, where~$\lambda$ is the genuine $\CFT_2$ coupling constant.\footnote{%
  Recall, for example, that in $\AdS_5/\CFT_4$ $h\sim\sqrt{\lambda}$ it is while in $\AdS_4/\CFT_3$ it is $h\sim\lambda$.%
} %
This prevents us from determining whether the $O(h^1)$ contribution to $\theta(p,q)$ in the equation above comes with an integral power of $\lambda$ as one would expect in a weakly-coupled planar limit.

Integrable spin chains with long-range interactions have been systematically considered in ref.~\cite{Beisert:2005wv} where, among other things, the weak-coupling expansion of the dressing factor and of some conserved charged was analysed. Interestingly, it appears that the weak-coupling properties of the $\AdS_3$ spin chain differ from those found in that general analysis. This is not entirely surprising, since in contrast to ref.~\cite{Beisert:2005wv}, the $\psu(1,1|2)^2$ spin chain consists of both a left-moving and a right-moving 
sector. Something similar happens in the study of alternating spin chains~\cite{Bargheer:2009xy}, where novel operators that do not exist for the homogeneous spin chains of ref.~\cite{Beisert:2005wv} modify the structure of the dressing factor found there.
We can then conclude that the~$\AdS_3/\CFT_2 $ weakly-coupled spin-chain dynamics is substantially different from the ones we are familiar with.

Let us now investigate how these contributions modify the Bethe ansatz picture. The form of the right hand side of eq.~\eqref{eq:weakcoupl-phases} is reminiscent of a gauge-dependent shift of the length, as discussed at the end of section~\ref{sec:nesting}. Indeed, if in the Bethe equations appeared \eg~only left excitations and therefore only~$\theta(p,q)$ we could use the level-matching to reabsorb such factor in a shift of~$\ell$ as in~\eqref{eq:BAframefactors}, at least 
in absence of winding.  However, as soon as left \emph{and} right excitations are present such a procedure is impossible, because the correction due to~$\widetilde{\theta}(p,q)$ has opposite sign than the one of~$\theta(p,q)$, see~\eqref{eq:weakcoupl-phases}. What is more, such terms are non-integer, and it is easy to see that they would appear in the identification of the global charges~\eqref{eq:dynkinlab}. In particular, the Dynkin label~$r_{2}$ corresponds to~$\su(2)$ and as such should be an integer---but the phase contribution would contain anomalous terms.

At the moment we have no resolution for such a puzzle. It is tempting to say that perhaps the dressing factors should be modified by a solution of the homogeneous crossing equations, in such a way to remove such contributions.%
\footnote{%
In the case of~$\AdS_5/\CFT_4$, for instance, there exist a crossing-symmetric phase that reproduces the leading large-$h$ behaviour, but fails to match the weak-coupling structure of the~$\CFT_4$~\cite{Beisert:2006ib}.
}
 However one should bear in mind that these $r=1$ terms appear also in independent string calculations~\cite{Beccaria:2012kb,Abbott:2013ixa}, as we will discuss in the next section. This once again points to some subtlety in interpreting these results, and in fact our whole construction, at weak coupling.

\section{Comparisons}
\label{sec:comparisons}
In this section we will look first at explicit S-matrix elements obtained from string theory. These include tree-level,  one-loop and two-loop results. While the tree-level terms can be found directly in full generality, evaluating the one-loop ones requires either going to particular kinematic regions or using additional tricks such as unitarity techniques.%
\footnote{%
A novel perturbative result recently appeared in ref.~\cite{Sundin:2014sfa}. There, fully-fledged one-loop calculations are performed in the near-BMN limit of~$\AdS_3/\CFT_2 $. Just like the ones that we will discuss more in detail hereafter, those results agree with our proposals for S~matrix and dressing factors.
}

For this reason, to investigate the one-loop structure of the dressing factors sometimes a different approach is more useful: one can focus on a specific classical solution of the string NLSM and compute the one-loop corrections to its energy, whence the dressing factors can be reverse engineered. This can be done using the classical integrability properties the we described in section~\ref{sec:integrability}, and in particular the spectral curve. In the same way, one can obtain \emph{finite-gap equations} that capture part of the semiclassical spectrum.

\subsection*{Tree-level S~matrix}
To expand the S~matrix of section~\ref{sec:stringsmat} at tree level it suffices to use the AFS~phase~\eqref{eq:AFS-xpxm} and to plug in~\eqref{eq:BMN-expansion}. As discussed in section~\ref{sec:BAE}, we should take into account gauge terms of the form~$e^{i\,a\,(p\,\delta D_q-q\,\delta D_p)}$.

In ref.~\cite{Hoare:2013pma} Hoare and Tseytlin (HT), among other things, wrote down a tree level S~matrix for all the massive excitations of~$\AdS_3\times\S^3\times\T^4$, from a suitable truncation of the~$\AdS_5\times\S^5$ S~matrix%
\footnote{%
Such a truncation is possible at tree level, but as we have seen the two matrices should be genuinely different at higher orders in a loop expansion.}.
 It is immediate to see that the scattering processes allowed in the HT S~matrix are those that are non-zero in the one we derived in section~\ref{sec:stringsmat}. In particular, both S~matrices are reflectionless, and both come from the tensor product of two $\su(1|1)^2_{\text{c.e.}}$ S~matrices. 
In fact we can match our results both with the full S~matrix~(4.1) and with each factor~(4.8) of~\cite{Hoare:2013pma}. We should note that the HT S~matrix is written in terms of~$\mathbf{R}$ rather than~$\mathbf{S}$ or~$\check{\mathbf{S}}$ which leads to some. Then, up to a change of basis, we find perfect agreement~\cite{Borsato:2013qpa} with the integrability results  of the previous chapters.

In ref.~\cite{Rughoonauth:2012qd, Sundin:2013ypa} the~$\AdS_3\times \S^3\times \S^3\times \S^1$ and~$\AdS_3\times \S^3\times \T^4$ S~matrices have been investigated perturbatively by Rughoonauth, Sundin and Wulff~(RSW) and by Sundin and Wulff~(SW).
At tree-level, the comparison is similar to the one above.
Since in SW some computations are performed for the more general $\AdS_3\times \S^3\times \S^3\times \S^1$ background, we should take the parameter $\alpha=1$ everywhere to recover the $\T^4$ background at tree level. Only a subset of the S~matrix elements has been computed. Still, this probes the left-left and left-right diagonal and non-diagonal scattering, yielding agreement with the integrability construction.

\subsection*{One- and two-loop S-matrix elements}
Sundin and Wulff~\cite{Sundin:2013ypa} also computed certain one-loop elements in the near-flat-space limit. These correspond to certain elements $\mathcal{A},\widetilde{\mathcal{A}},\mathcal{C},\widetilde{\mathcal{C}}$, some of which we encountered in chapter~\ref{ch:crossing}. They are defined by
\begin{equation}
\label{eq:stringframeS-elems}
\begin{aligned}
\mathcal{A}_{pq}=
\bra{\Phi^{\smallL}_{+{+}q}\Phi^{\smallL}_{+{+}p}} \check{\mathbf{S}}_{pq}
\ket{\Phi^{\smallL}_{+{+}p}\Phi^{\smallL}_{+{+}q}},
&\quad&
\widetilde{\mathcal{A}}_{pq}=
\bra{{\Phi}^{\smallR}_{+{+}q}\Phi^{\smallL}_{+{+}p}} \check{\mathbf{S}}_{pq}
\ket{\Phi^{\smallL}_{+{+}p}{\Phi}^{\smallR}_{+{+}q}},\\
\mathcal{C}_{pq}=
\bra{\Phi^{\smallL}_{-{-}q}\Phi^{\smallL}_{+{+}p}} \check{\mathbf{S}}_{pq}
\ket{\Phi^{\smallL}_{+{+}p}\Phi^{\smallL}_{-{-}q}},
&\quad&
\widetilde{\mathcal{C}}_{pq}=
\bra{{\Phi}^{\smallR}_{-{-}q}\Phi^{\smallL}_{+{+}p}} \check{\mathbf{S}}_{pq}
\ket{\Phi^{\smallL}_{+{+}p}{\Phi}^{\smallR}_{-{-}q}}.
\end{aligned}
\end{equation}
and read\footnote{The tree level expressions for $\mathcal{C}$ and $\widetilde{\mathcal{C}}$ ware not given in~\cite{Sundin:2013ypa}, but were communicated privately to us by the authors. We include them here for the sake of completeness.}
\begin{align}
\label{eq:1-loop-S-elements-1st}
\mathcal{A}_{pq} &= 1-\frac{i}{4h}\frac{p_-q_-(p_-+q_-)}{p_--q_-}\\
\nonumber
& \quad
+\frac{1}{32h^2}\frac{p_-^2q_-^2}{q_-^2-p_-^2}
\left(\frac{i}{\pi}(p_-+q_-)^2-\frac{2i}{\pi}\frac{q_-p_-(q_-+p_-)}{q_--p_-}\log\frac{q_-}{p_-}-\frac{(q_-+p_-)^3}{q_--p_-}\right),
\end{align}
\begin{align}\widetilde{\mathcal{A}}_{pq} &= 1-\frac{i}{4h}\frac{p_-q_-(p_--q_-)}{p_-+q_-}\\
\nonumber
& 
-\frac{1}{32h^2}\frac{p_-^2q_-^2}{q_-^2-p_-^2}
\left(\frac{i}{\pi}(p_--q_-)^2+\frac{2i}{\pi}\frac{q_-p_-(q_--p_-)}{q_-+p_-}\log\frac{q_-}{p_-}+\frac{(q_-^2+p_-^2)(q_--p_-)}{q_-+p_-}\right)\!\!,\\
\mathcal{C}_{pq} &= 1-\frac{i}{4h}p_-q_-\\
\nonumber
& 
+\frac{1}{32h^2}\frac{p_-^2q_-^2}{q_-^2-p_-^2}
\left(\frac{i}{\pi}(p_-+q_-)^2-\frac{2i}{\pi}\frac{q_-p_-(q_-+p_-)}{q_--p_-}\log\frac{q_-}{p_-}-\frac{(q_-^2+p_-^2)(q_-+p_-)}{q_--p_-}\right)\!\!,\\
\label{eq:1-loop-S-elements-last}
\widetilde{\mathcal{C}}_{pq} &= 1-\frac{i}{4h}p_-q_-\\
\nonumber
& \quad
-\frac{1}{32h^2}\frac{p_-^2q_-^2}{q_-^2-p_-^2}
\left(\frac{i}{\pi}(p_--q_-)^2+\frac{2i}{\pi}\frac{q_-p_-(q_--p_-)}{q_-+p_-}\log\frac{q_-}{p_-}+(q_-^2-p_-^2)\right),
\end{align}
where in contrast with~\cite{Sundin:2013ypa} we explicitly used the coupling constant~$h$ as a loop-counting parameter.  A first nontrivial requirement of our construction is that these elements satisfy the crossing equations of chapter~\ref{ch:crossing}, that read simply
\begin{equation}
\label{eq:stringframeS-crossing}
\mathcal{A}_{pq}\,\widetilde{\mathcal{A}}_{p{q}^{\mathsf{c}}}=1 ,\qquad\qquad \mathcal{C}_{p{q}^{\mathsf{c}}}\,\widetilde{\mathcal{C}}_{pq}=1 .
\end{equation}
It is easy to check that this is actually the case, observing that in light-cone coordinate crossing the second variable $q\to{q}^{\mathsf{c}}$ amounts to $q_-\to {q}^{\mathsf{c}}_-=-q_-$, and taking everywhere the upper branch of the logarithm~\cite{Borsato:2013hoa}.
Using the near-flat-space expansion of the dressing phases~\eqref{eq:NFSsol} we can check that we perfectly match these elements.
This is a first direct validation of the dressing factors presented in chapter~\ref{ch:crossing}.

A different approach to obtain higher-loop expressions is to use unitarity techniques (akin to the optical theorem) and exploit two-dimensional kinematics~\cite{Bianchi:2013nra, Engelund:2013fja}. Engelund, McKeown and Roiban~(EKR) have applied such techniques to~$\AdS^3\times\S^3\times\T^4$ in ref.~\cite{Engelund:2013fja}, obtaining near-BMN one- and two-loop results for the diagonal S-matrix elements. However, in such an approach only the~\emph{logarithmic} part of the elements is predicted unambiguously. Moreover, an $l$-loop calculation requires as input the matrix structure of~$\mathbf{S}$ at $(l-1)$-loop. At one-loop one can rely on the tree-level perturbative results above, but the two-loop calculation requires using the symmetries of the model to constrain~$\mathbf{S}$ as we did. Despite not being completely independent, the two-loop result is a very non-trivial consistency check of the S~matrix and dressing factors. Once again, the proposed all-loop S~matrix satisfies~it.

Yet another approach to computing the dressing factors is to obtain them from the semiclassical energy shifts to classical NSLM solutions. This approach was pioneered in~\cite{Beisert:2003xu,Beisert:2003ea} in the case of $\AdS_5\times\S^5$ and was later applied to $\AdS_3\times\S^3\times\T^4$ as well~\cite{Abbott:2012dd,Beccaria:2012kb, Beccaria:2012pm, Abbott:2013ixa}. In ref.~\cite{Beccaria:2012kb}, Beccaria, Levkovic-Maslyuk, Macorini and Tseytlin~(BLMMT) have computed the one-loop dressing phase. Their calculation pre-dated the all-loop proposal~\cite{Borsato:2013hoa} based on crossing symmetry. Up to different normalisation due to the choice of the expansion parameters resulting in a factor of $4\pi$, their results are given in terms of the expansion~\eqref{eq:theta-charges} by
\begin{equation}
\begin{aligned}
&c^{(1)}_{\BLMMT\, r,s}= c^{(1)}_{r,s}\,,\qquad\qquad
&&{\tilde c}^{(1)}_{\BLMMT\, r,s}= \tilde{c}^{(1)}_{r,s}\,,\\
&c^{(1)}_{\BLMMT\, 1,s}=
2 c^{(1)}_{1,s}
\,,\qquad\qquad
&&{\tilde c}^{(1)}_{\BLMMT\, 1,s}=
2 \tilde{c}^{(1)}_{1,s}\,.
\end{aligned}
\label{eq:BLMMT}
\end{equation}
The coefficients~${c}^{(1)}_{r,s}$ and~${\tilde c}^{(1)}_{r,s}$ do not match our proposal when~$r=1$. In fact, it is not hard to see~\cite{Borsato:2013qpa} that the resulting phases do not even satisfy the proposed crossing equations.

A resolution of this discrepancy was proposed by Abbott~\cite{Abbott:2013ixa}, who highlighted a subtlety in the definition of the charges~$\mathcal{Q}_{r}$ in the semiclassical derivation, as in refs.~\cite{Gromov:2007ky,Gromov:2008ie}. This results in a factor of $\tfrac{1}{2}$ in the normalisation of~$\mathcal{Q}_{1}$, and does not affect any higher charge. Taking this into account, Abbot showed that the semiclassical calculation perfectly matches our proposal for the dressing factors.
Later, two additional validations of this result were proposed. One was a direct one-loop calculation in the near-BMN expansion by Sundin~\cite{Sundin:2014sfa}, and one a unitarity-based calculation in the same regime by Bianchi and Hoare~\cite{Bianchi:2014rfa}. Both agreed with the expansion of the all-loop dressing factors proposed in~\cite{Borsato:2013hoa}.

\subsection*{Finite-gap equations}
The spectral curve $\Gamma(x)$~\eqref{eq:spectralcurve} encodes all the information about the conserved charges of the classical NLSM. The same is true for its eigenvalues $\{\gamma_{j}(x)\}$, or equivalently the ``quasi-momenta'' $\{p_j(x)\}$, given by~$\gamma_j(x)=e^{ip_j(x)}$. In the case where~$\Gamma(x)$ is an algebraic curve, \ie~it has finite genus, it is possible to write down integral equations for the discontinuities at its cuts to describe~$\Gamma(x)$---the \emph{finite-gap equations}. By solving those equations we can find the classical energy of a solution having a given set of charges.
The finite gap equations for~$\AdS_3\times\S^3\times\T^4$, were proposed in the seminal paper~\cite{Babichenko:2009dk} by Babichenko, Stefa\'nski and Zarembo, and we expect that they describe a suitable classical limit of the all-loop spectrum.

It is natural to expect that the finite-gap equations are a limit of the Bethe equations where the magnon momenta condense to form cuts~\cite{Kazakov:2004qf}. We introduce the densities%
\footnote{%
Here we find it convenient to work in terms of excitation numbers~$K_j$ rather than~$M_{j\,\smallL,\smallR}^{\I,\II}$. The relation between the two sets is given in eqs.~(\ref{eq:labelcartan1}--\ref{eq:labelcartan1}).
}
\begin{equation}
\rho_j(x)=\sum_{k=1}^{K_j}\frac{x^2}{x^2-1}\delta(x-x_{j,k}),\quad\quad j=1,2,3,\tilde{1},\tilde{2},\tilde{3},
\end{equation}
and take the excitation numbers to be large, $K_j\gg1$. By making use of the expansion~\eqref{eq:FG-expansion}, we find the following finite gap-equations
\begin{align}
\label{eq:FGlimit-beg}
&2\pi n_1 = - \int\frac{\rho_2(y)}{x-y}\de y- \int\frac{\rho_{\tilde{2}}(y)}{x-1/y}\frac{\de y}{y^2} \\
&\begin{aligned}
2\pi n_2&=-\frac{x}{x^2-1}2\pi\mathcal{E}-\int\frac{\rho_1(y)}{x-y}\de y + 2 \pint\frac{\rho_2(y)}{x-y}\de y -\int\frac{\rho_3(y)}{x-y}\de y\\
&\phantom{={}} +\int\frac{\rho_{\tilde{1}}(y)}{x-1/y}\frac{\de y}{y^2} + \int\frac{\rho_{\tilde{3}}(y)}{x-1/y}\frac{\de y}{y^2}+\frac{1}{x^2-1}\mathcal{M},
\end{aligned}
\\
&2\pi n_3= - \int\frac{\rho_2(y)}{x-y}\de y- \int\frac{\rho_{\tilde{2}}(y)}{x-1/y}\frac{\de y}{y^2}
\end{align}
\begin{align}
&2\pi n_{\tilde{1}}= \int\frac{\rho_{2}(y)}{x-1/y}\frac{\de y}{y^2}+\int\frac{\rho_{\tilde{2}}(y)}{x-y}\de y  \\
&\begin{aligned}
2\pi n_{\tilde{2}}&=-\frac{x}{x^2-1}2\pi\mathcal{E}-\int\frac{\rho_1(y)}{x-1/y}\frac{\de y}{y^2}  -\int\frac{\rho_3(y)}{x-1/y}\frac{\de y}{y^2}\\
&\phantom{={}} +\int\frac{\rho_{\tilde{1}}(y)}{x-y}\de y -2 \pint\frac{\rho_{\tilde{2}}(y)}{x-y}\de y + \int\frac{\rho_{\tilde{3}}(y)}{x-y}\de y+\frac{1}{x^2-1}\mathcal{M},
\end{aligned}
\\
&2\pi n_{\tilde{3}}= \int\frac{\rho_{2}(y)}{x-1/y}\frac{\de y}{y^2}+\int\frac{\rho_{\tilde{2}}(y)}{x-y}\de y ,
\label{eq:FGlimit-end}
\end{align}
where $\pint$ denotes the principal-value integral. Here $\mathcal{E}$ corresponds to the residue of the quasi-momentum and it is given by
\begin{equation}
\mathcal{E} = \frac{1}{2 \pi} (L -\epsilon_1 +2 \epsilon_2 -\epsilon_3 +\epsilon_{\tilde{1}} +\epsilon_{\tilde{3}}),
\end{equation}
where
\begin{equation}
\epsilon_j = \int \frac{\rho_j(x)}{x^2} \de x.
\end{equation}
The quantity $\mathcal{M}$ has the meaning of winding of the corresponding solutions and it is given by
\begin{equation}\label{eq:winding-finite-gap}
\mathcal{M}=\mathcal{P}_1+\mathcal{P}_3-\mathcal{P}_{\tilde{1}}+2 \mathcal{P}_{\tilde{2}}- \mathcal{P}_{\tilde{3}}=\mathcal{P}_1 - \mathcal{P}_{2} +\mathcal{P}_3-\mathcal{P}_{\tilde{1}}+ \mathcal{P}_{\tilde{2}}- \mathcal{P}_{\tilde{3}},
\end{equation}
where
\begin{equation}
\mathcal{P}_j = \int \frac{\rho_j (x)}{x} \de x.
\end{equation}
The last equality in~\eqref{eq:winding-finite-gap} is possible thanks to the level matching condition that reads
\begin{equation}
\mathcal{P}_2+\mathcal{P}_{\tilde{2}}=0.
\end{equation}
The finite-gap equations that we derived are apparently different but equivalent to the ones in~\cite{Babichenko:2009dk}. In fact, if we repeated the same construction performed there (see also refs.~ \cite{Zarembo:2010yz,Zarembo:2010sg}) with a different choice of the grading, such as~\eqref{eq:cartan}, we would have found precisely~(\ref{eq:FGlimit-beg}--\ref{eq:FGlimit-end}). While for the Bethe-Yang equations we had only four choices of grading, all of them involving different Cartan matrices in the left and right sectors, at the classical level the are more choices---one can change grading to each copy of~$\psu(1,1|2)$ independently. We can then conclude that the proposed all-loop construction is compatible with semiclassical integrability calculations too.

\section{Chapter summary}
In this chapter we have compared the all-loop integrability results, that relied on several assumptions, with independent ones. This comparison is a check of quantum integrability, but is also an important test for the dressing factors, that were not completely fixed by crossing invariance.
All of the independent calculations that we checked were performed on the \emph{string theory side}, that is at large values of~$h$. The proposal discussed in the previous chapters, including the dressing factors, reproduces all of them. This non-trivial matching holds at one-loop and puts strong requirements on the two-loop S~matrix elements too. Based on that, we are confident to say that we expect a quantum-integrable dynamics for the~$\AdS_3\times\S^3\times\T^4$ superstring, and we expect the $\psu(1|1)^4_{\text{c.e.}}$ symmetry and a discrete left-right symmetry to play a crucial role in it.

Still, it is highly desirable to test even further this construction results. In particular, we only have limited understanding of  the weakly-coupled~$h\ll1$ regime, and of how~$h(\lambda)$ scales in terms of the~$\CFT$ coupling~$\lambda$. Moreover, as we discussed in section~\ref{sec:expansions}, the dressing factors have a different structure compared to the ones of~$\AdS_5/\CFT_4$ or~$\AdS_4/\CFT_3$, that yields new puzzling features at weak coupling.

Let us conclude this chapter by reviewing some earlier and slightly different proposals for the S~matrix and dressing factors of~$\AdS_3\times\S^3\times\T^4$. The earliest proposal is due to David and Sahoo~\cite{David:2008yk,David:2010yg}, and it was derived from a study of the giant magnons and their symmetry properties. That proposal also relied on a $\su(1|1)^2_{\text{c.e.}}$ symmetry, but was restricted to left-left sector. As we have argued in chapter~\ref{ch:crossing}, such a restriction is not compatible with crossing symmetry. Indeed the structure of the dressing factors and crossing equations put forward there yielded a single dressing phase, which appears incompatible with later calculations~\cite{Beccaria:2012kb,Abbott:2013ixa, Sundin:2014sfa,Bianchi:2014rfa}.

In ref.~\cite{Babichenko:2009dk},  Babichenko, Stefa{\'n}ski and Zarembo (BSZ) conjectured a set of all-loop Bethe equations from the finite-gap equations. Those equations differed in two ways from the ones we derived from the all-loop S~matrix in chapter~\ref{ch:betheansatz}. Firstly they were written in a different grading, which as discussed in section~\ref{sec:comparisons} just amounts to a duality transformation. Secondly, some terms%
\footnote{%
Namely, the symmetric phases that appear in the couplings between left and right sectors in front of~$\widetilde{\sigma}_{pq}^2$, see eqs.~\eqref{eq:BA-2} and~\eqref{eq:BA-2b}. These are required for the unitarity of the S~matrix.%
}
 that are sub-leading in the finite-gap limit---and hence could not straightforwardly be reverse engineered---were not accounted for, which would lead to a different energy spectrum from the one of the construction we reviewed. After that, in ref.~\cite{Ahn:2012hw} Ahn and Bombardelli (AB) proposed an all-loop integrable S~matrix constructed so as to reproduce the BSZ conjecture for the Bethe equations. By construction, such an S~matrix does not match the one discussed here, in particular in the LR sector, and cannot be matched to the perturbative calculations described in this chapter.%
\footnote{%
As discussed more in detail in ref.~\cite{Borsato:2012ud}, the AB S~matrix is not invariant under the~\emph{extended} symmetry~$\su(1|1)^2_{\text{c.e.}}$, but only under~$\su(1|1)^2$, making it qualitatively different from the one discussed in chapter~\ref{ch:smatrix}.
}
Nonetheless, it is quite intriguing that such an S~matrix exists, and it would be interesting to investigate whether it describe an integrable worldsheet theory for some superstring background.

\chapter{Recent developments and new directions}
\label{ch:outlook}
Now that we have a good handle on the simplest aspects of~$\AdS_3/\CFT_2 $ integrability, in this chapter we will describe some directions that are now being investigated, which go beyond those aspects.
The most natural one in order to complete our treatment is the inclusion of massless excitations in the integrability description.

\section{Massless modes in \texorpdfstring{$\AdS_3\times\S^3\times\T^4$}{AdS3xS3xT4}}
\label{sec:massless}
We have seen in section~\ref{sec:bosonic1ord} that the bosonic string spectrum on~$\AdS_3\times\S^3\times\T^4$ features four fundamental massless excitations. By supersymmetry, these must be supplemented by four fermionic ones. Extending our S-matrix treatment to such excitations appears problematic at first sight. Massless excitations are characterised by the scaling of their dispersion relation at small momentum, of the form%
\footnote{%
More general definitions of massless (quasi)particles may be given, but this one will suffice for our purposes.
}
\begin{equation}
\omega(p)^2= c^2\, p^2+O(p^4)\,,
\end{equation}
which results in~$\omega(p)$ being a non-analytic function of~$p$ around zero. Consequently it is natural to distinguish between left- and right-movers \emph{on the worldsheet}, having respectively~$p>0$ and~$p<0$, and energy
\begin{equation}
E_{\text{left}}=+c\,p+O(p^3)\,,
\qquad
E_{\text{right}}=-c\,p+O(p^3)\,.
\end{equation}
This should not be confused with the notion of left- and right-movers \emph{in the dual CFT}, \ie~the ``L'' and ``R'' labels that were ubiquitous in the previous chapters.

In the familiar relativistic case, higher orders in~$p$ are absent, and the dispersion relation is linear. Therefore, the group velocity of a wave-packet is
\begin{equation}
v_{\text{rel}}=\frac{\partial \omega}{\partial p}=\pm c\,,
\end{equation}
\ie, massless relativistic particles move at the speed of light. Particles with the same worldsheet chirality then cannot scatter, regardless of the value of their  momentum. Still, a formal treatment of relativistic massless theories in terms of factorised scattering is possible~\cite{Zamolodchikov:1992zr,Fendley:1993wq,Fendley:1993xa, Fendley:1993jh}. To this end it is however necessary to introduce appropriate rapidity variables and take suitable limits.

In the case of our interest, however, factorised scattering appears to be \emph{simpler} than in the relativistic case. In fact, as argued in~\cite{Borsato:2014hja}, the dispersion relation of massless worldsheet particles should take the form
\begin{equation}
\label{eq:masslessdisp}
\omega(p)^2=4\,h^2\sin^2\big(\frac{p}{2}\big),
\end{equation}
from which we find the group velocity
\begin{equation}
v_{\text{non-rel}}=\pm\, h\,\cos\big(\frac{p}{2}\big).
\end{equation}
We see then that massless excitations that have different momenta also have different velocities, so that we can expect them to scatter in a way similar to massive excitations.
Starting from this intuition, in refs.~\cite{Borsato:2014exa,Borsato:2014hja} the fundamental S~matrix of massive and massless excitations was computed. Since ref.~\cite{Borsato:2014hja} gives a very detailed account of the derivation, here we will only outline its conceptual steps. The approach is similar to the one used in chapters~\ref{ch:sigmamodel} and~\ref{ch:smatrix}, relying on the off-shell symmetry algebra of the theory. As we have mentioned in chapter~\ref{ch:sigmamodel}, the coset formalism is not suitable for treating massive and massless excitations at the same time. The issue is that the coset~$\kappa$-gauge fixing---whereby all physical fermions are taken to live in the coset---is not compatible with light-cone gauge.
To overcome this, it is possible to use the Green-Schwarz superstring formalism, taking a $\kappa $~gauge where the massless fermions live on the torus. Then one can compute the Noether charges and Killing spinors of the $\AdS_3\times\S^3\times\T^4$ geometry to find the off-shell symmetries.

\begin{figure}
  \centering
  \begin{tikzpicture}[%
    box/.style={outer sep=1pt},
    Q node/.style={inner sep=1pt,outer sep=0pt},
    arrow/.style={-latex}
    ]%

    \node [box] (PhiM) at ( 0  , 1.5cm) {\small $\ket{\Phi^{\smallL}_{++}}$};
    \node [box] (PsiP) at (-1.5cm, 0cm) {\small $\ket{\Phi^{\smallL}_{-+}}$};
    \node [box] (PsiM) at (+1.5cm, 0cm) {\small $\ket{\Phi^{\smallL}_{+-}}$};
    \node [box] (PhiP) at ( 0  ,-1.5cm) {\small $\ket{\Phi^{\smallL}_{--}}$};

    \newcommand{\horshift}{0.09cm,0cm}
    \newcommand{\vershift}{0cm,0.10cm}
 
    \draw [arrow] ($(PhiM.west) +(\vershift)$) -- ($(PsiP.north)-(\horshift)$) node [pos=0.5,anchor=south east,Q node] {\scriptsize $\mathbf{Q}^{1\L}$};
    \draw [arrow] ($(PsiP.north)+(\horshift)$) -- ($(PhiM.west) -(\vershift)$) node [pos=0.5,anchor=north west,Q node] {};

    \draw [arrow] ($(PsiM.south)-(\horshift)$) -- ($(PhiP.east) +(\vershift)$) node [pos=0.5,anchor=south east,Q node] {};
    \draw [arrow] ($(PhiP.east) -(\vershift)$) -- ($(PsiM.south)+(\horshift)$) node [pos=0.5,anchor=north west,Q node] {\scriptsize $\overline{\mathbf{Q}}{}^{1\L}$};

    \draw [arrow] ($(PhiM.east) -(\vershift)$) -- ($(PsiM.north)-(\horshift)$) node [pos=0.5,anchor=north east,Q node] {};
    \draw [arrow] ($(PsiM.north)+(\horshift)$) -- ($(PhiM.east) +(\vershift)$) node [pos=0.5,anchor=south west,Q node] {\scriptsize $\overline{\mathbf{Q}}{}^{2\L}$};

    \draw [arrow] ($(PsiP.south)-(\horshift)$) -- ($(PhiP.west) -(\vershift)$) node [pos=0.5,anchor=north east,Q node] {\scriptsize $-\mathbf{Q}^{2\L}$};
    \draw [arrow] ($(PhiP.west) +(\vershift)$) -- ($(PsiP.south)+(\horshift)$) node [pos=0.5,anchor=south west,Q node] {};

    \draw [arrow] (PsiM) -- (PsiP) node [pos=0.6,anchor=south west,Q node] {\scriptsize $\mathbf{J}^{\ \alpha}_{\bullet}\;$};
    \draw [arrow] (PsiP) -- (PsiM);
  \end{tikzpicture}
\hspace{1.5cm}
  \begin{tikzpicture}[%
    box/.style={outer sep=1pt},
    Q node/.style={inner sep=1pt,outer sep=0pt},
    arrow/.style={-latex}
    ]%

    \node [box] (PhiM) at ( 0  , 1.5cm) {\small $\ket{\Phi^{\smallR}_{--}}$};
    \node [box] (PsiP) at (-1.5cm, 0cm) {\small $\ket{\Phi^{\smallR}_{+-}}$};
    \node [box] (PsiM) at (+1.5cm, 0cm) {\small $\ket{\Phi^{\smallR}_{-+}}$};
    \node [box] (PhiP) at ( 0  ,-1.5cm) {\small $\ket{\Phi^{\smallR}_{++}}$};

    \newcommand{\horshift}{0.09cm,0cm}
    \newcommand{\vershift}{0cm,0.10cm}
 
    \draw [arrow] ($(PhiM.west) +(\vershift)$) -- ($(PsiP.north)-(\horshift)$) node [pos=0.5,anchor=south east,Q node] {\scriptsize $\overline{\mathbf{Q}}{}^{1\R}$};
    \draw [arrow] ($(PsiP.north)+(\horshift)$) -- ($(PhiM.west) -(\vershift)$) node [pos=0.5,anchor=north west,Q node] {};

    \draw [arrow] ($(PsiM.south)-(\horshift)$) -- ($(PhiP.east) +(\vershift)$) node [pos=0.5,anchor=south east,Q node] {};
    \draw [arrow] ($(PhiP.east) -(\vershift)$) -- ($(PsiM.south)+(\horshift)$) node [pos=0.5,anchor=north west,Q node] {\scriptsize ${\mathbf{Q}}^{1\R}$};

    \draw [arrow] ($(PhiM.east) -(\vershift)$) -- ($(PsiM.north)-(\horshift)$) node [pos=0.5,anchor=north east,Q node] {};
    \draw [arrow] ($(PsiM.north)+(\horshift)$) -- ($(PhiM.east) +(\vershift)$) node [pos=0.5,anchor=south west,Q node] {\scriptsize 
    ${\mathbf{Q}}^{2\R}$};

    \draw [arrow] ($(PsiP.south)-(\horshift)$) -- ($(PhiP.west) -(\vershift)$) node [pos=0.5,anchor=north east,Q node] {\scriptsize $-\overline{\mathbf{Q}}{}^{2\R}$};
    \draw [arrow] ($(PhiP.west) +(\vershift)$) -- ($(PsiP.south)+(\horshift)$) node [pos=0.5,anchor=south west,Q node] {};
    
    \draw [arrow] (PsiM) -- (PsiP) node [pos=0.6,anchor=south west,Q node] {\scriptsize $\mathbf{J}^{\ \alpha}_{\bullet}\;$};
    \draw [arrow] (PsiP) -- (PsiM);
  \end{tikzpicture}
  \caption{The action of the supercharges on the massive excitations is supplemented by $\mathbf{J}_{\bullet}^{\ \a} $ from $\su(2)_{\bullet}$, acting on the massive fermions only. Note that for clarity we only depict the action of some of the supercharges, \ie~the ones that remain non-trivial on-shell.}
  \label{fig:representationK}
\end{figure}
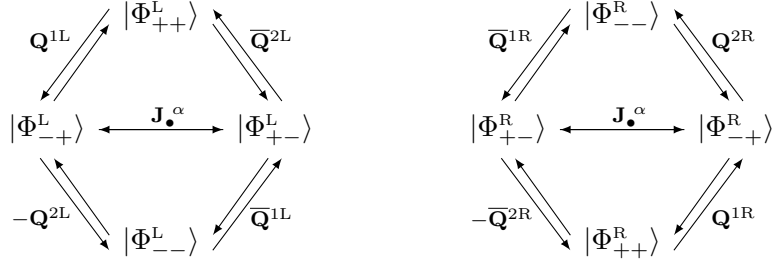
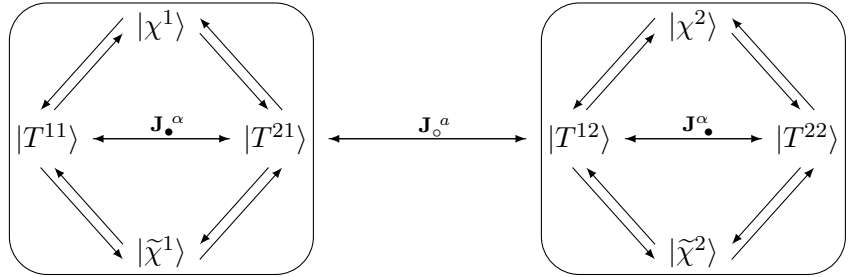
\begin{figure}
  \centering
  \begin{tikzpicture}[%
    box/.style={outer sep=1pt},
    Q node/.style={inner sep=1pt,outer sep=0pt},
    arrow/.style={-latex}
    ]%
\begin{scope}[xshift=-3.5cm]
    \node [box] (PhiM) at ( 0  , 1.5cm) {\small $\ket{\chi^{1}}$};
    \node [box] (PsiP) at (-1.5cm, 0cm) {\small $\ket{T^{11}}$};
    \node [box] (PsiM) at (+1.5cm, 0cm) {\small $\ket{T^{21}}$};
    \node [box] (PhiP) at ( 0  ,-1.5cm) {\small $\ket{\widetilde{\chi}^1}$};

    \newcommand{\horshift}{0.09cm,0cm}
    \newcommand{\vershift}{0cm,0.10cm}
 
    \draw [arrow] ($(PhiM.west) +(\vershift)$) -- ($(PsiP.north)-(\horshift)$) node [pos=0.5,anchor=south east,Q node] {};
    \draw [arrow] ($(PsiP.north)+(\horshift)$) -- ($(PhiM.west) -(\vershift)$) node [pos=0.5,anchor=north west,Q node] {};

    \draw [arrow] ($(PsiM.south)-(\horshift)$) -- ($(PhiP.east) +(\vershift)$) node [pos=0.5,anchor=south east,Q node] {};
    \draw [arrow] ($(PhiP.east) -(\vershift)$) -- ($(PsiM.south)+(\horshift)$) node [pos=0.5,anchor=north west,Q node] {};

    \draw [arrow] ($(PhiM.east) -(\vershift)$) -- ($(PsiM.north)-(\horshift)$) node [pos=0.5,anchor=north east,Q node] {};
    \draw [arrow] ($(PsiM.north)+(\horshift)$) -- ($(PhiM.east) +(\vershift)$) node [pos=0.5,anchor=south west,Q node] {};

    \draw [arrow] ($(PsiP.south)-(\horshift)$) -- ($(PhiP.west) -(\vershift)$) node [pos=0.5,anchor=north east,Q node] {};
    \draw [arrow] ($(PhiP.west) +(\vershift)$) -- ($(PsiP.south)+(\horshift)$) node [pos=0.5,anchor=south west,Q node] {};

    \draw [arrow] (PsiM) -- (PsiP) node [pos=0.6,anchor=south west,Q node] {\scriptsize $\mathbf{J}^{\ \alpha}_{\bullet}\;$};
    \draw [arrow] (PsiP) -- (PsiM);

        \draw[rounded corners=5mm] (-2cm,-1.8cm)rectangle (2cm,1.8cm);
\end{scope}
\begin{scope}[xshift=0cm]
    \draw [arrow] (-1.3cm,0cm) -- (1.3cm,0cm) node [Q node] at (0.1cm,0.15cm) {\scriptsize $\mathbf{J}^{\ a}_{\circ}$};
    \draw [arrow] (1.3cm,0cm) -- (-1.3cm,0cm);
\end{scope}
\begin{scope}[xshift=3.5cm]
    \node [box] (PhiM) at ( 0  , 1.5cm) {\small $\ket{\chi^{2}}$};
    \node [box] (PsiP) at (-1.5cm, 0cm) {\small $\ket{T^{12}}$};
    \node [box] (PsiM) at (+1.5cm, 0cm) {\small $\ket{T^{22}}$};
    \node [box] (PhiP) at ( 0  ,-1.5cm) {\small $\ket{\widetilde{\chi}^2}$};

    \newcommand{\horshift}{0.09cm,0cm}
    \newcommand{\vershift}{0cm,0.10cm}
 
    \draw [arrow] ($(PhiM.west) +(\vershift)$) -- ($(PsiP.north)-(\horshift)$) node [pos=0.5,anchor=south east,Q node] {};
    \draw [arrow] ($(PsiP.north)+(\horshift)$) -- ($(PhiM.west) -(\vershift)$) node [pos=0.5,anchor=north west,Q node] {};

    \draw [arrow] ($(PsiM.south)-(\horshift)$) -- ($(PhiP.east) +(\vershift)$) node [pos=0.5,anchor=south east,Q node] {};
    \draw [arrow] ($(PhiP.east) -(\vershift)$) -- ($(PsiM.south)+(\horshift)$) node [pos=0.5,anchor=north west,Q node] {};

    \draw [arrow] ($(PhiM.east) -(\vershift)$) -- ($(PsiM.north)-(\horshift)$) node [pos=0.5,anchor=north east,Q node] {};
    \draw [arrow] ($(PsiM.north)+(\horshift)$) -- ($(PhiM.east) +(\vershift)$) node [pos=0.5,anchor=south west,Q node] {};

    \draw [arrow] ($(PsiP.south)-(\horshift)$) -- ($(PhiP.west) -(\vershift)$) node [pos=0.5,anchor=north east,Q node] {};
    \draw [arrow] ($(PhiP.west) +(\vershift)$) -- ($(PsiP.south)+(\horshift)$) node [pos=0.5,anchor=south west,Q node] {};

    \draw [arrow] (PsiM) -- (PsiP) node [pos=0.6,anchor=south west,Q node] {\scriptsize $\mathbf{J}^{\alpha}_{\ \bullet}\;$};
    \draw [arrow] (PsiP) -- (PsiM);

        \draw[rounded corners=5mm] (-2cm,-1.8cm)rectangle (2cm,1.8cm);
\end{scope}
  \end{tikzpicture}
  \caption{The massless excitations sit in two~$\psu(1|1)^4_{\text{c.e.}} $ modules, with fermionic highest weight states. The two modules transform as a doublet under~$\su(2)_{\circ} $, and additionally the bosons~$T^{\alpha a}$ also carry a Greek index of~$\su(2)_{\bullet} $.}
  \label{fig:representationY}
\end{figure}

One then finds that the~$\psu(1|1)^4$ is supplemented by a bosonic~$\o(4)$ symmetry coming from the torus, with in particular an~$\so(4)=\su(2)_{\bullet}\oplus\su(2)_{\circ} $ symmetry. This may appear surprising, as one would na\"ively expect only~$\u(1)^4$ isometries from the torus. However, the presence of periodic boundary conditions on~$\T^4$ is probed only by winding modes, that in the decompactification limit, where the S~matrix is defined, are irrelevant.
The  massive fundamental excitations are charged under some of the torus isometries. In fact, the fermions~$\Phi_{\pm\mp}^{\smallL}$ and~$\Phi_{\pm\mp}^{\smallR}$ form two doublets under~$\su(2)_{\bullet}$, as depicted in figure~\ref{fig:representationK}. The massless excitations are also charged under~$\so(4)$, and we label them accordingly. Using Greek letters for fundamental and antifundamental~$\su(2)_{\bullet}$ indices, and Latin ones for~$\su(2)_{\circ}$, we can write the eight massless excitations as
\begin{equation}
\text{bosons:}\quad T^{\a a}\,,
\qquad\quad
\text{fermions:}\quad \chi^{a},\ \widetilde{\chi}^{a}\,,
\end{equation}
see figure~\ref{fig:representationY}. 
The fermions~$\chi^{a}$ and~$\widetilde{\chi}^{a}$ are distinguished by having opposite eigenvalues under the bosonic charge~$\mathbf{N}$ of eq.~\eqref{eq:bosonic-charges1}.
The~$\psu(1|1)^4 $ supercharges are also charged under~$\su(2)_{\bullet}$, forming four doublets
\begin{equation}
\mathbf{Q}^{\a \smallL}\,,\qquad
\overline{\mathbf{Q}}{}^{\phantom{\a} \smallL}_{\a}\,,\qquad
\mathbf{Q}^{\phantom{\a} \smallR}_{\a}\,,\qquad
\overline{\mathbf{Q}}{}^{\a \smallR}\,.
\end{equation}
This is the hidden $\su(2) $ action that we saw emerging in the massive sector from the tensor product structure when looking at the S-matrix elements in section~\ref{sec:stringsmat}.

The final ingredient for finding the S~matrix is understanding if and how the non-trivial central extension appears when going from the on-shell to the off-shell symmetry algebra. In the coset formulation we saw this easily because the light-cone coordinate~$x_{-}$ was%
\footnote{%
Recall that~$x_{-}$ is related to the total worldsheet momentum by~$\mathbf{P}=-\int \de \sigma\,x_{-}' $.
} 
neatly packaged into~$\Lambda(x_{\pm})$ as an exponential. Without reference to the coset representatives, the same terms naturally arise in the Green-Schwarz formulation, by requiring that the fermions are not charged under the light-cone isometries~\cite{Borsato:2014hja}, see also \eg~ref.~\cite{Alday:2005ww}. In fact one can see that the central extension takes the same form as in the massive sector. With these ingredients it is then straightforward to find the $2\to2 $ S~matrix, which for massless excitations is expressed in terms of modified Zukhovsky variables satisfying
\begin{equation}
\frac{x^+_p}{x^-_p}=e^{ip}\,,
\qquad
x^{+}_p\,x^{-}_p=1\,,
\qquad
x^{\pm}_p=e^{\pm \frac{i}{2}p}\, \text{sign}\left(\sin \frac{p}{2}\right),
\end{equation}
whereby we get the dispersion relation~\eqref{eq:masslessdisp} by the usual formula~\eqref{eq:dispersion-xpm}.
Unsurprisingly, since we are dealing with massless particles, a non-analyticity arises at $p=0$, whence a splitting into  left- and right-movers of the worldsheet can be introduced.
It is also interesting to observe that on symmetry grounds, unless the $\so(4) $ invariance of the S~matrix is broken, quantum effects cannot give a mass the massless particles---unlike what happens in other integrable theories such as the Gross-Neveu model~\cite{Gross:1974jv,Witten:1978qu}.

The worldsheet S~matrix again factors in a tensor-product structure, and depends on five distinct dressing phases,
\begin{equation}
\mathscr{S}^{\bullet\bullet}_{pq}\,,\qquad
\widetilde{\mathscr{S}}^{\bullet\bullet}_{pq}\,,\qquad
\mathscr{S}^{\bullet\circ}_{pq}\,,\qquad
\mathscr{S}^{\circ\bullet}_{pq}\,,\qquad
\mathscr{S}^{\circ\circ}_{pq}\,,
\end{equation}
where the first two phases are the massive-massive ones that we already studied in chapter~\ref{ch:crossing}, while the remaining three pertain to the massive-massless, massless-massive and massless-massless sectors, respectively. The requirement of crossing symmetry for all these phases can be written down straightforwardly, but a proposal for the latter three and a study of their properties is still lacking. To this end it may be useful to first obtain some insight in their perturbative form by performing worldsheet calculations, following ref.~\cite{Sundin:2013ypa} and perhaps taking advantage of the recent advances in unitarity techniques~\cite{Bianchi:2013nra, Engelund:2013fja}.

The proposal of the all-loop S~matrix for all fundamental excitations is a crucial step forward in~$\AdS_3/\CFT_2$ integrability, but should be supplemented by several related investigations. Among these, finding the Bethe-Yang equations and studying how the full~$\mathcal{N}=(4,4) $ infinite-dimensional symmetry is realised at the level of the asymptotic spectrum is probably the most important one. Once that has been understood, it should also be possible to formulate a string hypothesis and mirror TBA equations for this model, and to start to tackle finite-size physics, ``spectroscopy'', comparison with other approaches, and many other issues, some of which we will mention in the conclusions. We foresee many of these studies to be completed in the near future, rapidly advancing our understanding of this model.

\section{Integrability for the  \texorpdfstring{$\AdS_3\times\S^3\times\S^3\times\S^1$}{AdS3xS3xS3xS1} background}
\label{sec:ads3s3s3s1}
As discussed in the introduction, there is another~$\AdS_3 $ background (in fact, a family of backgrounds) that preserves the maximal amount of supersymmetry, \ie~16 supercharges. It is given by~$\AdS_3\times\S^3\times\S^3\times\S^1$ provided that the curvature radii of the two spheres~$R_{(1)}$ and~$R_{(2)}$ satisfy 
\begin{equation}
\frac{1}{R_{(1)}^2}+\frac{1}{R_{(2)}^2}
=\frac{1}{R_{\AdS}^2}\,,
\end{equation}
where~$R_{\AdS}$ is the radius of~$\AdS_3$. This gives an one-parameter family that can be labelled by~$\a$
\begin{equation}
\label{eq:alphadef}
  \alpha = \frac{R_{\AdS}^2}{R_{(1)}^2} = 1 - \frac{R_{\AdS}^2}{R_{(2)}^2} ,
\end{equation}
with $0<\alpha<1$. Clearly, up to exchanging the role of the two spheres, we can restrict to~$\tfrac{1}{2}\leq\alpha<1$, where the point~$\a=\tfrac{1}{2}$ is special in that the two spheres become identical there.
Another interesting configuration is the one where $\a\to1$, when~$R_{(1)}$ equates the AdS~radius and~$R_{(2)}$ blows up. In this limit the second sphere becomes flat, and decompactifies to~$\mathbbm{R}^3 $. Up to the fact that~$\mathbbm{R}^3 $ should be compactified back to~$\T^3 $, which can be done in several ways, we have basically obtained again the~$\AdS_3\times\S^3\times\T^4$ background. However, since this limit appears quite delicate, let us for the moment take~$\tfrac{1}{2}\leq\alpha<1$ and come back to it later.
\begin{figure}
  \centering
  \subfloat[\label{fig:dynkin-d21a-orig}]{
    \begin{tikzpicture}
      [
      thick,
      node/.style={shape=circle,draw,thick,inner sep=0pt,minimum size=5mm}
      ]
      \useasboundingbox (-1.1cm,-1.1cm) rectangle (1.6cm,1.1cm);

      \node (v1) at (-0.38cm,  0.65cm) [node] {};
      \node (v2) at ( 0.75cm,  0.00cm) [node] {};
      \node (v3) at (-0.38cm, -0.65cm) [node] {};

      \draw (v2.south west) -- (v2.north east);
      \draw (v2.north west) -- (v2.south east);

      \draw (v1) -- (v2);
      \draw (v2) -- (v3);
    \end{tikzpicture}
  }
  \hspace{2cm}
  \subfloat[\label{fig:dynkin-d21a-dual}]{
    \begin{tikzpicture}
      [
      thick,
      node/.style={shape=circle,draw,thick,inner sep=0pt,minimum size=5mm}
      ]
      \useasboundingbox (-1.1cm,-1.1cm) rectangle (1.6cm,1.1cm);

      \node (v1) at (-0.38cm,  0.65cm) [node] {};
      \node (v2) at ( 0.75cm,  0.00cm) [node] {};
      \node (v3) at (-0.38cm, -0.65cm) [node] {};

      \draw (v1.south west) -- (v1.north east);
      \draw (v1.north west) -- (v1.south east);

      \draw (v2.south west) -- (v2.north east);
      \draw (v2.north west) -- (v2.south east);

      \draw (v3.south west) -- (v3.north east);
      \draw (v3.north west) -- (v3.south east);

      \draw (v1) -- (v2);
      \draw (v2) -- (v3);
      \draw [double,double distance=3pt] (v3) -- (v1);
    \end{tikzpicture}
  }
  
  \caption{Two Dynkin diagrams for $\d21a$. Diagram~\protect\subref{fig:dynkin-d21a-dual} corresponds to a completely fermionic grading.}

  \label{fig:d21a-dynkin-diagrams}
\end{figure}

Once again the background has flat directions that give rise to massless excitations in its spectrum in light-cone gauge. Instead of the four massless modes from~$\T^4$, here we have two: one resides on the circle~$\S^1$, and one is shared by the two spheres%
\footnote{%
To preserve as much supersymmetry as possible, the light-cone geodesic runs trough the time coordinate in~$\AdS_3$ and both of the spheres. The massless excitation comes from a coordinate on the spheres that is orthogonal to such geodesic.
}. Once again, up to discarding the massless excitations, the string action can be written in terms of a coset~\cite{Babichenko:2009dk,Zarembo:2010yz,Zarembo:2010sg}
\begin{equation}
\label{eq:d21acoset}
\frac{D(2,1;\a)\times D(2,1;\a)}{SU(1,1)\times SU(2) \times SU(2)}\,,
\end{equation}
where~$D(2,1;\a)$ is the supergroup corresponding to the 	exceptional basic simple Lie superalgebra~$\d21a$, see refs.~\cite{Frappat:1996pb,Gavarini}. Here $\tfrac{1}{2}\leq\a<1$ is precisely the curvature ratio of eq.~\eqref{eq:alphadef}, and the case~$\a=\tfrac{1}{2}$ gives~$\mathfrak{d}(2,1;\tfrac{1}{2})\cong \mathfrak{osp}(4|2) $. The superalgebra~$\d21a$ can be written in two inequivalent gradings. In figure~\ref{fig:d21a-dynkin-diagrams} we illustrate the relative Dynkin diagrams.

\subsection*{Classical and perturbative aspects}
Clearly the coset~\eqref{eq:d21acoset} can be equipped with a~$\mathbbm{Z}_4$ automorphisms, whence classical integrability follows, in the sense that the equations of motion admit the Lax representation that we briefly discussed in section~\ref{sec:integrability}. However, this as usual requires some care in dealing with the massless modes, that are not automatically present in the coset construction and must be added by hand. This was shown to be possible by Babichenko, Stefa\'nski and Zarembo in ref.~\cite{Babichenko:2009dk} in a specific~$\kappa$-gauge fixing, and then extended by Sundin and Wulff~\cite{Sundin:2012gc} to an arbitrary $\kappa$~gauge.

A perturbative investigation of the spectrum in light-cone gauge yields again $8+8$ bosonic and fermionic fundamental excitations. They come in eight  supersymmetric doublets, with four different masses. We collect them in table~\ref{tab:excitations}.
There are~$4+4$ ``light'' excitations~$(\phi^{j}|\,\theta^{j}) $ of mass~$0<m<1$ and~$2+2$ ``heavy'' ones~$(\Phi^{j}|\,\Theta^{j}) $ of mass~$m=1$, plus the aforementioned massless modes~$(\varphi^{j}|\,\vartheta^{j}) $.

In ref.~\cite{Babichenko:2009dk}, besides establishing classical integrability for this background, the finite-gap equations for massive excitations were written down. In that context, it naturally appears that the heavy modes are actually composite, consisting of two light excitations of mass~$\a$ and~$1-\a$. From the point of view of the S~matrix, this is reflected in the presence of a tree-level light-light-heavy vertex.

In order to understand whether the heavy mode is truly composite one should analyse its (renormalised) two-point function. If this displays a pole at mass~$m=1$, then the heavy mode should be treated as a fundamental particle, and included in the asymptotic states. If the pole is replaced by a branch cut, then the mode should be treated as composite. This issue was investigated at one~loop in ref.~\cite{Sundin:2012gc}, where at least in the simpler case when~$\a=\tfrac{1}{2}$ it was established that the latter scenario occurs.
\begin{table}
  \centering
  \begin{tabular}{lcccc}
    \toprule
   Mass  & $m=\a $ & $m=1 $ & $m=1-\a $ & $m=0 $ \\
    \midrule
   Bosons     & $\phi^{1\smallL},\,\phi^{1\smallR}$ & $\Phi^{2\smallL},\,\Phi^{2\smallR}$ & $\phi^{3\smallL},\,\phi^{3\smallR}$ & $\varphi^{0},\,\varphi^{4}$ \\
    \bottomrule
  \end{tabular}
  \caption{%
    Fundamental bosonic excitations of the $\AdS_3\times\S^3\times\S^3\times\S^1$ in light-cone gauge. Fermions follow by supersymmetry. The massive modes are labelled (left-right) by the sign of a suitable target-space angular momentum, which can be interpreted as the chirality in the dual~$\CFT_2$---much like in $\AdS_3\times\S^3\times\T^4$. The massless excitation~$\varphi^{0}$ is shared between the two spheres, while~$\varphi^{4}$ lives on the circle~$\S^1$.
  }
  \label{tab:excitations}
\end{table}

\subsection*{Quantum integrability}

The investigation of integrability beyond the classical limit was initiated by Ohlsson~Sax and Stefa\'nski in ref.~\cite{OhlssonSax:2011ms}, by constructing a weakly-coupled spin chain with~$\d21a^2$ symmetry. This corresponds to the strongly-coupled regime in the NLSM, \ie~to the opposite of what was considered in ref.~\cite{Babichenko:2009dk}.
Even at weak coupling, the spin~chain appears substantially more complicated than the~$\psu(1,1|2)^2$ one, which we discussed in chapter~\ref{ch:spinchain}. The~$\d21a^2$ spin chain is given by the direct product of two \emph{alternating} spin chains.%
\footnote{%
Alternating spin chains were also investigated in the context of integrability for the ABJM theory~\cite{Minahan:2008hf,Gaiotto:2008cg}.
} Each alternating chain has its odd and even sites in the short (infinite-dimensional) $\d21a$ representations $(\tfrac{\alpha}{2};\tfrac{1}{2};0)$ and $(\tfrac{1-\alpha}{2};0;\tfrac{1}{2})$, respectively. At weak coupling, the alternating spin chains are decoupled, and correspond to left- and right-moving excitations, as it happened for~$\psu(1,1|2)^2 $. The setup is illustrated in figure~\ref{fig:altspinchain}.

In ref.~\cite{Borsato:2012ud} this picture was used to study the all-loop integrability properties of the chain. As we did in chapter~\ref{ch:spinchain}, the first step of this analysis was to pick a vacuum for the chain that preserved as much supersymmetry as possible, and study fundamental excitations on top of it. This resulted in $4+4$ excitations, the bosonic ones being
\begin{equation}
\ket{\phi^{1\smallL}}\,,\quad
\ket{\phi^{1\smallR}}\,,\qquad
\ket{\phi^{3\smallL}}\,,\quad
\ket{\phi^{3\smallR}}\,.
\end{equation}
These are precisely the light modes of table~\ref{tab:excitations}. Each of these excitations forms a doublet with its fermionic partner, and transforms under~$\su(1|1)^2_{\text{c.e.}}$ that is, under the very same algebra that we have analysed at length in the previous chapters. In fact, should we consider only~$\ket{\phi^{1\smallL}}$ and~$\ket{\phi^{1\smallR}}$, we would have the same algebra and representation as in section~\ref{sec:fullrepr}, up to rescaling the mass. This rescaling can be easily done in terms of modified Zhukovski variables
\begin{equation}
\label{eq:Zhukovskirescaled}
\frac{x^+_{j,p}}{x^-_{j,p}}=e^{ip}\,,
\qquad
x^+_{j,p}+\frac{1}{x^+_{j,p}}-x^-_{j,p}-\frac{1}{x^-_{j,p}}=\frac{2i\,m_j}{h}\,,\end{equation}
with $m_1=\a $ for~$\ket{\phi^{1\,\smallL,\smallR}} $ and~$m_3=1-\a $ for~$\ket{\phi^{3\,\smallL,\smallR}} $. Then the dispersion relation reads
\begin{equation}
E=\frac{h}{2i}
\left(x^{+}_{j,p}-\frac{1}{x^{+}_{j,p}}-x^{-}_{j,p}+\frac{1}{x^{-}_{j,p}}\right)
=
\sqrt{m_j^2+4h^2\sin^2\left(\frac{p}{2}\right)}\,,
\end{equation}
so that a large-$h$ expansion%
\footnote{%
Scaling as usual the momentum as~$1/h$.
}
 reproduces the spectrum of table~\ref{tab:excitations}. Furthermore, for the same reasons as in the previous chapters, it is natural to take the spin chain to have a discrete left-right symmetry.
\begin{figure}
  \centering
\begin{tikzpicture}
  \begin{scope}[xshift=-0.95cm]
    \coordinate (c0) at (-4.5cm,0);
    \coordinate (c1) at (-4cm,0);
    \coordinate (c2) at (+5.2cm,0);
    \coordinate (c3) at (+5.7cm,0);

    \coordinate (c0b) at (-4.5cm,1cm);
    \coordinate (c1b) at (-4cm,1cm);
    \coordinate (c2b) at (+5.2cm,1cm);
    \coordinate (c3b) at (+5.7cm,1cm);


    \node (ll) at (6cm,1.2cm)   {$\fixedspaceL{\text{right}}{\text{left}}$};
    \node (rr) at (6cm,0.2cm)   {$\fixedspaceL{\text{right}}{\text{right}}$};

    \node (sp10) at (-3.6cm,0cm) [spin1] {$\phantom{\phi}$};
    \node (sp11) at (-2.4cm,0cm)   [spin2] {$\phantom{\phi}$};
    \node (sp12) at (-1.2cm,0cm) [spin1] {$\phantom{\phi}$};
    \node (sp13) at (0cm,0cm)    [spin2] {$\phantom{\phi}$};
    \node (sp14) at (1.2cm,0cm)  [spin1] {$\phantom{\phi}$};
    \node (sp15) at (2.4cm,0cm)    [spin2] {$\phantom{\phi}$};
    \node (sp16) at (3.6cm,0cm)  [spin1] {$\phantom{\phi}$};
    \node (sp17) at (4.8cm,0cm)  [spin2] {$\phantom{\phi}$};

    \node (sp20) at (-3.6cm,1cm) [spin1] {$\phantom{\phi}$};
    \node (sp21) at (-2.4cm,1cm)   [spin2] {$\phantom{\phi}$};
    \node (sp22) at (-1.2cm,1cm) [spin1] {$\phantom{\phi}$};
    \node (sp23) at (0cm,1cm)    [spin2] {$\phantom{\phi}$};
    \node (sp24) at (1.2cm,1cm)  [spin1] {$\phantom{\phi}$};
    \node (sp25) at (2.4cm,1cm)    [spin2] {$\phantom{\phi}$};
    \node (sp26) at (3.6cm,1cm)  [spin1] {$\phantom{\phi}$};
    \node (sp27) at (4.8cm,1cm)  [spin2] {$\phantom{\phi}$};
    
    \draw [thick,dashed]  (c0) to (c1);
    \draw [thick]  (c1) to (sp10);
    \draw [thick]  (sp10) to (sp11);
    \draw [thick]  (sp11) to (sp12);
    \draw [thick]  (sp12) to (sp13);
    \draw [thick]  (sp13) to (sp14);
    \draw [thick]  (sp14) to (sp15);
    \draw [thick]  (sp15) to (sp16);
    \draw [thick]  (sp16) to (sp17);
    \draw [thick]  (sp17) to (c2);
    \draw [thick,dashed]  (c2) to (c3);

    \draw [thick,dashed]  (c0b) to (c1b);
    \draw [thick]  (c1b) to (sp20);
    \draw [thick]  (sp20) to (sp21);
    \draw [thick]  (sp21) to (sp22);
    \draw [thick]  (sp22) to (sp23);
    \draw [thick]  (sp23) to (sp24);
    \draw [thick]  (sp24) to (sp25);
    \draw [thick]  (sp25) to (sp26);
    \draw [thick]  (sp26) to (sp27);
    \draw [thick]  (sp27) to (c2b);
    \draw [thick,dashed]  (c2b) to (c3b);   

	\draw [thick,dotted,violet] (-4cm,-0.4cm) rectangle  (-2cm,1.4cm);
	\draw [thick,dotted,violet] (-1.6cm,-0.4cm) rectangle  (0.4cm,1.4cm);
	\draw [thick,dotted,violet] (0.8cm,-0.4cm) rectangle  (2.8cm,1.4cm);
	\draw [thick,dotted,violet] (3.2cm,-0.4cm) rectangle  (5.2cm,1.4cm);

  \end{scope}
\end{tikzpicture}

  \caption{Pictorial representation of the $\d21a^2$ alternating spin chains. We can think of it as composed of two (left and right) alternating chain. For each of these, odd (cyan) sites are in the~$(\tfrac{\a}{2};\tfrac{1}{2};0)$ representation of~$\d21a$, while even (pink) sites are in the~$(\tfrac{1-\a}{2};\tfrac{1}{2};0)$ representation. The dotted squares represent the combination of sites which we identify as fundamental excitations of the whole~$\d21a^2$ chain.}
  \label{fig:altspinchain}
\end{figure}

Despite the many similarities, let us point out two important differences between this spin chain and the~$\psu(1,1|2)^2$ one. Firstly, the present symmetry algebra~$\su(1|1)^2_{\text{c.e.}}$ is \emph{half} of the~$\psu(1|1)^4_{\text{c.e.}}$ algebra of chapter~\ref{ch:spinchain}. What earlier was only one tensor product factor, which we introduced for convenience, is now the full, physical symmetry algebra of the spin-chain excitations.  Secondly, even if when we restrict to a subset of particles of a given mass we just need to rescale the Zhukovski variables~\eqref{eq:Zhukovskirescaled}, the presence of different masses introduces novel features. In particular, scattering processes involving particles of different masses are new to this spin chain.

Using this spin-chain picture, an all-loop S~matrix was worked out in ref.~\cite{Borsato:2012ud}. We can summarise it as a block matrix
\begin{equation}
\mathbf{S}=\left(
\begin{array}{c|c}
\mathbf{S}_{11} & \mathbf{S}_{31}\\
\hline
\mathbf{S}_{13} & \mathbf{S}_{33}
\end{array}
\right),
\end{equation}
where~$\mathbf{S}_{jj}$ is given by rewriting the~$\su(1|1)^2$ S~matrix of chapter~\ref{ch:smatrix} using the Zhukovski variables~\eqref{eq:Zhukovskirescaled}, and~$\mathbf{S}_{jk}$ with~$j\neq k $ scatters particle of different masses. One finds that the mass quantum number is always transmitted,  \eg
\begin{equation}
\ket{\mathcal{X}_{j,p}\,\mathcal{Y}_{k,q}}
\longrightarrow \# \ket{\widetilde{\mathcal{Y}}_{k,q}\,\widetilde{\mathcal{X}}_{j,p}}\,,
\qquad j\neq k\,,
\end{equation}
where~$\mathcal{X}_{j,p},\, \mathcal{Y}_{k,q}$ are magnon excitations of definite mass and momentum and~$\widetilde{\mathcal{X}}_{j,p},\, \widetilde{\mathcal{Y}}_{k,q}$ are their scattering products.
We can further subdivide the mixed-mass scattering by chirality~\eg
\begin{equation}
\mathbf{S}_{13}=\left(
\begin{array}{c|c}
\mathbf{S}_{13}^{\smallLL} & \mathbf{S}_{13}^{\smallRL}\\
\hline
\mathbf{S}_{13}^{\smallLR} & \mathbf{S}_{13}^{\smallRR}
\end{array}
\right).
\end{equation}
We then find that each block is given by the corresponding one from chapter~\ref{ch:smatrix}, up to rescaling each of the two Zhukovski variables independently.

The resulting S~matrix satisfies the Yang-Baxter equation and, after imposing unitarity and left-right symmetry, depends on \emph{four} dressing factors
\begin{equation}
\mathscr{S}_{jj}\,,\qquad\widetilde{\mathscr{S}}_{jj}\,,
\qquad\qquad
\mathscr{S}_{jk}\,,\qquad\widetilde{\mathscr{S}}_{jk}\,.
\end{equation}
In reducing the number of the dressing factors to four, we assumed that the mass-dependence in \eg~$\mathscr{S}_{11}$ and~$\mathscr{S}_{33}$ comes from the same functional dependence on~$m$. Crossing symmetry relates the first two of dressing factors among themselves, and the last two among themselves. So far, there is no all-loop proposal for these dressing factors. Even if there have been some tree-level and one-loop worldsheet perturbative calculations that validate%
\footnote{%
There may arise some confusion concerning ref.~\cite{Sundin:2013ypa}, where the one-loop matrix elements found for the limiting case~$\a=1$ are compared with the all-loop S~matrix at~$\tfrac{1}{2}\leq\a<1$ and a mismatch is found. This is because, as we have seen, the S~matrix radically changes in that limit, and the notion of fundamental particles is different at~$\a<1$ and~$\a=1$, see also the next subsection.
Taking this into account, we can match the calculations of ref.~\cite{Sundin:2013ypa} with \emph{both} the~$\AdS_3\times\S^3\times\S^3\times\S^1$ and~$\AdS_3\times\S^3\times\T^4$ S~matrices, for appropriate values of~$\a$.
}
 the proposed all-loop S~matrix~\cite{Babichenko:2009dk, Rughoonauth:2012qd,Abbott:2012dd,Sundin:2013ypa}, the understanding  of this model is quite limited and calculations are more involved due to the presence of the additional parameter~$\a$.

\begin{figure}
  \centering
  
\begin{tikzpicture}
  \begin{scope}
    \coordinate (m) at (0cm,0cm);

    \node (v1L) at (-1.25cm,   1cm) [dynkin node] {};
    \node (v2L) at (-0.625cm,  0cm) [dynkin node] {};
    \node (v3L) at (-1.25cm,  -1cm) [dynkin node] {};

    \draw [dynkin line] (v2L.south west) -- (v2L.north east);
    \draw [dynkin line] (v2L.north west) -- (v2L.south east);

    \draw [dynkin line] (v1L) -- (v2L);
    \draw [dynkin line] (v2L) -- (v3L);
    
    \node (v1R) at (+1.25cm,   1cm) [dynkin node] {};
    \node (v2R) at (+0.625cm,  0cm) [dynkin node] {};
    \node (v3R) at (+1.25cm,  -1cm) [dynkin node] {};

    \draw [inverse line] [out=  0+45,in= 90] (v1L) to (v2R);
    \draw [inverse line] [out=  0-45,in=270] (v3L) to (v2R);
    \draw [inverse line] [out=180-45,in= 90] (v1R) to (v2L);
    \draw [inverse line] [out=180+45,in=270] (v3R) to (v2L);

    \draw [dynkin line] (v1R.south west) -- (v1R.north east);
    \draw [dynkin line] (v1R.north west) -- (v1R.south east);

    \draw [dynkin line] (v2R.south west) -- (v2R.north east);
    \draw [dynkin line] (v2R.north west) -- (v2R.south east);

    \draw [dynkin line] (v3R.south west) -- (v3R.north east);
    \draw [dynkin line] (v3R.north west) -- (v3R.south east);

    \draw [dynkin line] (v1R) -- (v2R);
    \draw [dynkin line] (v2R) -- (v3R);
    \draw [dynkin line,double,double distance=0.5pt] (v3R) -- (v1R);

    \draw [red phase] [out=270-20,in= 90+20] (v1L) to (v3L);
    \draw [blue phase] [out=270+20,in=270-20] (v3L) to (v3R);
    \draw [red phase] [out= 90-20,in=270+20] (v3R) to (v1R);
    \draw [blue phase] [out= 90+20,in= 90-20] (v1R) to (v1L);

    \draw [red phase] [out=0,in=180] (v1L) to (v3R);
    \draw [red phase] [out=0,in=180] (v3L) to (v1R);

    \draw [blue phase] [out=180,in= 90+20,loop] (v1L) to (v1L);
    \draw [blue phase] [out=180,in=270-20,loop] (v3L) to (v3L);
    \draw [blue phase] [out=  0,in= 90-20,loop] (v1R) to (v1R);
    \draw [blue phase] [out=  0,in=270+20,loop] (v3R) to (v3R);
  \end{scope}

  \begin{scope}[xshift=+3cm,yshift=-0.75cm]
    \draw [dynkin line]  (0cm,1.5cm) -- (1cm,1.5cm) node [anchor=west,black] {\small Dynkin links};
    \draw [inverse line] (0cm,1.0cm) -- (1cm,1.0cm) node [anchor=west,black] {\small Fermionic inversion symmetry links};
    \draw [blue phase]   (0cm,0.5cm) -- (1cm,0.5cm) node [anchor=west,black] {\small Dressing phases $\mathscr{S}_{jj}$ and $\widetilde{\mathscr{S}}_{jj}$};
    \draw [red phase]    (0cm,0.0cm) -- (1cm,0.0cm) node [anchor=west,black] {\small Dressing phases $\mathscr{S}_{jk}$ and $\widetilde{\mathscr{S}}_{jk}$};
  \end{scope}
\end{tikzpicture}

  \caption{The Dynkin diagram for $\d21a^2$ in mixed grading (see figure~\ref{fig:d21a-dynkin-diagrams}), with the various interaction terms of the Bethe ansatz indicated.
Note how now, besides dressing phases that couple nodes of the same mass (in blue), there appear phases coupling nodes of mass~$\a$ with ones of mass~$1-\a$ (in red).  
  }
  \label{fig:bethe-equations-d21a}
\end{figure}
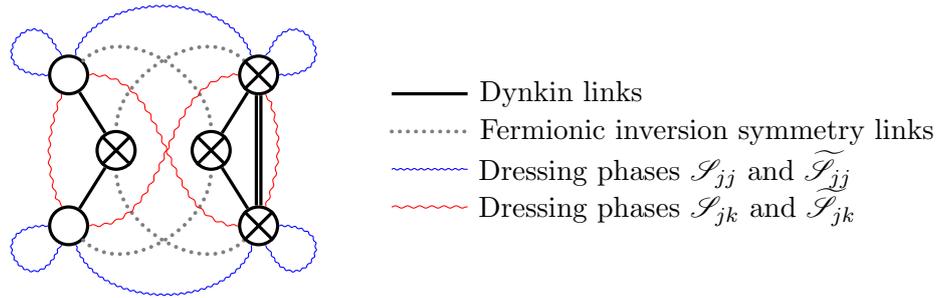

From the proposed S~matrix, the all-loop Bethe ansatz equations have been written in~\cite{Borsato:2012ss}, in a way much similar to what we did in section~\ref{sec:nesting}. Once again, the equations are written in a mixed grading for the two copies of~$\d21a$, and feature slightly different phases with respect to the early proposals that were reverse-engineered from the finite-gap equations, see refs.~\cite{Babichenko:2009dk,OhlssonSax:2011ms}. We schematically represent them in figure~\ref{fig:bethe-equations-d21a}.

\subsection*{Massless modes and \texorpdfstring{$\a\to1$}{alpha to 1} limit}
Once again, incorporating the massless excitations in the integrability description does not appear to be straightforward. Very recently, some progress has been made at the level of the finite-gap equations~\cite{Lloyd:2013wza}. The original set of finite-gap equations did not include any of the massless modes of the~$\AdS_3\times\S^3\times\S^3\times\S^1$ background, not even the one shared by the two spheres, which should be captured by the supercoset~\eqref{eq:d21acoset}. In ref.~\cite{Lloyd:2013wza} Lloyd and Stefa\'nski showed that such a mode can be accounted for by implementing the Virasoro constraints as suitable conditions on the residues  at the poles of the quasimomenta---in particular, as weaker conditions than the ones assumed in the earlier literature. Therefore, accounting for the massless mode does not result in additional cuts in the algebraic curve description, but rather in a modification of the residues of the quasimomenta, which contribute to the determination of the energy of the state. Such a prescription can be also employed to accommodate the massless mode of~$\S^1$, or the four modes of~$\T^4$. This is a promising advance, which hopefully will help including massless modes in general all-loop integrability scenarios.

Another interesting direction is to exploit this background to \emph{generate} massless modes. This can be done by taking the limit in which one of the spheres becomes flat,~$\a\to1$. At the level of~$\d21a$ this corresponds to a contraction of one of the~$\su(2)$ bosonic subalgebras. The limit is quite subtle, because this eliminates one of the simple roots, see figure~\ref{fig:dynkin-d21a-orig}, and introduces a novel fermionic one. At the level of the S~matrix, the limit is non-trivial too. We argued that the mass-one excitations~$\Phi^{2\,\smallL,\smallR}$ are composite in the case of~$\AdS_3\times\S^3\times\S^3\times\S^1$. In~$\AdS_3\times\S^3\times\T^4$, there are four \emph{fundamental} bosonic modes of mass one, two of which can be obtained from the one of mass~$\a$. The remaining two must be the once-composite modes~$\Phi^{2\,\smallL,\smallR}$.

All this makes the comparison not straightforward. However, in the limit of the weakly-coupled spin chain some progress could be made~\cite{Sax:2012jv}. There it was shown that the $\mathbf{R}$ matrix is regular in the limit~$\a\to1$, where its representations can be studied. The~$\d21a$ alternating spin chain does not quite go to the~$\psu(1,1|2)$ homogeneous one.
While the even sites are given by the~$(\tfrac{1}{2},\tfrac{1}{2})$ $\psu(1,1|2)$ module (see section~\ref{sec:spinchainrepr}), the odd ones are given by~$0\oplus0\oplus(\tfrac{1}{2},\tfrac{1}{2})$, \ie~also feature two singlets. These are related to the massless modes, and yield a degenerate vacuum. By writing down the corresponding Bethe ansatz equations and carefully studying their degeneracies it is possible to show that the degeneration is compatible with the chiral ring of the dual CFT---this is however quite subtle and requires considerations that go beyond the weakly-coupled regime.

\section{Mixed-flux backgrounds}
\label{sec:mixedflux}
Up to this point, our focus has been on integrable non-linear $\sigma $~models corresponding to superstring backgrounds supported by a pure Ramond-Ramond flux. This is in close analogy with the prototypical integrability example of the~$\AdS_5\times\S^5$ superstring. As we mentioned in the introduction, however, $\AdS_3$ backgrounds are special in that they can also be supported by a Neveu-Schwarz-Neveu-Schwarz flux,  or by a combination of NSNS and RR fluxes.
As discussed, the pure NSNS theory is in a way simpler to quantise and, at least in the case of~$\AdS_3\times\S^3\times\T^4$, is well-understood~\cite{Giveon:1998ns,Maldacena:2000hw, Maldacena:2000kv, Maldacena:2001km,Gaberdiel:2011vf}%
\footnote{%
The $\AdS_3\times\S^3\times\S^3\times\S^1$ background has also been studied, but remains more obscure~\cite{Elitzur:1998mm,FigueroaO'Farrill:2000ei, Gukov:2004ym,Tong:2014yna}.
}. Such a case can in fact be described as a supersymmetric extension of the~$SL(2)\times SU(2) $ Wess-Zumino-Witten (WZW) model, whence the free-string spectrum can be found by CFT techniques.
The mixed-flux case has also been considered~\cite{Berkovits:1999im,Rahmfeld:1998zn,Pesando:1998wm}, but until recently the solution of the relative spectral problem appeared to be out of reach.

\subsection*{Classical integrability}
In ref.~\cite{Cagnazzo:2012se}, Cagnazzo and Zarembo investigated such mixed flux backgrounds. In general, their Lagrangian takes the form
\begin{equation}
\mathcal{L}= -\frac{h}{2}\Big(\gamma^{\a\b}G_{\mu\nu}(X)+q\,\eps^{\a\b}B_{\mu\nu}(X)\Big)\,\partial_{\a} X^{\mu}\partial_{\b} X^{\nu}
+\text{fermions}\,.
\end{equation}
In a coset formulation such as the one of chapter~\ref{ch:sigmamodel}, this can be written as
\begin{equation}
\label{eq:NSSNcoset}
S=-\frac{h}{2}\Big(\int \de^2\sigma \gamma^{\a\b}\str(A^{(2)}_{\a}A^{(2)}_{\b})
+\frac{q}{3}\int_{\mathcal{B}}\!\!\! \de^3\sigma \eps^{\a\b\gamma} \str(A^{(2)}_{\a}A^{(2)}_{\b}A^{(2)}_{\gamma})\Big)
+\text{fermions}\,.
\end{equation}
Note that the last term is non-local, as it is a topological integral over a three-dimensional manifold~$\mathcal{B}$ such that its boundary~$\partial\mathcal{B}$ is the string worldsheet. Locally its integrand is a total derivative, so that the equations of motion can be consistently written in terms of worldsheet fields only. 
The choice~$q=0$ gives back the pure RR case, while~$q=1$ corresponds to the pure NSNS one. Interestingly, the string-theoretical S~duality exchanges the RR and NSNS fluxes, which corresponds to sending~$q\to\sqrt{1-q^2} $ in this description. However, S~duality is not realized perturbatively, so that the relative invariance will not necessarily be manifest in the string spectrum.
Using an appropriate extension of~\eqref{eq:NSSNcoset} to include fermions, it was shown in ref.~\cite{Cagnazzo:2012se} that the corresponding classical theory is integrable for a suitable choice of the couplings. Such a choice is precisely the same that guarantees $\kappa$~symmetry as well as conformal invariance. In fact, the same picture holds at the classical level also when including the massless modes and considering the~$\AdS_3\times\S^3\times\S^3\times\S^1 $ background.

\subsection*{Towards quantum integrability}
Following this realisation, Hoare and Tseytlin started investigating the integrability of the quantum theory in the S-matrix approach that we described in chapter~\ref{ch:smatrix}. In ref.~\cite{Hoare:2013pma}, they computed the tree-level S~matrix for~$q\neq0$, focusing on the simplest case of massive excitations on~$\AdS_3\times\S^3\times\T^4 $. In that case it turned out that the $q$-dependence was quite mild, and amounted to a relatively simply modification of some of the S-matrix elements and in particular of the tree-level dispersion relation
\begin{equation}
\label{eq:tree-levelNSNSdisp}
\sqrt{p^2+1}\quad\longrightarrow\quad
\sqrt{(p\pm q)^2+1-q^2}\,,
\end{equation}
where the sign~$\pm$ should be chosen suitably for each the S-matrix element. Building on that, in ref.~\cite{Hoare:2013ida} they proposed an all-loop, mixed-flux S~matrix in terms of a modification of the one proposed in ref.~\cite{Borsato:2013qpa}. The ``matrix part'' of~$\mathbf{S}$ should then be given by replacing the usual%
\footnote{%
The form in which we write the definition of the Zhukovski parameters here is equivalent to~\eqref{eq:zhukovski-def}, as it can be seen by using the definition of~$E(p)$ in terms of~$x^{\pm}$.
}
definition of the Zukhovski variables
\begin{equation}
e^{ip}=\frac{x^+(p)}{x^-(p)}\,,
\qquad
E(p)+1=i\,h\,\big(x^{-}(p)+x^{+}(p)\big),
\end{equation}
with the~$q$-dependent ones
\begin{equation}
e^{ip}=\frac{x^+_{\pm}(p)}{x_{\pm}^-(p)}\,,
\qquad
E_{\pm}(p)+1\pm 2 hq\,\sin \tfrac{p}{2}=ih\,\sqrt{1-q^2}\,\big(x_{\pm}^{-}(p)+x_{\pm}^{+}(p)\big),
\end{equation}
where we in fact introduced two pairs of Zhukovski parameters~$x^{\pm}_{\pm}$, each pair depending on one modified dispersion relation~$E_{\pm}$, with%
\footnote{%
Special care is needed to consider the limit~$q\to1$.
}
\begin{equation}
\label{eq:NSNSdispers}
E_{\pm}(p)^2=
\big(1\pm 2 hq\,\sin\frac{p}{2}\big)^2+ 4h^2(1-q^2)\sin^2\big(\frac{p}{2}\big).
\end{equation}
This conjecture for the S~matrix is compatible with the tree-level one, and with the off-shell symmetry algebra of the mixed flux theory, which however does not fix the dispersion relation entirely. The dispersion relation can essentially be narrowed down to~\eqref{eq:NSNSdispers} by requiring periodicity in~$p$, which is what happens in the pure-RR~case. As we mentioned in the introduction, one can think of this periodicity as coming from a discretisation of the worldsheet whence the spin-chain picture emerges---which is well established in several instances of integrable (pure-RR) string backgrounds.

To further validate this proposal, Hoare, Stepanchuk and Tseytlin~\cite{Hoare:2013lja} have investigated solitonic solutions for the mixed flux theory. Studying the dyonic giant magnon solution, they  were able to determine that the dispersion relation in that case takes the form
\begin{equation}
\label{eq:correctNSNSdispers}
E_{\pm}(p)^2=
(1\pm hq\,p)^2+ 4h^2(1-q^2)\sin^2\big(\frac{p}{2}\big),
\end{equation}
that in fact is incompatible with the earlier conjecture~\eqref{eq:NSNSdispers}, despite going to the same tree-level expression~\eqref{eq:tree-levelNSNSdisp}.
Now the $q$-linear term in brackets spoils periodicity in the momentum~$p$---a completely new and somewhat unexpected feature. Note that this also contradicts the na\"ive expectation of periodicity from a discretisation of the worldsheet that we discussed in the introduction and is at odds with a spin-chain description.
This last picture is also validated from the study of the bound-state dispersion relation~\cite{Hoare:2013lja}, and is consistent with the study of finite-size giant magnons later performed in ref.~\cite{Babichenko:2014yaa}.

Very recently, the complete all-loop S~matrix for the mixed-flux theory has been proposed from the study of the off-shell symmetry algebra~\cite{Lloyd:2014bsa}. Also in the massless and mixed-mass sector one finds that, modulo some subtelties, the ratios of the S-matrix elements can be expressed simply in terms of the pure-RR one with deformed Zhukovski variables. Note also that this last study also predict a dispersion relation as in eq.~\eqref{eq:correctNSNSdispers}. Still, some further investigation may be desirable, which may come through a two-loop calculation%
\footnote{%
In the present case, a simpler  near-flat-space calculation may not be sufficient to unveil the form of the dispersion relation, and a subtler near-BMN calculation would be necessary, perhaps taking advantage of the unitarity techniques that we mentioned in chapter~\ref{ch:comparison}.
}
for the dispersion relation, in the spirit of~\cite{Klose:2007rz,Murugan:2012mf}, that for these backgrounds has not been performed yet. Perhaps also due to the non-trivial analytic structure of the dispersion---implying in particular the absence of the familiar rapidity torus---the dressing factors of the theory seem to be very intricate. Not only they cannot be found by plugging deformed Zhukovski variables in the pure-RR dressing factors---which in any case was not to be expected on general grounds---but also their leading-order expressions appear quite are quite a bit more involved than in the pure-RR theory~\cite{Babichenko:2014yaa}.

In conclusion, there are still some puzzles for mixed-flux integrability even in the simplest case of~$\AdS_3\times\S^3\times\T^4 $, but we are  witnessing significant progress in the subject and we expect more in the next several months.
We find this particularly exciting because it may offer a way to understand the relation between the notion of solvability in terms of the worldsheet S~matrix (the pure-RR case) and by representation theory of chiral algebras (pure-NSNS). Such a structure has been partially uncovered in the case of relativistic massless integrable  theories that we mentioned in section~\ref{sec:massless}. In the case of the massless sine-Gordon model,  it was possible to  see how the higher charges from integrability fit into the Virasoro algebra, thus relating the two descriptions, see ref.~\cite{Fendley:1993jh}. Whether something similar will happen in the $\AdS_3/\CFT_2$ setup and what role is played by the dual $\CFT_2$ is a very interesting question.

\section{Conclusions and outlook}
In this chapter we have reviewed the most promising and relevant lines of research in integrability for  $\AdS_3/\CFT_2$. As we have seen, rapid progress is being made in including massless modes into the integrability machinery, see refs.~\cite{Sax:2012jv,Lloyd:2013wza, Borsato:2014exa,Borsato:2014hja}. More general backgrounds such as~$\AdS_3\times\S^3\times\S^3$~\cite{Borsato:2012ud, Borsato:2012ss} or the ones supported by mixed fluxes~\cite{Cagnazzo:2012se,Hoare:2013ida,Hoare:2013lja, Hoare:2013pma,Babichenko:2014yaa,Bianchi:2014rfa} are also being actively investigated, and show indications of being integrable.
All this offers a unique playground to gain a new understanding of integrability for massless (non-relativistic) S~matrices, of the relation between the charges of the Zamolodchikov-Faddeev algebra with the Virasoro ones appearing in the~$\CFT_2$ as well as in the worldsheet~$\CFT$, and of string dualities.

Once the massless modes issue has been finally cleared, it  would also be very interesting to further investigate the relations between the integrability construction and the symmetric-product orbifold CFT which is the dual to~$\AdS^3\times\S^3\times\T^4 $ strings~\cite{Maldacena:1997re,Giveon:1998ns,deBoer:1998pp, Maldacena:2000hw, Maldacena:2000kv, Maldacena:2001km, Kutasov:1999xu,Larsen:1999uk,Seiberg:1999xz}. Quite likely, such a study would be most fruitful in a spin-chain framework such as the one of ref.~\cite{Pakman:2009mi}. Repeating that study for the~$\AdS_3\times\S^3\times\S^3\times\S^1$ background could then shed some new light on the quite obscure $\CFT_2$ dual arising in that case~\cite{Elitzur:1998mm,FigueroaO'Farrill:2000ei, Gukov:2004ym,Tong:2014yna}.

We should also mention a related but somewhat more distant field of investigation, that is the study of \emph{higher-spin theories} in~$\AdS_3 $ backgrounds. Non-trivial theories involving fields of spin higher than two can be formulated on three-dimensional Minkowski space and on~$\AdS_{d}$ spaces. Such fields may be thought as arising from the tensionless limit of string theory~\cite{Sundborg:2000wp,witten:talk, Sezgin:2002rt} whereby the tower of string states becomes massless. A special feature of~$\AdS_3$ is that there (like it happens for gravity~\cite{Achucarro:1987vz,Witten:1988hc}) higher-spin theories can be described as Chern-Simons theories~\cite{Blencowe:1988gj}.
In this way, it was possible to show that asymptotically they enjoy a~$\mathcal{W}_\infty(\lambda)$ infinite-dimensional symmetry~\cite{Campoleoni:2010zq,Henneaux:2010xg}, which also characterises minimal models in~$\CFT_2$~\cite{Zamolodchikov:1985wn}, see ref.~\cite{Bouwknegt:1992wg} for a review. This provided a new approach to holograpic dualities, and lead to rapid developments~\cite{Gaberdiel:2010pz,Gaberdiel:2011wb, Candu:2012jq,Henneaux:2012ny}---see also ref.~\cite{Gaberdiel:2012uj} for a review---including a recent study of higher spins in the~$\AdS_3\times\S^3\times\S^3\times\S^1$ background~\cite{Gaberdiel:2013vva}, whereby the same~$\d21a$ representations constructed in~\cite{OhlssonSax:2011ms} and used in \cite{Borsato:2012ud,Borsato:2012ss} emerge~\cite{Gaberdiel:IGST}.
While a comparison of higher-spin theory results with the S-matrix integrability ones has not been performed yet, we expect these two approaches to make contact in the future, perhaps through the investigation of the Yangian symmetries of the theory---such an investigation has been recently initiated in ref.~\cite{Pittelli:2014ria}. This may lead to a more quantitative understanding of the tensionless limit---at least at the level of the free string spectrum---adding to the recent progresses in flat space~\cite{Sagnotti:2010at} and in the context of~$\AdS_4/\CFT_3 $~\cite{Chang:2012kt}.

Another interesting direction which should be explored in due course is to consider \emph{deformations} of the~$\AdS_3$ backgrounds. This road has been taken in the case of $\AdS_5/\CFT_4$, where a plethora of deformations has been shown to preserve integrability, such as orbifolds~\cite{Kachru:1998ys,Lawrence:1998ja} and T-duality-shift-T-duality (TsT) deformations~\cite{Lunin:2005jy,Frolov:2005dj} of the background, and quantum deformations (in the sense of quantum groups) of the underlying symmetry algebra~\cite{Beisert:2008tw,Beisert:2011wq, Hoare:2011wr,deLeeuw:2011jr} or of the NLSM Poisson structure~\cite{Delduc:2013qra}, which also result in an integrable NLSM closely related to the $\AdS_5\times\S^5 $ one~\cite{Arutyunov:2013ega,Arutynov:2014ota}.\footnote{A more systematic classification of integrable deformations of AdS/CFT has been recently initiated, see in particular ref.~\cite{Kawaguchi:2014qwa}.} For a review of the spectral problem in these contexts see refs.~\cite{Zoubos:2010kh,vanTongeren:2013gva}. The rationale in these approaches is to get rid of as much manifest symmetry as possible in an effort to get closer to more realistic theories.  This amounts to breaking R-symmetry and supersymmetry in the geometric deformations of the background, and even conformal (and Lorentz) symmetry in the case of quantum deformations.
What could be new and very interesting in the case of~$\AdS_3$ is that integrable deformations of that sort may include BTZ-like black hole geometries, which as we remarked in the introduction are in fact locally isometric to~$\AdS_3$. In fact,  it appears that the quotient that yields the BTZ background imposes boundary conditions that are compatible with classical integrability, so that the classical string theory is integrable there~\cite{David:2011iy,David:2012aq}. This raises high hopes of putting an integrability handle on such geometries, which would be extremely interesting.

All of these avenues appear very exciting, and unique to three-dimensional gravity theories and their dual two-dimensional CFTs. We are confident that in the near future we will witness substantial progress along many of them.

\renewcommand\thechapter{A}
\chapter{Appendices}
\fancyhf{}
\fancyhead[LE,RO]{\thepage}
\fancyhead[LO]{\nouppercase{Appendices}}
\fancyhead[RE]{\nouppercase{Appendix \rightmark}}
\renewcommand{\headrulewidth}{0.5pt}

\section{Generalities of the~\texorpdfstring{$\psu(1,1|2)$}{psu(1,1|2)} superalgebra}
\label{app:psu112}
We have given the (anti)commutation relations of the superalgebra $\psu(1,1|2)$ in section~\ref{sec:supercoset}.  In this appendix we will collect some useful additional notions about it, see also ref.~\cite{Frappat:1996pb}.

\subsection*{Serre-Chevalley bases}
Superalgebras have in general several inequivalent Dynkin diagrams, corresponding to different choices of simple roots. Each such choice gives a set of Cartan generators $\mathfrak{h}_i$, and corresponding raising and lowering operators $\mathfrak{e}_i$ and $\mathfrak{f}_i$, where the index $i$ takes values from $1$ to the rank of the algebra, which is $3$ for $\psu(1,1|2)$. These generators satisfy an algebra of the form
\begin{equation}
  \comm{\mathfrak{h}_i}{\mathfrak{h}_j} = 0 , \qquad
  \comm{\mathfrak{e}_i}{\mathfrak{f}_j} = \delta_{ij} \mathfrak{h}_j , \qquad
  \comm{\mathfrak{h}_i}{\mathfrak{e}_j} = + A_{ij} \mathfrak{e}_j , \qquad
  \comm{\mathfrak{h}_i}{\mathfrak{f}_j} = - A_{ij} \mathfrak{f}_j ,
\end{equation}
where $A_{ij}$ is the Cartan matrix.

Here we will mainly consider two gradings of $\psu(1,1|2)$. In the $\su(2)$ grading the simple roots are given by
\begin{equation}\label{eq:SC-basis-su2}
  \begin{aligned}
    \mathfrak{h}_1 &= +\mathbf{L}_3 - \mathbf{J}_3 , \qquad &
    \mathfrak{e}_1 &= +\mathbf{Q}_{+--} , \qquad &
    \mathfrak{f}_1 &= +\mathbf{Q}_{-++} , \\
    \mathfrak{h}_2 &= +2\,\mathbf{J}_3 , \qquad &
    \mathfrak{e}_2 &= +\mathbf{J}_+ , \qquad &
    \mathfrak{f}_2 &= +\mathbf{J}_- , \\
    \mathfrak{h}_3 &= +\mathbf{L}_3 - \mathbf{J}_3 , \qquad &
    \mathfrak{e}_3 &= +\mathbf{Q}_{+-+} , \qquad &
    \mathfrak{f}_3 &= -\mathbf{Q}_{-+-} .
  \end{aligned}
\end{equation}
This leads to the Cartan matrix
\begin{equation}\label{eq:Cartan-su2}
  \begin{pmatrix}
     0 & -1 &  0 \\
    -1 & +2 & -1 \\
     0 & -1 &  0
  \end{pmatrix}.
\end{equation}
The corresponding Dynkin diagram is shown in figure~\ref{fig:dynkin-su22}~\subref{fig:dynkin-su22-su}. In the construction of the coset of chapter~\ref{ch:sigmamodel} and in the spin-chain one of chapter~\ref{ch:spinchain}, we generally pick~$\psu(1,1|2)_{\L}$ to be in this grading. 
The Dynkin diagram for the $\sl(2)$ grading is shown in figure~\ref{fig:dynkin-su22}~\subref{fig:dynkin-su22-sl}. This is the grading that we generally use for~$\psu(1,1|2)_{\R}$.

There are also fermionic gradings of~$\psu(1,1|2)$, in which all three raising operators $\mathfrak{e}_i$ are odd. In particular we can choose them to be either
\begin{equation*}
  \mathbf{Q}_{+-+}, \quad \mathbf{Q}_{++-}, \quad \mathbf{Q}_{-++} \,, \qquad 
  \text{or} \qquad
  \mathbf{Q}_{-+-}, \quad \mathbf{Q}_{--+}, \quad \mathbf{Q}_{+--} \,.\,
\end{equation*}
This leads to the Cartan matrices
\begin{equation}\label{eq:Cartan-ferm}
    \begin{pmatrix}
     0 & +1 &  0 \\
    +1 & 0 & -1 \\
     0 & -1 &  0
  \end{pmatrix} , \qquad
  \text{and} \qquad
    \begin{pmatrix}
     0 & -1 &  0 \\
    -1 & 0 & +1 \\
     0 & +1 &  0
  \end{pmatrix},
\end{equation}
respectively, corresponding to the Dynkin diagram in figure~\ref{fig:dynkin-su22}~\subref{fig:dynkin-su22-fff}.

\begin{figure}
  \centering

  \subfloat[\label{fig:dynkin-su22-su}]{
    \begin{tikzpicture}
      [
      thick,
      node/.style={shape=circle,draw,thick,inner sep=0pt,minimum size=5mm}
      ]

      \useasboundingbox (-1.5cm,-1cm) rectangle (1.5cm,1cm);

      \node (v1) at (-1.1cm, 0cm) [node] {};
      \node (v2) at (  0.0cm, 0cm) [node] {};
      \node (v3) at (  1.1cm, 0cm) [node] {};

      \draw (v1.south west) -- (v1.north east);
      \draw (v1.north west) -- (v1.south east);

      \draw (v3.south west) -- (v3.north east);
      \draw (v3.north west) -- (v3.south east);

      \draw (v1) -- (v2);
      \draw (v2) -- (v3);

      \node at (v2.south) [anchor=north] {$+1$};
    \end{tikzpicture}
  }
  \hspace{1cm}
  \subfloat[\label{fig:dynkin-su22-fff}]{
    \begin{tikzpicture}
      [
      thick,
      node/.style={shape=circle,draw,thick,inner sep=0pt,minimum size=5mm}
      ]

      \useasboundingbox (-1.5cm,-1cm) rectangle (1.5cm,1cm);

      \node (v1) at (-1.1cm, 0cm) [node] {};
      \node (v2) at (  0.0cm, 0cm) [node] {};
      \node (v3) at (  1.1cm, 0cm) [node] {};

      \draw (v1.south west) -- (v1.north east);
      \draw (v1.north west) -- (v1.south east);

      \draw (v2.south west) -- (v2.north east);
      \draw (v2.north west) -- (v2.south east);

      \draw (v3.south west) -- (v3.north east);
      \draw (v3.north west) -- (v3.south east);

      \draw (v1) -- (v2);
      \draw (v2) -- (v3);

      \node at (v2.south) [anchor=north] {$\pm 1$};
    \end{tikzpicture}
  }
  \hspace{1cm}
  \subfloat[\label{fig:dynkin-su22-sl}]{
    \begin{tikzpicture}
      [
      thick,
      node/.style={shape=circle,draw,thick,inner sep=0pt,minimum size=5mm}
      ]

      \useasboundingbox (-1.5cm,-1cm) rectangle (1.5cm,1cm);

      \node (v1) at (-1.1cm, 0cm) [node] {};
      \node (v2) at (  0.0cm, 0cm) [node] {};
      \node (v3) at (  1.1cm, 0cm) [node] {};

      \draw (v1.south west) -- (v1.north east);
      \draw (v1.north west) -- (v1.south east);

      \draw (v3.south west) -- (v3.north east);
      \draw (v3.north west) -- (v3.south east);

      \draw (v1) -- (v2);
      \draw (v2) -- (v3);

      \node at (v2.south) [anchor=north] {$-1$};
    \end{tikzpicture}
  }
  
  \caption{Three Dynkin diagrams for $\psu(1,1|2)$.}
  \label{fig:dynkin-su22}
\end{figure}

In the $\sl(2)$ grading we have
\begin{equation}\label{eq:SC-basis-sl2}
  \begin{aligned}
    \widetilde{\mathfrak{h}}_1 &= -\mathbf{L}_3 + \mathbf{J}_3 , \qquad &
    \widetilde{\mathfrak{e}}_1 &= -\mathbf{Q}_{-++} , \qquad &
    \widetilde{\mathfrak{f}}_1 &= +\mathbf{Q}_{+--} , \\
    \widetilde{\mathfrak{h}}_2 &= +2\,\mathbf{L}_3 , \qquad &
    \widetilde{\mathfrak{e}}_2 &= +\mathbf{L}_- , \qquad &
    \widetilde{\mathfrak{f}}_2 &= -\mathbf{L}_+ , \\
    \widetilde{\mathfrak{h}}_3 &= -\mathbf{L}_3 + \mathbf{J}_3 , \qquad &
    \widetilde{\mathfrak{e}}_3 &= -\mathbf{Q}_{-+-} , \qquad &
    \widetilde{\mathfrak{f}}_3 &= -\mathbf{Q}_{+-+} ,
  \end{aligned}
\end{equation}
with the Cartan matrix
\begin{equation}\label{eq:Cartan-sl2}
  \begin{pmatrix}
     0 & +1 &  0 \\
    +1 & -2 & +1 \\
     0 & +1 &  0
  \end{pmatrix}.
\end{equation}

\subsection*{A continuous automorphism}
It is useful to note that~$\psu(1,1|2)$ admits an $\u(1)$-automorphism~$\mathbf{U}$, acting on the supercharges as
\begin{equation}
\label{eq:Uautomorphism}
  \comm{\mathbf{U}}{\mathbf{Q}_{a\kappa\pm}} = \pm \frac{1}{2} \mathbf{Q}_{a\kappa\pm}\, ,
\end{equation}
and commuting with the bosonic charges. If one thinks of $\psu(1,1|2)$ as a contraction of the exceptional superalgebra $\mathfrak{d}(2,1;\alpha)$ with $\alpha\to0$, the generator~$\mathbf{U}$ can be identified with with one of the generators of~$\su(2)\subset\mathfrak{d}(2,1;\alpha)$ that is contracted. This is described in more detail in ref.~\cite{OhlssonSax:2011ms}.

\subsection*{Supermatrix realisation}
Once the reality condition~\eqref{eq:algebrareality} is imposed, we are left with eight even independent supermatrices. Eliminating trace and supertrace, we can identify the six even generators as 
\begin{equation}
\begin{aligned}
&\mathbf{L}_{1}=\frac{1}{2}
\mbox{\scriptsize$\left(\begin{array}{cccc}
0&-i&0&0\\
i&0&0&0\\
0&0&0&0\\
0&0&0&0
\end{array}\right)$},
\qquad\qquad
&&\mathbf{J}_{1}=\frac{1}{2}
\mbox{\scriptsize$\left(\begin{array}{cccc}
0&0&0&0\\
0&0&0&0\\
0&0&0&i\\
0&0&i&0
\end{array}\right)$},
\\
&\mathbf{L}_{2}=\frac{1}{2}
\mbox{\scriptsize$\left(\begin{array}{cccc}
0&1&0&0\\
1&0&0&0\\
0&0&0&0\\
0&0&0&0
\end{array}\right)$},
\qquad
&&\mathbf{J}_{2}=\frac{1}{2}
\mbox{\scriptsize$\left(\begin{array}{cccc}
0&0&0&0\\
0&0&0&0\\
0&0&0&-1\\
0&0&1&0
\end{array}\right)$},
\\
&\mathbf{L}_{3}=\frac{1}{2}
\mbox{\scriptsize$\left(\begin{array}{cccc}
i&0&0&0\\
0&-i&0&0\\
0&0&0&0\\
0&0&0&0
\end{array}\right)$},
\qquad
&&\mathbf{J}_{3}=\frac{1}{2}
\mbox{\scriptsize$\left(\begin{array}{cccc}
0&0&0&0\\
0&0&0&0\\
0&0&i&0\\
0&0&0&-i
\end{array}\right)$},
\end{aligned}
\end{equation}
while the eight supercharges are
\begin{equation}
\begin{aligned}
&\mathbf{Q}^{1}_{11}=
\mbox{\scriptsize$\left(\begin{array}{cccc}
0&0&i&0\\
0&0&0&0\\
i&0&0&0\\
0&0&0&0
\end{array}\right)$},
\qquad\qquad
&&\mathbf{Q}^{2}_{11}=
\mbox{\scriptsize$\left(\begin{array}{cccc}
0&0&-1&0\\
0&0&0&0\\
1&0&0&0\\
0&0&0&0
\end{array}\right)$},\\
&\mathbf{Q}^{1}_{21}=
\mbox{\scriptsize$\left(\begin{array}{cccc}
0&0&0&i\\
0&0&0&0\\
0&0&0&0\\
i&0&0&0
\end{array}\right)$},
\qquad
&&\mathbf{Q}^{2}_{21}=
\mbox{\scriptsize$\left(\begin{array}{cccc}
0&0&0&-1\\
0&0&0&0\\
0&0&0&0\\
1&0&0&0
\end{array}\right)$},\\
&\mathbf{Q}^{1}_{12}=
\mbox{\scriptsize$\left(\begin{array}{cccc}
0&0&0&0\\
0&0&-i&0\\
0&i&0&0\\
0&0&0&0
\end{array}\right)$},
\qquad
&&\mathbf{Q}^{2}_{12}=
\mbox{\scriptsize$\left(\begin{array}{cccc}
0&0&0&0\\
0&0&1&0\\
0&1&0&0\\
0&0&0&0
\end{array}\right)$},\\
&\mathbf{Q}^{1}_{22}=
\mbox{\scriptsize$\left(\begin{array}{cccc}
0&0&0&0\\
0&0&0&-i\\
0&0&0&0\\
0&i&0&0
\end{array}\right)$},
\qquad
&&\mathbf{Q}^{2}_{22}=
\mbox{\scriptsize$\left(\begin{array}{cccc}
0&0&0&0\\
0&0&0&1\\
0&0&0&0\\
0&1&0&0
\end{array}\right)$}.
\end{aligned}
\end{equation}
While this choice of the generators satisfies~\eqref{eq:algebrareality}, it can be useful to consider complex combinations such as~$\mathbf{L}_{\pm}, \mathbf{J}_{\pm}$ and~$\mathbf{Q}_{\pm\pm\pm}$. In particular we have 
\begin{equation}
\mathbf{L}_{\pm}=\mathbf{L}_{1}\pm i\,\mathbf{L}_{2}\,,
\qquad
\mathbf{J}_{\pm}=\mathbf{J}_{1}\pm i\,\mathbf{J}_{2}\,.
\end{equation}

\subsection*{Two copies of $\psu(1,1|2)$}
We can take the representation for $\mathcal{M}$, \ie\ the one defining the left copy of~$\psu(1,1|2)$, to be given by the previous expressions. The matrix in the representation~$\widetilde{\mathcal{M}}$, instead, is in a representation that has opposite weights for the bosonic subalgebra. This can be realised by  sending
\begin{equation}
\label{eq:bosonflip}
\mathbf{L}_2\to-\mathbf{L}_2\,,
\quad
\mathbf{L}_3\to-\mathbf{L}_3\,,
\qquad
\mathbf{J}_2\to-\mathbf{J}_2\,,
\quad
\mathbf{J}_3\to-\mathbf{J}_3\,.
\end{equation}
In fact, this corresponds to exchanging raising with lowering operators~$\mathbf{L}_{\pm}$ and~$\mathbf{J}_{\pm}$ without affecting the commutation relations. The transformation straightforwardly extends to the whole superalgebra.

The transformation~\eqref{eq:bosonflip} implies that the bosonic subalgebras in each (left and right) copy of~$\psu(1,1|2)$ have  opposite notions of what are the positive bosonic roots. If we take this notion to be induced from the choice of simple roots, it follows that the two copies cannot be in the same grading. More specifically, if one copy is in the grading of fig.~\ref{fig:dynkin-su22-su} with Cartan matrix~\eqref{eq:Cartan-su2}, then the other is in the grading of fig.~\ref{fig:dynkin-su22-sl} with Cartan matrix~\eqref{eq:Cartan-sl2}. If one of them has the fully-fermion grading of fig.~\ref{fig:dynkin-su22-fff}, then the other does too, and they have Cartan matrices of opposite signs, see~\eqref{eq:Cartan-ferm}.

It is also interesting to notice that then the $\mathbbm{Z}_{4}$-automorphism $\Omega$ of eq.~\eqref{eq:automorphism} acts on a bosonic matrix as
\begin{equation}
\Omega(\mathbf{L}_j\oplus \mathbf{O})=
\mathbf{O}\oplus\widetilde{\mathbf{L}}_j\,,
\qquad
\Omega(\mathbf{J}_j\oplus \mathbf{O})=
\mathbf{O}\oplus\widetilde{\mathbf{J}}_j\,,
\end{equation}
\ie~exchanges the left and right bosonic subalgebras.

\section{Explicit expressions for the coset action}
\label{app:coset}
Here we will collect some explicit expressions that appear when evaluating the coset action of chapter~\ref{ch:sigmamodel} and expanding it.

\subsection*{Parametrisation of group elements}
We parametrise a group element by~\eqref{eq:gparam}. The light-cone coordinate~$\Lambda(t,\phi)$ has been defined in eq.~\eqref{eq:Lambdaparam}. The transverse coordinates~$\{x_k\}=(z_1,z_2,y_1,y_2)$ are given by
\begin{equation}
\label{eq:gxparam}
\begin{aligned}
g(x)=&\;i\,\frac{\mathbf{I}_z+z_1\Sigma_{1}+z_2\Sigma_{2}}{\sqrt{1-\frac{1}{4}|z|^2}}
+
i\,\frac{\mathbf{I}_y+ y_1\Sigma_{3}+y_2\Sigma_{4}}{\sqrt{1+\frac{1}{4}|y|^2}}\,,
\end{aligned}
\end{equation}
with
\begin{equation}
\label{eq:Sigmai-def}
\begin{aligned}
\mathbf{I}_z=\tfrac{1}{2}\big(\mathbf{I}+\Sigma_{+}\Sigma_{-}\big),
\qquad
&\Sigma_{1}=\mathbf{L}_{1}^{\smallL}-\mathbf{L}_{1}^{\smallR}\,,
\qquad
&\Sigma_{3}=\mathbf{J}_{1}^{\smallL}-\mathbf{J}_{1}^{\smallR}\,,
\\
\mathbf{I}_y=\tfrac{1}{2}\big(\mathbf{I}-\Sigma_{+}\Sigma_{-}\big),
\qquad
&\Sigma_{2}=\mathbf{L}_{2}^{\smallL}-\mathbf{L}_{2}^{\smallR}\,,
\qquad
&\Sigma_{4}=\mathbf{J}_{2}^{\smallL}-\mathbf{J}_{2}^{\smallR}\,.
\end{aligned}
\end{equation}

We have split the fermions in massive and massless ones. The former are $\theta^{1}_{\smallL},\theta^{2}_{\smallL}$ and $\theta^{1}_{\smallR},\theta^{2}_{\smallR}$ that in terms of a $\psu(1,1|2)^2$ matrix we write as
\begin{equation}
\label{eq:Psim-param}
{\Psi_{m}}=
\left({
\begin{array}{cccc}
0 & 0 & 0 & -{\theta}^1_{\smallL}\\
0 & 0 & \bar{\theta}_{2\smallL} & 0\\
0 & {\theta}^{2}_{\smallL} & 0 & 0\\
\bar{\theta}_{1\smallL} & 0 & 0 & 0
\end{array}
}\right)
\oplus
\left({
\begin{array}{cccc}
0 & 0 & 0 &-{\theta}^{2}_{\smallR}\\
0 & 0 & -\bar{\theta}_{1\smallR} & 0\\
0 & -{\theta}^1_{\smallR} & 0 & 0\\
\bar{\theta}_{2\smallR} & 0 & 0 & 0
\end{array}
}\right).
\end{equation}
The four massless fermions are instead
\begin{equation}
{\Psi_{l}}=
\left({
\begin{array}{cccc}
0 & 0 & {\eta}^{1}_{\smallL} & 0\\
0 & 0 & 0 & {\eta}^{2}_{\smallL}\\
-\bar{\eta}_{1\smallL} & 0 & 0 & 0\\
0 & \bar{\eta}_{2\smallL} & 0 & 0
\end{array}
}\right)
\oplus
\left({
\begin{array}{cccc}
0 & 0 & \bar{\eta}^{2}_{\smallR} & 0\\
0 & 0 & 0 & -{\eta}^{1}_{\smallR}\\
-\bar{\eta}_{2\smallR} & 0 & 0 & 0\\
0 & -\bar{\eta}_{1\smallR} & 0 & 0
\end{array}
}\right).
\end{equation}
In both cases, the bar denotes complex conjugation, \eg
\begin{equation}
\big({\theta}^{1}_{\smallL}\big)^{\dagger}= \bar{\theta}_{1\smallL}\,,
\end{equation}
ans so on. We can now parametrise the corresponding group element as
\begin{equation}
g(\Psi_{m,l})=1+\Psi_{m,l}+\frac{1}{2}\Psi_{m,l}^2\,.
\end{equation}
Note that one can think that such a parametrisation emerges equivalently from $g(\Psi)=\exp(\Psi)$ or $g(\Psi)=\Psi+\sqrt{1+\Psi^2}$, since the expansions terminate at~$O(\Psi^2)$.

Finally, the $\T^4$ elements can be found from exponentiating the corresponding generators,
\begin{equation}
g(X)=e^{i \sum_{k=1}^4X_k T^k}=\prod_{k=1}^4e^{i X_k T^k}\,,
\end{equation}
where we made it explicit that they all commute.

\subsection*{Explicit expressions for perturbative evaluation}
To define quantities related to the transverse bosonic fields $x_{k}$ let us introduce indices $a,s$ so that $x_{a}=(z_1,z_2)$ and $x_{s}=(y_1,y_2)$. Then the metric elements are
\begin{equation}
g_{\pm}=\left(
\frac{1}{\sqrt{4+|y|^2}}\pm
\frac{1}{\sqrt{4-|z|^2}}
\right),
\qquad
g_a=\frac{2i\, z_a}{\sqrt{4-|z|^2}}\,,
\quad
g_s=\frac{2i\, y_s}{\sqrt{4+|y|^2}}\,.
\end{equation}
It is useful to rewrite the parametrisation~\eqref{eq:gxparam} of~$g(x)$ in light-cone coordinates
\begin{equation}
g(x)=i\,g_{+}\mathbf{I}+i\,g_{-}\Sigma_8+g_k\Sigma_k\,,
\end{equation}
where $\mathbf{I}$ is the identity, $\Sigma_8=-\Sigma_{+}\Sigma_{-}$ and $\Sigma_{k}$ are given by~\eqref{eq:Sigmai-def}.
It is then easy to compute
\begin{equation}
g(x)^2=G_{+}\mathcal{I}+G_{-}\Sigma_{8}+G_{M}\Sigma_M\,,
\end{equation}
where now
\begin{equation}
 G_{\pm}=-\left(
\frac{4-|y|^2}{4+|y|^2}\pm
\frac{4+|z|^2}{4-|z|^2}
\right),
\qquad
G_a=-\frac{8\, z_a}{{4-|z|^2}}\,,
\quad
G_s=-\frac{8\, y_s}{{4+|y|^2}}\,.
\end{equation}
We can work out the expression of the momenta ~$\{p_k\}=(p^{z}_{1},p^{z}_{2},p^{y}_{1},p^{y}_{2})$ canonically conjugated to~$\{x_k\}=(z_1,z_2,y_1,y_2)$, in terms of the auxiliary fields~$\varpi_{k}$. They are
\begin{equation}
p_a=\frac{4\,\varpi_a}{4-|z|^2}\,,
\qquad
p_s=\frac{4\,\varpi_s}{4+|y|^2}\,.
\end{equation}

Let us now look at the Lagrangian more in detail.
The even and odd currents are given by
\begin{equation}
\begin{aligned}
A_{\text{even}}=&-\frac{i}{2}g^{-1}_{x}\left(\de x_{+}\Sigma_{+}\big(1+2\Psi_m^2\big)+\frac{1}{2}\de x_{-}\Sigma_{-}-i\big[\de \Psi_m,\Psi_m\big]\right)g_{x}-g^{-1}_{x}\de g_{x}\,,\\
A_{\text{odd}}=&-i\,g^{-1}_{x}\,\de x_{+}\Sigma_{+}\Psi_{m}g_{x}-g^{-1}_{x}\de \Psi_{m}g_{x}\,,
\end{aligned}
\end{equation}
where we  used that the fermion parametrisation~\eqref{eq:Psim-param} is quadratic at most.
The part of the even current that depends solely on the transverse bosonic coordinates is
\begin{equation}
A^{\perp}_{\text{even}}=-\frac{1}{2}g^{-1}_{x}\big[\dot{\Psi}_m,\Psi_m\big]g_{x}
-g^{-1}_{x}\dot{g}_{x}\,.
\end{equation}
Using these expressions it is easy to get to eq.~\eqref{eq:lagrangianevenodd}. Moreover we have
\begin{equation}
\str(\varpi g^{-1}_{x}\dot{g}_{x})=p_k\dot{x}_k\,.
\end{equation}
We can also further simplify
\begin{equation}
\begin{aligned}
\str(\varpi g^{-1}_x \big[\dot{\Psi}_m,\Psi_m\big] g_x)=
&
\frac{i}{4}\big(\varpi_{+}G_{+}-\varpi_{-} G_{-}\big)\str\big(\Sigma_{+}\big[\dot{\Psi}_m,\Psi_m\big]\big)\\
&+\frac{1}{4}\varpi_{k}\str\big(\Sigma_{k}g^{-1}_x\big[\dot{\Psi}_m,\Psi_m\big]g_x\big)\,.
\end{aligned}
\end{equation}

\subsubsection*{Solving the constraints}
We have from $\mathcal{C}_1=0$
\begin{equation}
\label{eq:c1fermions}
-x'_{-}=p_{k}x'_{k}+\frac{1}{4}\varpi_{k} \str\left(\Sigma_{k}g^{-1}_{x}\big[\Psi'_{m},\Psi_m\big]g_{x}\right)+\frac{i}{8}\,\str\left(\Sigma_{+}\big[\Psi'_{m},\Psi_m\big]\right)\,,
\end{equation}
which we can expand as
\begin{equation}
-x_{-}'=p_{k}x_{k}'-\frac{i}{4}\str \big(\Sigma_{+}\Psi_{m}\Psi_{m}'\big)+\dots\,,
\end{equation}
where the ellipsis indicate higher order terms.
The quadratic constraint is
\begin{equation}
\mathcal{C}_{2}=\frac{1}{G_{+}}\big(\varpi_{-}G_{-}+1\big)\varpi_{-}+
|\varpi_{k}|^2+h\,\str\big((A^{(2)}_1)^2\big)\,,
\end{equation}
where
\begin{equation}
\begin{aligned}
A^{(2)}_1=&
-\frac{i}{4}x_{-}'\big(g^{-1}_x\Sigma_{-}g_{x}
-\Omega(g^{-1}_x \Sigma_{-}g_{x})\big)-\frac{1}{2}\big(g^{-1}_x g_x'
+g_{x}'g^{-1}_x \big)\\
&-\frac{1}{4}\left(g^{-1}_x \big[\Psi_m',\Psi_m\big] g_x
-\Omega\big(g^{-1}_x \big[\Psi_m',\Psi_m\big] g_{x}\big)\right)\,.
\end{aligned}
\end{equation}
By dropping higher order terms, from $\mathcal{C}_2=0$ we have
\begin{equation}
\varpi_{-}=-\frac{1}{2}|\varpi_{k}|^2-\frac{1}{2}|x_{k}'|^2+\dots
\end{equation}
Dropping the total $\tau$-derivative $\dot{x}_{-}$, the Lagrangian reads
\begin{equation}
\label{eq:expandedLagrangian}
\mathcal{L}_2=\mathbf{p}_{-}+p_{k}\dot{x}_k
-\frac{i}{4}\str\big(\Sigma_{+}\Psi_{m}\dot{\Psi}_{m}\big)
-\frac{i}{4}\eps^{\a\b}\str\big( A^{\text{odd}}_{\a}\,\Omega(A^{\text{odd}}_{\b})\big)
+\dots
\,,
\end{equation}
with
\begin{equation}
\mathbf{p}_{-}=\varpi_{-}-\frac{1}{2}\big(|x_k|^2+\str\Psi_m^2\big)+\dots\,.
\end{equation}
For the purpose of expanding the Lagrangian and the Noether charge~\eqref{eq:noethercharge} in powers of the fields  it is useful to note that at leading order it is
\begin{equation}
A^{\text{odd}}_{0}=-i\Psi_m-\dot{\Psi}_m+\dots\,,
\qquad
A^{\text{odd}}_{1}=-\Psi_m'+\dots\,.
\end{equation}
When plugged into the Lagrangian~\eqref{eq:expandedLagrangian}, and of using the explicit form of $\Psi_m$, this gives contributions of the form
\begin{equation}
-\frac{i}{4}\eps^{\a\b}\str\big( A^{\text{odd}}_{\a}\,\Omega(A^{\text{odd}}_{\b})\big)=
\frac{1}{2}\str\big(\Sigma_{+}\Psi_{m} \,\Omega(\Psi_{m}')\big)
-\frac{i}{2}\big(\dot{\Psi}_{m}\,\Omega(\Psi_{m}')\big)+\dots\,,
\end{equation}
where the double-derivative terms all vanish after two integrations by parts. By these expressions it is easy to derive the Lagrangian and Hamiltonian of the main text~\eqref{eq:quadrH-matrix}.

To derive some of these results is useful to use the identities
\begin{equation}
\Sigma_{\pm}g^{-1}_x=g_{x}\Sigma_{\pm}\,,
\qquad
\Sigma_{\pm}\Psi_{m}=\mp \Psi_{m}\Sigma_{\pm}\,.
\end{equation}

\section{Quadratic charges in terms of the fields}
\label{app:quadraticcharges}
In order  to compactly write down that form  of the symmetry generators, it is useful to introduce complex bosonic coordinates
\begin{equation}
X^{1}_{\smallL}=z_{1}+i\,z_{2}\,,
\quad
X^{2}_{\smallL}=y_{1}+i\,y_{2}\,,
\qquad
{X}^{1}_{\smallR}=z_{1}-i\,z_{2}\,,
\quad
{X}^{2}_{\smallR}=y_{1}-i\,y_{2}\,,
\end{equation}
and conjugate momenta
\begin{equation}
P_{1\,\smallL}=p^{z}_{1}-ip^{z}_{2}\,,
\quad
P_{2\,\smallL}=p^{y}_{1}-ip^{y}_{2}\,,
\quad
P_{1\,\smallR}=p^{z}_{1}+ip^{z}_{2}\,,
\quad
P_{2\,\smallR}=p^{y}_{1}+ip^{y}_{2}\,.
\end{equation}
It is clear that $(X^{j}_{\smallL})^{\dagger}=X^{j}_{\smallR}$ and $(P_{j\,\smallL})^{\dagger}=P_{j\,\smallR}$.
We will raise and lower indices by
\begin{equation}
X_{j\,\smallL}=\delta_{jk}X^{k}_{\smallL}\,,
\quad
X_{j\,\smallR}=\delta_{jk}X^{k}_{\smallR}\,,
\qquad
P^{j}_{\smallL}=\delta^{jk}P_{k\,\smallL}\,,
\quad
P^{j}_{\smallR}=\delta^{jk}P_{k\,\smallR}\,.
\end{equation}
Similarly, if necessary, we will raise an lower the fermion indices by
\begin{equation}
\theta_{j\,\smallL,\smallR}=\delta_{jk}\theta^{j}_{\smallL,\smallR}\,,
\qquad
\bar{\theta}^{j}_{\smallL,\smallR}=\delta^{jk}\bar{\theta}_{j\,\smallL,\smallR}\,.
\end{equation}
The quadratic Lagrangian becomes
\begin{equation}
\mathcal{L}_2=P_{j\,\smallL}\dot{X}^{j}_{\smallL}+P_{j\,\smallR}\dot{X}^{j}_{\smallR}
+i\,\bar{\theta}_{j\,\smallL}\dot{\theta}^{j\,\smallL}
+i\,\bar{\theta}_{j\,\smallR}\dot{\theta}^{j\,\smallR}
-\mathcal{H}_2\,.
\end{equation}

In particular, this implies that the upon quantisation the non-vanishing commutators are
\begin{equation}
\big[X^{j}_{\smallL}(\sigma),P_{k\,\smallL}(\tilde{\sigma})\big]=i\, 
\delta^{j}_{k}\,\delta(\sigma-\tilde{\sigma})\,,
\quad
\big[X^{j}_{\smallR}(\sigma),P_{k\,\smallR}(\tilde{\sigma})\big]=i\, 
\delta^{j}_{k}\,\delta(\sigma-\tilde{\sigma})\,,
\end{equation}
compatibly with the ones we originally found~(\ref{eq:PB-boson}--\ref{eq:PB-fermion}).

The form of the bosonic charges is then
\begin{equation}
\label{eq:quadraticbos}
\begin{aligned}
&\mathbf{H}=\int \de \sigma\bigg(\frac{1}{2}P^{j}_{\smallL}P_{j\,\smallR}
+2X^{j}_{\smallL}{}'X_{j\,\smallR}'
+2X^{j}_{\smallL}{}X_{j\,\smallR}
+\theta^{j}_{\smallL}\theta_{j\smallR}'
-\bar{\theta}^{j}_{\smallL}\bar{\theta}_{j\smallR}'
+\bar{\theta}_{j\,\smallL}{\theta}^{j}_{\smallL}
+\bar{\theta}_{j\,\smallR}{\theta}^{j}_{\smallR}\bigg),\\
&\mathbf{M}=\int \de \sigma\left(
\bar{\theta}_{j\,\smallL}{\theta}^{j}_{\smallL}
-\bar{\theta}_{j\,\smallR}{\theta}^{j}_{\smallR}
+i P_{j\,\smallL}X^{j}_{\smallL}-i P_{j\,\smallR}X^{j}_{\smallR}
\right),
\\
&\mathbf{N}=\int \de \sigma\,i\left(
P_{1\,\smallL}X^{1}_{\smallL}
-P_{2\,\smallL}X^{2}_{\smallL}
- P_{1\,\smallR}X^{1}_{\smallR}
+ P_{2\,\smallR}X^{2}_{\smallR}
\right).
\end{aligned}
\end{equation}
The four supercharges~$\mathbf{Q}^{j\,\L,\R}$ read
\begin{equation}
\label{eq:charges-fields}
\begin{aligned}
&\mathbf{Q}^{1\L}=
e^{-\frac{i}{4}\pi} \int \de \sigma
e^{\frac{i}{2}x_{-}}
\bigg(
\frac{i}{2}P_{2\smallR}\,\bar{\theta}_{1\smallL}
-X_{\smallL}^{2}\,\theta_{\smallR}^{1}{}'
+X_{\smallL}^{2}\,\bar{\theta}_{1\smallL}\\
&\hspace{7cm}
-\frac{i}{2}P_{1\smallL}\,\theta_{2\smallL}
+X_{\smallR}^{1}\,\bar{\theta}_{2\smallR}'
+X_{\smallR}^1\,\theta_{\smallL}^2
\bigg),
\\
&\mathbf{Q}^{1\R}=
e^{-\frac{i}{4}\pi}\int \de \sigma
e^{\frac{i}{2}x_{-}}
\bigg(
\frac{i}{2}P_{2\smallL}\,\bar{\theta}_{1\smallR}
-X_{\smallR}^{2}\,\theta_{\smallL}^{1}{}'
+X_{\smallR}^{2}\,\bar{\theta}_{1\smallR}\\
&\hspace{7cm}
-\frac{i}{2}P_{1\smallR}\,\theta_{\smallR}^2
+X_{\smallL}^{1}\,\bar{\theta}_{2\smallL}'
+X_{\smallL}^1\,\theta_{\smallR}^2
\bigg),
\\
&\mathbf{Q}^{2\L}=
e^{+\frac{i}{4}\pi}\int \de \sigma
e^{\frac{i}{2}x_{-}}
\bigg(
\frac{i}{2}P_{1\smallL}\,{\theta}^{1}_{\smallL}
+X_{\smallR}^{1}\,\bar{\theta}_{1\smallR}'
-X_{\smallR}^{1}\,{\theta}^{1}_{\smallL}\\
&\hspace{7cm}
+\frac{i}{2}P_{2\smallR}\,\bar{\theta}_{2\smallL}
+X_{\smallL}^{2}\,{\theta}^{2}_{\smallR}{}'
+X_{\smallL}^2\,\bar{\theta}_{2\smallL}
\bigg),
\\
&\mathbf{Q}^{2\R}=
e^{+\frac{i}{4}\pi}\int \de \sigma
e^{\frac{i}{2}x_{-}}
\bigg(
\frac{i}{2}P_{1\smallR}\,{\theta}^{1}_{\smallR}
+X_{\smallL}^{1}\,\bar{\theta}_{1\smallL}'
-X_{\smallL}^{1}\,{\theta}^{1}_{\smallR}\\
&\hspace{7cm}
+\frac{i}{2}P_{2\smallL}\,\bar{\theta}_{2\smallR}
+X_{\smallR}^{2}\,{\theta}^{2}_{\smallL}{}'
+X_{\smallR}^2\,\bar{\theta}_{2\smallR}
\bigg),
\end{aligned}
\end{equation}
and the remaining four can be found by taking the complex conjugate of these expressions. Notice how left and right supercharges differ only by exchanging the labels~$\bigL\leftrightarrow\bigR$. The factors $e^{\pm\frac{i}{4}\pi} $ is inserted so that the charges take a simpler expression in terms of oscillators.

\subsection*{From field to oscillators}
We can define bosonic lowering operators
\begin{equation}
\begin{aligned}
a_{\smallL}^{--}(p)
=&
\frac{1}{\sqrt{2\pi}}
\int \frac{\de \sigma}{\sqrt{\omega_p}}
\left(
\omega_p X_{\smallL}^{1}(\sigma)
+\frac{i}{2}
P_{1\smallR}(\sigma)
\right)\,e^{-ip\sigma}\,,
\\
a_{\smallR}^{--}(p)
=&
\frac{1}{\sqrt{2\pi}}
\int \frac{\de \sigma}{\sqrt{\omega_p}}
\left(
\omega_p X_{\smallR}^{1}(\sigma)
+\frac{i}{2}
P_{1\smallL}(\sigma)
\right)\,e^{-ip\sigma}\,,
\\
a_{\smallL}^{++}(p)
=&
\frac{1}{\sqrt{2\pi}}
\int \frac{\de \sigma}{\sqrt{\omega_p}}
\left(
\omega_p X_{\smallL}^{2}(\sigma)
+\frac{i}{2}
P_{2\smallR}(\sigma)
\right)\,e^{-ip\sigma}\,,
\\
a_{\smallR}^{++}(p)
=&
\frac{1}{\sqrt{2\pi}}
\int \frac{\de \sigma}{\sqrt{\omega_p}}
\left(
\omega_p X_{\smallR}^{2}(\sigma)
+\frac{i}{2}
P_{2\smallL}(\sigma)
\right)\,e^{-ip\sigma}\,,
\end{aligned}
\end{equation}
with the raising operator being the conjugate of these. Similarly, for fermions we have
\begin{equation}
\begin{aligned}
a_{\smallL}^{-+}(p)
=&
\frac{e^{+\frac{i}{4}\pi}}{\sqrt{2\pi}}
\int \frac{\de \sigma}{\sqrt{\omega_p}}
\left(
f_p\,\theta^{1}_{\smallL}(\sigma)
+i\,g_p \bar{\theta}_{1\smallR}(\sigma)
\right)\,e^{-ip\sigma}\,,
\\
a_{\smallR}^{-+}(p)
=&
\frac{e^{+\frac{i}{4}\pi}}{\sqrt{2\pi}}
\int \frac{\de \sigma}{\sqrt{\omega_p}}
\left(
f_p\,\theta^{1}_{\smallR}(\sigma)
+i\,g_p \bar{\theta}_{1\smallL}(\sigma)
\right)\,e^{-ip\sigma}\,,
\\
a_{\smallL}^{+-}(p)
=&
\frac{e^{-\frac{i}{4}\pi}}{\sqrt{2\pi}}
\int \frac{\de \sigma}{\sqrt{\omega_p}}
\left(
f_p\bar{\theta}_{2\smallL}(\sigma)
-i\,g_p\theta^2_{\smallR}(\sigma)
\right)\,e^{-ip\sigma}\,,
\\
a_{\smallR}^{+-}(p)
=&
\frac{e^{-\frac{i}{4}\pi}}{\sqrt{2\pi}}
\int \frac{\de \sigma}{\sqrt{\omega_p}}
\left(
f_p\bar{\theta}_{2\smallR}(\sigma)
-i\,g_p\theta^2_{\smallL}(\sigma)
\right)\,e^{-ip\sigma}\,.
\end{aligned}
\end{equation}
Once we drop the $x_-$-dependence from~\eqref{eq:charges-fields}, and by making use of the explicit form of the wave-function parameters~(\ref{eq:bosonicparam}--\ref{eq:fermionicparam}) with mass $m=1$, the charges can be recast in the form~(\ref{eq:bosonic-charges1}--\ref{eq:supercharges1}).

\section{Dualisation of the Bethe ansatz equations}
\label{app:duality}

While performing the nesting procedure in chapter~\ref{ch:betheansatz}, we had to chose a set of level-I excitations. Here we will show that the different choices we could make lead there, to equivalent equations (\ie~that describe the same spectrum), by performing a fermionic duality~\cite{Essler:1992nk,Beisert:2005fw} of the Bethe equations.

\subsection*{Duality for the~$\su(1|1)^2_{\text{c.e.}}$ Bethe equations}
Let us start from the simpler case we treated in section~\ref{sec:nesting}. We will work in the spin-chain frame for definiteness.

The idea is that one can write a new set of equations equivalent to the previous one, in which all the auxiliary roots $y, \tilde{y}$ are replaced by a dual set of auxiliary roots $\widehat{y}, \widehat{\tilde{y}}$.
We will see that the resulting equations have the same form as the original ones, up to exchanging the left and right sectors, which makes it clear that they correspond to the choice~$V^{\I}_B$ in~\eqref{eq:BA-excit-set}.

Let us define the polynomial $\mathscr{P}(\xi)$ as
\begin{equation}
\mathscr{P}(\xi) = \prod_{j=1}^{M^{\I}_{\smallL}} (\xi - x_j^+)   \prod_{j=1}^{M^{\I}_{\smallR}} (\xi - \frac{1}{ \tilde{x}_j^-})    - \prod_{j=1}^{M^{\I}_{\smallL}} (\xi - x_j^-)    \prod_{j=1}^{M^{\I}_{\smallR}} (\xi- \frac{1}{ \tilde{x}_j^+}) .
\end{equation}
This is a polynomial of degree $n=M^{\I}_{\smallL} + M^{\I}_{\smallR} -1$. Then the Bethe equations for the left and right auxiliary excitations can be rewritten respectively as
\begin{equation}
\mathscr{P}(y)=0\,,
\qquad\qquad
 \mathscr{P}(1/\tilde{y})=0\,.
\end{equation}
Another zero of the polynomial is at $\xi =0$. The equation $\mathscr{P}(0)=0$ is equivalent to the level-matching condition. We denote the remaining zeros by $\widehat{y}$ and $\widehat{\tilde{y}}$, because they correspond to the dualisation of respectevely left and right excitations. We can thus rewrite~$\mathscr{P}(\xi)$ as
\begin{equation}
\mathscr{P}(\xi) = \xi \prod_{j=1}^{M^{\II}_{\smallL}} (\xi - y_j)  \prod_{j=1}^{\widehat{M}^{\II}_{\smallL}} (\xi - \widehat{y}_j) \prod_{j=1}^{M^{\II}_{\smallR}} (\xi - \frac{1}{ \tilde{y}_j})  \prod_{j=1}^{\widehat{M}^{\II}_{\smallR}} (\xi - \frac{1}{ \widehat{\tilde{y}}_j})
\end{equation}
for $\widehat{M}^{\II}_{\smallL}=M^{\I}_{\smallL} -K^{\II}_{\smallL} -1$ and $\widehat{M}^{\II}_{\smallR}=M^{\I}_{\smallR} -K^{\II}_{\smallR} -1$ roots.
Let us now dualise the Bethe equations for level-I excitations.
We write the expression $\mathscr{P}(x_k^+)/\mathscr{P}(x_k^-)$ using the two possible representations for the polynomial, getting the equation
\begin{equation}
\begin{aligned}
 \frac{x_k^+}{x_k^-} \prod_{j=1}^{M^{\II}_{\smallL}} \frac{x_k^+ - y_j}{x_k^- - y_j}  \prod_{j=1}^{\widehat{M}^{\II}_{\smallL}} \frac{x_k^+ - \widehat{y}_j}{x_k^- - \widehat{y}_j} \prod_{j=1}^{M^{\II}_{\smallR}} \frac{x_k^+ -1/ \tilde{y}_j}{x_k^- - 1/ \tilde{y}_j}  \prod_{j=1}^{\widehat{M}^{\II}_{\smallR}} \frac{x_k^+ - 1/ \widehat{\tilde{y}}_j}{x_k^- - 1/ \widehat{\tilde{y}}_j} = 
\qquad\qquad \qquad \qquad 
 \\
 - \prod_{j=1}^{M^{\I}_{\smallL}} \frac{x_k^+ - x_j^-}{x_k^- - x_j^+}   \prod_{j=1}^{M^{\I}_{\smallR}} \frac{x_k^+ - 1/ \tilde{x}_j^+}{x_k^- -1/ \tilde{x}_j^-} \,,
\end{aligned}
\end{equation}
that becomes
\begin{equation}
\begin{aligned}
 \left(\frac{x_k^+}{x_k^-}\right)^{-1-M^{\II}_{\smallR}-\widehat{M}^{\II}_{\smallR}+ M^{\I}_{\smallR}}  
\prod_{\substack{j = 1\\j \neq k}}^{M^{\I}_{\smallL}} \frac{x_k^+ - x_j^-}{x_k^- - x_j^+}  
\prod_{j=1}^{M^{\II}_{\smallL}} \frac{x_k^- - y_j}{x_k^+ - y_j} 
\prod_{j=1}^{M^{\II}_{\smallR}} \frac{1 - \frac{1}{x_k^- \tilde{y}_j}}{1 -\frac{1}{x_k^+ \tilde{y}_j}} =
\qquad\qquad\qquad
 \\
  \prod_{j=1}^{\widehat{M}^{\II}_{\smallL}} \frac{x_k^+ - \widehat{y}_j}{x_k^- - \widehat{y}_j}   
\prod_{j=1}^{\widehat{M}^{\II}_{\smallR}} \frac{1 - \frac{1}{x_k^+ \widehat{\tilde{y}}_j}}{1 - \frac{1}{x_k^- \widehat{\tilde{y}}_j}} 
\prod_{j=1}^{M^{\I}_{\smallR}} \frac{1 -\frac{1}{x_k^- \tilde{x}_j^-}}{1 - \frac{1}{x_k^+ \tilde{x}_j^+}} \,,
\end{aligned}
\end{equation}
where the exponent of $x_k^+/x_k^-$ is in fact 0. With the help of this substitution, the Bethe equation for left excitations can thus be rewritten as
\begin{equation}
      \left(\frac{x_k^+}{x_k^-}\right)^{\ell} = 
      \prod_{\substack{j = 1\\j \neq k}}^{M^{\I}_{\smallL}} \mathscr{S}^{\smallLL}_{kj}
	 \prod_{j=1}^{\widehat{M}^{\II}_{\smallL}} \frac{x_k^+ - \widehat{y}_j}{x_k^- - \widehat{y}_j}
   \prod_{j=1}^{M^{\I}_{\smallR}} \sqrt{\frac{1- \frac{1}{x_k^- \tilde{x}_j^-}}{1- \frac{1}{x_k^+ \tilde{x}_j^+}}} \mathscr{S}^{\smallLR}_{k\tilde{\jmath}}
	\prod_{j=1}^{\widehat{M}^{\II}_{\smallR}} \frac{1 - \frac{1}{x_k^+ \widehat{\tilde{y}}_j}}{1 - \frac{1}{x_k^- \widehat{\tilde{y}}_j}} \,.
\end{equation}
This equation has precisely the same form as the original equation for the right excitations~\eqref{eq:BAE11:Right}.

Similarly, using $\mathscr{P}(1/\tilde{x}_k^+)/\mathscr{P}(1/\tilde{x}_k^-)$, one can obtain the dualised Bethe equation for type $\tilde{1}$ excitations
\begin{equation}
 \left(\frac{\tilde{x}_k^+}{\tilde{x}_k^-}\right)^{\ell} =
      \prod_{\substack{j = 1\\j \neq k}}^{M^{\I}_{\smallR}} \frac{\tilde{x}_k^+ - \tilde{x}_j^-}{\tilde{x}_k^- - \tilde{x}_j^+} \mathscr{S}^{\smallRR}_{\tilde{k}\tilde{\jmath}}
      \prod_{j=1}^{\widehat{M}^{\II}_{\smallR}} \frac{\tilde{x}_k^- - \widehat{\tilde{y}}_j}{\tilde{x}_k^+ - \widehat{\tilde{y}}_j}
     \prod_{j=1}^{M^{\I}_{\smallL}} \sqrt{\frac{1- \frac{1}{\tilde{x}_k^+ x_j^+}}{1- \frac{1}{\tilde{x}_k^- x_j^-}}} \mathscr{S}^{\smallRL}_{\tilde{k}j}
      \prod_{j=1}^{\widehat{M}^{\II}_{\smallL}} \frac{1 - \frac{1}{\tilde{x}_k^- \widehat{y}_j}}{1- \frac{1}{\tilde{x}_k^+ \widehat{y}_j}}\,,
\end{equation}
which has the same form as the original Bethe equations for type 1 excitations.

We have found a new set of Bethe equations, with the same form of the original ones up to exchanging left with right excitations and  substituting $(y,K_2)$ and $(\tilde{y},K_{\tilde{2}})$ by $(\widehat{y},\widehat{K}_2)$ and~$(\widehat{\tilde{y}},\widehat{K}_{\tilde{2}})$, respectively.

\subsection*{Duality for the $\psu(1|1)^4_{\text{c.e.}}$ Bethe equations}

As argued in section \ref{sec:BAE}, there are four different possible gradings in which we can write the all-loop Bethe equations for the~$\psu(1|1)^4_{\text{c.e.}}$ S~matrix.
Once again, these are related by dualities. 
Labelling the excitation numbers~$K_j$ by the corresponding Cartan elements as in eqs.~(\ref{eq:labelcartan1}--\ref{eq:labelcartan2}), let us define again the polynomial
\begin{equation}
\mathscr{P}(\xi)= \prod_{j=1}^{K_2} (\xi - x_j^+) \prod_{j=1}^{K_{\tilde{2}}} (\xi - \frac{1}{\tilde{x}_j^-}) -\prod_{j=1}^{K_2}(\xi - x_j^-) \prod_{j=1}^{K_{\tilde{2}}}(\xi- \frac{1}{ \tilde{x}_j^+})\,,
\end{equation}
of degree  $n=K_2+K_{\tilde{2}}-1$.
The Bethe equations for auxiliary roots $y_1,y_3,y_{\tilde{1}},y_{\tilde{3}}$ can be written respectively as
\begin{equation}
\mathscr{P}(y_1)=0, \qquad \mathscr{P}(y_3)=0, \qquad \mathscr{P}(1/y_{\tilde{1}})=0, \qquad \mathscr{P}(1/y_{\tilde{3}})=0.
\end{equation}
We can choose to dualise either the auxiliary roots $y_1, y_{\tilde{1}}$ or $y_3, y_{\tilde{3}}$. In the first case we consider a set of dual $\widehat{K}_1,\widehat{K}_{\tilde{1}}$ roots such that $K_1+ \widehat{K}_1=K_2 -1$ and $K_{\tilde{1}}+\widehat{K}_{\tilde{1}}=K_{\tilde{2}}-1$. The polynomial can thus be rewritten as
\begin{equation}
  \mathscr{P}(\xi) =  \xi  \prod_{j=1}^{K_1} (\xi - y_{1,j}) \prod_{j=1}^{\widehat{K}_1} (\xi - \widehat{y}_{1,j})  \prod_{j=1}^{K_{\tilde{1}}} \left(\xi - \frac{1}{y_{\tilde{1},j}}\right)  \prod_{j=1}^{\widehat{K}_{\tilde{1}}} \left(\xi - \frac{1}{\widehat{y}_{\tilde{1},j}}\right) 
\end{equation}
As we did in the previous subsection, we can evaluate the quantities~$\frac{\mathscr{P}(x^+_k)}{\mathscr{P}(x^-_k)}$ and~$\frac{\mathscr{P}(1/\tilde{x}^-_k)}{\mathscr{P}(1/\tilde{x}^+_k)}$. The resulting identities, plugged into the Bethe ansatz equations, allow us to rewrite them as
\begin{align}\label{eq:BA-ferm-first}
    1 &= 
    \prod_{j=1}^{K_2} \frac{\widehat{y}_{1,k} - x_j^-}{\widehat{y}_{1,k} - x_j^+}
    \prod_{j=1}^{K_{\tilde{2}}} \frac{1 - \frac{1}{\widehat{y}_{1,k} \tilde{x}_j^+}}{1- \frac{1}{\widehat{y}_{1,k} \tilde{x}_j^-}} , \\
    \begin{split}
      \left(\frac{x_k^+}{x_k^-}\right)^{\ell} &=
      \prod_{\substack{j = 1\\j \neq k}}^{K_2} \frac{1- \frac{1}{x_k^+ x_j^-}}{1- \frac{1}{x_k^- x_j^+}} \sigma^2(x_k,x_j)
      \prod_{j=1}^{\widehat{K}_1} \frac{x^+_k - \widehat{y}_{1,j}}{x^-_k - \widehat{y}_{1,j}}  
	\prod_{j=1}^{K_3} \frac{x_k^- - y_{3,j}}{x_k^+ - y_{3,j}}
      \\ &\phantom{\ = \ }\times
      \prod_{j=1}^{K_{\tilde{2}}}  \frac{1- \frac{1}{x_k^+ \tilde{x}_j^-}}{1- \frac{1}{x_k^- \tilde{x}_j^+}} \widetilde{\sigma}^2(x_k,\tilde{x}_j)
      \prod_{j=1}^{\widehat{K}_{\tilde{1}}} \frac{1 - \frac{1}{x^+_k \widehat{y}_{\tilde{1},j}}}{1 - \frac{1}{x^-_k \widehat{y}_{\tilde{1},j}}} 
	\prod_{j=1}^{K_{\tilde{3}}} \frac{1 - \frac{1}{x_k^- y_{\tilde{3},j}}}{1- \frac{1}{x_k^+ y_{\tilde{3},j}}},
    \end{split} \\
     1 &= 
    \prod_{j=1}^{K_2} \frac{y_{3,k} - x_j^+}{y_{3,k} - x_j^-}
    \prod_{j=1}^{K_{\tilde{2}}} \frac{1 - \frac{1}{y_{3,k} \tilde{x}_j^-}}{1- \frac{1}{y_{3,k} \tilde{x}_j^+}} , 
\end{align}
\begin{align}
    1 &= 
    \prod_{j=1}^{K_{\tilde{2}}} \frac{\widehat{y}_{\tilde{1},k} - \tilde{x}_j^+}{\widehat{y}_{\tilde{1},k} - \tilde{x}_j^-}
    \prod_{j=1}^{K_2} \frac{1 - \frac{1}{\widehat{y}_{\tilde{1},k} x_j^-}}{1- \frac{1}{\widehat{y}_{\tilde{1},k} x_j^+}} , \\
    \begin{split}
      \left(\frac{\tilde{x}_k^+}{\tilde{x}_k^-}\right)^{\ell} &=
      \prod_{\substack{j = 1\\j \neq k}}^{K_{\tilde{2}}} \frac{1- \frac{1}{\tilde{x}_k^+ \tilde{x}_j^-}}{1- \frac{1}{\tilde{x}_k^- \tilde{x}_j^+}} \sigma^2(\tilde{x}_k,\tilde{x}_j)
      \prod_{j=1}^{\widehat{K}_{\tilde{1}}} \frac{\tilde{x}^-_k - \widehat{y}_{\tilde{1},j}}{\tilde{x}^+_k - \widehat{y}_{\tilde{1},j}} 
	 \prod_{j=1}^{K_{\tilde{3}}} \frac{\tilde{x}_k^+ - y_{\tilde{3},j}}{\tilde{x}_k^- - y_{\tilde{3},j}}
      \\ &\phantom{\ = \ }\times
      \prod_{j=1}^{K_2} \frac{1- \frac{1}{\tilde{x}_k^+ x_j^-}}{1- \frac{1}{\tilde{x}_k^- x_j^+}} \widetilde{\sigma}^2(\tilde{x}_k,x_j)
      \prod_{j=1}^{\widehat{K}_1} \frac{1 - \frac{1}{\tilde{x}^-_k \widehat{y}_{1,j}}}{1 -\frac{1}{\tilde{x}^+_k \widehat{y}_{1,j}}} 
	\prod_{j=1}^{K_{3}} \frac{1 - \frac{1}{\tilde{x}_k^+ y_{3,j}}}{1- \frac{1}{\tilde{x}_k^- y_{3,j}}} ,
    \end{split} \\
	\label{eq:BA-ferm-last}
     1 &= 
    \prod_{j=1}^{K_{\tilde{2}}} \frac{y_{\tilde{3},k} - \tilde{x}_j^-}{y_{\tilde{3},k} - \tilde{x}_j^+}
    \prod_{j=1}^{K_2} \frac{1 - \frac{1}{y_{\tilde{3},k} x_j^+}}{1- \frac{1}{y_{\tilde{3},k} x_j^-}} .
\end{align}
The above equations are the ones that we would have found if we had chosen $\Phi^{\smallL}_{-{+}},{\Phi}^{\smallR}_{+{-}}$ to be the fields composing the level-II vacuum.
Similarly, one could have dualised the auxiliary roots $y_3, y_{\tilde{3}}$ and obtained the Bethe equations  corresponding to the choice of $\Phi^{\smallL}_{+{-}},{\Phi}^{\smallR}_{-{+}}$ in the level-II vacuum. We do not write them, since they are equal to the ones written above after exchanging 1 and 3, as expected.
Two consecutive dualisations, \ie~dualising   $y_1, y_{\tilde{1}}$ and then $y_3, y_{\tilde{3}}$ (in any order) give Bethe equations  corresponding to the choice of $\Phi^{\smallL}_{-{-}},{\Phi}^{\smallR}_{+{+}}$ in the level-II vacuum. They are equal to the Bethe equations derived in section \ref{sec:BAE} after exchanging L$\leftrightarrow $R.


\backmatter

\fancyhf{}
\fancyhead[LE,RO]{\thepage}
\fancyhead[LO]{\nouppercase{Bibliography}}
\fancyhead[RE]{\nouppercase{Bibliography}}
\renewcommand{\headrulewidth}{0.5pt}

\bibliographystyle{nb}
\addcontentsline{toc}{chapter}{Bibliography}
\bibliography{thesis}

\cleardoublepage

\end{document}